\documentclass[a4paper,11pt]{article}
\usepackage{jheppub}
\usepackage{mathrsfs}
\usepackage{amsfonts}
\usepackage{setspace}
\usepackage{cellspace}
\usepackage{amsmath,amssymb,bm}
\allowdisplaybreaks
\usepackage[colorlinks=true,linkcolor=blue]{hyperref}
\usepackage{xcolor}
\usepackage{epsfig}
\usepackage{slashed}
\usepackage{caption}
\usepackage{hhline,multirow,tabularx}  
\usepackage{dcolumn}    
\usepackage{url}        
\usepackage{braket}
\definecolor{cyan}{rgb}{0.0, 1.0, 1.0}
\definecolor{applegreen}{rgb}{0.55, 0.71, 0.0}
\definecolor{arylideyellow}{rgb}{0.91, 0.84, 0.42}
\definecolor{bananayellow}{rgb}{1.0, 0.88, 0.21}
\definecolor{burlywood}{rgb}{0.87, 0.72, 0.53}
\definecolor{buff}{rgb}{0.94, 0.86, 0.51}
\definecolor{blond}{rgb}{0.98, 0.94, 0.75}
\definecolor{bisque}{rgb}{1.0, 0.89, 0.77}
\definecolor{bananamania}{rgb}{0.98, 0.91, 0.71}
\definecolor{apricot}{rgb}{0.98, 0.81, 0.69}
\definecolor{almond}{rgb}{0.94, 0.87, 0.8}
\usepackage[export]{adjustbox}
\usepackage{titlesec}
\titleclass{\subsubsubsection}{straight}[\subsubsection]
\newcounter{subsubsubsection}[subsubsection]
\renewcommand\thesubsubsubsection{\thesubsubsection.\arabic{subsubsubsection}}
\titleformat{\subsubsubsection}
  {\normalfont\normalsize\bfseries}{\thesubsubsubsection}{1em}{}
\titlespacing*{\subsubsubsection}{0pt}{1.5ex plus .2ex minus .2ex}{1ex plus .2ex}

\title{Asymptotic Long-Distance Expansion of Euclidean Correlators in Lattice Parton Applications}
\author[a,b]{Xiangdong Ji}
\author[c]{Yizhuang Liu}
\author[b]{Yushan Su}

\affiliation[a]{T. D. Lee Institute, Shanghai 201210, China}
\affiliation[b]{Department of Physics, University of Maryland, College Park, MD 20742, USA}
\affiliation[c]{Institute of Theoretical Physics and Mark Kac Center for Complex Systems Research, Jagiellonian University, 30-348 Kraków, Poland}

\emailAdd{xji@umd.edu}
\emailAdd{yizhuang.liu@uj.edu.pl}
\emailAdd{ysu12345@umd.edu}

\abstract{Bilinear Euclidean quark and gluon correlators with Wilson links have been used widely for applications of large-momentum effective field theories to computing non-perturbative collinear and soft parton physics. Due to color confinement, these correlators decay exponentially at large spatial distances, a behavior crucial for computing momentum-space Fourier transformations with controlled errors from lattice QCD data. Using heavy-quark effective theory reduction, dispersive analysis, Lorentz symmetry, and heavy-flavor spectra, we determine the leading and next-to-leading asymptotic behaviors and relate the expansion parameters to binding energies of heavy-flavor hadrons. We demonstrate the results through two-loop calculations in $\phi^3$ theory and from the perspective of locality and analyticity. We also study the impact of the asymptotic analysis on realistic lattice QCD data and demonstrate reliable error estimates. 
}
\date{\today}

\begin{document}
\maketitle
\flushbottom

\section{Introduction}
Parton distribution functions (PDFs) describe the momentum distributions of quarks and gluons in the hadrons moving at an asymptotically-large  momentum~\cite{Bjorken:1969ja}. They are important Standard Model (SM) inputs in the search for beyond SM physics as well as crucial non-perturbative functions for probing the inner structures of hadrons. They can be extracted from experimental data through global fit, such as ATLASpdf21~\cite{ATLAS:2021vod}, NNPDF4.0~\cite{NNPDF:2021njg}, MSHT20~\cite{Bailey:2020ooq}, CT18~\cite{Hou:2019efy}, ABMP16~\cite{Alekhin:2017kpj}, HERAPDF2.0~\cite{H1:2015ubc}, and JR14~\cite{Jimenez-Delgado:2014twa}. This paper primarily focuses on first-principles theoretical calculations of PDFs, which have the potential to improve the precision of beyond SM searches, and, in some cases, to provide important information on hadronic structure that cannot be or has not been accessed experimentally. 

In Ref.~\cite{Ji:2024oka}, partons are considered as effective degrees of freedom for high-energy scattering processes in the infinite momentum limit or light-cone frame, where the PDFs are defined using a lightlike-separated quark bilinear operator~\cite{Collins_2023}, which has a direct momentum-density or particle number interpretation, yielding a clear and elegant physical picture. Since the known perturbative dependences of the high-energy scales are factorized out, such as the photon virtuality $Q^2$ in the deep inelastic scattering (DIS) or the invariant mass $M$ in the Drell-Yan (DY) process, PDFs are universal across multiple experiments, thereby enhancing the predictive power of quantum chromodynamics (QCD). 

Despite their elegance and universality, it is difficult to calculate parton physics from the first principles of QCD, such as lattice QCD~\cite{Wilson:1974sk}. One challenge stems from the real-time dependence of light-cone correlators, which cannot be directly simulated on the Euclidean lattice QCD. Another challenge involves light-front singularities, additional ultraviolet (UV) divergences that arise in the infinite-momentum limit and require appropriate regularization and renormalization schemes. One well-studied approach is light-front quantization~\cite{Dirac:1949cp,Brodsky:1997de}, where systematically regularizing and renormalizing the light-front singularities remains a difficult task to this day. 

The proposal of large momentum effective theory (LaMET)~\cite{Ji:2013dva,Ji:2014gla} is an important step toward resolving those challenges. LaMET~\cite{Ji:2013dva,Ji:2014gla,Ji:2020ect,Ji:2024oka} is an effective field theory (EFT) to directly calculate $x$-dependent parton physics with controlled precision, under the large momentum expansion and perturbative matching of Euclidean correlators, which are available from lattice QCD. The light-front singularities are naturally regularized by the finite large momentum, whose dependence can be analyzed within perturbation theory. Since its proposal, LaMET has a wide range of applications, including calculations of quark collinear PDFs~\cite{Xiong:2013bka,Lin:2014zya,Alexandrou:2015rja,Chen:2016utp,Alexandrou:2016jqi,Alexandrou:2018pbm,Chen:2018xof,Lin:2018pvv,LatticeParton:2018gjr,Alexandrou:2018eet,Liu:2018hxv,Chen:2018fwa,Izubuchi:2018srq,Izubuchi:2019lyk,Shugert:2020tgq,Chai:2020nxw,Lin:2020ssv,Fan:2020nzz,Gao:2021hxl,Gao:2021dbh,Gao:2022iex,Su:2022fiu,LatticeParton:2022xsd,Gao:2022uhg,Chou:2022drv,Gao:2023lny,Gao:2023ktu,Chen:2024rgi,Holligan:2024umc,Holligan:2024wpv,Ji:2024hit,Zhang:2024omt,Zhang:2024wyq,Gao:2024fbh,Chu:2025jsi,Zhang:2025mer,Miller:2025wgr,Chen:2025cxr}, gluon collinear PDFs~\cite{Zhang:2018diq,Fan:2018dxu,Chowdhury:2024ymm,Good:2024iur,Good:2025daz,NieMiera:2025mwj,Chen:2025xww,NieMiera:2025vcx}, light-meson distribution amplitudes (DAs)~\cite{Zhang:2017bzy,Chen:2017gck,Zhang:2020gaj,Hua:2020gnw, LatticeParton:2022zqc,Gao:2022vyh,Holligan:2023rex,Baker:2024zcd,Cloet:2024vbv,Ji:2025mvk,Xiong:2025obq,Ling:2025olz}, heavy-meson DAs~\cite{Xu:2022guw,Hu:2023bba,LatticeParton:2024zko,Han:2024cht,Deng:2024dkd,Han:2024fkr,Guo:2025obm}, baryon DAs~\cite{Deng:2023csv,Han:2023xbl,Han:2023hgy,Han:2024ucv,LatticeParton:2024vck,LPC:2025jvd,Zhang:2025npd}, transverse-momentum-dependent (TMD) distributions~\cite{Ji:2014hxa,Shanahan:2019zcq,Shanahan:2020zxr,Zhang:2020dbb,Ji:2021znw,LatticePartonLPC:2022eev,Liu:2022nnk,Zhang:2022xuw,Deng:2022gzi,Zhu:2022bja,LatticePartonCollaborationLPC:2022myp,Rodini:2022wic,Shu:2023cot,LatticeParton:2023xdl,delRio:2023pse,LatticePartonLPC:2023pdv,Alexandrou:2023ucc,Avkhadiev:2023poz,Zhao:2023ptv,Avkhadiev:2024mgd,Bollweg:2024zet,Spanoudes:2024kpb,LatticeParton:2024mxp,Alexandrou:2025xci,Bollweg:2025ecn,Tan:2025ofx,Bollweg:2025iol,LatticeParton:2025eui,Xie:2025rrw}, generalized parton distributions (GPDs)~\cite{Ji:2015qla,Chen:2019lcm,Alexandrou:2019dax,Lin:2020rxa,Alexandrou:2020zbe,Lin:2021brq,Scapellato:2022mai,Bhattacharya:2022aob,Bhattacharya:2023nmv,Bhattacharya:2023jsc,Lin:2023gxz,Holligan:2023jqh,Ding:2024saz,Chu:2025jsi,Holligan:2025baj,Chu:2025kew,Bhattacharya:2025yba}, and double parton distributions (DPDs)~\cite{Zhang:2023wea,Jaarsma:2023woo,Reitinger:2024ulw}. Recent reviews on LaMET can be found in Refs.~\cite{Cichy:2018mum,Ji:2020ect}. Differences between LaMET expansion for direct parton calculations and short-distance factorization for moments of partons have been discussed in Ref. \cite{Ji:2022ezo}. 

In the past few years, considerable efforts have been made to control the precision of LaMET calculations, at each step of analysis procedure, including lattice data generation~\cite{Bali:2016lva,Gao:2023lny,Zhao:2023ptv,Bhattacharya:2023jsc,Zhang:2025hyo}, renormalization~\cite{Ji:2017oey,Ishikawa:2017faj,Green:2017xeu,Constantinou:2017sej,Stewart:2017tvs,Alexandrou:2017huk,Chen:2017mzz,Ji:2020brr,LatticePartonLPC:2021gpi,Zhang:2023bxs,Alexandrou:2023ucc,Spanoudes:2024kpb}, Fourier transform (FT)~\cite{Ji:2020brr,Gao:2021dbh,Chen:2025cxr,Xiong:2025obq,Ling:2025olz} and perturbative matching~\cite{Li:2020xml,Chen:2020ody,Ji:2025mvk,Su:2022fiu,Holligan:2023rex,Zhang:2023bxs,Gao:2021hxl,Ji:2023pba,Liu:2023onm,Ji:2024hit,Cloet:2024vbv,Holligan:2025baj}. At present, the uncertainties in the last three steps are well under control, amounting to only a few percent in the moderate-$x$ region. However, more effort is needed in lattice data generation, especially for the excited state control of nucleon matrix elements, perhaps by adopting the methods in Refs.~\cite{Blossier:2009kd,Barca:2025det,Wang:2025nsd,Abbott:2025yhm,Lyu:2025lnd}. Please also check Ref.~\cite{Zhao:2025oto} for a brief review on LaMET precision control. 

The focus of this paper is the precision control of the Fourier transform (FT), especially for the large distance behavior. One step in LaMET analysis is the Fourier transform of the lattice correlator from coordinate space to momentum space, which formally requires correlator values over an infinite range of spatial separations~\cite{Ji:2020brr}. In principle, lattice data should decay exponentially at large distances (e.g. $\sim 1 \, {\rm fm}$), implying that contributions from the long-distance tail (e.g. $ \gtrsim 1 \, {\rm fm}$) to FT should be negligible and that the FT precision should be well controlled. In practice, lattice data become increasingly noisy at large distances and cannot be directly used for FT. To control the FT precision, we adopt the large-distance asymptotic analysis, which discards noisy data at large distances, performs the physical extrapolation of the correlators to large distances, and computes the FT using the extrapolated form. 

The discussion therefore turns to the theoretical understanding of the large-distance asymptotic behavior of Euclidean correlators for physical extrapolation, as well as to the quantification of the corresponding FT uncertainty arising from asymptotic analysis. In Ref.~\cite{Ji:2020brr}, motivated by the Regge and threshold limits of momentum-space PDFs, a phenomenological parametrization of the large-distance asymptotic form was proposed and has since been widely adopted in LaMET calculations. Ref.~\cite{Gao:2021dbh} subsequently provided a derivation of the asymptotic form based on heavy quark effective theory (HQET) and dispersive analysis, under the assumptions of constant form factors and the validity of analytic continuation. That work also derived an upper bound for the FT uncertainty by approximating the extrapolated function as a flat curve within one oscillatory period. More recently, Ref.~\cite{Chen:2025cxr} argued that LaMET is fundamentally a forward-problem formalism, which is systematically improvable without encountering an ill-posed inverse problem. It further emphasized that a physics-based asymptotic extrapolation yields the most reliable estimates of FT error. It also noted that there are and will be more high-quality lattice data for robust asymptotic analysis. These points directly responded to the concerns from Refs.~\cite{Dutrieux:2025jed,Dutrieux:2025axb}. 

This work extends and complements the previous papers in several key aspects, leading to more sophisticated derivations and results on large-distance asymptotic forms as well as FT precision estimates. First, we carefully study the HQET reduction of the coordinate space propagator into the gauge link, which involves additional exponential behavior related to the heavy quark pole mass $m_Q$, and polynomial $z$-dependence from the ``kinematic energy term" $D_{\perp}^2/(2 m_Q)$. Second, we explore in the dispersive analysis the form factor connectedness~\cite{Jaffe:1983hp}, which affects the phases of the large-distance asymptotic forms, thereby explaining why the lattice data for quasi-DA matrix elements exhibit stronger oscillations than most quasi-PDFs. Third, we further investigate the exponential decay of spacelike correlators from the viewpoints of locality and analyticity. We prove explicit lower bounds for the decay speed consistent with the dispersive analysis. Fourth, we perform two-loop calculations in the $\phi^3$ theory as supporting evidence for the dispersive analysis. Fifth, we provide all-order mathematical structures for large-distance asymptotic expansions, which are series of exponential decays, each exponential decay multiplied by a series of polynomial dependences. We also discuss the accuracy-counting principles for truncation up to a given order for practical lattice data analysis. Finally, we derive upper bounds on the FT uncertainty by considering three phases that primarily affect different end-point regions. 

Thanks to all the efforts mentioned above, we now have a solid theoretical foundation for large-distance asymptotic forms, grounded in the fundamental properties of quantum field theory. Moreover, we accurately determine the large-distance behaviors, including the physical origins of mass gaps in the exponential decay, the powers of polynomial dependences, and the overall phases containing the external momentum. Last but not least, we have sophisticated FT precision control for both moderate $y$ range and endpoint regions. Under the current lattice status, the FT uncertainty is just a few percent in the moderate $y$ range and slightly larger but still under control at the endpoints. 

The rest of the paper is organized as follows. Sec.~\ref{sec:summary} summarizes the main results in this paper for the convenience of lattice practitioners. In Sec.~\ref{sec:theory}, we present the theoretical derivations of large distance asymptotic forms, based on HQET reduction, dispersive analysis, Lorentz symmetry, and heavy-flavor spectra, taking unpolarized quasi-PDF as an example. In Sec.~\ref {sec:numerics}, we implement the derived large distance behaviors for the Fourier transform of lattice data to demonstrate controlled precision. In Sec.~\ref{sec:spacelike}, we further investigate the exponential decay of spacelike correlators from the perspective of locality and analyticity, where we prove the existence of exponential decay under various conditions and derive the corresponding lower bounds for the decay speed. We conclude this paper in Sec.~\ref{sec:conclusion}. Some supporting evidence and applications to other quasi-correlators are provided in the appendices.

\section{Summary of main results}\label{sec:summary}
We summarize the large-distance expansions of various quasi-correlators up to leading asymptotics (LA) and next-to-leading asymptotics (NLA). They are derived using the methods in Sec.~\ref{sec:theory}, including HQET reduction in Sec.~\ref{sec:HQETred}, dispersive analysis in Sec.~\ref{sec:Disana}, Lorentz symmetry in Sec.~\ref{sec:Lorentz}, and accuracy-counting principles in Sec.~\ref{sec:AccuCount}. 
\begin{itemize}
\item Pion quark quasi-PDF matrix element $\tilde{h}\left(z,P^z\right) =\langle \pi(P) | \bar{\psi}(z) U(z,0) \gamma^t \psi(0) | \pi(P)  \rangle$ where $\psi$ and $\bar{\psi}$ denote the quark and anti-quark fields separated by a spatial distance $z$, $U(z,0)$ is a straight Wilson link to preserve the gauge invariance, and $| \pi(P)  \rangle$ is an external pion state with four-momentum $P=\left(\sqrt{m_{\pi}^2+P_z^2},0,0,P^z\right)$. This matrix element is expanded at $z \rightarrow \infty$ with fixed large $P^z$, up to LA and NLA, respectively, 
\begin{align}
\tilde{h}^{\rm LA}\left(z,P^z\right) = 
\left[A_2 e^{i \, {\rm sign}(z) \, \phi_2} +  A_1 e^{i \, {\rm sign}(z) \, \phi_1} e^{-i P^z z} + A_3 e^{i \, {\rm sign}(z) \, \phi_3} e^{i P^z z} \right] e^{-\Lambda |z|}  \ ,
\end{align}
\begin{align}
\tilde{h}^{\rm NLA}\left(z,P^z\right) 
& = \left[A_2 e^{i \, {\rm sign}(z) \, \phi_2} +  A_1 e^{i \, {\rm sign}(z) \, \phi_1} e^{-i P^z z} + A_3 e^{i \, {\rm sign}(z) \, \phi_3} e^{i P^z z} \right] e^{-\Lambda |z|} \nonumber\\
& + \left[A'_2 e^{i \, {\rm sign}(z) \, \phi'_2} +  A'_1 e^{i \, {\rm sign}(z) \, \phi'_1} e^{-i P^z z} + A'_3 e^{i \, {\rm sign}(z) \, \phi'_3} e^{i P^z z} \right] \frac{e^{-\Lambda |z|}}{|z|} \ ,
\end{align}  
where $A_i$, $A'_i$, $\phi_i$, $\phi'_i$ and $\Lambda$ are real parameters to be fitted with lattice data. 

If $\psi$ and $\bar{\psi}$ are light quarks, the mass gap $\Lambda$ is related to the binding energy of a vector or pseudoscalar heavy-light meson ($Q \bar{q}$), which is $\sim 500$ MeV estimated using physical states as discussed in Sec.~\ref{sec:AccuCount}. On lattice, the mass gap may deviate from the physical value up to $O(\Lambda_{\rm QCD})$ due to unphysical pion mass, discretization effect, or the choice of renormalon scheme. So one can fit it from lattice data under a physics-motivated constraint such as $[500\,{\rm MeV}-O(\Lambda_{\rm QCD})$, $500\,{\rm MeV}+O(\Lambda_{\rm QCD})]$. If $\psi$ and $\bar{\psi}$ are heavy quarks, the mass gap can be estimated analogously from the spectrum of the corresponding heavy–heavy meson. 

For valence combination, the parameters are constrained by isospin symmetry: $\phi_2=\phi'_2=0$, $\phi_3 = -\phi_1$, $\phi'_3 = -\phi'_1$, $A_1=A_3$, and $A'_1 = A'_3$. For sea quarks (including light and heavy quarks),  $A_1 = A_3 =A'_1 = A'_3 = 0$ since the corresponding mass gaps are associated with four-quark states with the combined quantum numbers of a light meson and a heavy meson, and are therefore larger than those of a single heavy meson. 

The general mathematical structure of the above results can be found in Eq.~(\ref{eq:qPDFLzGen}), the accuracy counting for the pion valence PDF in Sec.~\ref{sec:AccuCount}, and the numerical tests in Sec.~\ref{sec:pionval}.

\item Nucleon quark unpolarized quasi-PDF $\tilde{h}\left(z,P^z\right) =\langle N(P) | \bar{\psi}(z) U(z,0) \gamma^t \psi(0) | N(P)  \rangle$,
\begin{align}
\tilde{h}^{\rm LA}\left(z,P^z\right) = 
A_2 e^{i \phi_2 \, {\rm sign}(z)} e^{-\Lambda |z|}  \ ,
\end{align}
\begin{align}
\tilde{h}^{\rm NLA}\left(z,P^z\right) = 
\left[A_2 e^{i \phi_2 \, {\rm sign}(z)} +  \frac{A'_2 e^{i \phi'_2 \, {\rm sign}(z)}}{|z|} \right] e^{-\Lambda |z|}  \ ,
\end{align}
where the mass gap $\Lambda$ here has a similar physical origin as the pion quark quasi-PDF case, so does the $\Lambda$ in all the other asymptotic forms in this section. The above formats apply to valence and sea quarks. More detailed discussions can be found in Eq.~(\ref{eq:qPDFLzGen}), Sec.~\ref{sec:AccuCount}, and Sec.~\ref{sec:protonupol}.

\item Nucleon quark transversity quasi-PDF $$\tilde{h}\left(z,P^z\right) =\langle N(P S_{\perp})| \bar{\psi}(z) \gamma^t \gamma^{\perp} \gamma^5 U(z,0) \psi(0) | N(P S_{\perp}) \rangle \ ,$$ and helicity quasi-PDF $$\tilde{h}\left(z,P^z\right) =\langle N(P S_{L}) | \bar{\psi}(z) \gamma^z \gamma^5 U(z,0) \psi(0) | N(P S_{L}) \rangle \ .$$
Their large distance asymptotic expansions truncated up to LA and NLA are,
\begin{align}
\tilde{h}^{\rm LA}\left(z,P^z\right) = 
A_2 e^{i \phi_2 \, {\rm sign}(z)} e^{-\Lambda |z|}  \ ,
\end{align}
\begin{align}
\tilde{h}^{\rm NLA}\left(z,P^z\right) = 
\left[A_2 e^{i \phi_2 \, {\rm sign}(z)} +  \frac{A'_2 e^{i \phi'_2 \, {\rm sign}(z)}}{|z|} \right] e^{-\Lambda |z|}  \ .
\end{align}
Theoretical derivations can be found in Appendix~\ref{sec:PolPDF} and numerical tests on nucleon quark transversity quasi-PDF in Sec.~\ref{sec:protontran}. 

\item Meson quasi-DA $\tilde{h}\left(z,P^z\right) = \langle M(P) | \bar{\psi}_1(z) U(z,0) \gamma^z \gamma^5 \psi_{2}(0) | \Omega \rangle$, 
\begin{align}
\tilde{h}^{\rm LA}\left(z,P^z\right) = 
\left[A_1 e^{i \phi_1 \, {\rm sign}(z)} e^{-i z P^z} + A_2 e^{i \phi_2 \, {\rm sign}(z)} \right]e^{-\Lambda |z|}  \ ,
\end{align}
\begin{align}
&\tilde{h}^{\rm NLA}\left(z,P^z\right) = 
\left[ \left(A_1 e^{i \phi_1 \, {\rm sign}(z)} e^{-i z P^z} + A_2 e^{i \phi_2 \, {\rm sign}(z)} \right) \right. \nonumber\\
&\left.+ \left(A'_1 e^{i \phi'_1 \, {\rm sign}(z)} e^{-i z P^z} + A'_2 e^{i \phi'_2 \, {\rm sign}(z)} \right) \frac{1}{|z|} \right] e^{-\Lambda |z|}  \ .
\end{align}
If both $\psi_1$ and $\psi_2$ are light quarks with similar masses, all terms should be kept. In particular, $A_1=A_2$, $\phi_1=-\phi_2$, $A'_1=A'_2$ and $\phi'_1=-\phi'_2$ for pion quasi-DA. If $\psi_1$ is a light quark and $\psi_2$ is a heavy quark, $A_1 = A'_1 = 0$. If $\psi_1$ is a heavy quark and $\psi_2$ is a light quark, $A_2 = A'_2 = 0$. In all the situations mentioned here, $\Lambda$ is the binding energy of a pseudoscalar heavy-light meson, which also appears in the pion quark quasi-PDF case. Theoretical derivations can be found in Appendix~\ref{sec:DA} and numerical tests on pion quasi-DA in Sec.~\ref{sec:pionDA}.

\item Pion quark quasi-GPD $\tilde{h}\left(z,P^z,P'^z\right)=\langle \pi(P) | \bar{\psi}(z) U(z,0) \gamma^t  \psi(0) | \pi(P') \rangle$,
\begin{align}
\tilde{h}^{\rm LA}\left(z,P^z,P'^z\right) =
&\left[A_1 e^{i \phi_1 \, {\rm sign}(z)} e^{-i z P^z} + A_3 e^{i \phi_3 \, {\rm sign}(z)} e^{i z P'^z}  \right. \nonumber\\
&\left.+A_2 e^{i \phi_2 \, {\rm sign}(z)} + \tilde{A}_2 e^{i \tilde{\phi}_2 \, {\rm sign}(z)} e^{-i(P^z-P'^z)z}  \right] e^{-\Lambda |z|} \ ,
\end{align}
\begin{align}
\tilde{h}^{\rm NLA}\left(z,P^z,P'^z\right) =
&\left[A_1 e^{i \phi_1 \, {\rm sign}(z)} e^{-i z P^z}  + A_3 e^{i \phi_3 \, {\rm sign}(z)} e^{i z P'^z} \right. \nonumber\\
&\left.+A_2 e^{i \phi_2 \, {\rm sign}(z)} + \tilde{A}_2 e^{i \tilde{\phi}_2 \, {\rm sign}(z)} e^{-i(P^z-P'^z)z}\right]e^{-\Lambda |z|} \nonumber\\ 
&+\left[A'_1 e^{i \phi'_1 \, {\rm sign}(z)} e^{-i z P^z} + A'_3 e^{i \phi'_3 \, {\rm sign}(z)} e^{i z P'^z} \right. \nonumber\\
&\left.+A'_2 e^{i \phi'_2 \, {\rm sign}(z)} + \tilde{A}'_2 e^{i \tilde{\phi}'_2 \, {\rm sign}(z)} e^{-i(P^z-P'^z)z} \right] \frac{e^{-\Lambda |z|}}{|z|} \ .
\end{align}
For valence combination, all terms are preserved. For sea quarks (including light and heavy quarks),  $A_1 = A_3 = A'_1 = A'_3 = 0$ for the same reason as the pion quasi-PDF case. More details can be found in Appendix~\ref{sec:GPD}.

\item Nucleon quark quasi-GPD, including the unpolarized case $$\tilde{h}\left(z,P^z,P'^z\right)=\langle N(P) | \bar{\psi}(z) U(z,0) \gamma^t  \psi(0) | N(P') \rangle \ , $$ and polarized case $$\tilde{h}\left(z,P^z,P'^z\right)=\langle N(P S)| \bar{\psi}(z) U(z,0) \gamma^z \gamma^5 \psi(0) | N(P' S')\rangle \ .$$
The large distance expansions up to LA and NLA are, 
\begin{align}
\tilde{h}^{\rm LA}\left(z,P^z,P'^z\right) = 
\left[ A_2 e^{i \phi_2 \, {\rm sign}(z)} + \tilde{A}_2 e^{i \tilde{\phi}_2 \, {\rm sign}(z)} e^{-i(P^z-P'^z)z} \right] e^{-\Lambda |z|}  \ ,
\end{align}
\begin{align}
&\tilde{h}^{\rm NLA}\left(z,P^z,P'^z\right) = 
\left[ A_2 e^{i \phi_2 \, {\rm sign}(z)} + \tilde{A}_2 e^{i \tilde{\phi}_2 \, {\rm sign}(z)} e^{-i(P^z-P'^z)z} \right.\nonumber\\ 
&\left.+ \left( A'_2 e^{i \phi'_2 \, {\rm sign}(z)} + \tilde{A}'_2 e^{i \tilde{\phi}'_2 \, {\rm sign}(z)} e^{-i(P^z-P'^z)z} \right) \frac{1}{|z|}  \right] e^{-\Lambda |z|}  \ .
\end{align}
More discussions can be found in Appendix~\ref{sec:GPD}. 
\end{itemize}

Using the above large distance expansions and following the procedure and criteria in Sec.~\ref{sec:fitcriteria}, we demonstrate the controlled precision of asymptotic analysis, through the analytically-derived upper error bound in Eq.~(\ref{eq:errbud})\footnote{It applies to all the quasi-correlators discussed in Sec.~\ref{sec:summary} except for GPD.}, and numerical tests among various lattice quasi-correlators in Sec.~\ref{sec:applications}. With the current quality of lattice data, the uncertainty from asymptotic analysis is just a few percent compared to the center value at $O(1)$ in the moderate $y$ range. The precision can even be improved in the near future with recently proposed techniques~\cite{Gao:2023lny,Zhao:2023ptv,Zhang:2025hyo}.

\section{Large distance behavior of unpolarized quark quasi-PDF}\label{sec:theory}
In this section, we derive the large distance asymptotic behavior\footnote{In LaMET applications, the large distance behavior is referred to as the expansion at $z \rightarrow \infty$ for fixed large $P^z$.} of the unpolarized quark quasi-PDF matrix element~\cite{Ji:2013dva,Ji:2014gla,Cichy:2018mum,Ji:2020ect,Ji:2024oka},
\begin{align}\label{eq:qPDF}
\langle H(P) | \bar{\psi}(z) U(z,0) \gamma^t \psi(0) | H(P) \rangle \ ,
\end{align}
where $| H(P) \rangle$ denotes a boosted hadron with momentum $P=\left(\sqrt{m_H^2+P_z^2},0,0,P^z\right)$ and mass $m_{H}$, where $P^z$ is much larger than $m_{H}$. $\psi,\bar{\psi}$ are quark and anti-quark fields separated by the spatial distance $z$, and $U(z,0)$ is a spacelike Wilson line to ensure the gauge invariance.

Understanding this correlator’s large-distance behavior is essential because it provides a physical constraint for controlling the precision of the Fourier transformation. The quasi-correlator Eq.~(\ref{eq:qPDF}) is one of the key objects in LaMET for parton physics calculations. It can be calculated non-perturbatively in lattice QCD, renormalized using standard self-renormalization and hybrid schemes~\cite{Ji:2020brr,LatticePartonLPC:2021gpi}, Fourier transformed to momentum space, and finally perturbatively matched to the unpolarized quark light-cone PDF. The Fourier transform, however, formally requires input over an infinite spatial range. In practice, lattice data at large separations (e.g. $\gtrsim 1$ fm) suffer from exponentially deteriorating signal-to-noise ratios, making them unusable directly. Our goal in this section is therefore to derive the theoretical large-distance asymptotic form, which will serve as a physical constraint enabling precision control of the Fourier transformation in the next section.

Our methods are summarized here and will be detailed in the following subsections. First, we represent the gauge link with dynamical heavy quarks through the coordinate-space HQET reduction. Then, a dispersive analysis along the timelike separation with physical states is performed to obtain the large-time asymptotic form. Next, using Lorentz symmetry, this form is converted to the spacelike separation, which corresponds to the Euclidean correlator simulated in lattice QCD. More theoretical justifications for the exponential decay in the spacelike region will be given in Sec.~\ref{sec:spacelike}. Finally, we discuss the accuracy-counting principles for the large-distance expansions, including the definitions of leading asymptotics (LA) and next-to-leading asymptotics (NLA). The methods here apply to many other quasi-correlators shown in Appendices~\ref{sec:PolPDF},~\ref{sec:DA},~\ref{sec:GPD} and~\ref{sec:Gluon}.

\subsection{Matching between heavy quark and gauge link}\label{sec:HQETred}
In this subsection, we represent the gauge link with dynamical heavy quarks (not auxiliary heavy quarks), for the convenience of dispersive analysis in the next subsection. 

A natural approach to deriving the large-distance expansion is dispersive analysis, which inserts a complete basis of intermediate particles into the correlator and relates the mass gaps of exponential decays to the spectrum of these particles. However, it is challenging to perform dispersive analysis directly with quasi-PDF because of the gauge link -- one has to insert colored states, which are not the physical states of QCD. To avoid this issue, we represent the gauge link with dynamical heavy quarks. For the convenience of follow-up discussions, we will start from a general situation that applies to either spacelike (as in Eq.~(\ref{eq:qPDF})) or timelike separation. 

According to the heavy quark effective theory (HQET)~\cite{IsgurWise1989,IsgurWise1990,EichtenHill1990,Georgi1990,Grinstein1990,Neubert1994,ManoharWise2000,FalkLuke1992,Bigi1993,Mannel:1991mc,Neubert:1996wg}, the momentum space heavy quark with multiple gluon insertions is eikonalized into a gauge link along its worldline. As shown in Appendix~\ref{sec:MatQGL}, we derive a novel reduction formula for the coordinate space heavy quark propagator $Q(x) \bar{Q}(0)$ in the heavy pole mass limit $m_{Q} \rightarrow +\infty$, 
\begin{align}\label{eq:MatQGLorg}
Q(x) \bar{Q}(0) \rightarrow U(x,0) D_{Q}(x^2,m_{Q}^2) \frac{i \slashed{x} + \sqrt{-x^2}}{2 \sqrt{-x^2}} \left[ 1 + O\left( \frac{1}{m_{Q} \sqrt{-x^2}}, \frac{\Lambda_{\rm QCD}}{m_{Q}} \right) \right] \ ,
\end{align}
where $x$ is a Lorentz coordinate that can be either timelike or spacelike. When $x$ is timelike, we take $x=(t,0,0,0)$ with $t>0$. When it is spacelike, we use $x=(0,0,0,z)$. $U(x,0)$ is a straight-line gauge link connecting $0$ to $x$ and encodes the color structure of the heavy quark propagator. The scalar function $D_{Q}(x^2,m_{Q}^2)$ contains the kinematic structure,
\begin{align}\label{eq:DQ}
D_{Q}(x^2,m_{Q}^2)  = \frac{m_{Q}^3}{2\sqrt{2} \pi^{3/2}} \frac{e^{-m_{Q} \sqrt{-x^2}}}{\left(m_{Q}\sqrt{-x^2}\right)^{3/2}} \ .
\end{align}
The prescription of $\sqrt{-x^2}$ depends on the time ordering. For non time-ordered correlator (Wightman functions), one needs $\sqrt{-(t-i 0^{+})^2+\vec{x}^2}$, while for time-ordered correlator one needs $\sqrt{-x^2+i0^{+}}$~\cite{Streater:1989vi,Jost:1965yxu,Kravchuk:2021kwe}. These two prescriptions coincide for $t>0$. When $-x^2>0$, the correlator is within the domain of analyticity and no prescription is needed.   

Therefore, the coordinate space current-current correlator,
\begin{align}\label{eq:MqQQq}
M_{\Gamma_1\Gamma_2}(x,P,\mu) \equiv  \langle H(P) | \bar{\psi}(x) \Gamma_1 Q(x) \, \bar{Q}(0) \Gamma_2 \psi(0) | H(P) \rangle_c \ ,   
\end{align}
can be matched onto the gauge link object,
\begin{align}\label{eq:hqUq}
\tilde{h}_{\Gamma_1\Gamma_2}\left(x,P,\mu\right) \equiv   \langle H(P) | \bar{\psi}(x) \Gamma_1 U(x,0) \frac{i \slashed{x} + \sqrt{-x^2}}{2 \sqrt{-x^2}} \Gamma_2 \psi(0) | H(P) \rangle_c  \ ,
\end{align}
in the heavy quark pole mass expansion $m_{Q} \rightarrow +\infty$, 
\begin{align}\label{eq:MatQGL}
M_{\Gamma_1\Gamma_2}(x,P,\mu) = H_{\Gamma_1\Gamma_2,\hat{x}}\left(\frac{m_{Q}^2} {\mu^2}\right) D_{Q}(x^2,m_{Q}^2) \tilde{h}_{\Gamma_1\Gamma_2}\left(x,P,\mu\right) \left[ 1 + O\left( \frac{1}{m_{Q} \sqrt{-x^2}}, \frac{\Lambda_{\rm QCD}}{m_{Q}} \right) \right] \ .
\end{align}
Here $\Gamma_1$ and $\Gamma_2$ are Dirac Gamma matrices, and the kinematic scalar factor $D_Q$ is defined in Eq.~(\ref{eq:DQ}). The hard coefficient $H_{\Gamma_1\Gamma_2,\hat{x}}\left(m_{Q}^2/\mu^2\right)$ is perturbatively calculable and depends on the gauge link direction $\hat{x} \equiv i x/\sqrt{-x^2}$. All matrix elements are renormalized in the $\overline{\rm MS}$ scheme at scale $\mu$\footnote{In the following discussions, we suppress the explicit scale $\mu$ dependence for simplicity.}. The subscript ``$c$" indicates the connected contribution,
\begin{align}\label{eq:connected}
&\langle H(P) | \bar{\psi}(x) \Gamma_1 Q(x) \, \bar{Q}(0) \Gamma_2 \psi(0) | H(P) \rangle_c \nonumber\\
&\equiv \langle H(P) | \bar{\psi}(x) \Gamma_1 Q(x) \, \bar{Q}(0) \Gamma_2 \psi(0) | H(P) \rangle
- \langle H(P) | H(P) \rangle \langle \Omega | \bar{\psi}(x) \Gamma_1 Q(x) \, \bar{Q}(0) \Gamma_2 \psi(0) | \Omega \rangle \ .
\end{align}
Here $| \Omega \rangle$ denotes the QCD vacuum. This matching relation shows that the current-current correlator $M_{\Gamma_1\Gamma_2}(x,P)$ and the quasi-PDF matrix element $\tilde{h}_{\Gamma_1\Gamma_2}(x,P)$ exhibit different exponential and polynomial large distance behaviors. These differences are captured by the kinematic factor $D_Q$ in Eq.~(\ref{eq:DQ}), indicating that the gauge link is a more efficient approach for transporting quantum correlations over large distances. Intuitively, the heavy-quark propagator transports quantum correlations via diffusion in all directions, whereas the gauge link is more like a direct transport along its path. 

The coordinate-space HQET reduction formulas in Eqs.~(\ref{eq:MatQGLorg}) and~(\ref{eq:MatQGL}) are justified in the timelike case through an expansion-by-region analysis in Appendix~\ref{sec:MatQGLexp} and perturbative calculations up to one-loop in Appendix~\ref{sec:MatQGL1loop}. It is crucial to keep the ``kinematic energy term" $D_{\perp}^2/(2 m_Q)$, which provides the correct polynomial-$t$ dependence. Given the HQET reduction in the timelike region, in Appendix~\ref{sec:reductionspace}, we use techniques in complex analysis to show that the reduction also holds in the spacelike region.

Different choices of $\Gamma_1,\Gamma_2$ correspond to different partonic polarizations. In the following subsections, we focus on the unpolarized combination, 
\begin{align}
M_{\{\Gamma_x,I\}}(x,P) = M_{\Gamma_x,I}(x,P)+M_{I,\Gamma_x}(x,P) \ ,
\end{align}
where the gamma matrix $\Gamma_x$ is defined as,
\begin{align}
\Gamma_x \equiv 
\begin{cases}\label{eq:Gammax}
\gamma^z \ , \quad x=(t,0,0,0) \\
\gamma^t \ , \quad x=(0,0,0,z)
\end{cases} \ .
\end{align}
We will begin with the timelike separation and convert the result to the spacelike case at the end. According to Eq.~(B4) of Ref.~\cite{Gao:2021dbh}, this gamma matrix construction involves only a single Lorentz scalar function, which simplifies the conversion between timelike and spacelike correlators, as will be discussed in Sec.~\ref{sec:Lorentz}. Finally, in the spacelike case, $M_{\{\Gamma_x,I\}}(x,P)$ can be matched to the standard quasi-PDF matrix element $\langle H(P) | \bar{\psi}(z) U(z,0) \gamma^t \psi(0) | H(P) \rangle$.

\subsection{Dispersive analysis considering form factor connectedness}\label{sec:Disana}
In this subsection, we start with timelike separation for clear physical interpretations and derive the large-time behavior using dispersive analysis, considering three types of form-factor representations regarding connectedness. 

To analyze the large time behaviors ($x=(t,0,0,0)$ with $t\rightarrow +\infty$) of Eq.~(\ref{eq:MqQQq}) for the unpolarized case, we insert a complete basis of eigenstates of QCD Hamiltonian,
\begin{align}\label{eq:FFrep}
M_{\{\Gamma_x,I\}}(x,P)  = \int d \Gamma_{X}(k) e^{i (P^t-k^t) t}  \langle H(P) | \bar{\psi} \gamma^z Q |X(k)\rangle \langle X(k)| \bar{Q}  \psi | H(P) \rangle |_{\rm c} + (\gamma^z \leftrightarrow I) \ ,
\end{align}
where $X(k)$ denotes a multi-particle state,
\begin{align}\label{eq:Xmps}
X(k) \equiv  n_1(k_1) n_2(k_2) ... n_N(k_N)\ ,
\end{align}
where $n_i(k_i)$ is a single particle state with species $n_i$ and momentum $k_i$, and the total momentum $k=\sum_{i \in X} k_{i}$. The notation ``$i \in X$" means $i$ is a single particle in the multi-particle state $X$. The Lorentz indices and permutation signatures are omitted for simplicity. $\int d \, \Gamma_X(k)$ is the Lorentz invariant phase space,
\begin{align}\label{eq:PhaseSpace}
\int d \, \Gamma_X(k) \equiv \frac{1}{S_X} \Pi_{i \in X} \int \frac{d^3 \vec{k_i}}{(2\pi)^3} \frac{1}{2\sqrt{\vec{k}_{i}^2+m_{i}^2}} \ ,
\end{align}
where $S_X$ is the symmetry factor for the identical particles, and $m_{i}$ is the mass of the single particle state ``$i$".

\begin{figure}
    \centering
    \includegraphics[width=0.4\linewidth]{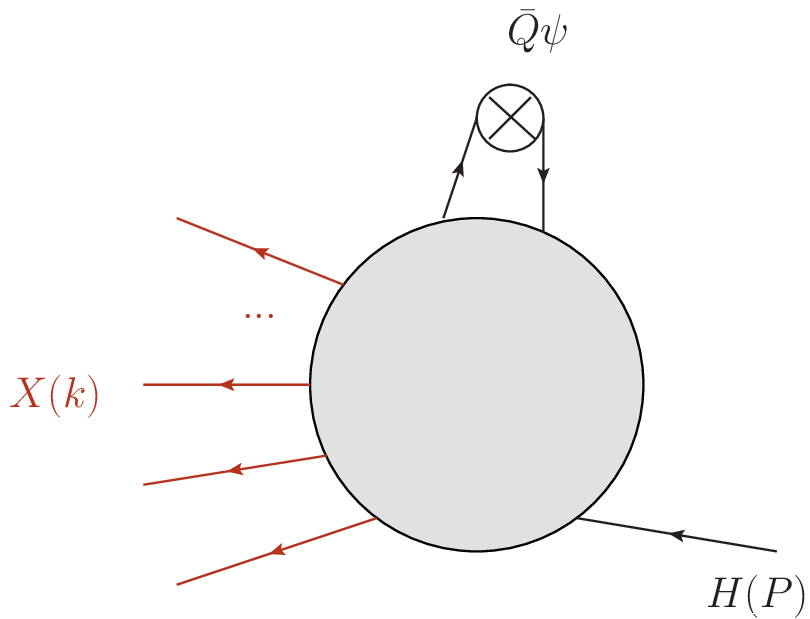}
    \quad\quad
    \includegraphics[width=0.44\linewidth]{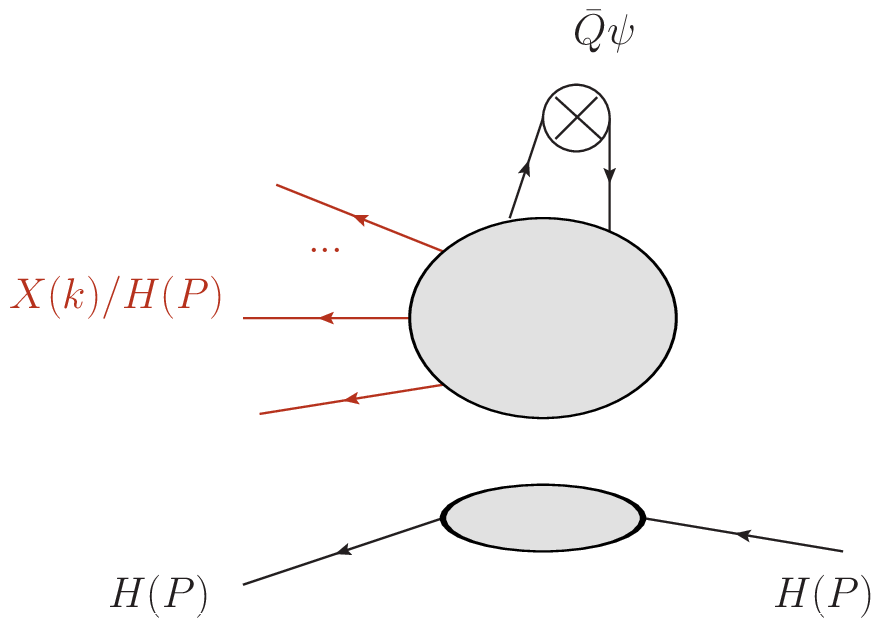}
    \caption{Connected and disconnected contributions to the form factor $\langle X(k) | \bar{Q}  \psi | H(P) \rangle$. The left panel denotes the diagram $\langle X(k) | \bar{Q}  \psi | H(P) \rangle_{\rm C}$ where $\bar{Q}  \psi$ and $| H(P) \rangle$ are connected. The right panel represents the contribution $\langle X(k) | \bar{Q}  \psi | H(P) \rangle_{\rm DC}$ for $\bar{Q}  \psi$ and $| H(P) \rangle$ being disconnected. $H(P)$ is the external hadron of quasi-correlator and $X(k)$ is the inserted multi-particle state.}
    \label{fig:FFCDC}
\end{figure}
\begin{figure}
    \centering
    \includegraphics[width=0.74\linewidth]{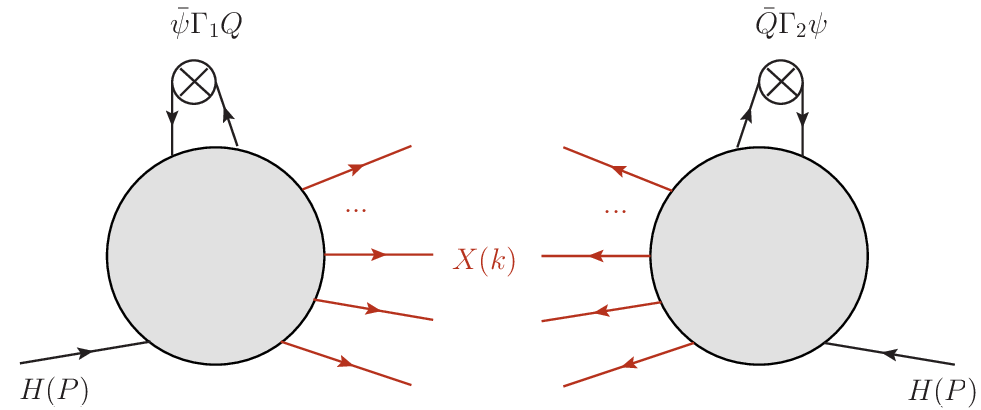}
    \caption{Form factor representation $M_X$, where both form factors are connected. For the unpolarized case, we sum up $(\Gamma_1,\Gamma_2)=(\gamma^z,I)$ and $(\Gamma_1,\Gamma_2)=(I,\gamma^z)$. }
    \label{fig:Corre_C_C}
\end{figure}
\begin{figure}
    \centering
    \includegraphics[width=0.74\linewidth]{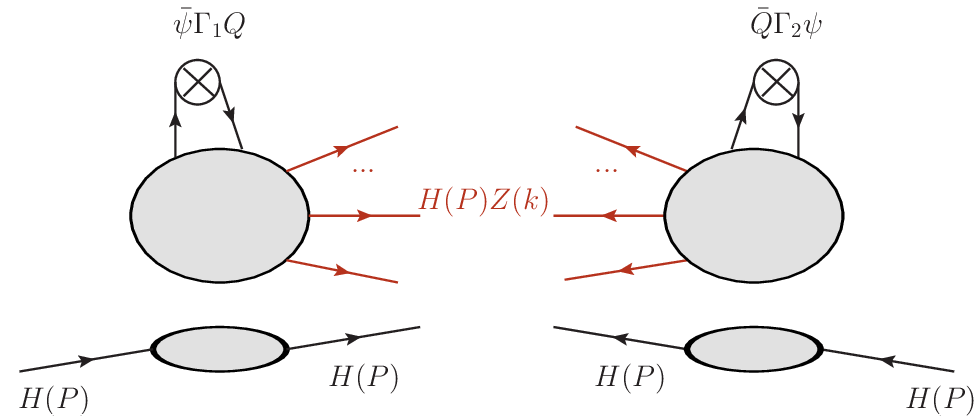}
    \caption{Form factor representation $M_Z$, where both form factors are disconnected. }
    \label{fig:Corre_DC_DC}
\end{figure}
\begin{figure}
    \centering
    \includegraphics[width=0.475\linewidth]{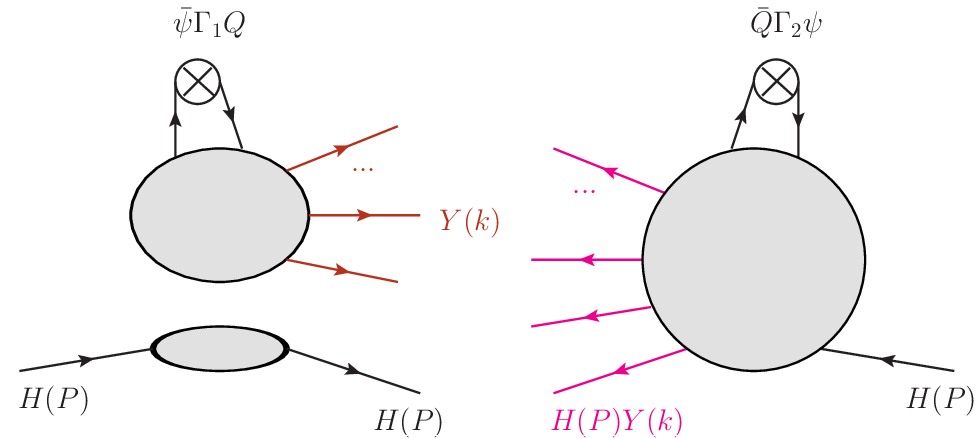}
    \raisebox{6.18\height}{$+$}
    \includegraphics[width=0.475\linewidth]{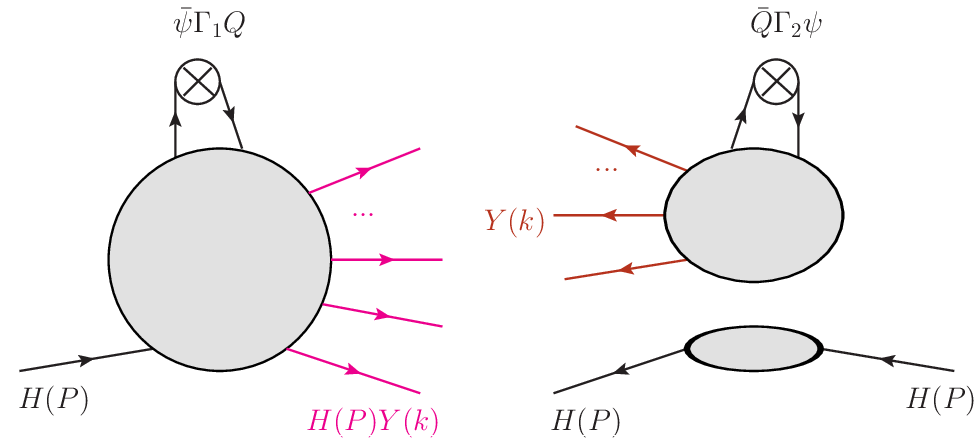}
    \caption{Form factor representation $M_Y$, where one form factor is disconnected, and the other one is connected. There are two terms in this representation. }
    \label{fig:Corre_DC_C_C_DC}
\end{figure}

Although the full correlator $M_{\{\Gamma_x,I\}}(x,P)$ is connected, the form factors contain disconnected contributions~\cite{JAFFE1983205,Collins_2023},
\begin{align}
\langle X(k) | \bar{Q}  \psi | H(P) \rangle
= \langle X(k) | \bar{Q}  \psi | H(P) \rangle_{\rm C} + \langle X(k) | \bar{Q}  \psi | H(P) \rangle_{\rm DC} \ ,
\end{align}
where the matrix element $\langle X(k) | \bar{Q}  \psi | H(P) \rangle_{\rm DC}$ is defined as $\bar{Q}  \psi$ and $| H(P) \rangle$ being disconnected, shown as the right panel of Fig.~\ref{fig:FFCDC},
\begin{align}\label{eq:FFDC}
\langle X(k) | \bar{Q}  \psi | H(P) \rangle_{\rm DC}
&\equiv \sum_{n \in X} \langle X(k)/n(q) | \, \bar{Q}  \psi \, | \Omega \rangle \, \langle n(q) | H(P)\rangle \nonumber\\
&= \sum_{n \in X} \langle X(k)/n(q) | \, \bar{Q}  \psi \, | \Omega \rangle \, (2\pi)^3 2 \sqrt{\vec{P}^2+m_H^2} \delta^3(\vec{q}-\vec{P}) \delta_{n,H} \ ,
\end{align}
where $X(k)/n(q)$ denotes the state $X(k)$ eliminating the single particle $n(q)$. The matrix element $\langle X(k) | \bar{Q}  \psi | H(P) \rangle_{\rm C}$ is the remaining part where $\bar{Q}  \psi$ and $| H(P) \rangle$ are connected, denoted as the left panel of Fig.~\ref{fig:FFCDC}. Eq.~(\ref{eq:FFrep}) is decomposed based on form factor connectedness,
\begin{align}\label{eq:Mcla}
M_{\{\Gamma_x,I\}}(x,P) = M_{X}(x,P) + M_{Y}(x,P) + M_{Z}(x,P) \ .
\end{align}
In $M_{X}(x,P)$, both form factors are connected, as shown in Fig.~\ref{fig:Corre_C_C}. In $M_{Y}(x,P)$, one form factor is connected while the other is disconnected, as depicted in Fig.~\ref{fig:Corre_DC_C_C_DC}. Finally, in $M_{Z}(x,P)$, both form factors are disconnected, as illustrated in Fig.~\ref{fig:Corre_DC_DC}.

Start from $M_{X}(x,P)$ in Fig.~\ref{fig:Corre_C_C},
\begin{align}
&M_{X}(x,P) 
= e^{i P^t t} \sum_{X} \int d \, \Gamma_X(k) \, e^{-i k^t  t} \langle H(P) | \bar{\psi} \gamma^z Q |X(k)\rangle_{\rm C} \langle X(k)| \bar{Q}  \psi | H(P) \rangle_{\rm C}  + (\gamma^z \leftrightarrow I) \ .
\end{align}
Express it in terms of the multiplication of single-particle states,
\begin{align}
&M_{X}(x,P) = e^{i P^t t} \sum_{X} \frac{1}{S_X} \left( \Pi_{i \in X} \int \frac{d^3 \vec{k_i}}{(2\pi)^3} \frac{1}{2\sqrt{\vec{k}_{i}^2+m_{i}^2}} \, e^{-i \sqrt{\vec{k}_{i}^2+m_{i}^2} t} \right) \nonumber\\
&\quad \times \langle H(P) | \bar{\psi} \gamma^z Q | \Pi_{i \in X} n_i(k_i) \rangle_{\rm C} \langle \Pi_{i \in X} n_i(k_i) | \bar{Q}  \psi | H(P) \rangle_{\rm C}  + (\gamma^z \leftrightarrow I) \ .
\end{align}  
See the meaning of ``$i \in X$" below Eq.~(\ref{eq:Xmps}). The large-time asymptotic form can be calculated using the stationary-phase analysis~\cite{stein2003complex}. In the $t\rightarrow+\infty$ limit, the phase factor $e^{-i \sqrt{\vec{k}_{i}^2+m_{i}^2} t}$ is highly oscillatory, leading to strong cancellations in the phase space integral, except for a domain near the stationary-phase point $\vec{k}_{i}^2\sim 0$, where we perform the expansion, 
\begin{align}\label{eq:MXSP}
&M_{X}(x,P)_{t \rightarrow +\infty}  = e^{i P^t t} \sum_{X} \frac{1}{S_X} \left( \Pi_{i \in X} \int \frac{d^3 \vec{k_i}}{(2\pi)^3} \frac{1}{2 m_i} \, e^{-i m_i t} e^{-i \frac{\vec{k}_i^2}{2 m_i} t} \right) \nonumber\\
&\quad \times \langle H(P) | \bar{\psi} \gamma^z Q | \Pi_{i \in X} n_i( m_i \hat{x} ) \rangle_{\rm C} \langle \Pi_{i \in X} n_i( m_i \hat{x} ) | \bar{Q}  \psi | H(P) \rangle_{\rm C} \nonumber\\
&\quad \times\left[1 + O\left(k_i^z\right) + O\left(\vec{P}\cdot \vec{k}_i\right) + O\left(\vec{k}_i\cdot \vec{k}_j\right)_{j\neq i} + O\left(\vec{k}_i^2\right)\right]  + (\gamma^z \leftrightarrow I) \nonumber\\
& = e^{i P^t t} \sum_{X} \frac{1}{S_X} \left( \Pi_{i \in X} \frac{1}{2 m_i}  \frac{m_i^3}{2\sqrt{2} \pi^{3/2}}  \frac{e^{-i m_i t}}{(i m_i t)^{3/2}}  \right) \nonumber\\
&\quad \times \langle H(P) | \bar{\psi} \gamma^z Q | \Pi_{i \in X} n_i( m_i \hat{x} ) \rangle_{\rm C} \langle \Pi_{i \in X} n_i( m_i \hat{x} ) | \bar{Q}  \psi | H(P) \rangle_{\rm C} \left[1+O\left(\frac{1}{t}\right)\right]  + (\gamma^z \leftrightarrow I) \ .
\end{align}
See the definition of $\hat{x}$ below Eq.~(\ref{eq:MatQGL}). The result has a clear physical interpretation, with a simple power counting rule. The dominant contribution arises from the momentum mode $k_i\sim \left(m_i,\sqrt{m_i/t},\sqrt{m_i/t},\sqrt{m_i/t}\right)$, which is almost like a particle at rest except for slight spatial quantum fluctuations. The time component gives the exponential behavior $e^{-i m_i t}$, which is the phase for a particle at rest, and the spatial components provide the polynomial dependence $1/t^{3/2}$, which is related to the size of quantum fluctuations. 

The form factors in Eq.~(\ref{eq:MXSP}) imply that the mass gaps $m_X\equiv\sum_{i\in X} m_i$ in the large time expansion can be identified with the spectrum of states projected by the common quantum numbers of $\bar{Q} \psi | H(P) \rangle$ and $\bar{Q} \gamma^z \psi | H(P) \rangle$. 

The higher order correction is $O\left(1/t\right)$ suppressed instead of $O\left(1/\sqrt{t}\right)$, as a consequence of rotational symmetry. In the expansion around the stationary phase point $\vec{k}_{i}^2\sim 0$, there are $O\left(k_i^z\right)$, $O\left(\vec{P}\cdot \vec{k}_i\right)$ and $O\left(\vec{k}_i\cdot \vec{k}_j\right)_{j\neq i}$ corrections originating from the form factors. In particular, $O\left(k_i^z\right)$ and $O\left(\vec{P}\cdot \vec{k}_i\right)$ could potentially lead to $O\left(1/\sqrt{t}\right)$ by power counting. However, these corrections vanish in the phase space integral because the factor $e^{-i \frac{\vec{k}_i^2}{2 m_i} t}$ is isotropic. In contrast, the $O\left(\vec{k}_i^2\right)$ term survives, yielding the $O(1/t)$ correction. Generally speaking, only corrections with even powers of spatial momenta contribute, leading to an integer-power series in $t$. 

For $M_{Z}(x,P)$ in Fig.~\ref{fig:Corre_DC_DC} where both form factors are disconnected, integrating out the delta functions from Eq.~(\ref{eq:FFDC}) yields,
\begin{align}
&M_{Z}(x,P) 
= e^{-i P^t t} \sum_{Z} \int d \, \Gamma_Z(k) \, e^{-i k^t  t} \langle \Omega | \bar{\psi} \gamma^z Q |H(P) Z(k)\rangle \langle H(P) Z(k) | \bar{Q}  \psi | \Omega \rangle  + (\gamma^z \leftrightarrow I) \ .
\end{align}
Following a similar logic as $M_{X}(x,P)$, we obtain the large time asymptotic form,
\begin{align}\label{eq:MZSP}
&M_{Z}(x,P)_{t \rightarrow +\infty} = e^{-i P^t t} \sum_{Z} \frac{1}{S_Z} \left( \Pi_{i \in Z} \frac{1}{2 m_i}  \frac{m_i^3}{2\sqrt{2} \pi^{3/2}}  \frac{e^{-i m_i t}}{(i m_i t)^{3/2}}  \right) \nonumber\\
&\quad \times \langle \Omega | \bar{\psi} \gamma^z Q | H(P) \Pi_{i \in Z} n_i( m_i \hat{x} ) \rangle \langle H(P) \Pi_{i \in Z} n_i( m_i \hat{x} )  | \bar{Q}  \psi | \Omega \rangle \left[1+O\left(\frac{1}{t}\right)\right]  + (\gamma^z \leftrightarrow I) \ .
\end{align}
The result is similar to $M_X$ except for the overall phase factor and the physical origins of mass gaps. In $M_X$, the external hadron $H(P)$ appears as an incoming state, giving rise to the phase factor $e^{i P^t t}$. In contrast, for $M_Z$, after integrating out the disconnected contributions, the external hadron $H(P)$ effectively appears as an outgoing state, leading to the phase factor $e^{-i P^t t}$. The mass gap $m_Z \equiv \sum_{i\in Z} m_i$ in $M_Z$ is related to the state $Z$ projected by the common quantum numbers of $\bar{Q} \psi | \bar{H}(P) \rangle$ and $\bar{Q} \gamma^z \psi | \bar{H}(P) \rangle$, where $\bar{H}(P)$ denotes the anti-particle of $H(P)$. 

\begin{figure}
    \centering
    \includegraphics[width=0.475\linewidth]{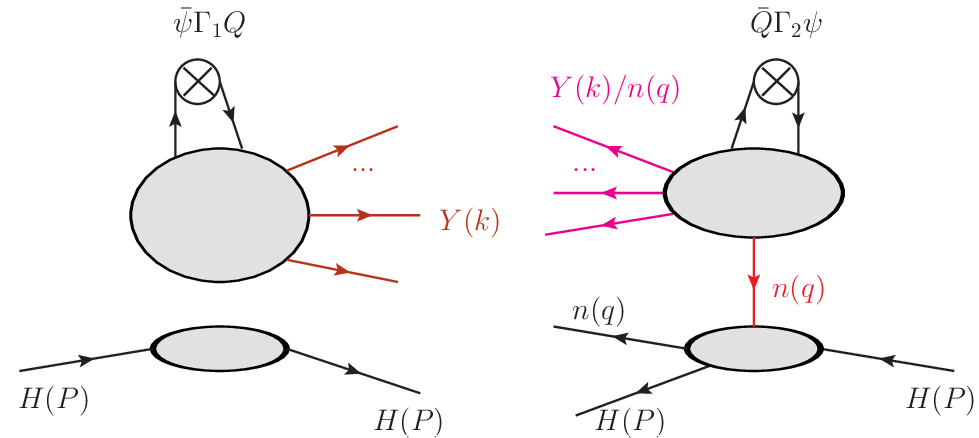}
    \raisebox{6.18\height}{$+$}
    \includegraphics[width=0.475\linewidth]{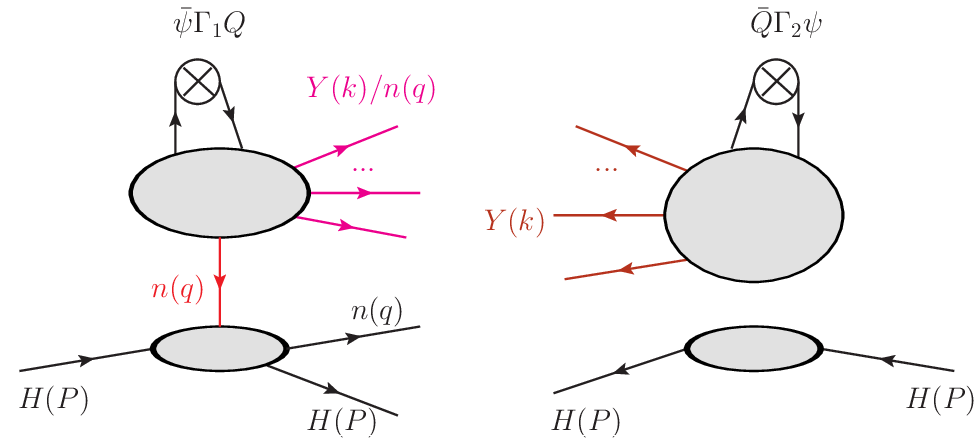}
    \caption{One-particle resonance diagram $M_{Y,fs}$ (or called forward singularity diagram), belonging to $M_Y$ in Fig.~\ref{fig:Corre_DC_C_C_DC}. The multi-particle state $Y(k)$ contains the single particle $n(q)$, and see the definition of $Y(k)/n(q)$ below Eq.~(\ref{eq:FFDC}). For the unpolarized case, we sum up $(\Gamma_1,\Gamma_2)=(\gamma^z,I)$ and $(\Gamma_1,\Gamma_2)=(I,\gamma^z)$. For moving intermediate states, allowed by the quantum numbers of the operators, the propagator {\color{red} $n(q)$} and the inserted single-particle $n(q)$ can be of the same species. Constrained by momentum conservation, the propagator {\color{red} $n(q)$} is on-shell. }
    \label{fig:Corre_DC_C_C_DC_sg}
\end{figure}

As for $M_{Y}(x,P)$ in Fig.~\ref{fig:Corre_DC_C_C_DC}, where one form factor is connected and the other one is not, integrating out the delta function in Eq.~(\ref{eq:FFDC}) leads to
\begin{align}\label{eq:MYorg}
&M_{Y}(x,P) =  \sum_{Y} \int d \, \Gamma_{Y}(k) \, e^{-i k^t t} 
\left[ \langle \Omega | \bar{\psi} \gamma^z Q | Y(k) \rangle \langle H(P) Y(k) | \bar{Q}  \psi | H(P) \rangle_{\rm C} \right. \nonumber\\
&\left. \quad + \langle H(P) | \bar{\psi} \gamma^z Q | H(P) Y(k) \rangle_{\rm C} \langle Y(k) | \bar{Q}  \psi | \Omega \rangle \right] + (\gamma^z \leftrightarrow I)  \ .
\end{align}
The form factors $\langle H(P) Y(k) | \bar{Q}  \psi | H(P) \rangle_{\rm C}$ and $\langle H(P) | \bar{\psi} \gamma^z Q | H(P) Y(k) \rangle_{\rm C}$ contain divergences associated with one-particle resonance poles~\cite{Lin:2021umz}. Fig.~\ref{fig:Corre_DC_C_C_DC_sg} displays the one-particle resonance diagram $M_{Y,fs}$, as part of the contributions to the form factor representation $M_Y$. Since $\bar{Q} \psi$ and $\bar{Q} \gamma^z \psi$ have the common quantum numbers to interpolate the same state as long as $\vec{k} \neq 0$~\cite{Zhang:2025hyo}, the species of the propagator {\color{red} $n(q)$} can be the same as the inserted single-particle $n(q)$. Constrained by momentum conservation, the propagator momentum {\color{red} $q$} is the same as this inserted on-shell state $n(q)$. Therefore, the propagator {\color{red} $n(q)$} is on-shell, and the form factor becomes divergent. 

\begin{figure}
    \centering
    \includegraphics[width=0.475\linewidth]{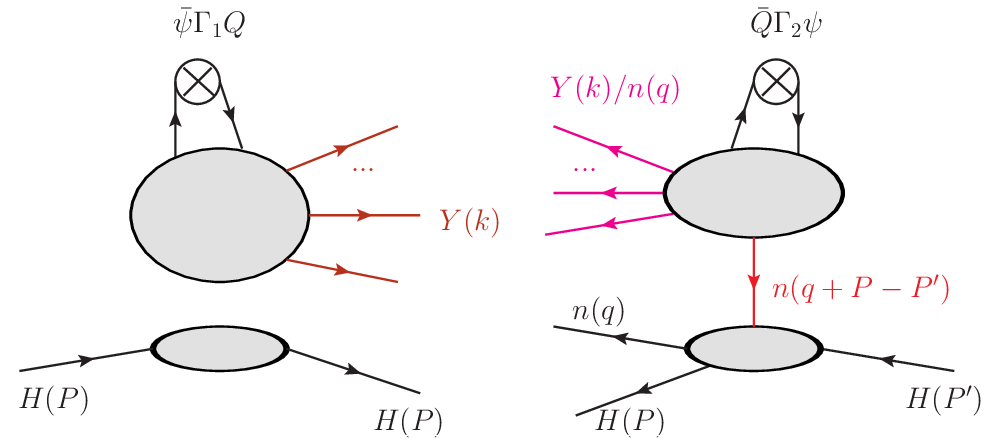}
    \raisebox{6.18\height}{$+$}
    \includegraphics[width=0.475\linewidth]{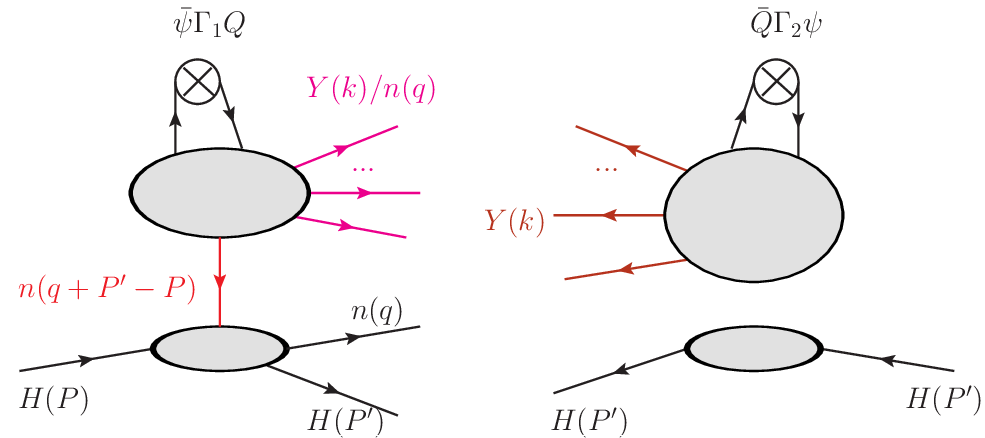}
    \caption{Off-forward regulator for the forward singularity. The propagators {\color{red} $n(q+P-P')$} and {\color{red}  $n(q+P'-P)$} are off-shell under the off-forwardness. }
    \label{fig:Corre_DC_C_C_DC_sg_reg}
\end{figure}

To study this divergence, we introduce the off-forward regulator, which has been discussed in Sec.~6.11.6 in Ref.~\cite{Collins_2023}, 
\begin{align}
\langle H(P) | \bar{\psi}(x) \gamma^z Q(x)  \bar{Q}(0) \psi(0) | H(P) \rangle_{\rm c} \rightarrow \langle H(P) | \bar{\psi}(x) \gamma^z Q(x)  \bar{Q}(0) \psi(0) | H(P') \rangle_{\rm c} \ ,
\end{align}
and the one-particle resonance diagram $M_{Y,fs}$ becomes Fig.~\ref{fig:Corre_DC_C_C_DC_sg_reg}. The propagator {\color{red} $n(q+P-P')$} or {\color{red} $n(q+P'-P)$} now becomes off-shell, which means the off-forwardness regularizes the divergence. Since each term becomes singular in the forward limit $P'\rightarrow P$, this divergence is called the {\it forward singularity} in our language. Using the methods in Ref.~\cite{Lin:2021umz}, the forward singularity diagram $M_{Y,fs}$ can be reduced into the single-particle propagator, two-particle forward scattering amplitude $T(n(q),H(P))$\footnote{In this work, the amplitude $T$ does not include the Dirac delta function of momentum conservation, which differs from the convention used in Ref.~\cite{Peskin:1995ev}.}, and wavefunction amplitudes, in the domain $P' \sim P$,
\begin{align}\label{eq:MYoff}
&M_{Y,fs}(x,P,P') =  \sum_{Y} \sum_{n(q) \in Y} \int d \, \Gamma_{Y}(k) \, e^{-i k^t t} 
\left[ \langle \Omega | \bar{\psi} \gamma^z Q | Y(k) \rangle \frac{ T(n(q),H(P)) \, \langle Y(k) | \bar{Q}  \psi | \Omega \rangle}{(q+P-P')^2 - m_n^2 + i0^{+}}  \right. \nonumber\\
&\left.  + e^{i \left( P^t-P'^t \right) t} \frac{\langle \Omega | \bar{\psi} \gamma^z Q | Y(k) \rangle \, T(n(q),H(P))}{(q+P'-P)^2 - m_n^2 - i0^{+}} \langle Y(k) | \bar{Q}  \psi | \Omega \rangle \right]+O\left[\left(P'^t-P^t\right)^{0}\right] + (\gamma^z \leftrightarrow I)  \ .
\end{align}
Here $Y(k)$ denotes a multi-particle state containing a single particle $n(q)$ with on-shell momentum $q^2=m_n^2$. The notation $O\left[\left(P'^t-P^t\right)^{0}\right]$ represents part of the regular contributions in the forward limit, while the other parts are retained explicitly in the propagators and phase $e^{i (P^t-P'^t)t}$ for later convenience.  

We provide a heuristic argument for the cancellation of the forward singularity between the two terms in Eq.~(\ref{eq:MYoff}) or Fig.~\ref{fig:Corre_DC_C_C_DC_sg_reg}. The propagators {\color{red}$n(q+P-P')$} or {\color{red}$n(q+P'-P)$} are factorized out from the remaining parts labeled as $g_1$ and $g_2$,
\begin{align}\label{eq:MYsgcancel}
&M_{Y,fs}(x,P,P') = \int d^3 \vec{q} \left[\frac{g_1(P,P',\vec{q},t)}{(q+P-P')^2 - m_n^2 + i0^{+}} + \frac{g_2(P,P',\vec{q},t)}{(q+P'-P)^2 - m_n^2 - i0^{+} }\right] \nonumber\\
&= \int d^3 \vec{q} \frac{1}{E_{\vec{q}} + P^t - P'^t - E_{\vec{q}+\vec{P}-\vec{P}'}+i0^{+}} \nonumber\\
&\quad \times \left[ \frac{g_1(P,P',\vec{q},t)}{E_{\vec{q}} + P^t - P'^t + E_{\vec{q}+\vec{P}-\vec{P}'}} - \frac{g_2(P,P',\vec{q}-\vec{P}'+\vec{P},t)}{E_{\vec{q}+\vec{P}-\vec{P}'}  + P'^t - P^t + E_{\vec{q}} }\right] \ ,
\end{align}
where $E_{\vec{q}} \equiv \sqrt{\vec{q}^2+m_n^2}$. Due to the symmetric structures of the diagrams, $g_1$ and $g_2$ are the same in the forward limit $P' \rightarrow P$. Provided that $g_1$ and $g_2$ are analytic in the neighborhood of $P'^t \sim P^t$, the above expression is finite in the $P'\rightarrow P$ limit, which means the forward singularity is canceled. 

Although the forward singularity is canceled, it could have an impact on the polynomial-$t$ dependence. After taking the forward limit, the forward singularity diagram $M_{Y,fs}(x,P)$ becomes
\begin{align}\label{eq:MYsgLint}
&M_{Y,fs}(x,P) =  \sum_{Y} \sum_{n(q) \in Y} \int d \, \Gamma_{Y}(k) \, e^{-i k^t t} 
\left[ \frac{i (P^t -P'^t) {\color{red} t}}{(q+P'-P)^2 - m_n^2} \right]_{P' \rightarrow P} \nonumber\\
&\times \langle \Omega | \bar{\psi} \gamma^z Q | Y(k) \rangle \, T(n(q),H(P)) \, \langle Y(k) | \bar{Q}  \psi | \Omega \rangle \left[1+O\left(\frac{\left(P'^t-P^t\right)^{0}}{t}\right)\right] + (\gamma^z \leftrightarrow I)  \ ,
\end{align}
where the factor ${\color{red} t}$ arises from the expansion of the phase $e^{i \left( P^t-P'^t \right) t}$ in Eq.~(\ref{eq:MYoff}) in the forward limit. The notation $O\left(\left(P'^t-P^t\right)^{0}/t\right)$ denotes other regular contributions without the linear $t$ enhancement. 

However, the above linear-$t$ enhancement does not contribute to the leading asymptotics of the large time expansion, which corresponds to the zero-momentum intermediate states, that cannot be simultaneously projected by the operators $\bar{Q} \psi$ and $\bar{Q} \gamma^z \psi$. To demonstrate this, we calculate the large time limit of Eq.~(\ref{eq:MYsgLint}) using stationary phase analysis as before, 
\begin{align}\label{eq:MYsfLt}
&M_{Y,fs}(x,P)_{t\rightarrow+\infty} =  \sum_{Y} \sum_{n(q) \in Y} \frac{1}{S_Y} \left( \Pi_{i \in Y} \frac{1}{2 m_i}  \frac{m_i^3}{2\sqrt{2} \pi^{3/2}}  \frac{e^{-i m_i t}}{(i m_i t)^{3/2}}  \right) \nonumber\\
&\times T(n(m_n \hat{x}),H(P)) \left[ \frac{i (P^t -P'^t) t}{(m_n \hat{x}+P'-P)^2 - m_n^2} \right]_{P'^t \rightarrow P^t} \nonumber\\
&\times \langle \Omega | \bar{\psi} \gamma^z Q | \Pi_{i \in Y} n_i( m_i \hat{x} ) \rangle \,  \langle \Pi_{i \in Y} n_i( m_i \hat{x} ) | \bar{Q}  \psi | \Omega \rangle  \nonumber\\
&\times \left[1+O\left(\frac{1}{t}\right)\right] + (\gamma^z \leftrightarrow I) \ ,
\end{align}
where the intermediate states $Y$ are at rest, which can be determined from the quantum numbers of the rest states, as shown in Tab.~6.1 in Ref.~\cite{Gattringer:2010zz}. Based on the form factor $\langle \Pi_{i \in Y} n_i( m_i \hat{x} ) | \bar{Q}  \psi | \Omega \rangle$, $Y$ has the quantum number of a scalar meson. According to $\langle \Omega | \bar{\psi} \gamma^z Q | \Pi_{i \in Y} n_i( m_i \hat{x} ) \rangle$, $Y$ has the quantum number of a vector meson. As a result, $Y$ does not exist, and the linear-$t$ enhanced contribution vanishes. 

Therefore, the large time behavior of $M_Y$ can be calculated using similar methods as $M_X$ or $M_Z$,
\begin{align}\label{eq:MY}
&M_{Y}(x,P)_{t\rightarrow+\infty} =  \sum_{Y} \frac{1}{S_Y} \left( \Pi_{i \in Y} \frac{1}{2 m_i}  \frac{m_i^3}{2\sqrt{2} \pi^{3/2}}  \frac{e^{-i m_i t}}{(i m_i t)^{3/2}}  \right) \nonumber\\
&\times \left[ \langle \Omega | \bar{\psi} \gamma^z Q | \Pi_{i \in Y} n_i( m_i \hat{x} ) \rangle \langle H(P) \Pi_{i \in Y} n_i( m_i \hat{x} ) | \bar{Q}  \psi | H(P) \rangle_{\rm C} \right. \nonumber\\
&\left. \quad + \langle H(P) | \bar{\psi} \gamma^z Q | H(P) \Pi_{i \in Y} n_i( m_i \hat{x} ) \rangle_{\rm C} \langle \Pi_{i \in Y} n_i( m_i \hat{x} ) | \bar{Q}  \psi | \Omega \rangle \right] \nonumber\\
&\times \left[1+O\left(\frac{1}{t}\right)\right]+ (\gamma^z \leftrightarrow I)  \ .
\end{align}
Compared to $M_X$ and $M_Z$, there is no overall phase $e^{i P^t t}$ or $e^{-i P^t t}$. The mass gap $m_Y \equiv \sum_{i\in Y} m_i$ has a different physical origin, related to a state $Y$ projected by the quantum numbers of $\bar{Q}  \psi | \Omega \rangle$ or $\bar{Q} \gamma^z \psi | \Omega \rangle$.

The derived large-time asymptotic forms here are supported by the perturbative calculations up to two-loop in $\phi^3$ theory in Appendix~\ref{sec:phi3pert}, including the methods to determine the physical origins of mass gaps, the influence of form factor connectedness on the phases, and the power-counting of polynomial-$t$ dependence for $M_X$ and $M_Z$. Moreover, the cancellation and impact of forward singularities have been demonstrated in $\phi^3$ theory through perturbative calculations in Appendices~\ref{sec:phi3FWSG} and~\ref{sec:CalMets} and all-order discussions in Appendix~\ref{sec:phi3FWSGsplit}. 

The discussion of forward singularities in this subsection can be straightforwardly generalized to other “forward” quasi-correlators, in particular for utilizing Eq.~(\ref{eq:MYsfLt}) to determine whether the large time expansion of $Y$ terms exhibits the enhanced polynomial dependence. Starting from a general operator displacement in a forward matrix element,
\begin{align}\label{eq:GOdis}
    \langle H(P) | O_1(x) O_2(0) | H(P) \rangle \ ,
\end{align}
the factor $\langle \Omega | \bar{\psi} \gamma^z Q | \Pi_{i \in Y} n_i( m_i \hat{x} ) \rangle \,  \langle \Pi_{i \in Y} n_i( m_i \hat{x} ) | \bar{Q}  \psi | \Omega \rangle$ appearing in Eq.~(\ref{eq:MYsfLt}) is replaced by
\begin{align}\label{eq:replace}
    \langle \Omega | O_1 | \Pi_{i \in Y} n_i( m_i \hat{x} ) \rangle \,  \langle \Pi_{i \in Y} n_i( m_i \hat{x} ) | O_2 | \Omega \rangle \ .
\end{align}
Consequently, if $O_1$ and $O_2$ have overlap with the same rest-frame intermediate state, as in the examples of $\phi^3$ theory discussed in Appendix~\ref{sec:phi3} and the gluon PDF analyzed in Appendix~\ref{sec:Gluon}, the polynomial dependence of the $Y$ term is enhanced in the large time expansion. By contrast, when $O_1$ and $O_2$ do not share overlap with the same rest state, as in the unpolarized quark PDF considered in this subsection and the polarized cases studied in Appendix~\ref{sec:PolPDF}, such an enhancement does not occur. We emphasize that these examples correspond to the specific choices of Dirac gamma matrices commonly employed in lattice QCD calculations. For alternative choices, different conclusions may follow; for instance, the correlator $\langle H(P) | \bar{\psi}(z) U(z,0) \gamma^z \psi(0) | H(P) \rangle$ exhibits polynomial enhancement compared to $\langle H(P) | \bar{\psi}(z) U(z,0) \gamma^t \psi(0) | H(P) \rangle$.

Here we should comment that the forward-singularities discussed here can be perceived as the effects on the coordinate space structure functions, of those singularities in the multi-particle form factors, that occur when the momentum of the out-going particles coincide with the incoming ones. In 2D integrable QFTs, these singularities were sometimes called ``kinematical singularities''~\cite{Bajnok:2006ze} and were quite crucial when solving the multi-particle form factors~\cite{Karowski:1978vz,Berg:1978sw,Kirillov:1988zk,Smirnov:1992vz,Balog:1994np,Babujian:2013roa}. See~\cite{Babujian:1998uw} for a nice introduction to such singularities. Their roles on the large distance behaviors of the structure functions, on the other hand, were not very well known in literature yet.

\subsection{Conversion to Euclidean correlator using Lorentz symmetry}\label{sec:Lorentz}
In this subsection, we summarize the large-time asymptotic expansions derived in the previous subsection and convert the results to the spacelike separation using Lorentz symmetry. 

Combining $M_X$ in Eq.~(\ref{eq:MXSP}), $M_Y$ in Eq.~(\ref{eq:MY}), and $M_Z$ in Eq.~(\ref{eq:MZSP}), the large-time asymptotic expansion of the current-current correlator takes the form,
\begin{align}\label{eq:Mexp}
M_{\{\Gamma_x,I\}}(x,P)_{\sqrt{-x^2} \rightarrow \infty}
=&  \sum_{X,l=0} \frac{e^{i P \cdot x} e^{-m_X\sqrt{-x^2}}}{\left(\sqrt{-x^2}\right)^{\frac{3}{2}N_X+l}} \, C_{X,l}\left(P, \left\{k_j = \frac{i x m_j}{\sqrt{-x^2}} \bigg| j\in X\right\}\right) \nonumber\\
& + \sum_{Y,l=0} \frac{e^{-m_Y\sqrt{-x^2}}}{\left(\sqrt{-x^2}\right)^{\frac{3}{2}N_Y+l}} \, C_{Y,l}\left(P, \left\{k_j = \frac{i x m_j}{\sqrt{-x^2}} \bigg| j\in Y\right\}\right) \nonumber\\
& + \sum_{Z,l=0} \frac{e^{-i P \cdot x} e^{-m_Z\sqrt{-x^2}}}{\left(\sqrt{-x^2}\right)^{\frac{3}{2}N_Z+l}} \, C_{Z,l}\left(P, \left\{k_j = \frac{i x m_j}{\sqrt{-x^2}} \bigg| j\in Z\right\}\right) \ ,
\end{align}
where $m_X$ is the mass of the state $X$ projected by common quantum numbers of $\bar{Q} \psi | H(P) \rangle$ and $\bar{Q} \gamma^z \psi | H(P) \rangle$, $m_Y$ is for the state $Y$ with quantum numbers of $\bar{Q}  \psi | \Omega \rangle$ or $\bar{Q} \gamma^z \psi | \Omega \rangle$, and $m_Z$ is for the state $Z$ with common quantum numbers of $\bar{Q} \psi | \bar{H}(P) \rangle$ and $\bar{Q} \gamma^z \psi | \bar{H}(P) \rangle$. The quantities $N_X$, $N_Y$, and $N_Z$ denote the numbers of single-particles in the corresponding multi-particle states $X$, $Y$, and $Z$, respectively. The coefficients $C_{X,l}$, $C_{Y,l}$, and $C_{Z,l}$ depend on the momenta of the external hadron and inserted particles. 

The asymptotic expansion in Eq.~(\ref{eq:Mexp}) is written in a Lorentz invariant form, which corresponds to a single Lorentz scalar function by construction in Eq.~(\ref{eq:Gammax}). The phase factors $e^{i P \cdot x}$ and $e^{-i P \cdot x}$ are Lorentz invariant. In the exponential factors, the separation necessarily appears as the Lorentz-invariant quantity $\sqrt{-x^2}$, since the associated mass gaps are Lorentz invariant. See the discussions on the prescription of $\sqrt{-x^2}$ below Eq.~(\ref{eq:DQ}). We also write the polynomial dependence in terms of $\sqrt{-x^2}$. While alternative parameterizations of the polynomial factors are possible, any such differences can be absorbed into a redefinition of the coefficients. These coefficients depend on the momenta of the inserted particles and can, in general, be expanded along the “velocity” $i x /\sqrt{-x^2}$ like moving HQET. 

Based on HQET reduction in Eq.~(\ref{eq:MatQGL}), the asymptotic expansion of quasi-PDF matrix element is,
\begin{align}\label{eq:hexp}
\tilde{h}_{\{\Gamma_x,I\}}(x,P)_{\sqrt{-x^2} \rightarrow \infty} 
= & \sum_{X,l=0} \frac{e^{i P \cdot x}e^{-(m_X-m_Q)\sqrt{-x^2}}}{\left(\sqrt{-x^2}\right)^{\frac{3}{2}(N_X-1)+l}} \, a_{X,l}\left(P, \left\{k_j = \frac{i x m_j}{\sqrt{-x^2}} \bigg| j\in X\right\}\right) \nonumber\\
&+ \sum_{Y,l=0} \frac{e^{-(m_Y-m_Q)\sqrt{-x^2}}}{\left(\sqrt{-x^2}\right)^{\frac{3}{2}(N_Y-1)+l}} \, a_{Y,l}\left(P, \left\{k_j = \frac{i x m_j}{\sqrt{-x^2}} \bigg| j\in Y\right\}\right) \nonumber\\
&+ \sum_{Z,l=0} \frac{e^{-i P \cdot x}e^{-(m_Z-m_Q)\sqrt{-x^2}}}{\left(\sqrt{-x^2}\right)^{\frac{3}{2}(N_Z-1)+l}} \, a_{Z,l}\left(P, \left\{k_j = \frac{i x m_j}{\sqrt{-x^2}} \bigg| j\in Z\right\}\right) \ ,
\end{align}
which differs from $M_{\{\Gamma_x,I\}}(x,P)_{\sqrt{-x^2} \rightarrow \infty}$ by exponential and polynomial dependences in the kinematic factor $D_{Q}$ in Eq.~(\ref{eq:DQ}).

Using Lorentz symmetry and setting $x=(0,0,0,z)$, we obtain the large-distance expansion for the spacelike correlator in $\overline{\rm MS}$ scheme,
\begin{align}\label{eq:qPDFLzGen}
&\tilde{h}_{\{\gamma^t,I\}}(z,P^z)  
=   \langle H(P) | \bar{\psi}(z) U(z,0) \gamma^t  \psi(0) | H(P) \rangle_c   \nonumber\\
&{\overset{z\rightarrow \infty}{=}}   
 \sum_{X,l=0} \frac{e^{-i P^z z} e^{-\Lambda_X |z| }}{|z|^{\frac{3}{2}(N_X-1)+l}} \, A_{X,l}\left(P^z\right) \, e^{i \, {\rm sign}(z) \, \phi_{X,l}(P^z)} \nonumber\\
&\quad + \sum_{Y,l=0} \frac{e^{-\Lambda_Y |z|}}{|z|^{\frac{3}{2}(N_Y-1)+l}} \, A_{Y,l}\left(P^z\right) \, e^{i \, {\rm sign}(z) \, \phi_{Y,l}(P^z)} \nonumber\\
&\quad + \sum_{Z,l=0} \frac{e^{i P^z z} e^{-\Lambda_Z |z|}}{|z|^{\frac{3}{2}(N_Z-1)+l}} \, A_{Z,l}\left(P^z\right) \, e^{i \, {\rm sign}(z) \, \phi_{Z,l}(P^z)} \ ,
\end{align}
where $\Lambda_X=m_X-m_Q$ is the binding energy of the state $X$ projected by the common quantum numbers of $\bar{Q} \psi | H(P) \rangle$ and $\bar{Q} \gamma^z \psi | H(P) \rangle$, $\Lambda_Y=m_Y-m_Q$ is for the state $Y$ with quantum numbers of $\bar{Q}  \psi | \Omega \rangle$ or $\bar{Q} \gamma^z \psi | \Omega \rangle$, and $\Lambda_Z=m_Z-m_Q$ is for the state $Z$ of the common quantum numbers of $\bar{Q} \psi | \bar{H}(P) \rangle$ and $\bar{Q} \gamma^z \psi | \bar{H}(P) \rangle$.\footnote{Because of Lorentz invariance of mass gaps, we determine the spectra in the spacelike case using the quantum numbers of physical states in the timelike case. This approach is more convenient than studying the spacelike case directly, which involves unphysical states with imaginary spatial momenta shown in Eq.~(\ref{eq:hexp}), although such an approach is in principle possible via complex Lorentz transformations~\cite{Streater:1989vi}.} 

These binding energies $\Lambda_i$ are the $O\left(\Lambda_{\rm QCD}\right)$ corrections in the HQET expansion of physical masses $m_i$ of heavy-flavor systems, 
\begin{align}\label{eq:PhyMass}
m_i = m_Q(\tau) + \Lambda_i(\tau) + O\left(\frac{\Lambda_{\rm QCD}^2}{m_Q}\right) \ ,
\end{align}
where $\tau$ denotes the renormalon regularization scheme of the heavy quark pole mass~\cite{Pineda:2001zq,Hoang:2008yj,Hoang:2009yr,Ayala:2019hkn}. While the individual components $m_Q(\tau)$ and $\Lambda_i(\tau)$ depend on $\tau$, the physical masses $m_i$ are renormalon-scheme independent. In lattice QCD calculations, the heavy quark pole mass $m_Q$ is effectively replaced by the lattice cutoff $1/a$. In this case, the binding energies appear as the $O\left(\Lambda_{\rm QCD}\right)$ corrections in the small $a$ expansion of bare mass gaps $m^{B}_i(a)$ of quasi-correlators,  
\begin{align}\label{eq:BareMass}
m^{B}_i(a) = \left[\frac{\sum_{n=0} r_{n} \alpha_{\rm lat}^{n+1}(a)}{a}\right](\tau) + \Lambda_i(\tau) + O\left(a \Lambda_{\rm QCD}^2\right) \ .    
\end{align}
The linear divergences depend on the renormalon scheme $\tau$~\cite{Braun:2018brg,Zhang:2023bxs}, and its $\tau$-dependence is compensated by that of the binding energies. Under the same choice of renormalon scheme $\tau$ (as well as physical pion mass and other possible conditions in lattice setup), these binding energies are expected be consistent between nature and lattice QCD. In the following discussions, we will adopt the principal value (PV) prescription for $\tau$~\cite{Zhang:2023bxs}.

The polynomial powers in $z$ are affected by $N_X$, $N_Y$ and $N_Z$ defined below Eq.~(\ref{eq:Mexp}). In the spacelike separation $x=(0,0,0,z)$, the coefficients $a_{X,l}$, $a_{Y,l}$ and $a_{Z,l}$ appearing in Eq.~(\ref{eq:hexp}) generally acquire phases that depend on the sign of $z$. These phases are made explicit in Eq.~(\ref{eq:qPDFLzGen}) through the factors $e^{i \, {\rm sign}(z) \, \phi_{X,l}(P^z)}$, $e^{i \, {\rm sign}(z) \, \phi_{Y,l}(P^z)}$ and $e^{i \, {\rm sign}(z) \, \phi_{Z,l}(P^z)}$. Both the amplitudes $A_i$ and phases $\phi_i$ depend on the external momentum $P^z$ in general. 

Eq.~(\ref{eq:qPDFLzGen}) applies directly to the hybrid renormalized matrix elements~\cite{Ji:2020brr}, since the hybrid scheme differs from the $\overline{\rm MS}$ scheme only by an overall constant factor at large distances. Eq.~(\ref{eq:qPDFLzGen}) can also be generalized to bare matrix elements if one replaces the parameters $\{\Lambda_X,\Lambda_Y,\Lambda_Z\}$ into the bare mass gaps $\{m^B_{X}(a),m^B_{Y}(a),m^B_{Z}(a)\}$. However, it does not apply to the so-called ratio method~\cite{Orginos:2017kos}, because the leading exponential decay cancels in the ratio of the large momentum to the zero momentum matrix elements. The follow-up discussions in this paper are based on the $\overline{\rm MS}$ or hybrid scheme~\cite{Ji:2020brr}.  

The connection between timelike and spacelike correlators using Lorentz symmetry has been verified up to two-loop in $\phi^3$ theory in Eqs.~(\ref{eq:MLx(1)phi3}) and~(\ref{eq:MLx(2)phi3}). 

\subsection{Accuracy-counting of large distance expansion}\label{sec:AccuCount}
In this subsection, we discuss the accuracy-counting principles of the large-distance expansion in Eq.~(\ref{eq:qPDFLzGen}), and define the leading asymptotics (LA) and the next-to-leading asymptotics (NLA). These discussions are closely connected to the data analysis in the next section, where the large-distance expansion is truncated up to a certain accuracy. 

Eq.~(\ref{eq:qPDFLzGen}) is a double expansion structure, which is a series of exponential-$z$ decays, each of which is multiplied by a series of polynomial-$z$ dependences. The accuracy in the large distance limit ($z \rightarrow \infty$) should be counted in this way: 
\begin{itemize}
\item Combine the three spectra $\{\Lambda_X,\Lambda_Y,\Lambda_Z\}$, and rank the mass gaps in this joint spectrum, $\Lambda_0<\Lambda_1<\Lambda_2<\Lambda_3<...$, which provides the leading exponential decay $e^{-\Lambda_0 |z|}$, next-to-leading exponential decay $e^{-\Lambda_1 |z|}$, next-to-next-to-leading exponential decay $e^{-\Lambda_2 |z|}$, etc. An illustrative example is shown below,
\begin{align}
\tilde{h}_{\{\gamma^t,I\}}(z,P^z) {\overset{z\rightarrow \infty}{=}}  
&e^{-\Lambda_0 |z|} \left(a_{0,0}+\frac{a_{0,1}}{|z|}+\frac{a_{0,2}}{z^2}+...\right) \nonumber\\
&+  e^{-\Lambda_1 |z|} \left(a_{1,0}+\frac{a_{1,1}}{|z|}+\frac{a_{1,2}}{z^2}+...\right) \nonumber\\
&+e^{-\Lambda_2 |z|} \left(a_{2,0} + \frac{a_{2,1}}{|z|}+\frac{a_{2,2}}{z^2}+...\right) \nonumber\\
&+ ...
\end{align}
Note that the powers of the polynomial-$z$ dependences could differ among real cases according to the discussions around Eqs.~(\ref{eq:GOdis}) and~(\ref{eq:replace}). The phase factors, irrelevant to the discussions here, are absorbed into those coefficients. The momentum $P^z$ dependence in those coefficients is omitted for simplicity. 
\item Count the polynomial accuracy within each exponential decay. For example, in the leading exponential decay, count the leading polynomial accuracy $a_{0,0}$, next-to-leading polynomial accuracy $a_{0,1}/|z|$, etc.
\item One cannot achieve the next-to-leading exponential accuracy until the polynomial accuracy in the leading exponential decay is entirely under control. This accuracy-counting principle is similar to the operator product expansion (OPE)~\cite{Ayala:2019uaw}, where one cannot achieve power accuracy until the logarithmic accuracy in the leading twist is fully under control.  
\item Focus on the leading exponential decay (associated with the smallest mass gap) and count its polynomial accuracy, where we call the leading polynomial accuracy the {\it leading asymptotics} (LA) and the next-to-leading polynomial accuracy the {\it next-to-leading asymptotics} (NLA). 
\end{itemize}

It is convenient to use physical heavy hadron masses as references when applying the above accuracy-counting principles to lattice QCD quasi-correlators, particularly for identifying the smallest mass gap that governs the leading exponential decay used to construct LA and NLA, provided that the following conditions are met: 
\begin{itemize}
\item The higher power corrections $O(\Lambda_{\rm QCD}^2/m_Q)$ in Eq.~(\ref{eq:PhyMass}) and $O(\Lambda_{\rm QCD}^2 a)$ in Eq.~(\ref{eq:BareMass}) are negligible. To satisfy this condition, $m_Q$ and $1/a$ should be large enough compared to $\Lambda_{\rm QCD}$. Under current lattice techniques, $a=0.03-0.1 \, {\rm fm}$, much smaller than $1/\Lambda_{\rm QCD} \sim 0.7 - 1$ fm, thereby suppressing the power corrections $O(\Lambda_{\rm QCD}^2 a)$ in Eq.~(\ref{eq:BareMass}). As for heavy-flavor systems in nature, we will choose the spectra of bottom mesons and baryons in Tabs.~\ref{tab:Bmesons} and~\ref{tab:Bbaryons} as references, since the bottom quark pole mass is estimated as $m_Q \sim 4.8$ GeV~\cite{Ayala:2019hkn}, large enough to suppress the power corrections $O(\Lambda_{\rm QCD}^2/m_Q)$ in Eq.~(\ref{eq:PhyMass}). 
\item The unphysical pion masses in lattice setups do not alter the ranking of the mass gaps, or some mass gaps are pretty close to each other so that their differences can be ignored. To estimate the effects of unphysical pion masses, we will compare the bottom hadrons with $s$ and $u/d$ quarks. This is reasonable since the kaon and pion masses in nature are $\sim 500$ MeV and $\sim 130$ MeV, respectively, which roughly cover the pion mass variation range $\sim 100-400$ MeV in practical lattice simulations. 
\end{itemize}

\begin{table}[]
    \centering
    \begin{tabular}{|c|c|c|c|c|}
     \hline
     State & $I(J^{P})$ & $\Gamma$ & Particle and Mass  \\
     \hline
     Scalar ($u/d$) & $1/2(0^{+})$ & $I,\gamma^t$ & $B^{*}_{0}$ (5756 MeV) \\
     Pseudo-scalar ($u/d$) & $1/2(0^{-})$ & $\gamma^5,\gamma^t\gamma^5$ & $B^{\pm}$ ($5279.41\pm0.07$ MeV)  \\
     Vector ($u/d$) & $1/2(1^{-})$ & $\gamma^i,\gamma^t\gamma^i$ &  $B^{*}$ ($5324.75\pm0.20$ MeV) \\
     Axial or Tensor ($u/d$) & $1/2(1^{+})$ & $\gamma^i\gamma^5$ or $\gamma^i\gamma^j$ & $B_1$ ($5726.0^{+2.5}_{-2.7}$ MeV) \\
     Pseudo-scalar ($s$) & $0(0^{-})$ & $\gamma^5,\gamma^t\gamma^5$ & $B_s^{0}$ ($5366.91\pm0.11$ MeV)  \\
     Vector ($s$) & $0(1^{-})$ & $\gamma^i,\gamma^t\gamma^i$ &  $B_s^{*}$ ($5415.4\pm1.4$ MeV) \\
    \hline
    \end{tabular}
    \caption{The B mesons of various spins and parities. The ``$\Gamma$" here denotes the Dirac structures in the meson interpolators $\bar{b} \Gamma q$ for particles at rest~\cite{Gattringer:2010zz}, where $q=u/d$ or $s$. The Dirac structures for moving particles can be obtained using the Lorentz boost. ``$i$" and ``$j$" denote the spatial indices. The masses of pseudo-scalar, vector, and axial mesons are obtained from Particle Data Group~\cite{ParticleDataGroup:2024cfk}. The mass of the scalar meson is calculated using heavy meson chiral perturbation theory in Ref.~\cite{Cheng:2017oqh}. We only present the lowest mass for each category. }
    \label{tab:Bmesons}
\end{table}
Take the pion valence quasi-PDF matrix element as an example, 
\begin{align}
\left\langle \pi(P) \left| \, \left[ \bar{u}(z) U(z,0) \gamma^t u(0) - \bar{d}(z) U(z,0) \gamma^t d(0) \right] \, \right| \pi(P) \right\rangle \ ,
\end{align}
which is a particular case of Eq.~(\ref{eq:qPDFLzGen}). Based on the quantum numbers listed below Eq.~(\ref{eq:qPDFLzGen}) and the masses of B mesons in Tab.~\ref{tab:Bmesons}, the lightest $Y$ corresponds to a heavy-light vector meson ($Q\bar{q}$, $1/2(1^{-})$, 5325 MeV if estimated using $B^{*}$), and $X$ or $Z$ is related to a heavy-light pseudo-scalar meson ($Q\bar{q}$, $1/2(0^{-})$, 5279 MeV if estimated using $B^{\pm}$). The two masses differ by only a small amount $\sim 46$ MeV. On lattice, due to the unphysical pion masses, those values could differ from the physical values, by $\sim 87$ MeV if estimated using the difference between $B_s^{0}$ and $B^{\pm}$. Therefore, we are unsure whether lattice artifacts could alter the ranking of these two mass gaps. On the other hand, this fine mass split effect would not affect the practical data analysis and precision control, as we will see in the next section. Therefore, our strategy is to treat the two mass gaps as the same when constructing the LA and NLA, and keep all relevant terms to be fitted with lattice data,
\begin{align}
\tilde{h}^{\rm LA}\left(z,P^z\right) = 
\left[ A_2 + 2 A_1 \cos\left(\phi-z P^z\right) \right]e^{-\Lambda |z|}  \ ,
\end{align}
\begin{align}
\tilde{h}^{\rm NLA}\left(z,P^z\right) 
= \left[ A_{2} + \frac{A'_2}{|z|} + 2 A_1 \cos\left(\phi-z P^z\right) + \frac{2 A'_1  \cos\left(\phi'-z P^z\right)}{|z|} \right] e^{-\Lambda |z|} \ ,
\end{align}    
where all the parameters $A_2$, $A_1$, $\phi$, $\Lambda$, $A'_2$, $A'_1$ and $\phi'$ are real numbers. The $A_2$ and $A'_2$ terms arise from the state $Y$, where the phases vanish due to the isospin symmetry. The combination of $X$ and $Z$ terms gives the $A_1$ and $A'_1$ terms here. The mass gap $\Lambda \sim 400-600 \, {\rm MeV}$ if estimated using vector and pesudo scalar B-meson masses $\sim5279-5415$ MeV and bottom quark pole mass in PV prescription $\sim4836$ MeV~\cite{Ayala:2019hkn}. 

\begin{table}[]
    \centering
    \begin{tabular}{|c|c|}
    \hline
    $I(J^{P})$    & Particle and Mass\\
    \hline
    $0(1/2^{+})$  & $\Lambda^0_b$ ($5619.57\pm0.16$ MeV) \\
    $0(1/2^{-})$  & $\Lambda_b(5912)^{0}$ ($5912.16\pm 0.16$ MeV) \\
    $0(3/2^{+})$  & $\Lambda_b(6146)^{0}$ ($6146.2\pm 0.4$ MeV) \\
    $0(3/2^{-})$  & $\Lambda_b(5920)^{0}$ ($5920.07\pm 0.16$ MeV) \\
    $1(1/2^{+})$  & $\Sigma_b^{+}$ ($5810.56\pm 0.25$ MeV) \\
    $1(3/2^{+})$  & $\Sigma_b^{*+}$ ($5830.32\pm 0.27$ MeV) \\
    $1/2(1/2^{+})$  & $\Xi_b^{0}$ ($5791.7 \pm 0.4$ MeV) \\
    $1/2(3/2^{+})$  & $\Xi_b(5945)^0$ ($5952.3 \pm 0.6$ MeV) \\
    \hline
    \end{tabular}
    \caption{The bottom baryons of various isospins, spins, and parities obtained from Particle Data Group~\cite{ParticleDataGroup:2024cfk}. For each set of quantum numbers, only the lightest particle is presented. }
    \label{tab:Bbaryons}
\end{table}
Consider the isovector quark quasi-PDF of an unpolarized nucleon,
\begin{align}
\left\langle N(P) \left| \, \left[ \bar{u}(z) U(z,0) \gamma^t u(0) - \bar{d}(z) U(z,0) \gamma^t d(0) \right] \, \right| N(P) \right\rangle \ ,
\end{align}
as another example of Eq.~(\ref{eq:qPDFLzGen}). The leading exponential decay can be determined by comparing the quantum numbers below Eq.~(\ref{eq:qPDFLzGen}) and the masses of bottom quark systems in Tabs.~\ref{tab:Bmesons} and~\ref{tab:Bbaryons}. The smallest mass gap of the $Y$ terms is $\Lambda\sim 0.5 \, {\rm GeV}$, the binding energy of a heavy-light vector meson ($Q \bar{q}$, $1^{-}$), following a similar logic as the pion valence PDF. The smallest mass gaps of $X$ and $Z$ terms are related to binding energies of heavy baryon ($Q qq$, $1/2^{+}$) and heavy anti-pentaquark ($Q \bar{q} \bar{q}\bar{q}\bar{q}$), respectively, which are $\sim 0.9 \, {\rm GeV}$ and $\sim 1.5 \, {\rm GeV}$, if estimated using the average of $\Lambda^0_b\sim 5.6$ GeV and $\Xi^0_b\sim 5.8$ GeV in Tab.~\ref{tab:Bbaryons} and the baryon-meson molecule $\sim 6.3$ GeV, subtracted by the bottom quark pole mass in the ``PV-prescription" $\sim4.8$ GeV~\cite{Ayala:2019hkn}. At large distances, the $X$ and $Z$ terms are exponentially suppressed compared to the $Y$ terms (e.g. $e^{-(0.9-0.5) {\rm GeV}\times 1.5 {\rm fm}}=0.0476$ and $e^{-(1.5-0.5) {\rm GeV}\times 1.5 {\rm fm}}=0.0005$). Therefore, the LA and NLA formulas are the $Y$ terms,
\begin{align}
\tilde{h}^{\rm LA}\left(z,P^z\right) = 
A e^{i \phi \, {\rm sign}(z)}  e^{-\Lambda |z|}  \ ,
\end{align}
\begin{align}
\tilde{h}^{\rm NLA}\left(z,P^z\right) = 
\left[A e^{i \phi \, {\rm sign}(z)}  + \frac{ A' e^{i \phi' \, {\rm sign}(z)} }{|z|} \right] e^{-\Lambda |z|}  \ ,
\end{align}
where the parameters $A$, $\phi$, $\Lambda$, $A'$ and $\phi'$ are real numbers.

The methods discussed in Sec.~\ref{sec:theory} are generalizable to many other quasi-correlators, such as the proton quark polarized PDF in Appendix~\ref{sec:PolPDF}, meson DA in Appendix~\ref{sec:DA}, quark GPD in Appendix~\ref{sec:GPD}, and gluon PDF in Appendix~\ref{sec:Gluon}.

\section{Precision control of large distance asymptotic analysis}\label{sec:numerics}
In this section, we apply the previously obtained large distance asymptotic forms as physical constraints on the Fourier transform (FT) of the lattice data to control its precision, a process called large distance asymptotic analysis. At present, the uncertainty from this asymptotic analysis is only a few percent in the intermediate $y$ region, and slightly larger but still well controlled near the endpoints. 

\subsection{General discussions of asymptotic analysis}
In this subsection, we discuss the general procedure and criteria for asymptotic analysis and derive upper error bounds for FT using analytical methods.  

\subsubsection{Procedure and criteria}\label{sec:fitcriteria}
The procedure of asymptotic analysis, 
\begin{itemize}
    \item Determine the large distance asymptotic forms $\tilde{h}_{\rm asym}(z,P^z)$ up to certain accuracy (e.g. LA or NLA) according to the discussions in Sec.~\ref{sec:AccuCount}.
    \item Fit the asymptotic form $\tilde{h}_{\rm asym}\left(z,P^z\right)$ with lattice data $\tilde{h}_{\rm dat}(z,P^z)$ within $[z_{\rm min}, z_{\rm max}]$. 
    \item Perform the Fourier transform (FT) using lattice data $\tilde{h}_{\rm dat}(z,P^z)$ for $|z|<z_L$ and fitted asymptotic form $\tilde{h}_{\rm asym}(z,P^z)$ for $|z|>z_L$:
    \begin{align}\label{eq:FT}
    \tilde{f}(y,P^z) = P^z \int_{|z|<z_L} \frac{d z}{2\pi} \, e^{i z P^z y} \tilde{h}_{\rm dat}(z,P^z) + P^z \int_{|z|>z_L} \frac{d z}{2\pi} \, e^{i z P^z y} \tilde{h}_{\rm asym}(z,P^z) \ . 
    \end{align}
\end{itemize}
General guidelines for choosing fit ranges, parameter settings, and truncation orders of asymptotic forms,
\begin{itemize}
\item For lattice data to be reliably described by the asymptotic expansion, $z_{\rm min}$ should be large enough compared to the mass gap $\Lambda$ of exponential decay,
\begin{align}
 z_{\rm min}  \gtrsim \frac{1}{\Lambda} \ .
\end{align}
This is confirmed by a simple example in $\phi^3$ theory shown in Fig.~\ref{fig:AEphi3}. For most light-quark bilinear quasi-correlators, $\Lambda$ can be estimated as the binding energy of heavy-light meson $\sim 400-600 \, {\rm MeV}$ as discussed in Sec.~\ref{sec:AccuCount}, and the appropriate $z$ for asymptotic analysis should be no less than $\sim 0.5 \, {\rm fm}$. 

\item For $z_{\rm min} \sim 2/\Lambda$, $\sim 3/\Lambda$, or larger, the LA alone could provide a good description of the lattice data. As illustrated in Fig.~\ref{fig:AEphi3}, the deviation between LA and the full result is only a few percent at $z\sim 2/m$. However, when $z_{\rm min} \sim 1/\Lambda$, the inclusion of NLA may be necessary to achieve comparable precision.

\item The mass gaps $\Lambda$ appearing in the asymptotic form can be loosely constrained around their physical expectations to improve the stability and robustness of the fits.

\item The switching point $z_L$ can be chosen between $z_{\rm min}$ and $z_{\rm max}$ to ensure a smooth transition from lattice data to asymptotic form. 
\end{itemize}

\subsubsection{FT precision in the moderate $y$ range}\label{sec:FTerrmy}
Following the above procedure and criteria, we derive the upper error bounds of FT from asymptotic analysis for the moderate $y$ range, which is related to the uncertainty of asymptotic forms at $z_L$, that originates from the statistical uncertainty of the fit with data, and systematic uncertainty from the omitted higher order terms in the large distance expansion. 

For the convenience of our discussions, we first consider the $Y$ terms, that appear in Eq.~(\ref{eq:qPDFLzGen}) or other quasi correlators in Appendices~\ref{sec:PolPDF},~\ref{sec:DA}, and~\ref{sec:Gluon}, up to arbitrary exponential and polynomial accuracy\footnote{For simplicity, the dependence on $P^z$ in the function $\tilde{h}$ is suppressed in this and the following subsubsections.}, 
\begin{align}
\tilde{h}_{Y}(z) = \sum_{Y,l} \frac{e^{-\Lambda_Y |z|} A_{Y,l} \, e^{i \, {\rm sign}(z) \, \phi_{Y,l}}}{|z|^{\frac{3}{2}(N_Y-1)+l}}  \ .
\end{align}
The contribution from this extrapolated form to quasi-PDF is,
\begin{align}
&\tilde f_{Y}(y,P^z) = P^z \int_{|z|>z_L} \frac{d z}{2\pi} e^{i y P^z z} \tilde{h}_{Y}(z) 
\nonumber\\
&=  \sum_{Y,l} \frac{A_{Y,l} P^z}{2 \pi z_L^{\frac{3}{2}(N_Y-1)+l-1}}  \left[e^{i \phi _{Y,l}} E_{\frac{3}{2}(N_Y-1)+l}\left(z_L \left(-i P^z y+\Lambda_{Y}\right)\right) \right. \nonumber\\
&\left.\quad\quad\quad\quad + e^{-i \phi _{Y,l}} E_{\frac{3}{2}(N_Y-1)+l}\left(z_L \left(i P^z y+\Lambda_{Y}\right)\right)\right] \ ,
\end{align}
where $E$ denotes the exponential integral function. 
The large momentum limit gives,
\begin{align}\label{eq:fYLM}
&\tilde f_{Y}(y,P^z)_{P^z \rightarrow\infty} = \frac{-1}{\pi  y} \sum_{Y,l} \frac{A_{Y,l} e^{-z_L \Lambda_{Y}}}{z_L^{\frac{3}{2}(N_Y-1)+l}} \left[\cos \left(\phi_{Y,l}\right) \sin \left(P^z y z_L\right)+\sin \left(\phi_{Y,l}\right) \cos \left(P^z y z_L\right) \right] \nonumber\\
&=-\frac{{\rm Re}\left[\tilde{h}_Y(z_L)\right] \sin \left(P^z y z_L\right) + {\rm Im}\left[\tilde{h}_Y(z_L)\right] \cos \left(P^z y z_L\right)}{\pi y} \ .
\end{align}
Part of the errors of $\tilde f_{Y}(y,P^z)_{P^z \rightarrow\infty}$ arise from the statistical uncertainties from the fit, denoted as $\delta^{\rm stat}{\rm Re}\tilde{h}_{Y}(z_L)$ and $\delta^{\rm stat}{\rm Im}\tilde{h}_{Y}(z_L)$,
\begin{align}\label{eq:errbudstatY}
\delta^{\rm stat} \tilde f_{Y}(y,P^z)_{P^z \rightarrow\infty} &= \frac{1}{\pi |y|} \sqrt{ v^{T} K v} < \frac{1}{\pi |y|} \sqrt{ {\rm Tr}[K]} \nonumber\\
&= \frac{1}{\pi |y|} \sqrt{\left(\delta^{\rm stat}{\rm Re}\tilde{h}_{Y}(z_L)\right)^2 + \left(\delta^{\rm stat}{\rm Im}\tilde{h}_{Y}(z_L)\right)^2} \ ,
\end{align}
where $K$ is the covariance matrix of ${\rm Re}\left[\tilde{h}_{Y}(z_L)\right]$ and ${\rm Im}\left[\tilde{h}_{Y}(z_L)\right]$, and the vector $v = (\sin \left(P^z y z_L\right),\cos \left(P^z y z_L\right))$. The other part comes from the omitted higher order terms for truncating the large distance expansion up to certain accuracy, denoted as $\delta^{\rm sys}{\rm Re}\tilde{h}_Y(z_L)$ and $\delta^{\rm sys}{\rm Im}\tilde{h}_Y(z_L)$,
\begin{align}\label{eq:errbudsysY}
\delta^{\rm sys} \tilde f_{Y}(y,P^z)_{P^z \rightarrow\infty} &= \frac{1}{\pi |y|} 
\left| (\delta^{\rm sys}{\rm Re}\tilde{h}_Y(z_L),\delta^{\rm sys}{\rm Im}\tilde{h}_Y(z_L)) \cdot \left(\sin \left(P^z y z_L\right), \cos \left(P^z y z_L\right)\right) \right| \nonumber\\
&< \frac{1}{\pi |y|} \sqrt{\left(\delta^{\rm sys}{\rm Re}\tilde{h}_{Y}(z_L)\right)^2 + \left(\delta^{\rm sys}{\rm Im}\tilde{h}_{Y}(z_L)\right)^2} \ ,
\end{align}
where the first line is written in the form of vector product. Since the statistical and systematic uncertainties are irrelevant to each other, the total uncertainty is estimated as,
\begin{align}\label{eq:errbudY}
&\delta \tilde f_{Y}(y,P^z)_{P^z \rightarrow\infty} = \sqrt{\left(\delta^{\rm stat} \tilde f_{Y}(y,P^z)_{P^z \rightarrow\infty}\right)^2 + \left(\delta^{\rm sys} \tilde f_{Y}(y,P^z)_{P^z \rightarrow\infty}\right)^2} \nonumber\\
&< \frac{1}{\pi |y|} \sqrt{\left(\delta^{\rm stat}{\rm Re}\tilde{h}_{Y}(z_L)\right)^2 + \left(\delta^{\rm stat}{\rm Im}\tilde{h}_{Y}(z_L)\right)^2 + \left(\delta^{\rm sys}{\rm Re}\tilde{h}_{Y}(z_L)\right)^2 + \left(\delta^{\rm sys}{\rm Im}\tilde{h}_{Y}(z_L)\right)^2} \ .
\end{align}
In the above derivation, we utilize the upper bounds that $\sin^2\left(P^z y z_L\right), \cos^2\left(P^z y z_L\right) \leq 1$. If one keeps those triangular functions, one obtains the oscillation period in the error, $\Delta y = \pi/(P^z z_L)$. 

Then, the above logic is generalized to all three types of terms in Eq.~(\ref{eq:qPDFLzGen}) or other quasi correlators, 
\begin{align}
\tilde{h}_{\rm asym}\left(z\right) = 
\tilde{h}_{X}(z)  + \tilde{h}_{Y}(z) 
+ \tilde{h}_{Z}(z)  \ ,
\end{align}
where
\begin{align}
\tilde{h}_{X}(z) =  \sum_{X,l=0} \frac{e^{-i P^z z} e^{-\Lambda_X |z| }}{|z|^{\frac{3}{2}(N_X-1)+l}} \, A_{X,l} \, e^{i \, {\rm sign}(z) \, \phi_{X,l}}   \ ,
\end{align}
\begin{align}
\tilde{h}_{Z}(z) =  \sum_{Z,l=0} \frac{e^{i P^z z} e^{-\Lambda_Z |z|}}{|z|^{\frac{3}{2}(N_Z-1)+l}} \, A_{Z,l} \, e^{i \, {\rm sign}(z) \, \phi_{Z,l}}   \ ,
\end{align}
whose FT in the large $P^z$ limit is,
\begin{align}
&\tilde f_{\rm asym}(y,P^z)_{P^z \rightarrow\infty} 
= P^z \int_{|z|>z_L} \frac{d z}{2\pi} e^{i y P^z z} \tilde{h}_{\rm asym}\left(z\right) \bigg|_{P^z \rightarrow \infty} \nonumber\\
=&-\frac{{\rm Re}\left[\tilde{h}_X(z_L)\right] \sin \left(P^z y z_L\right) + {\rm Im}\left[\tilde{h}_X(z_L)\right] \cos \left(P^z y z_L\right)}{\pi (y-1)} \nonumber\\
&-\frac{{\rm Re}\left[\tilde{h}_Y(z_L)\right] \sin \left(P^z y z_L\right) + {\rm Im}\left[\tilde{h}_Y(z_L)\right] \cos \left(P^z y z_L\right)}{\pi y} \nonumber\\
&-\frac{{\rm Re}\left[\tilde{h}_Z(z_L)\right] \sin \left(P^z y z_L\right) + {\rm Im}\left[\tilde{h}_Z(z_L)\right] \cos \left(P^z y z_L\right)}{\pi (y+1)}\ ,
\end{align}
which mainly contributes to different end-point regions in momentum space. To estimate the error of $\tilde f_{\rm asym}$, one can first evaluate the contribution from each term individually following similar logics as Eqs.~(\ref{eq:errbudstatY}) and~(\ref{eq:errbudsysY}), 
\begin{align}\label{eq:errbudstatX}
\delta^{\rm stat} \tilde f_{X}(y,P^z)_{P^z \rightarrow\infty}  <  \frac{1}{\pi |y-1|} \sqrt{\left(\delta^{\rm stat}{\rm Re}\tilde{h}_{X}(z_L)\right)^2 + \left(\delta^{\rm stat}{\rm Im}\tilde{h}_{X}(z_L)\right)^2} \ ,
\end{align}
\begin{align}\label{eq:errbudstatZ}
\delta^{\rm stat} \tilde f_{Z}(y,P^z)_{P^z \rightarrow\infty}  <  \frac{1}{\pi |y+1|} \sqrt{\left(\delta^{\rm stat}{\rm Re}\tilde{h}_{Z}(z_L)\right)^2 + \left(\delta^{\rm stat}{\rm Im}\tilde{h}_{Z}(z_L)\right)^2} \ ,
\end{align}
\begin{align}\label{eq:errbudsysX}
\delta^{\rm sys} \tilde f_{X}(y,P^z)_{P^z \rightarrow\infty} < \frac{1}{\pi |y-1|} \sqrt{\left(\delta^{\rm sys}{\rm Re}\tilde{h}_{X}(z_L)\right)^2 + \left(\delta^{\rm sys}{\rm Im}\tilde{h}_{X}(z_L)\right)^2} \ ,
\end{align}
\begin{align}\label{eq:errbudsysZ}
\delta^{\rm sys} \tilde f_{Z}(y,P^z)_{P^z \rightarrow\infty} < \frac{1}{\pi |y+1|} \sqrt{\left(\delta^{\rm sys}{\rm Re}\tilde{h}_{Z}(z_L)\right)^2 + \left(\delta^{\rm sys}{\rm Im}\tilde{h}_{Z}(z_L)\right)^2} \ ,
\end{align}
which share similar mathematical structures as Eqs.~(\ref{eq:errbudstatY}) and~(\ref{eq:errbudsysY}) except for the overall $y$-dependent factors. 
Then, the most conservative error bounds for total statistical and systematic uncertainties are estimated as,
\begin{align}\label{eq:errbudstat}
&\delta^{\rm stat} \tilde f_{\rm asym}(y,P^z)_{P^z \rightarrow\infty} < \delta^{\rm stat} \tilde f_{X}(y,P^z)_{P^z \rightarrow\infty} + \delta^{\rm stat} \tilde f_{Y}(y,P^z)_{P^z \rightarrow\infty} + \delta^{\rm stat} \tilde f_{Z}(y,P^z)_{P^z \rightarrow\infty} \ ,
\end{align}
\begin{align}\label{eq:errbudsys}
&\delta^{\rm sys} \tilde f_{\rm asym}(y,P^z)_{P^z \rightarrow\infty} < \delta^{\rm sys} \tilde f_{X}(y,P^z)_{P^z \rightarrow\infty} + \delta^{\rm sys} \tilde f_{Y}(y,P^z)_{P^z \rightarrow\infty} + \delta^{\rm sys} \tilde f_{Z}(y,P^z)_{P^z \rightarrow\infty} \ ,
\end{align}
and the total uncertainty is,
\begin{align}\label{eq:errbud}
&\delta \tilde f_{\rm asym}(y,P^z)_{P^z \rightarrow\infty} < \sqrt{\left(\delta^{\rm stat} \tilde f_{\rm asym}(y,P^z)_{P^z \rightarrow\infty}\right)^2 + \left(\delta^{\rm sys} \tilde f_{\rm asym}(y,P^z)_{P^z \rightarrow\infty}\right)^2} \ ,    
\end{align}
which treats statistical and systematic uncertainties as independent error budgets. 

Therefore, the FT precision from asymptotic analysis is completely under control in the moderate $y$ range, which depends on the error of the asymptotic form at $z_L$, that is further decomposed into the statistical uncertainty from the fit with lattice data, and systematic uncertainty from the omitted higher order terms in the large distance expansion. As will be shown in most numerical tests with current lattice data using $z_L \sim 0.8 - 1 \, {\rm fm}$, the leading asymptotics mainly comes from $\tilde{h}_{Y}(z_L)$\footnote{This is the case for most quasi-correlators except for pion DA where both $X$ and $Y$ terms contribute the similar amounts to the leading asymptotics. However, the uncertainty in pion DA at moderate $y$ is still a few percent.}, whose statistical uncertainty is
\begin{align}
&\delta^{\rm stat}{\rm Re}\tilde{h}_{Y}(z_L)\sim \delta^{\rm stat}{\rm Im}\tilde{h}_{Y}(z_L) \sim 0.05 \ ,
\end{align}
and the systematic uncertainties from the omitted higher order terms scale as, 
\begin{align}
&\delta^{\rm sys}{\rm Re}\tilde{h}_{X}(z_L) 
\sim \delta^{\rm sys}{\rm Im}\tilde{h}_{X}(z_L) 
\sim \delta^{\rm sys}{\rm Re}\tilde{h}_{Y}(z_L) 
\sim \delta^{\rm sys}{\rm Im}\tilde{h}_{Y}(z_L) \nonumber\\
&\sim \delta^{\rm sys}{\rm Re}\tilde{h}_{Z}(z_L) 
\sim \delta^{\rm sys}{\rm Im}\tilde{h}_{Z}(z_L) 
\sim 0.01 \ .
\end{align}
The error budget can be estimated using Eq.~(\ref{eq:errbud}),
\begin{align}
\delta \tilde f_{\rm asym}(y,P^z)_{P^z \rightarrow\infty} < \sqrt{\frac{0.05^2+0.05^2}{(\pi y)^2} + \left(\frac{\sqrt{2}\times 0.01}{\pi|y-1|}+\frac{\sqrt{2}\times 0.01}{\pi|y|}+\frac{\sqrt{2}\times 0.01}{\pi|y+1|}\right)^2} \ ,
\end{align}
which gives $\delta \tilde f_{\rm asym}(y=0.5,P^z)_{P^z \rightarrow\infty} < 0.05$, which is just a few percent compared to most quasi-PDFs at $O(1)$ for $y=0.5$. This precision can be improved in the near future with recently proposed techniques~\cite{Gao:2023lny,Zhao:2023ptv,Zhang:2025hyo}.

\subsubsection{FT precision in the end point regions}\label{sec:FTerrendpoint}
The error estimate in Eq.~(\ref{eq:errbud}) does not apply to the end-point regions, such as $y\rightarrow -1$, $0$, or $1$, because the large momentum limit $P^z \rightarrow \infty$ does not commute with the end-point limits $y\rightarrow -1$, $0$, or $1$. In this subsubsection, we derive the FT error bounds for the end-point regions, which are slightly larger than the moderate $y$ range but still under control.   

Keep only the leading exponential decay for each term for simplicity,
\begin{align}
\tilde{h}_{\rm asym}\left(z\right) = 
\tilde{h}_{X}(z)  + \tilde{h}_{Y}(z) 
+ \tilde{h}_{Z}(z)  \ ,
\end{align}
where 
\begin{align}
\tilde{h}_{X}(z)  =  \sum_{l=0} \frac{e^{-i P^z z} e^{-\Lambda_X |z| }}{|z|^{\frac{3}{2}(N_X-1)+l}} \, A_{X,l} \, e^{i \, {\rm sign}(z) \, \phi_{X,l}}   \ ,
\end{align}
\begin{align}
\tilde{h}_{Y}(z) = \sum_{l} \frac{e^{-\Lambda_Y |z|} A_{Y,l} \, e^{i \, {\rm sign}(z) \, \phi_{Y,l}}}{|z|^{\frac{3}{2}(N_Y-1)+l}}  \ ,
\end{align}
\begin{align}
\tilde{h}_{Z}(z) =  \sum_{l=0} \frac{e^{i P^z z} e^{-\Lambda_Z |z|}}{|z|^{\frac{3}{2}(N_Z-1)+l}} \, A_{Z,l} \, e^{i \, {\rm sign}(z) \, \phi_{Z,l}}   \ .
\end{align}
Its contribution to Fourier transformation is,
\begin{align}
&\tilde f_{\rm asym}(y,P^z) = P^z \int_{|z|>z_L} \frac{d z}{2\pi} e^{i y P^z z} \tilde{h}_{\rm asym}(z)  \ .
\end{align}
First take the end-point limits $y \rightarrow -1$ or $0$ or $1$, and then perform the large $z_L$ and $P^z$ expansions,
\begin{align}
\tilde f_{\rm asym}(1,P^z)  = \frac{{\rm Re}\left[e^{i z_L P^z}\tilde{h}_{X}(z_L)\right]}{\pi} \frac{P^z}{\Lambda_X} \left(1+O\left(\frac{1}{z_L},\frac{1}{P^z}\right)\right) \ ,
\end{align}
\begin{align}
\tilde f_{\rm asym}(0,P^z)  = \frac{{\rm Re}\left[\tilde{h}_{Y}(z_L)\right]}{\pi} \frac{P^z}{\Lambda_Y} \left(1+O\left(\frac{1}{z_L},\frac{1}{P^z}\right)\right) \ ,
\end{align}
\begin{align}
\tilde f_{\rm asym}(-1,P^z)  = \frac{{\rm Re}\left[ e^{-i z_L P^z}\tilde{h}_{Z}(z_L)\right]}{\pi} \frac{P^z}{\Lambda_Z} \left(1+O\left(\frac{1}{z_L},\frac{1}{P^z}\right)\right)  \ .
\end{align}
Therefore, the asymptotic uncertainties at the end-point regions are\footnote{The symbol ``$\lesssim$" means the power corrections in the large $z_L$ and $P^z$ expansions are omitted in the error estimate.},
\begin{align}
\delta\tilde f_{\rm asym}(1,P^z)  \lesssim  &\frac{P^z}{\Lambda_X \pi} \left[ \left(\delta^{\rm sys}{\rm Re}\left[e^{i z_L P^z}\tilde{h}_{X}(z_L)\right]\right)^2 + \left(\delta^{\rm stat}{\rm Re}\left[e^{i z_L P^z}\tilde{h}_{X}(z_L)\right]\right)^2 \right.\nonumber\\
&\left. + \left({\rm Re}\left[e^{i z_L P^z}\tilde{h}_{X}(z_L)\right]\frac{\delta \Lambda_{X}}{\Lambda_{X}}\right)^2 \right]^{1/2} \ ,
\end{align}
\begin{align}
\delta\tilde f_{\rm asym}(0,P^z)  \lesssim  \frac{P^z}{\Lambda_Y \pi} \sqrt{ \left(\delta^{\rm sys}{\rm Re}\tilde{h}_{Y}(z_L)\right)^2 + \left(\delta^{\rm stat}{\rm Re}\tilde{h}_{Y}(z_L)\right)^2 + \left({\rm Re}\left[\tilde{h}_{Y}(z_L)\right]\frac{\delta \Lambda_{Y}}{\Lambda_{Y}}\right)^2 } \ ,
\end{align}
\begin{align}
\delta\tilde f_{\rm asym}(-1,P^z)  \lesssim  &\frac{P^z}{\Lambda_Z \pi}  \left[ \left(\delta^{\rm sys}{\rm Re}\left[ e^{-i z_L P^z}\tilde{h}_{Z}(z_L)\right]\right)^2 + \left(\delta^{\rm stat}{\rm Re}\left[ e^{-i z_L P^z}\tilde{h}_{Z}(z_L)\right]\right)^2 \right.\nonumber\\ 
&\left. + \left({\rm Re}\left[ e^{-i z_L P^z}\tilde{h}_{Z}(z_L)\right]\frac{\delta \Lambda_{Z}}{\Lambda_{Z}}\right)^2 \right]^{1/2}  \ .
\end{align}
Compared to the errors in the moderate $y$ range in Eq.~(\ref{eq:errbud}), the errors in the end-point regions are enhanced by $O(P^z/2\Lambda)$ but still finite and under control. At current lattice status for $P^z \sim 2 \, {\rm GeV}$ and $\Lambda \sim 0.5 \, {\rm GeV}$, this enhancement factor is $\sim 2$.\footnote{In most cases tested later, the major uncertainty arises from the $Y$ term, corresponding to the endpoint uncertainty near $y\sim 0$. In the matching process, the propagation of the $y\sim 0$ uncertainty to the moderate $x$ range is negligible according to the discussions in Ref.~\cite{Chen:2025cxr}, which means the moderate $x$-range targeted in LaMET analysis is almost not contaminated by the slightly larger uncertainty near $y\sim 0$.}

\subsection{Applications to realistic lattice data}\label{sec:applications}
In this subsection, we apply the asymptotic analysis discussed in Sec.~\ref{sec:fitcriteria} to realistic lattice data to demonstrate controlled precision using numerical methods. In all the tested cases, the uncertainty from asymptotic analysis is just a few percent in the moderate $y$ range. 

\subsubsection{Pion valence PDF}\label{sec:pionval}
\begin{figure}
    \centering
    \includegraphics[width=0.49\linewidth]{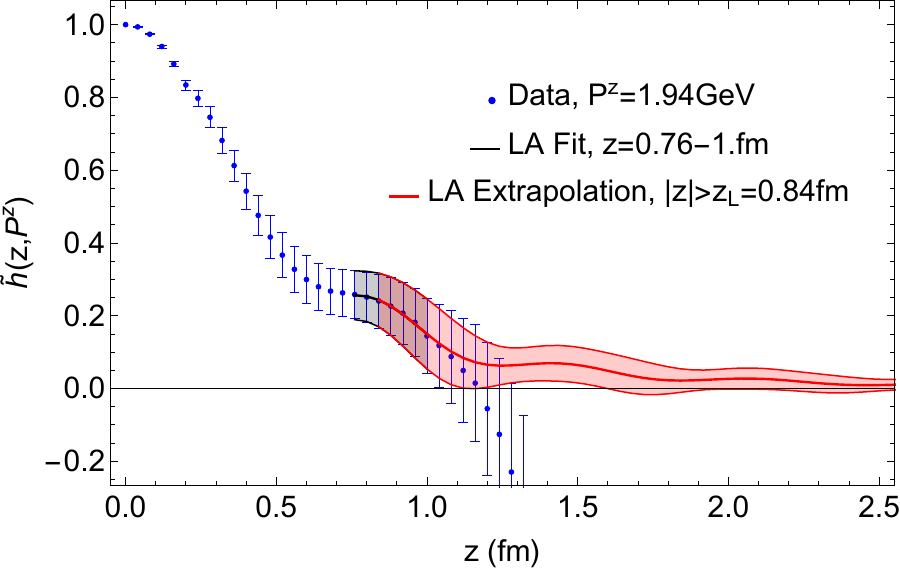}
    \includegraphics[width=0.49\linewidth]{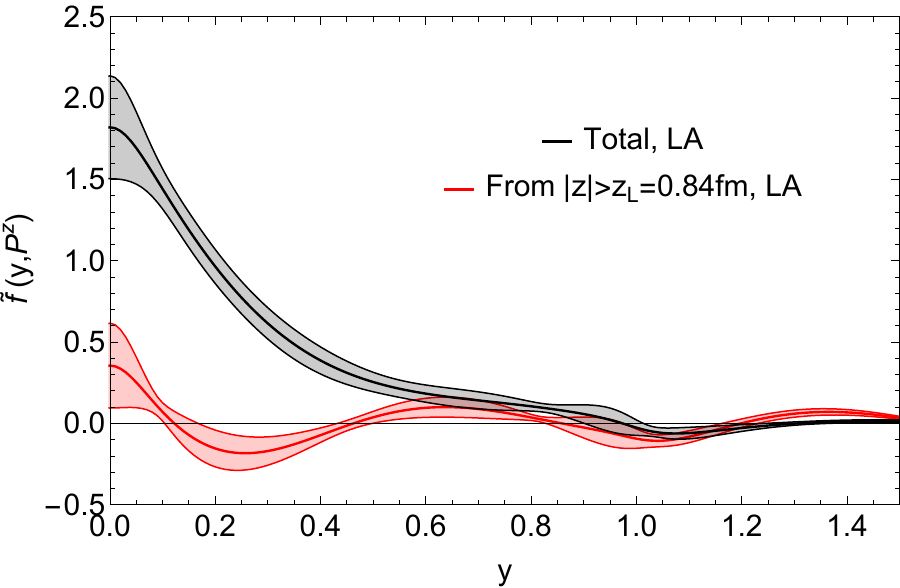}
    \caption{The leading asymptotic (LA) analysis for pion valence PDF. The left panel shows the renormalized matrix element in coordinate space, where the blue points are renormalized lattice data, the black band covers the fit range, and the red band denotes the extrapolated result. The right panel displays the quasi-PDF in momentum space obtained from FT, where the red band only includes the contribution from the extrapolated form, and the black band denotes the total result. The bands only include the statistical uncertainty. }
    \label{fig:pionfitextra}
\end{figure}

We conduct numerical studies of the large distance extrapolation and Fourier transformation on lattice QCD matrix elements for the valence quark distribution of a pion at lattice spacing $a$ = 0.04\,fm and momentum $P^{z}$ = 1.94 GeV. These matrix elements have been calculated by the BNL/ANL collaboration and analyzed in the papers~\cite{Izubuchi:2019lyk,Gao:2020ito,Gao:2021hxl,Gao:2021dbh,Gao:2022iex,Gao:2022ytj,Zhang:2023bxs,Ji:2024hit,Zhao:2025oto}. The renormalization of the lattice data is performed using the hybrid scheme~\cite{Ji:2020brr} with the PV prescription following the method in Ref.~\cite{Zhang:2023bxs}, shown as blue data points in the left panel of Fig.~\ref{fig:pionfitextra}. The signal-to-noise ratio becomes worse for larger distances, which forbids the direct Fourier transformation of the data. To control the FT precision, we perform the large distance asymptotic analysis following Sec.~\ref{sec:fitcriteria}. 

According to the discussions in Sec.~\ref{sec:AccuCount}, the leading asymptotics (LA) is,
\begin{align} 
\tilde{h}^{\rm LA}\left(z,P^z\right) = 
\left[ A_2 + 2 A_1 \cos\left(\phi-z P^z\right) \right]e^{-\Lambda |z|}   \ .
\end{align}
We fit it with lattice data. As discussed in Sec.~\ref{sec:AccuCount}, the mass gap is estimated as $\Lambda \sim 0.5 \, {\rm GeV}$, corresponding to the binding energy of a heavy-light meson under the PV prescription. We choose the fit range as $z=0.76-1$ fm, where the lower bound is roughly $2/\Lambda$, making it suitable for LA analysis. The mass gap $\Lambda$ is treated as a fit parameter, loosely constrained as $\Lambda \in [0.2,0.8] \, {\rm GeV}$ to ensure the stability of the fit. The quantities $A_2$, $A_1$, and $\phi$ are taken as free parameters in the fit. The fitted result is shown as the black band in the left panel of Fig.~\ref{fig:pionfitextra}, which agrees well with lattice data, with $\chi^2/{\rm dof} = 1.46$. The fitted values are $\Lambda =  0.48(25) \, {\rm GeV}$, $\tilde{h}^{\rm LA}_Y\left(z_L,P^z\right) = 0.20(09)$ and $\tilde{h}^{\rm LA}_X\left(z_L,P^z\right) = \tilde{h}^{\rm LA}_Z\left(z_L,P^z\right) = 0.023(17)$.

We then perform the FT, with the results shown in the right panel of Fig.~\ref{fig:pionfitextra}. The black curve represents the total result given in Eq.~(\ref{eq:FT}), while the red curve shows the contribution arising solely from the region $|z|>z_L$ using the fitted LA. In the moderate $y$ range, the contribution from the extrapolated form takes a small amount of the total result; for example, -0.07 compared to 0.39 at $y=0.4$, and 0.04 compared to 0.25 at $y=0.5$. 

\begin{figure}
    \centering
    \includegraphics[width=0.49\linewidth]{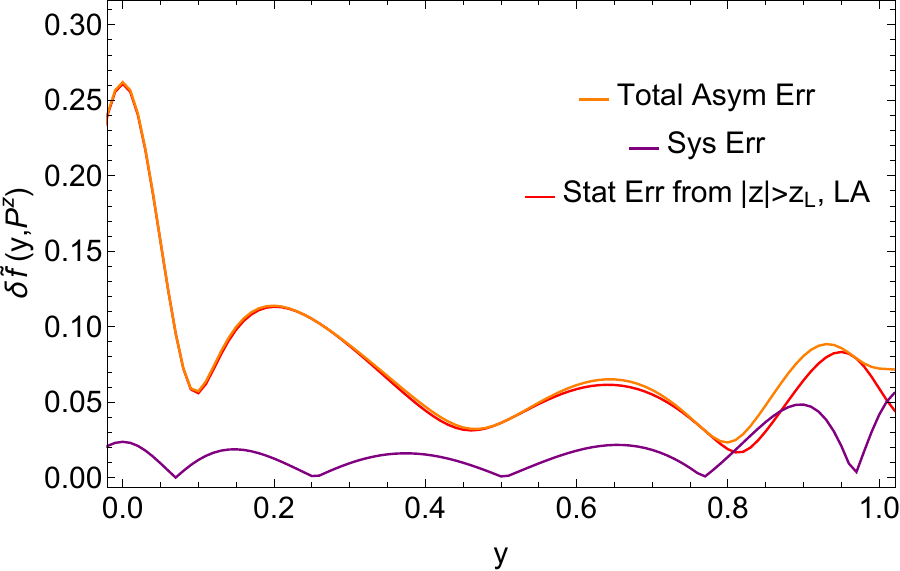}
    \includegraphics[width=0.49\linewidth]{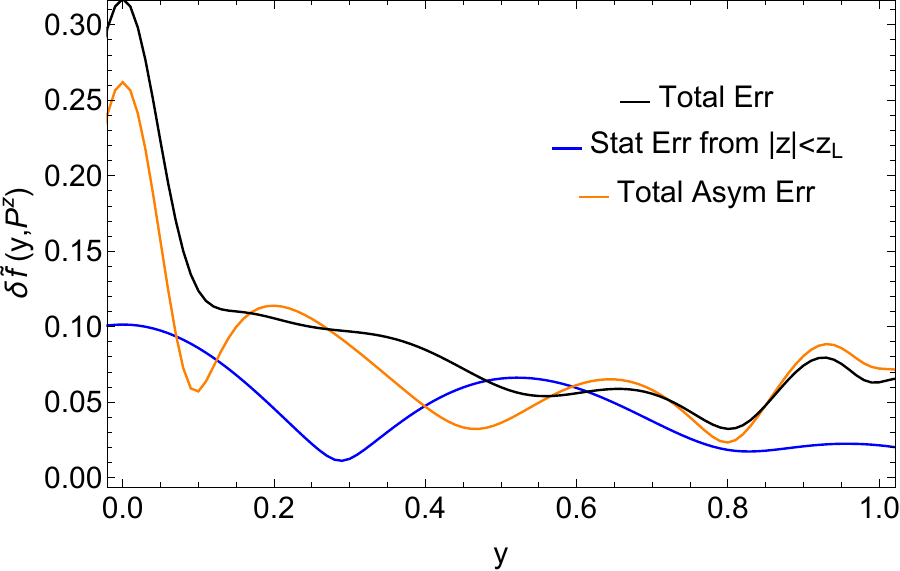}
    \caption{Error budgets of FT for pion valence quasi-PDF calculated using numerical methods. The left panel displays the uncertainties from the large distance asymptotic analysis, corresponding to $|z|>z_L$. The red curve originates from the statistical uncertainty of LA, which comes from the fit to lattice data. The purple curve represents the systematic uncertainty from the omitted higher-order terms in the truncated large distance expansion, estimated as the difference between LA and NLA results. The right panel contains all the sources of FT uncertainties, where the orange curve is the total uncertainty from asymptotic analysis, and the blue curve denotes the statistical uncertainty propagated from the data in the region $|z|<z_L$. The black curve is the total FT uncertainty. }
    \label{fig:pionerrorbudget}
\end{figure}

The red curve in the left panel of Fig.~\ref{fig:pionerrorbudget} represents the momentum-space statistical uncertainty propagated from the LA\footnote{Since the LA parameters are fitted from lattice data, the statistical uncertainty from the data is propagated to the LA, which is then transmitted to the momentum-space quasi-PDF after FT.}, which is the uncertainty of the red band in the right panel of Fig.~\ref{fig:pionfitextra}. This uncertainty is slightly larger near $y\sim0$, but well under control in the moderate $y$ range (e.g., $\delta^{\rm stat}\tilde f_{\rm asym}(y=0.5) = 0.036$). The curve exhibits local minima that are spaced by $\Delta y \sim 0.36$.

The above error estimates from numerical methods are consistent with the theoretical discussions in Secs.~\ref{sec:FTerrmy} and~\ref{sec:FTerrendpoint}. The analytical upper error bound derived in Eq.~(\ref{eq:errbudstat}) is\footnote{The symbol ``$\lesssim$" means the omitted power corrections in the large $P^z$ expansion could slightly influence the error estimate (e.g. $\sim10\%$ of the total error in the moderate $y$ range for the tested pion valence PDF data). Due to their small impacts, they have not been discussed in detail for simplicity.},
\begin{align}
&\delta^{\rm stat} \tilde f_{\rm asym}(y,P^z)  \lesssim \frac{\delta^{\rm stat} \tilde{h}^{\rm LA}_{Y}\left(z_L,P^z\right)}{\pi |y|}  
+ \frac{\delta^{\rm stat} \tilde{h}^{\rm LA}_{X}\left(z_L,P^z\right)}{\pi |y-1|}  
+ \frac{\delta^{\rm stat} \tilde{h}^{\rm LA}_{Z}\left(z_L,P^z\right)}{\pi |y+1|}  \nonumber\\
&=\frac{0.09}{\pi |y|} + \frac{0.017}{\pi |y-1|}  + \frac{0.017}{\pi |y+1|} \ ,
\end{align}
which gives $\delta^{\rm stat} \tilde f_{\rm asym}(y=0.5) \lesssim 0.072$ compatible with the numerical calculation. The theoretical derivation predicts an increased error as $y \rightarrow 0$, which is consistent with the trend in the numerical result displayed as the red curve in the left panel of Fig.~\ref{fig:pionerrorbudget}. The oscillation period from theoretical derivation in Eq.~(\ref{eq:fYLM}) is $\Delta y= \pi/(P^z z_L) \sim 0.38$, close to the numerical result.  

\begin{figure}
    \centering
    \includegraphics[width=0.49\linewidth]{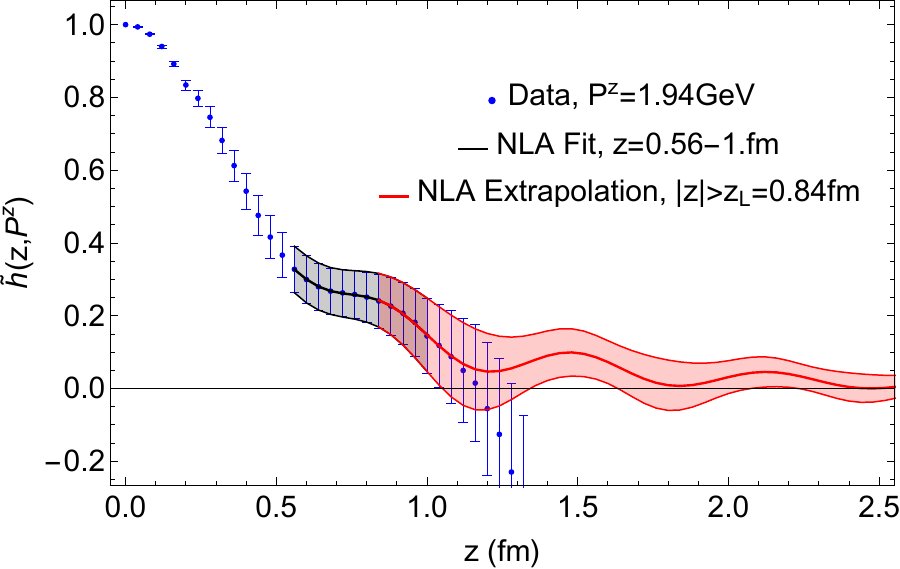}
    \includegraphics[width=0.49\linewidth]{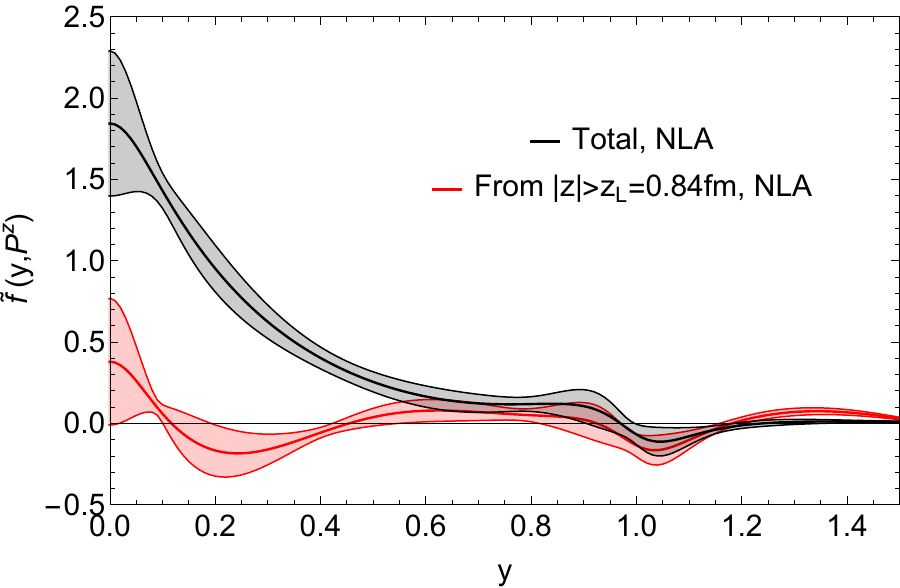}
    \caption{Similar to Fig.~\ref{fig:pionfitextra} except for next-to-leading asymptotic (NLA) analysis.}
    \label{fig:pionNLAfitextra}
\end{figure}
To estimate the systematic uncertainty from the omitted higher order terms in the truncated large distance expansion, we fit lattice data with the following NLA form discussed in Sec.~\ref{sec:AccuCount},
\begin{align}
\tilde{h}^{\rm NLA}\left(z,P^z\right) 
= \left[ A_{2} + \frac{A'_2}{|z|} + 2 A_1 \cos\left(\phi-z P^z\right) + \frac{2 A'_1  \cos\left(\phi'-z P^z\right)}{|z|} \right] e^{-\Lambda |z|} \ .
\end{align}    
The fit range is chosen as $[0.56,1]\, {\rm fm}$, where the lower bound is slightly smaller than the LA case but larger than $\sim 1/\Lambda$. The constraints $|A'_2|/z_L<|A_2|$ and $|A'_1|/z_L<|A_1|$ are set during the fit because the NLA correction should be smaller than LA. As shown in the left panel of Fig.~\ref{fig:pionNLAfitextra}, the fitted results agree well with data. The right panel of Fig.~\ref{fig:pionNLAfitextra} displays the quasi-PDF in momentum space. The difference between the LA and NLA results in momentum space is considered as the systematic uncertainty, shown as the purple curve in the left panel of Fig.~\ref{fig:pionerrorbudget}, where the value at $y=0.4$ is just 0.015. That means the omitted higher-order terms are negligible in our calculations. 

On the other hand, the upper bound of systematic uncertainty can be estimated using the theoretically-derived error bound in Eq.~(\ref{eq:errbudsys}). The omitted higher-order terms in LA can be estimated using the fit parameters of NLA,
\begin{align}
&\delta^{\rm sys} {\rm Re} \tilde h_{X}(z_L) = \delta^{\rm sys} {\rm Re} \tilde h_{Z}(z_L) = \frac{\left| A'_1 \cos[\phi'-z_L P^z] \right|}{z_L} e^{-\Lambda z_L} = 0.04(5) \ ,\nonumber\\
&\delta^{\rm sys} {\rm Re} \tilde h_{Y}(z_L) = \frac{|A'_2|}{z_L} e^{-\Lambda z_L}  = 0.05(16) \ .
\end{align}
Plug the above values into Eq.~(\ref{eq:errbudsys}),
\begin{align}
\delta^{\rm sys} \tilde{f}_{\rm asym}(y,P^z) \lesssim \frac{0.05}{\pi |y|} + \frac{0.04}{\pi |y-1|} + \frac{0.04}{\pi |y+1|} \ ,
\end{align}
which gives $\delta^{\rm sys} \tilde{f}_{\rm asym}(y=0.4,P^z) \lesssim 0.07$, which is a very loose bound compared to the error estimate using numerical methods. 

We compare various origins of FT uncertainties in Fig.~\ref{fig:pionerrorbudget}, calculated using numerical methods. The left panel shows the errors from asymptotic analysis, which are well under control in the moderate $y$ range. The systematic uncertainty is much smaller compared to the statistical one. The right panel compares asymptotic-analysis uncertainty from $|z|>z_L$ and statistical uncertainty from $|z|<z_L$. Near $y\sim 0$, asymptotic uncertainty makes a dominant contribution to the total error. In the moderate $y$ range, the two error budgets are similarly small in our test, and the total error is well controlled.

\subsubsection{Proton quark unpolarized PDF}\label{sec:protonupol}
The proton quark unpolarized quasi-PDF matrix elements are generated by Lattice Parton Collaboration (LPC) and have been used to study the ``EMC effect" of a deuteronlike dibaryon system~\cite{Chen:2024rgi}. We choose the isovector combination with $P^z= 1.85 \, {\rm GeV}$ generated on C32P29 ensemble~\cite{CLQCD:2023sdb,CLQCD:2024yyn}. The data are renormalized under hybrid scheme~\cite{Ji:2020brr} and converted to the ``PV-prescription" for mass renormalon regularization~\cite{Zhang:2023bxs}. Using these data, we demonstrate the controlled precision of FT following the asymptotic analysis in Sec.~\ref{sec:fitcriteria}.

The LA and NLA fit formulas for the isovector quark quasi-PDF of an unpolarized nucleon are discussed in Sec.~\ref{sec:AccuCount},
\begin{align}
\tilde{h}^{\rm LA}\left(z,P^z\right) = 
A e^{i \phi \, {\rm sign}(z)} e^{-\Lambda |z|}  \ ,
\end{align}
\begin{align}
\tilde{h}^{\rm NLA}\left(z,P^z\right) = 
\left[A e^{i \phi \, {\rm sign}(z)}  +  \frac{A' e^{i \phi' \, {\rm sign}(z)}}{|z|} \right] e^{-\Lambda |z|}  \ .
\end{align}
The fit parameters $A$, $\phi$, $\Lambda$, $A'$ and $\phi'$ are real numbers. Compared to the $Y$ terms in the pion valence PDF case, they contain a phase dependent on the sign of $z$. 

\begin{figure}
    \centering
    \includegraphics[width=0.49\linewidth]{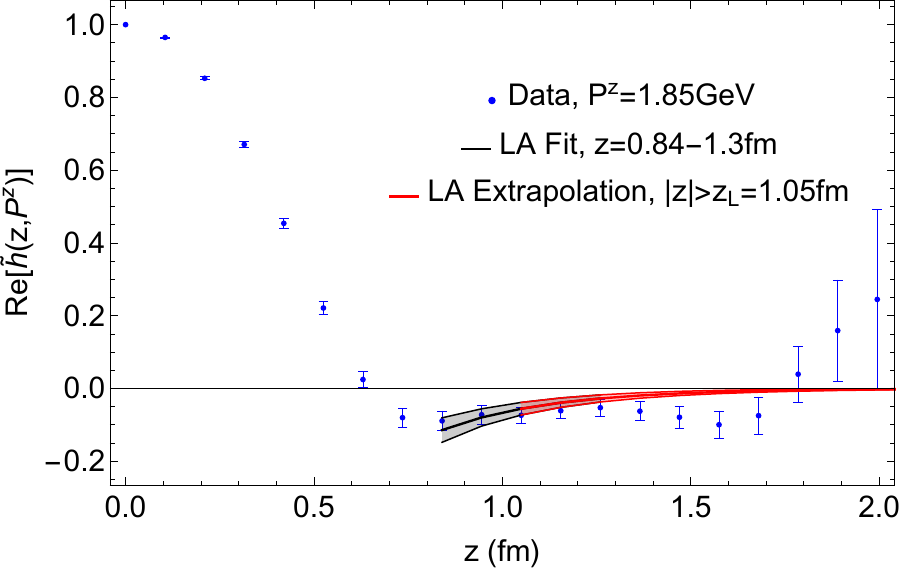}
    \includegraphics[width=0.49\linewidth]{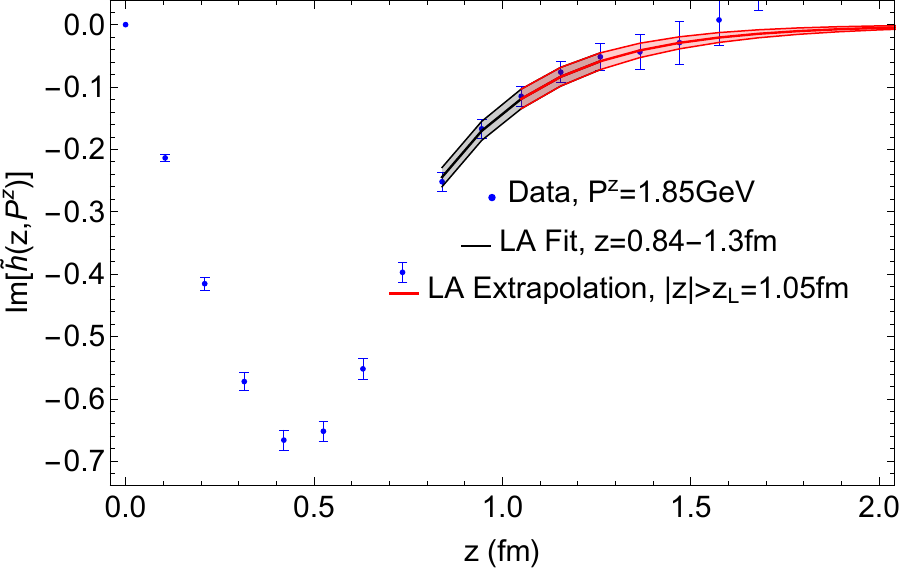}
    \caption{The leading asymptotic (LA) analysis for unpolarized nucleon quasi-PDF matrix element. The left panel shows the real part of the renormalized matrix element in coordinate space, where the blue points are renormalized lattice data, the black band covers the fit range, and the red band denotes the extrapolated result. The right panel displays the imaginary part.}
    \label{fig:upolNfitextra}
\end{figure}

\begin{figure}
    \centering
    \includegraphics[width=0.49\linewidth]{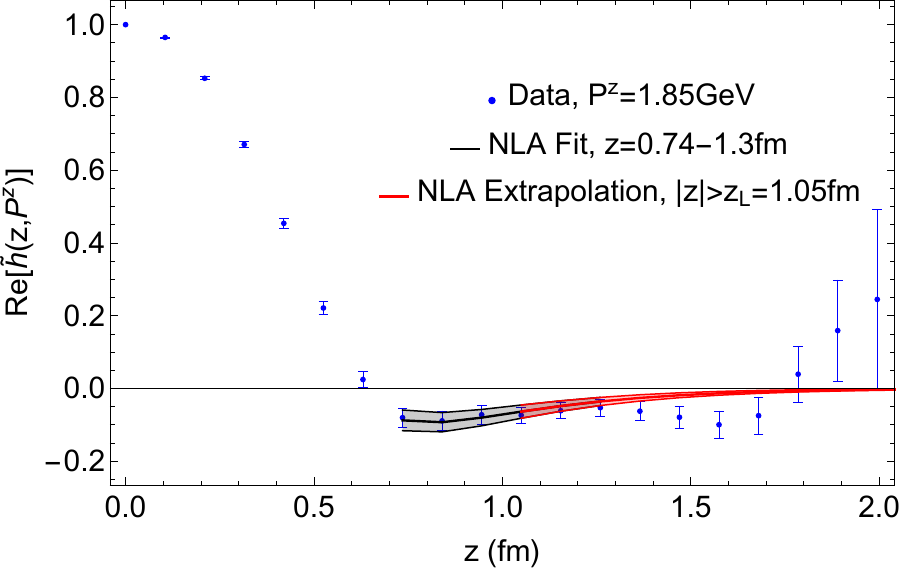}
    \includegraphics[width=0.49\linewidth]{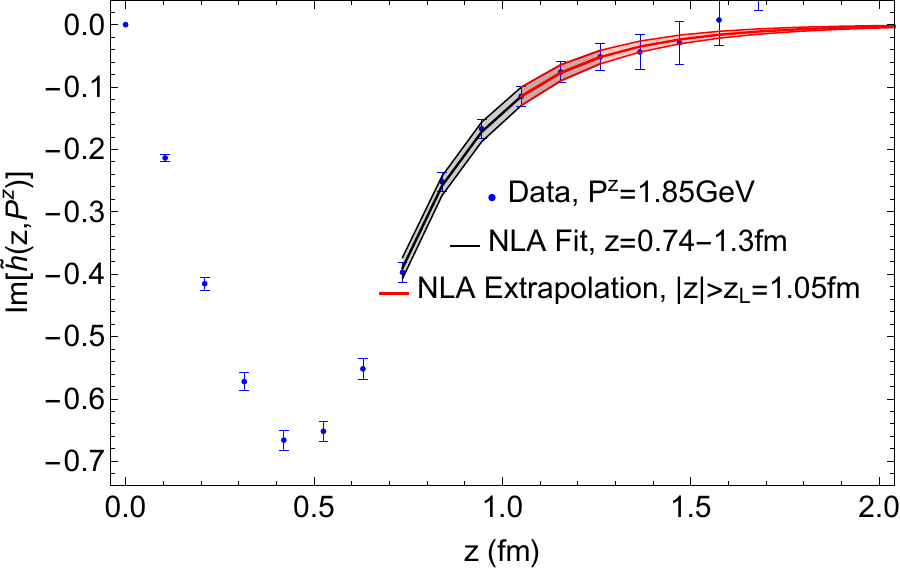}
    \caption{Similar to Fig.~\ref{fig:upolNfitextra} except for the next-to-leading asymptotic (NLA) analysis.}
    \label{fig:upolNfitextraNLA}
\end{figure}
Figs.~\ref{fig:upolNfitextra} and~\ref{fig:upolNfitextraNLA} display the fits with lattice data, where the mass gap $\Lambda$ is loosely bounded within [0.01,0.99] GeV to improve the stability, and the constraint $|A'|<|A| z_L/2$ is set in the NLA case required by the asymptotic expansion. Both LA and NLA forms agree well with data.

\begin{figure}
    \centering
    \includegraphics[width=0.49\linewidth]{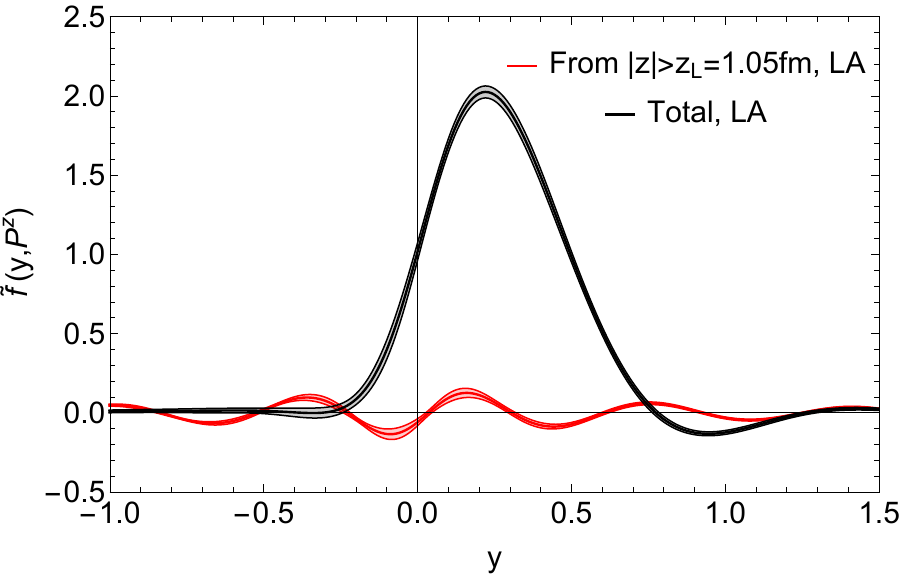}
    \includegraphics[width=0.49\linewidth]{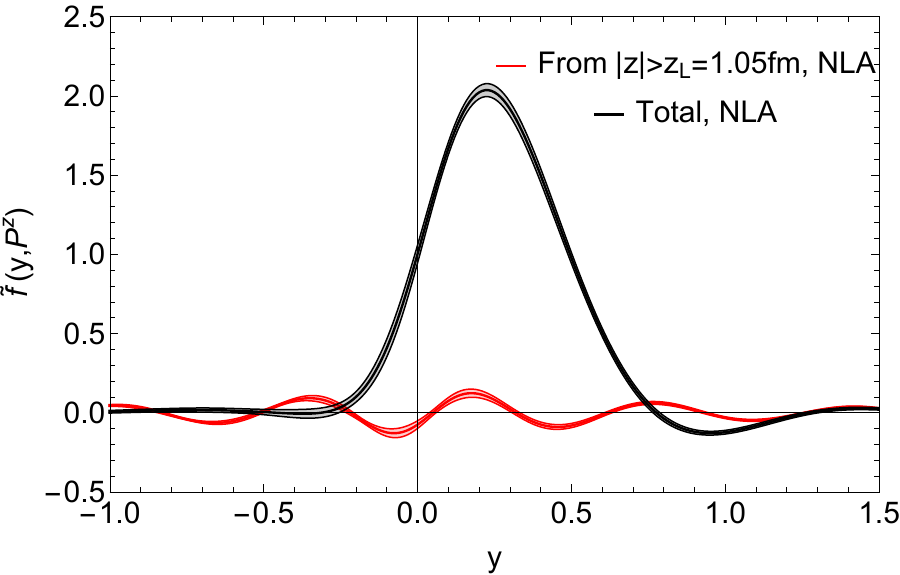}
    \caption{The unpolarized nucleon quasi-PDF in momentum space obtained from FT, where the black curve denotes the total result in Eq.~(\ref{eq:FT}), and the red curve only includes the contribution from the asymptotic form. The bands only contain the statistical uncertainty. The left panel comes from LA analysis in Fig.~\ref{fig:upolNfitextra} and the right panel is for NLA analysis in Fig.~\ref{fig:upolNfitextraNLA}.}
    \label{fig:upolNfitextramom}
\end{figure}
Fig.~\ref{fig:upolNfitextramom} shows the unpolarized nucleon quasi-PDF in momentum space obtained from FT. The contributions from asymptotic analysis only take a small amount of the total results (e.g. -0.072 compared to 0.985 at $y=0.5$). 

\begin{figure}
    \centering
    \includegraphics[width=0.49\linewidth]{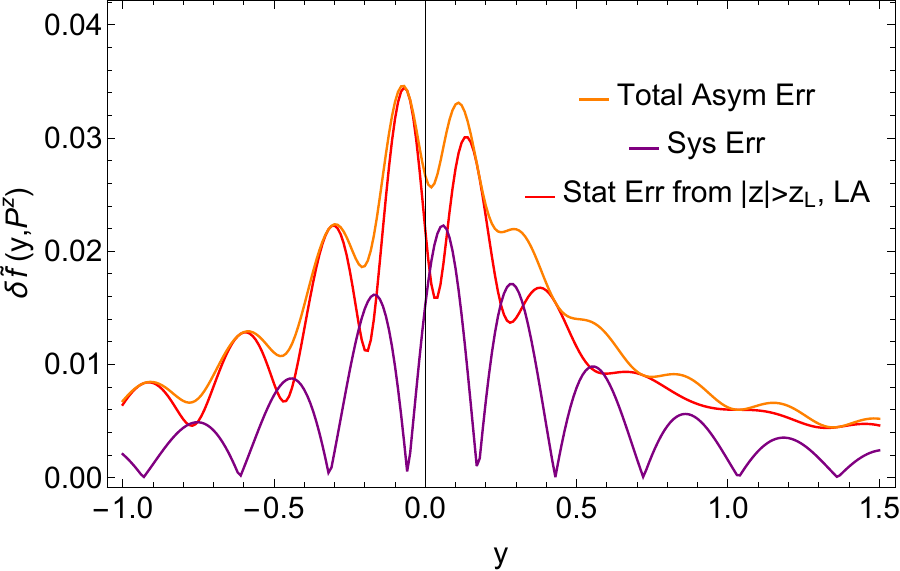}
    \includegraphics[width=0.49\linewidth]{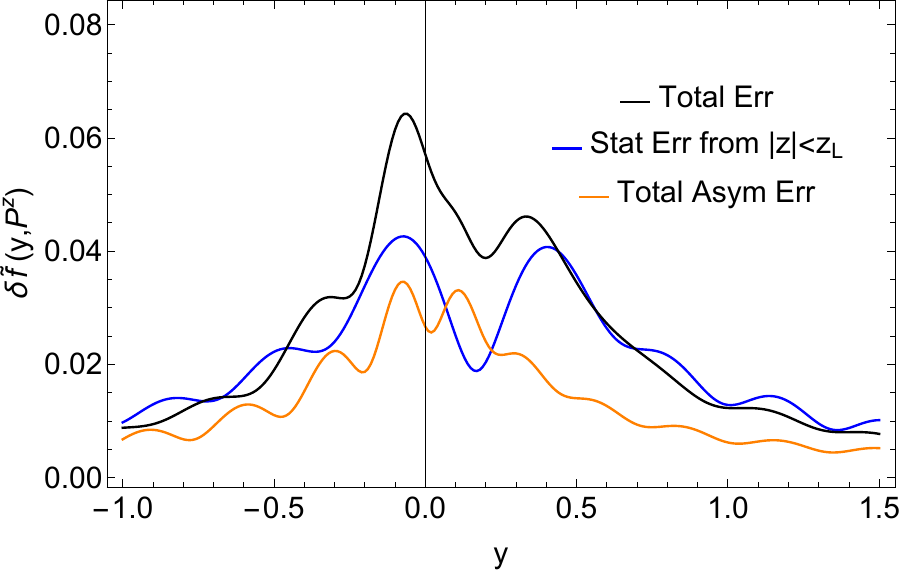}
    \caption{Error budgets of FT for unpolarized proton quasi-PDF calculated using numerical methods. The left panel displays the uncertainties from the large distance asymptotic analysis, corresponding to $|z|>z_L$. The red curve originates from the statistical uncertainty of LA, which comes from the fit to lattice data. The purple curve represents the systematic uncertainty from the omitted higher-order terms in the truncated large distance expansion, estimated as the difference between LA and NLA results. The right panel contains all the sources of FT uncertainties, where the orange curve is the total uncertainty from asymptotic analysis, and the blue curve denotes the statistical uncertainty propagated from the data in the region $|z|<z_L$. The black curve is the total FT uncertainty. }
    \label{fig:upolNerrbud}
\end{figure}
We compare various sources of FT uncertainties in Fig.~\ref{fig:upolNerrbud}. The left panel displays the uncertainties from asymptotic analysis, which is slightly larger near $y\sim 0$ (e.g. 0.026 at $y=0$) and well under control in the moderate $y$ range (e.g. 0.014 at $y=0.5$). The error size is much smaller than the pion valence PDF case because the data here are more precise at large distance, providing smaller statistical error in the asymptotic analysis. The right panel indicates that at most $y$ the asymptotic error from $|z|>z_L$ is smaller than the statistical error from $|z|<z_L$, though both of them are just a few percent compared to the center values of PDF.

\subsubsection{Proton quark transversity PDF}\label{sec:protontran}
The proton transversity quasi-PDF matrix elements are calculated by Lattice Parton Collaboration (LPC) and presented in Ref.~\cite{LatticeParton:2022xsd} with a careful study of continuum, physical pion mass and infinite momentum extrapolations. We take $P^z=1.82$ GeV generated on CLS H102 ensemble~\cite{Bruno:2014jqa} as an example to demonstrate the large distance asymptotic analysis following Sec.~\ref{sec:fitcriteria}. 

The asymptotic expansions of proton quark transversity PDF have been derived in Appendix~\ref{sec:PolPDF}, where the LA and NLA fit formulas are, 
\begin{align}
\tilde{h}^{\rm LA}\left(z,P^z\right) = 
A e^{i \phi \, {\rm sign}(z)} e^{-\Lambda |z|}  \ ,
\end{align}
\begin{align}
\tilde{h}^{\rm NLA}\left(z,P^z\right) = 
\left[A e^{i \phi \, {\rm sign}(z)} + \frac{A' e^{i \phi' \, {\rm sign}(z)}}{|z|} \right] e^{-\Lambda |z|}  \ .
\end{align}
Similar to the unpolarized proton PDF case, the LA does not contain the polynomial-$z$ enhancement because the quantum numbers of the operators forbid the forward singularity.

\begin{figure}
    \centering
    \includegraphics[width=0.49\linewidth]{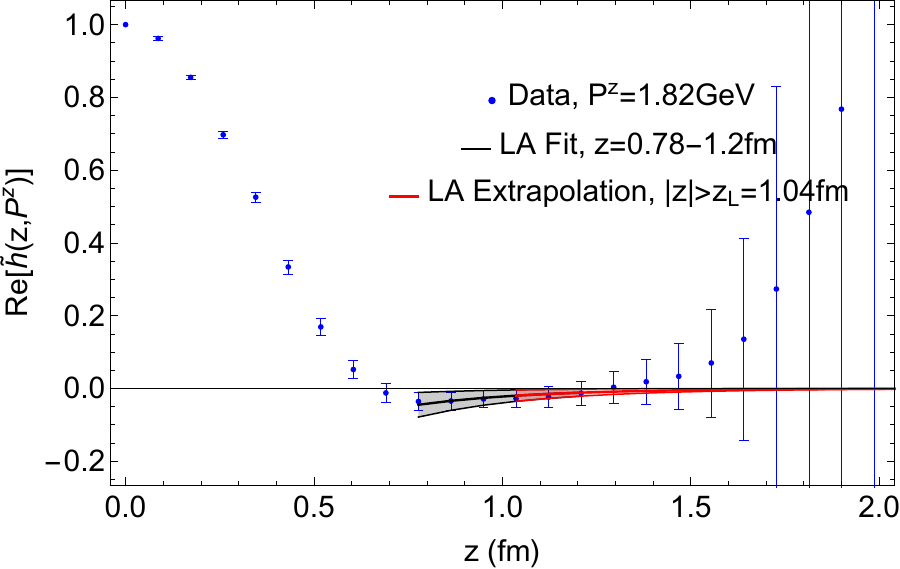}
    \includegraphics[width=0.49\linewidth]{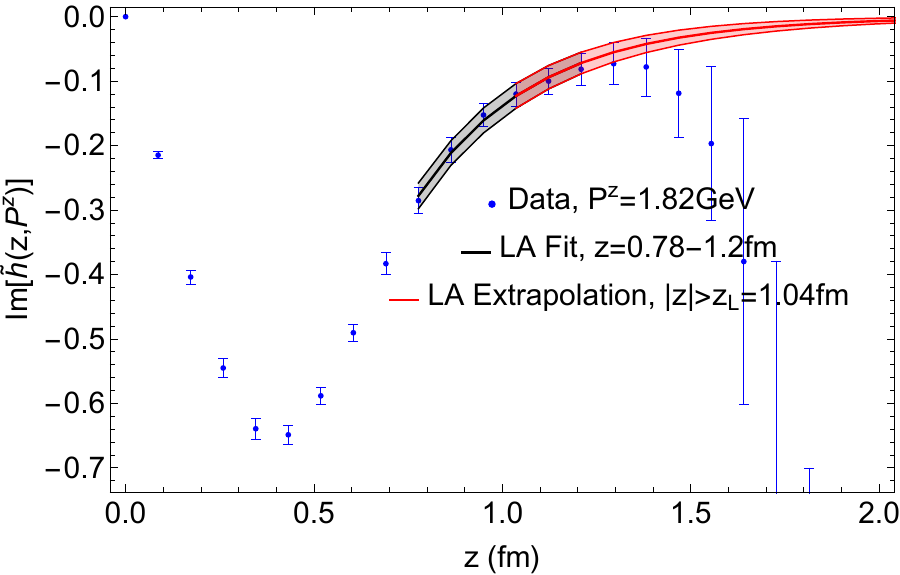}
    \caption{Similar to Fig.~\ref{fig:upolNfitextra} except for nucleon transversity quasi-PDF matrix element.}
    \label{fig:tranNfitextra}
\end{figure}

\begin{figure}
    \centering
    \includegraphics[width=0.49\linewidth]{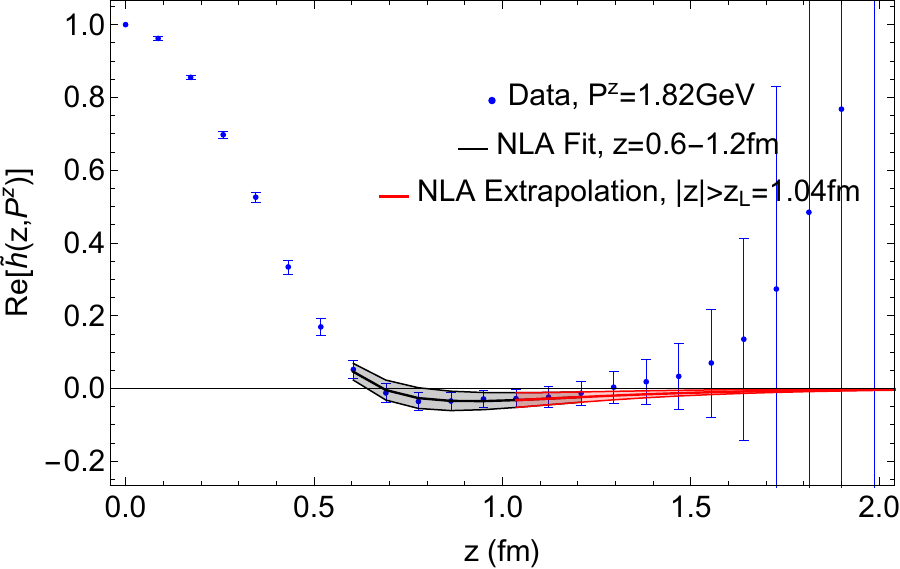}
    \includegraphics[width=0.49\linewidth]{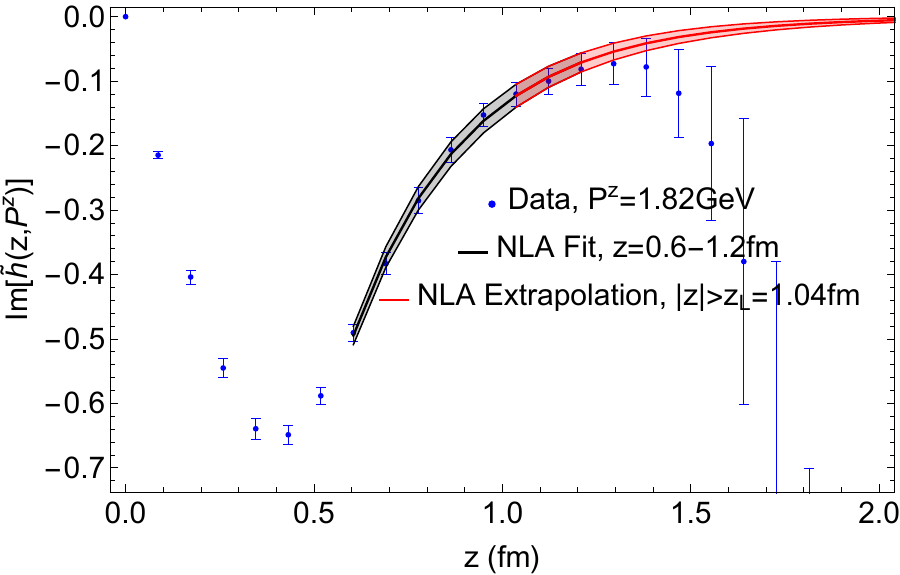}
    \caption{Similar to Fig.~\ref{fig:tranNfitextra} except for the next-to-leading asymptotic (NLA) analysis.}
    \label{fig:tranNfitextraNLA}
\end{figure}

Following the procedure and criteria in Sec.~\ref{sec:fitcriteria}, we fit the asymptotic forms with lattice data shown in Figs.~\ref{fig:tranNfitextra} and~\ref{fig:tranNfitextraNLA}. The mass gap $\Lambda$ is loosely bounded within [0.01,0.99] GeV to improve the stability, and the constraint $|A'|/z_L<|A|/2$ is set in the NLA case. The fitted results agree well with lattice data. Particularly, the NLA fit can even describe the non-monotonic behavior of lattice data.

\begin{figure}
    \centering
    \includegraphics[width=0.49\linewidth]{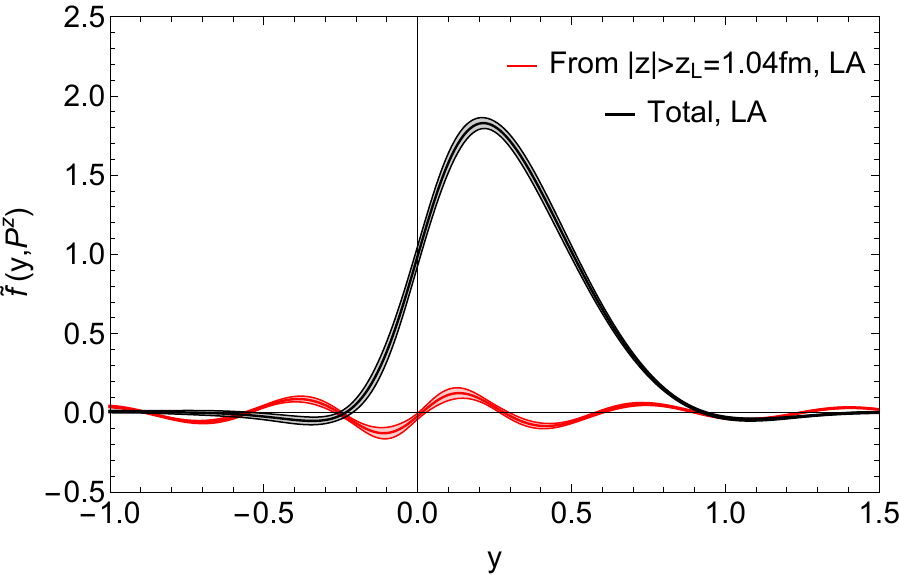}
    \includegraphics[width=0.49\linewidth]{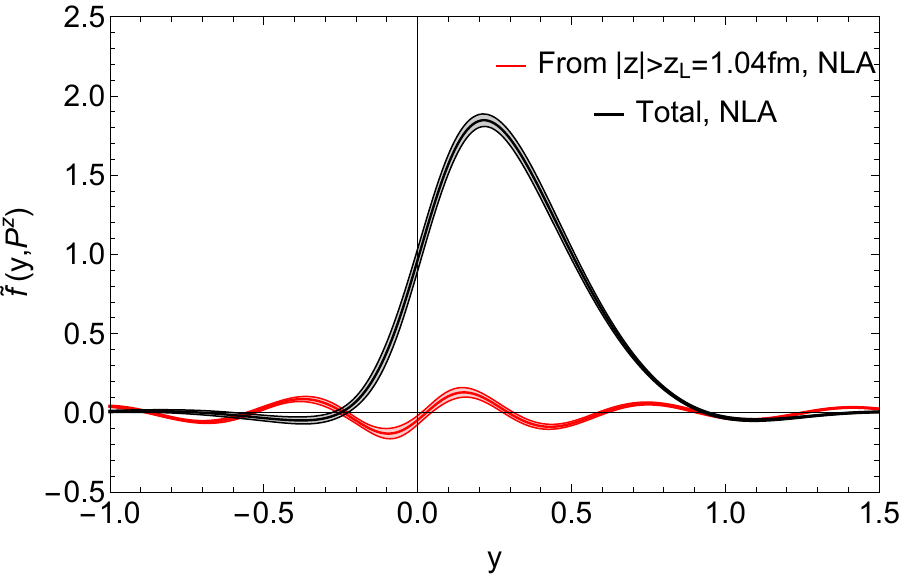}
    \caption{Similar to Fig.~\ref{fig:upolNfitextramom} except for the nucleon transversity quasi-PDF.}
    \label{fig:tranNfitextramom}
\end{figure}

Fig.~\ref{fig:tranNfitextramom} shows the momentum space quasi-PDF obtained through FT. Again, the large distance asymptotic form only contributes a small percent to the total result. 

\begin{figure}
    \centering
    \includegraphics[width=0.49\linewidth]{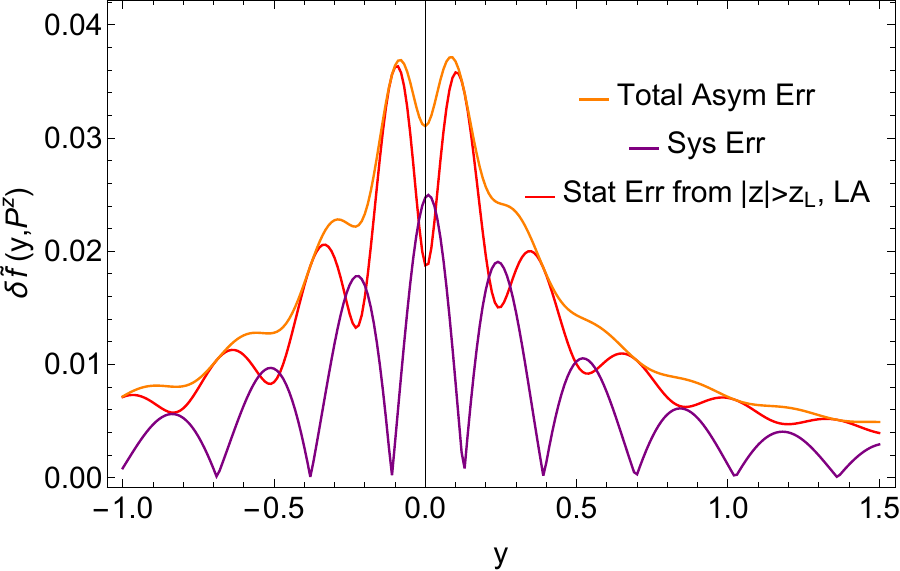}
    \includegraphics[width=0.49\linewidth]{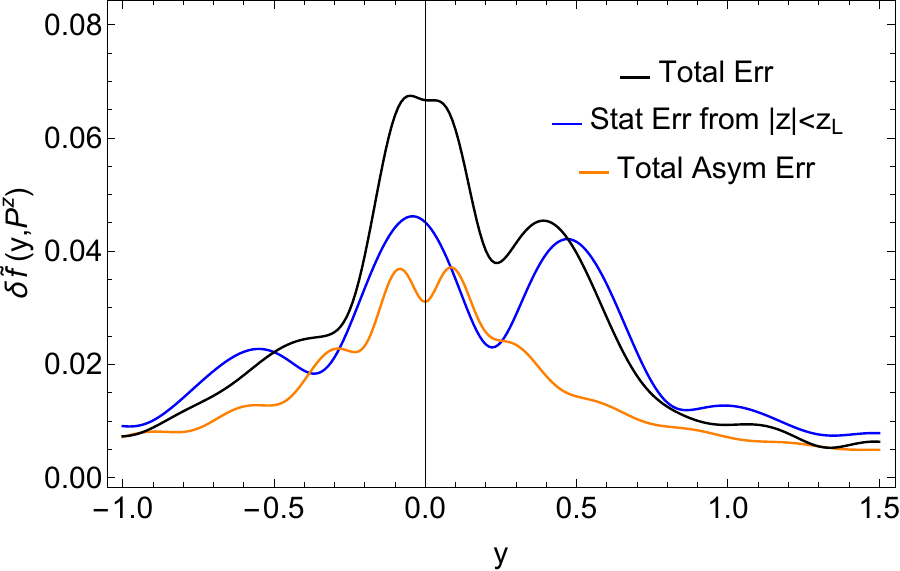}
    \caption{Similar to Fig.~\ref{fig:upolNerrbud} except for nucleon transversity quasi-PDF. }
    \label{fig:tranNerrbud}
\end{figure}

Fig.~\ref{fig:tranNerrbud} includes various FT uncertainties. The asymptotic analysis has a controlled precision with a few percent uncertainty compared to quasi PDF at $O(1)$ (e.g. 0.015 compared to 1.0 at $y=0.5$).

\subsubsection{Pion DA}\label{sec:pionDA}
The pion quasi-DA matrix elements are generated by LPC and studied in Ref.~\cite{LatticeParton:2022zqc} with hybrid renormalization, continuum extrapolation and infinite momentum limit. We choose $P^z = 1.72$ GeV on a09m130 ensemble~\cite{Follana:2006rc,MILC:2012znn} to study the asymptotic analysis in Sec.~\ref{sec:fitcriteria}.
 
Appendix~\ref{sec:DA} shows the large distance asymptotic forms for the meson quasi-DA matrix elements. Constrained by the isospin symmetry of pion quasi-DA, the LA and NLA asymptotic forms are, 
\begin{align}
\tilde{h}^{\rm LA}\left(z,P^z\right) = 
\left[A e^{i \phi \, {\rm sign}(z)} e^{-i z P^z} + A e^{-i \phi \, {\rm sign}(z)} \right] e^{-\Lambda |z|}  \ ,
\end{align}
\begin{align}
&\tilde{h}^{\rm NLA}\left(z,P^z\right) = 
\left[ \left(A e^{i \phi \, {\rm sign}(z)} e^{-i z P^z} + A e^{-i \phi \, {\rm sign}(z)} \right) \right. \nonumber\\
&\left.+ \left(A' e^{i \phi' \, {\rm sign}(z)} e^{-i z P^z} + A' e^{-i \phi' \, {\rm sign}(z)} \right) \frac{1}{|z|} \right] e^{-\Lambda |z|} \ .
\end{align}
These fit formulas exhibit stronger oscillations than the nucleon quasi-PDF cases. 

\begin{figure}
    \centering
    \includegraphics[width=0.49\linewidth]{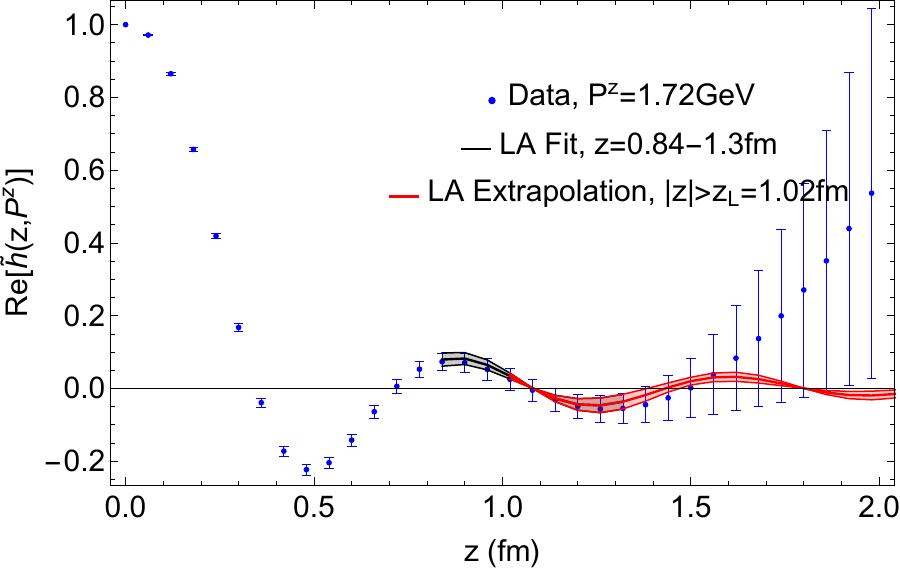}
    \includegraphics[width=0.49\linewidth]{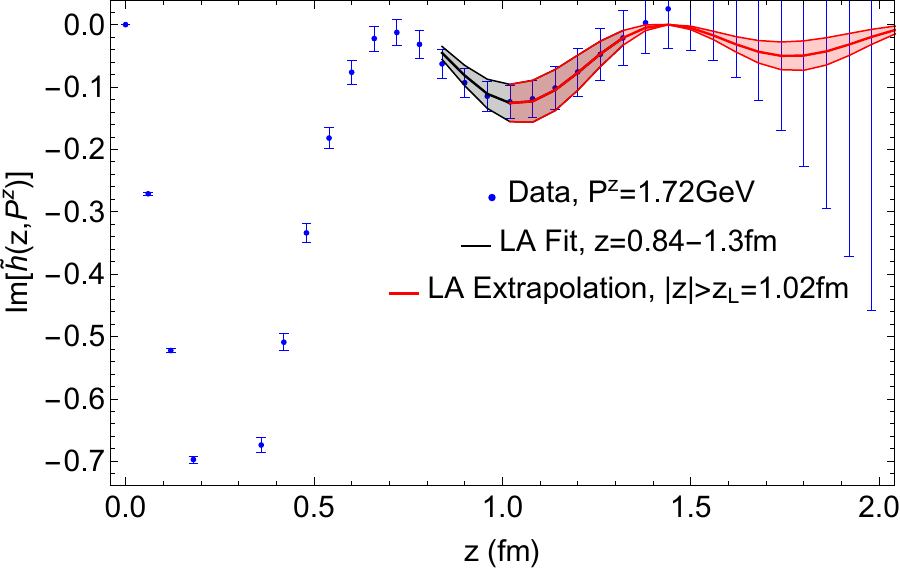}
    \caption{Similar to Fig.~\ref{fig:upolNfitextra} except for pion quasi-DA matrix element.}
    \label{fig:pionDAfitextra}
\end{figure}

\begin{figure}
    \centering
    \includegraphics[width=0.49\linewidth]{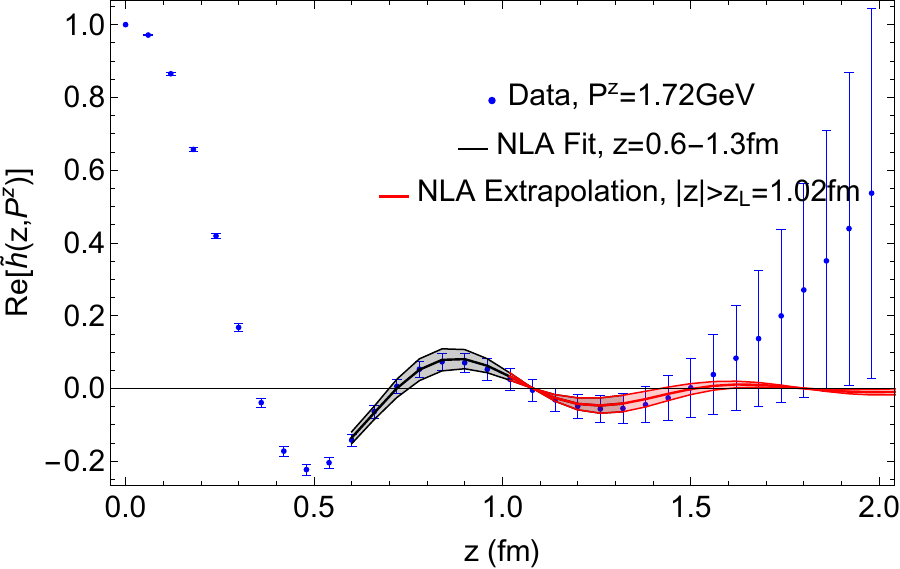}
    \includegraphics[width=0.49\linewidth]{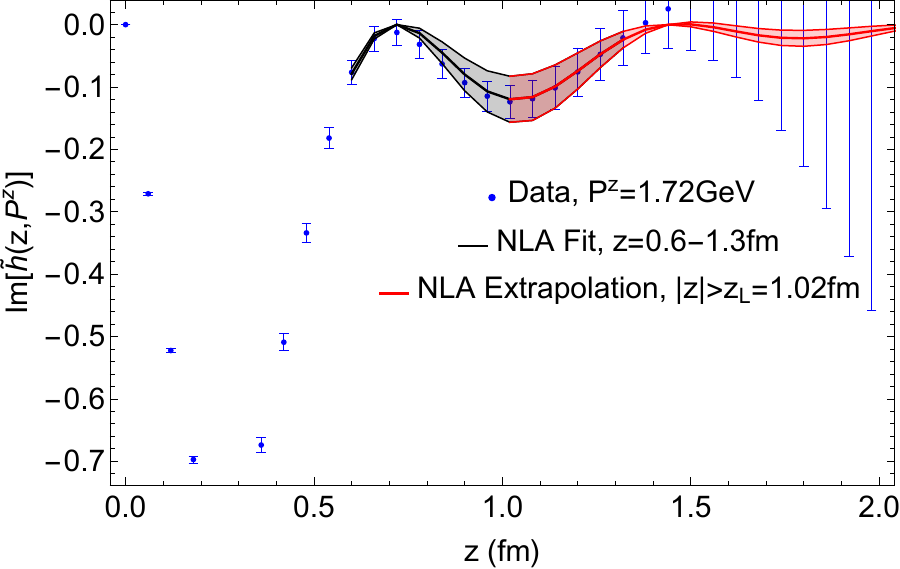}
    \caption{Similar to Fig.~\ref{fig:pionDAfitextra} except for the next-to-leading asymptotic (NLA) analysis.}
    \label{fig:pionDAfitextraNLA}
\end{figure}
In Figs.~\ref{fig:pionDAfitextra} and~\ref{fig:pionDAfitextraNLA}, we fit the asymptotic forms with lattice data. The mass gap $\Lambda$ is loosely bounded within [0.2,0.8] GeV to improve the stability, and the constraint $|A'|/z_L<|A|/2$ is set in the NLA case. Both LA and NLA forms agree well with lattice data, especially for the oscillatory behavior. 

\begin{figure}
    \centering
    \includegraphics[width=0.49\linewidth]{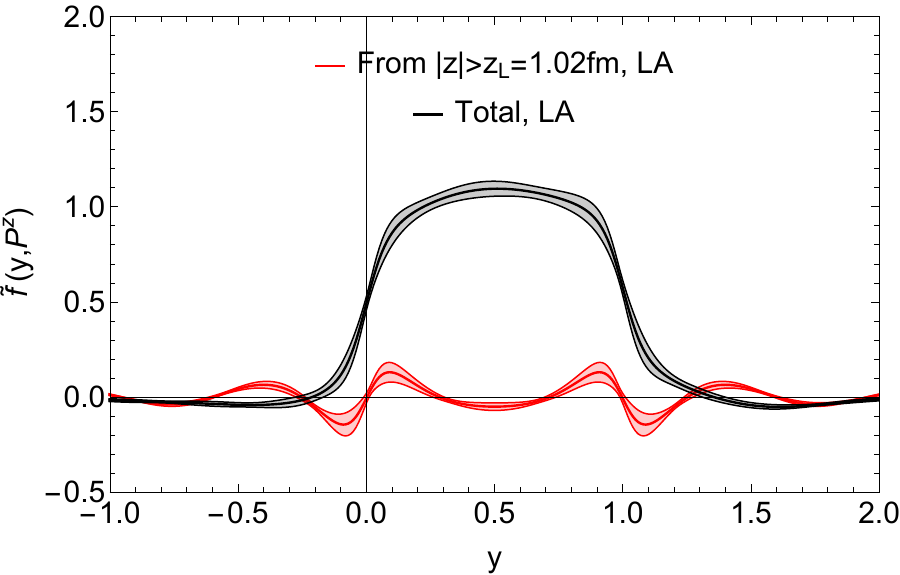}
    \includegraphics[width=0.49\linewidth]{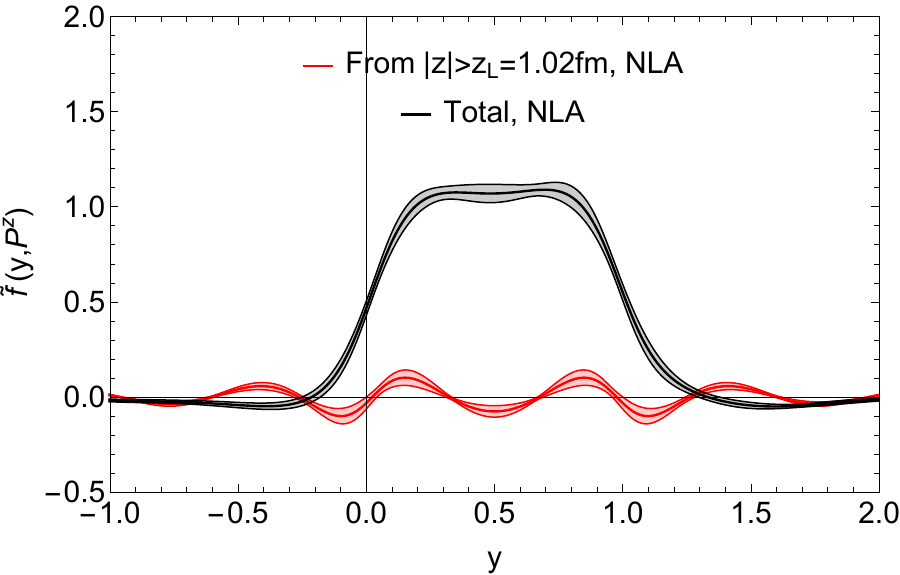}
    \caption{Similar to Fig.~\ref{fig:upolNfitextramom} except for the pion quasi-DA.}
    \label{fig:pionDAfitextramom}
\end{figure}
Fig.~\ref{fig:pionDAfitextramom} displays the quasi-DA in momentum space obtained from FT. The contribution from large distance asymptotic analysis is negligible compared to the total result. 

\begin{figure}
    \centering
    \includegraphics[width=0.49\linewidth]{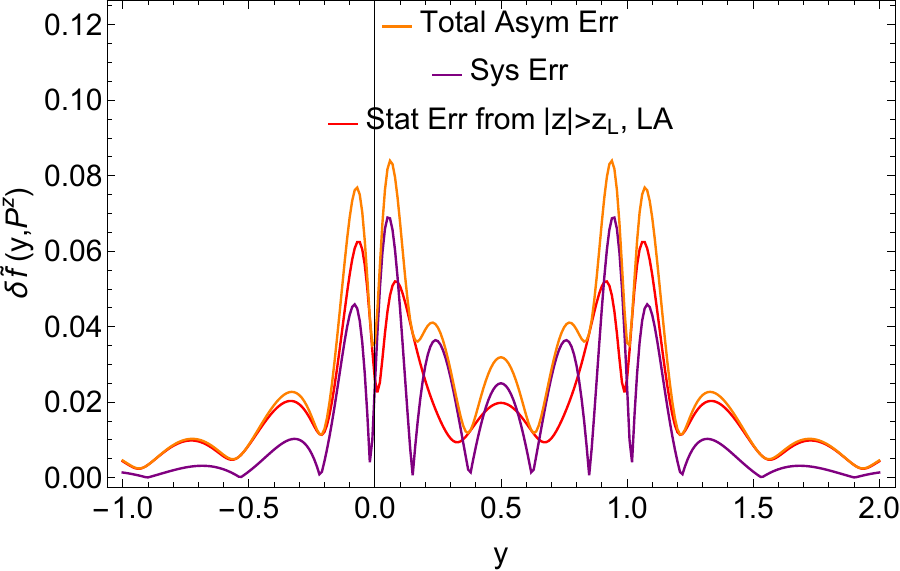}
    \includegraphics[width=0.49\linewidth]{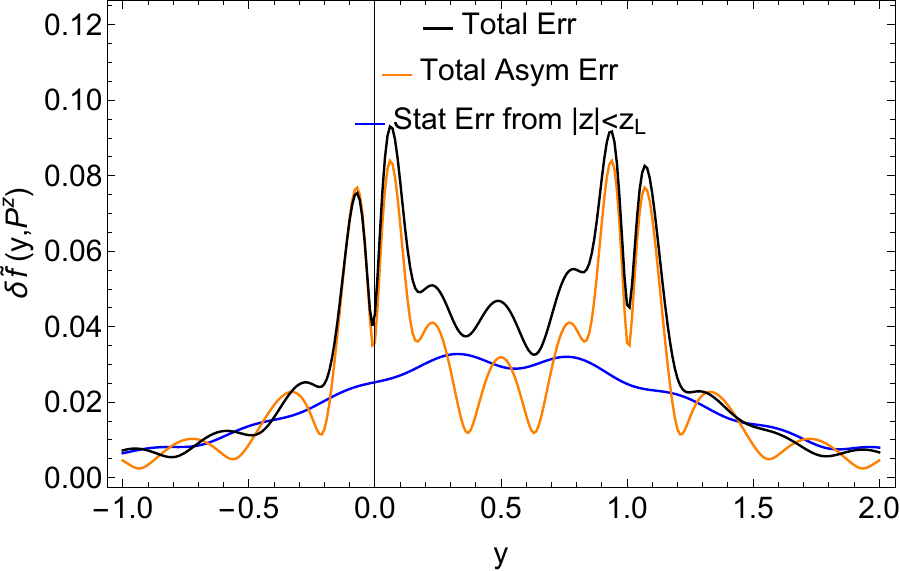}
    \caption{Similar to Fig.~\ref{fig:upolNerrbud} except for pion quasi-DA. }
    \label{fig:pionDAerrbud}
\end{figure}
Fig.~\ref{fig:pionDAerrbud} shows various contributions to FT uncertainties. The total asymptotic error is well controlled in the moderate $y$ range (e.g. $0.03$ at $y=0.5$ compared to the center value $1.06$).

\section{Exponential decay of space-like correlators from locality} \label{sec:spacelike}
In this section, we provide further justification for the existence of exponential decay of spacelike hadronic matrix elements, such as Eqs.~(\ref{eq:MqQQq}) and~(\ref{eq:hqUq}), from the perspective of {\it locality} and analyticity.

Here are the basic assumptions throughout this section. First, on the spectrum, on top of the ground state $|\Omega\rangle$, there are massive one-particle states $|\vec{k}\rangle$, followed by the two-particle states $|\vec{k}_1;\vec{k}_2\rangle$ and higher. We choose the external state $|\vec{p}\rangle$ to be of mass $m$, and we assume that the minimal mass of the allowed one-particle intermediate states after applying the operator is $M\ge m$. For the allowed two-particle states and higher, we assume that there is a finite gap $\Delta m>0$ with respect to the external state. Second, for the operator $\phi(t,\vec{x})$, we assume that it is a local quantum field with moderate regularity properties (temperedness and power-law bound). And, $\phi(t,\vec{x})$ commutes with $\phi(0,0)$ at space-like separations $t^2-|\vec{x}|^2<0 $, a property called {\it locality}. For correlators of $\phi$, the temperedness combined with positivity of energy leads to the analyticity when imaginary time is introduced, and {\it locality} allows different operator orderings to be glued together and leads to the spacelike analyticity. 

We start with correlators at zero external momentum without the one-particle intermediate states, such as the scalar-scalar correlator in the $\phi^4$ type/Ising type theories. We provide an extremely general proof for the exponential decay of such matrix elements, in a way that does not require the full Lorentz symmetry. It generalizes the arguments in~\cite{Fredenhagen:1984bz} for the vacuum two-point correlator. 

We then include one-particle intermediate states, with the help of spectral representations of one-particle form factors and their locality completions inspired by one-loop Feynman diagrams. For the locality completed one-particle contributions to the matrix elements, our approach establishes the exponential decay at an arbitrary external momentum, and leads to sharp bounds depending on the initial and final state masses, and the thresholds in the one particle form factors. 

Finally, based on the first two subsections, we generalize the arguments to an arbitrary external momentum using spectral representations of the full correlator in the forward limit. To support the argumentation, we provide an all-order proof for the existence of the spectral representation as well as the exponential decay in the scalar $\phi^3$/$\phi^4$ theories, based on parametric representations. As a by-product of the analysis, we derive a general formula for the forward-singularity contribution to the large distance asymptotics.

For simplicity of the discussion, we choose the two scalar operators in the correlator to be the same, but the proof can be easily generalized to different operators, including the operator displacements in Eq.~(\ref{eq:MqQQq}).

\subsection{Decay of correlators for zero external momentum without one-particle intermediate states}\label{sec:decayzerowoone}
\begin{figure}
    \centering
    \includegraphics[width=0.618\linewidth]{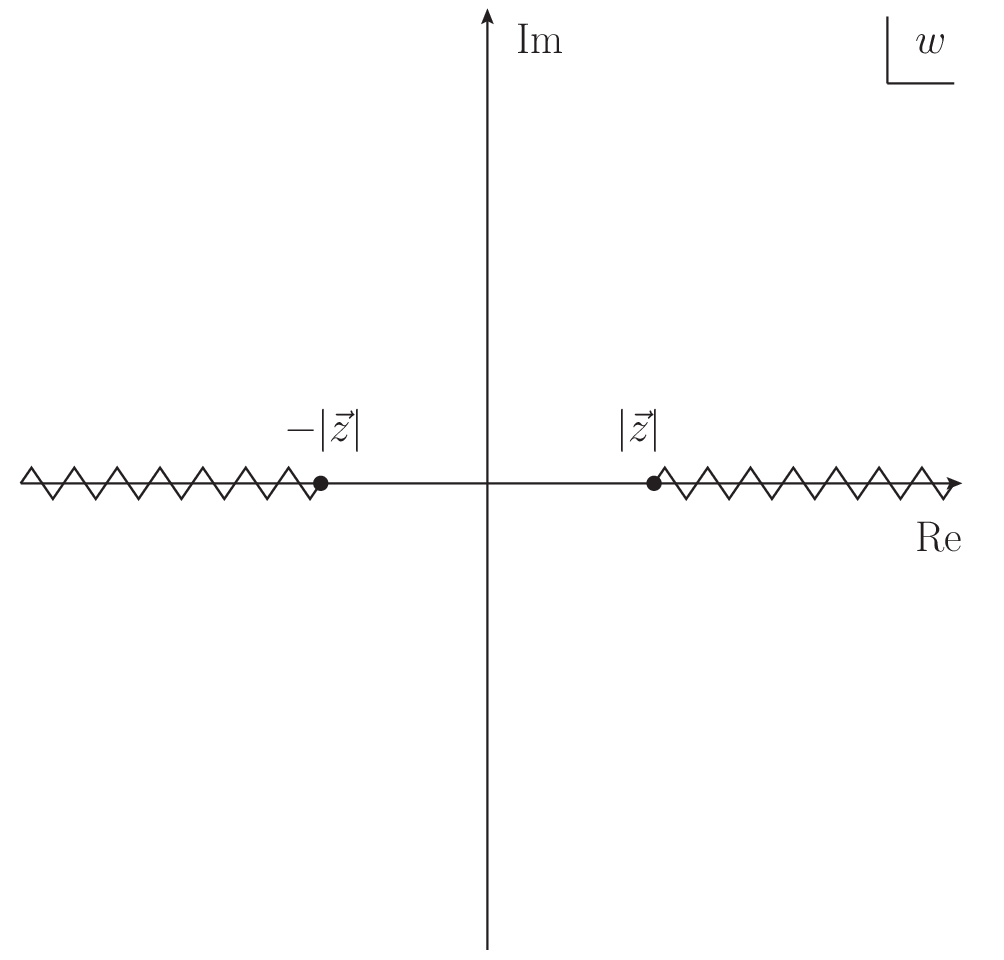}
    \caption{The analytic structure of $G_>(w,\vec{z})$ in Eq.~(\ref{eq:decomfplus}) and $G_<(w,\vec{z})$ in Eq.~(\ref{eq:decomfminus}) in the complex plane of $w$, where $w$ and $\vec{z}$ are time and spatial separations in the correlator, respectively. These functions are analytic except for the branch cuts ${\rm Im}[w]=0 \cap {\rm Re}[w]^2>\vec{z}^2$. 
    Our goal is to prove the exponential decay bound $\sim e^{-\Delta m |\vec{z}|}$ in the spatial correlator in Eq.~(\ref{eq:largezerop}), related to the origin $w=0$. 
    We start by constructing the analytic functions $G_>(w,\vec{z})$ in Eq.~(\ref{eq:decomfplus}) and $G_<(w,\vec{z})$ in Eq.~(\ref{eq:decomfminus}), in the lower (${\rm Im}[w]<0$) and upper (${\rm Im}[w]>0$) half-planes, respectively, where the analyticity is guaranteed by the temperedness and power-law bound of the spectral function in the high-energy limit. 
    Because of locality (permutation symmetry of the two scalar operators) on the spacelike edge (${\rm Im}[w]=0 \cap {\rm Re}[w]^2<\vec{z}^2$) by construction in Eq.~(\ref{eq:Crossout}), the two functions have identical boundary values on this edge, which means $G_>(w,\vec{z})$ and $G_<(w,\vec{z})$ can be combined into a single analytic function based on the Edge-of-the-wedge theorem. 
    Finally, the exponential decay bound $\sim e^{-\Delta m |\vec{z}|}$ at $w=0$ can be proved using conformal mapping and Jensen's formula as discussed in Ref.~\cite{Fredenhagen:1984bz}, which is about transferring the exponential decay along the imaginary axis to the origin using sub-harmonic property of norm logarithms.}
    \label{fig:w_plane}
\end{figure}

We first consider zero external momentum without one-particle intermediate states and prove the existence of exponential decay based on locality. The scalar operator $\phi(t,\vec{x})$ has a wave function normalization $ \sqrt{Z}=\langle \Omega|\phi(0)|\vec{p}\rangle$ in the one-particle states. See Fig.~\ref{fig:w_plane} for a summary of the logic chain in this subsection. 

We now consider the following matrix element at zero momentum of the external particle, and separated at a spacelike separation $(0,\vec{z})$
\begin{align}
f(t=0,\vec{z})&=\langle \vec{p}=0|\phi(0,\vec{z})\phi(0)|\vec{p}=0\rangle \nonumber \\ &\equiv \frac{1}{2}\lim_{\vec{p}\rightarrow0}\bigg(\langle \vec{p}|\phi(0,\vec{z})\phi(0)|-\vec{p}\rangle+\langle -\vec{p}|\phi(0,\vec{z})\phi(0)|\vec{p}\rangle\bigg)  \ . 
\end{align}
We would like to prove: at large $|\vec{z}|$, the above decays exponentially on top of the vacuum contribution and its crossing:
\begin{align}
f(t=0,\vec{z})-2Z={\cal O}(e^{-\Delta m|\vec{z}|}) \ . \label{eq:largezerop}
\end{align}
To start, we insert the resolution of the identity
\begin{align}
{\cal I}=P_{\Omega}+P_{1}+P_{n\ge 2} \ ,
\end{align}
between the two operators $\phi(0,\vec{z})$ and $\phi(0)$. The $P_{\Omega}$ projects the ground state, the $P_1$ projects the one-particle states, and $P_{n\ge 2}$ is the projection to the multi-particle states. In this subsection, we do not consider $P_1$ and will leave it to the next subsection. Thus one has
\begin{align}
&f(t=0,\vec{z}) = Z+\frac{1}{2}\lim_{\vec{p}\rightarrow0}\bigg(\langle \vec{p}|\phi(0,\vec{z})P_{n\ge 2}\phi(0)|-\vec{p}\rangle+\langle -\vec{p}|\phi(0,\vec{z})P_{n\ge 2}\phi(0)|\vec{p}\rangle\bigg) \ . 
\end{align}
However, if only the vacuum contribution were separated out, then the remaining part would not be symmetric under the operator permutation once real-time dependence is introduced later. To maintain the permutation property, in the two-particle projection, we further separate out the crossed channel for the vacuum projection\footnote{Even when this channel's existence is unclear, one can still separate it out manually at the level of the correlator, without changing the spectral properties. The asymptotics Eq.~(\ref{eq:largezerop}) in fact shows that such a cross channel must exist.}
\begin{align}\label{eq:Crossout}
P_{n\ge 2}=|\vec{p};-\vec{p}\rangle\langle \vec{p};-\vec{p}|+P_{n\ge 2}' \ .
\end{align}
This way, the remaining matrix element with $P_{n\ge 2}'$ is still symmetric under the operator permutation in the spacelike region. 

After this, we perform the analytical continuation by introducing the {\it time component} $w=t-it_E$ as a complex variable. More precisely, we define the following analytic functions in the lower and upper half-planes of $w$, respectively, 
\begin{align}
f_>(t-it_E,\vec{z})&=e^{im(t- i t_E)}\langle\vec{p}=0|\phi(0,\vec{z})e^{-iH(t-it_E)}\phi(0,0)|\vec{p}=0\rangle\nonumber \\ 
&=2Z\cos m(t-it_E)+\langle \vec{p}=0|\phi(t-it_E,\vec{z}) \phi(0)|\vec{p}=0\rangle_{P_{n \ge2}'} , t_E>0 \ , \\
f_<(t-it_E,\vec{z})&=2Z\cos m(t-it_E)+\langle \vec{p}=0|\phi(0)\phi(t-it_E,\vec{z}) |\vec{p}=0\rangle_{P_{n \ge2}'} \ , t_E<0 \ . 
\end{align}
The subscript ``$P'_{n \ge2}$" means inserting the state projection operator $P'_{n \ge2}$ between the two scalar operators. Clearly, they can be written in the following forms with $w=t-it_E$
\begin{align}
&f_>(w,\vec{z})=2Z\cos mw +G_>(w,\vec{z}) , \ {\rm Im}(w)<0 \ , \label{eq:decomfplus}\\
&f_<(w,\vec{z})=2Z\cos mw +G_<(w,\vec{z}), \  {\rm Im}(w)>0 \ .\label{eq:decomfminus}
\end{align}
Because the functions $G_>(w,\vec{z})$ and $G_<(w,\vec{z})$ receive only the two-particle contributions or above, the following bounds hold in each of the two function's domains
\begin{align}\label{eq:boundim}
&|G_>(w,\vec{z})| \le Ce^{-\Delta m|{\rm Im}(w)|} \ ,  \ | G_<(w,\vec{z})| \le C'e^{-\Delta m|{\rm Im}(w)|} \ , 
\end{align}
where $\Delta m$ is the mass gap between the inserted two or higher-particle states and the external state. Here, $C$ and $C'$ denote the pre-coefficients that can be polynomial in $w$ and $\vec{z}$. Moreover, at $t_E=0$, when $|t|<|\vec{z}|$, the operators are separated spacelikely and the boundary values of the two analytic functions agree
\begin{align}
&f_{>}(w,\vec{z})=f_{<}(w,\vec{z}) \rightarrow  \  G_>(w,\vec{z})=G_<(w,\vec{z}); \  {\rm Im}(w)=0,  \  |{\rm Re}(w)|<|\vec{z}| \ .
\end{align}
See Fig.~\ref{fig:w_plane} for the analytic structure of $G_>(w,\vec{z})$ and $G_<(w,\vec{z})$. The situation is then identical to the vacuum correlator~\cite{Fredenhagen:1984bz}: two analytic functions $G_>(w,\vec{z})$ and $G_<(w,\vec{z})$ in $w$, defined initially on the lower and upper half-planes and decay exponentially in the imaginary directions,  can be combined together into a single analytic function, with a branch cut on the real axis along $(-\infty,-|\vec{z}|)\bigcup (|\vec{z}|,\infty)$. Under these conditions and very moderate regularity assumptions on the behavior near the branch cut and at large real times, the values of the analytic functions $G_{>}(w,\vec{z})$,  $G_{<}(w,\vec{z})$ at $w=0$ can be bounded rigorously\footnote{Intuitively, this reflects that the norm-logarithms of analytic functions also behave like harmonic functions: if on average the norm-logarithms are very negative, then so at the individual points. Presence of zeros can only make the norm-logarithms more negative. } using the method in~\cite{Fredenhagen:1984bz} as
\begin{align}\label{eq:Causalitybound}
\bigg|G_>(0,\vec{z})\bigg|=\bigg|G_<(0,\vec{z})\bigg|={\cal O}(e^{-\Delta m|\vec{z}|}) \ . 
\end{align}
Combining with the $2Z$ terms in the decompositions Eq.~(\ref{eq:decomfplus}) and Eq.~(\ref{eq:decomfminus}), one obtains the desired bound Eq.~(\ref{eq:largezerop}). In fact, the $e^{-\Delta m|\vec{z}|}$ speed is sharp and can not be improved further: it is saturated exactly by the ``forward singularity'' contributions discussed before. 

\subsection{Including one-particle intermediate states: spectral representation and locality completion}
The discussions in the previous subsection are for correlators without one-particle intermediate states in the form-factor expansions. This subsection will focus on the one-particle intermediate state. If the mass $M$ of the lightest intermediate one-particle is heavier than the mass $m$ of the external state, such as Eq.~(\ref{eq:MqQQq}), the arguments in the last subsection can be directly generalized to the one-particle intermediate state, which leads to a decay at least of the speed $e^{-(M-m)|z|}$. However, we hope to derive a more constrained bound using more detailed properties, such as Lorentz invariance and spectral representations of on-shell form factors, in addition to locality. The derivations in this subsection also apply to non-zero external momentum or $M=m$ cases.

We introduce the spectral representation for the on-shell scalar form factor
\begin{align}
\langle \vec{k}|\phi(0)|\vec{p}\rangle=\int_{s_0}^{\infty}\frac{\rho(s)ds}{(k-p)^2-sm^2+i0} \ . 
\end{align}
Here $p^2=m^2$, $k^2=M^2$ are the on-shell momenta of the initial and final states. The $\rho(s)$ is the spectral function, which is required to be tempered with power-law bounds, and $s_0>0$ is the  threshold for the form factor in the time-like channel. For this to hold, we require $\phi$ to be free from vacuum condensates. This way, one has the following general form for the one-particle contribution
\begin{align}
\langle &\vec{p}|\phi(\vec{z})\phi(0)|\vec{p}\rangle\nonumber_{P_1}= \\ 
&\int_{s_0}^{\infty}  ds_1ds_2\int \frac{d^3\vec{k{}}}{(2\pi)^3}\frac{1}{2\sqrt{\vec{k}^2+M^2}}\frac{\rho(s_1)\rho(s_2)e^{i\vec{k}\cdot \vec{z}-i\vec{p}\cdot \vec{z}}}{((k-p)^2-s_1m^2+i0)((k-p)^2-s_2m^2-i0)} \ . \label{eq:onepform}
\end{align}
For $p^2=m^2\ll k^2=M^2$, the momentum transfer $(k-p)^2$
could become above threshold and create singularities that could potentially prevent the exponential decay. The traditional approach is to notice that these singularities are in the final states (the intermediate states between $\phi(\vec{z})$ and $\phi(0)$) and cancel after summing over the cuts~\cite{Collins_2023}. 

Below, we will show in general that the locality also implies the cancellation of such singularities. The trick is to add additional contributions that are two-particle and higher at the level of the spectral condition, yet complete the one-particle contribution to be permutation symmetric. The minimal locality completion of the one-particle contribution is essentially the addition of two one-loop Feynman diagrams ($z=(t,\vec{z})$)
\begin{align}
&\langle\vec{p}|\phi(t>0,\vec{z})\phi(0)|\vec{p}\rangle\nonumber_{P_1^c}=\int_{s_0}^{\infty}ds_1ds_2 \rho(s_1)\rho(s_2) \bigg(I_{t}(s_1,s_2,z,\vec{p})+I_t(s_1,s_2,-z,\vec{p})\bigg) \ , \\
&I_t(s_1,s_2,z,\vec{p})\nonumber \\ 
&=\int\frac{d^4k}{(2\pi)^4}\frac{i}{k^2-M^2+i0}\frac{e^{iE_pt-itk^0+i\vec{k}\cdot \vec{z}-i\vec{p}\cdot \vec{z}}}{((k-p)^2-s_1m^2+i0)((k-p)^2-s_2m^2+i0)} \ . 
\end{align}
For $t<0$, one needs the following for the Wightman correlator
\begin{align}
&\langle\vec{p}|\phi(t<0,\vec{z})\phi(0)|\vec{p}\rangle\nonumber_{P_1^c}=\bigg(\langle\vec{p}|\phi(-t,-\vec{z})\phi(0)|\vec{p}\rangle\nonumber_{P_1^c}\bigg)^{\dagger} \ . 
\end{align}
For another operator ordering, one needs
\begin{align}
\langle\vec{p}|\phi(0)\phi(t,\vec{z})|\vec{p}\rangle\nonumber_{P_1^c}=\langle\vec{p}|\phi(-t,-\vec{z})\phi(0)|\vec{p}\rangle\nonumber_{P_1^c} \ . 
\end{align}
It is now clear that the $\langle \vec{p}|\phi(t,\vec{z})\phi(0)| \vec{p}\rangle_{P_1^c}$ can be analytically continued with $t\rightarrow t-it_E, t_E>0$, while $\langle \vec{p}|\phi(0)\phi(t,\vec{z})| \vec{p}\rangle_{P_1^c}$ can be continued with $t_E<0$. Namely, we can define
\begin{align}
&G_>^{(1)}(w=t-it_E,\vec{z})=\langle \vec{p}|\phi(t-it_E,\vec{z})\phi(0)|\vec{p} \rangle_{P_1^c}, \ t_E>0 \ , \\
&G_<^{(1)}(w=t-it_E,\vec{z})=\langle \vec{p}|\phi(0)\phi(t-it_E,\vec{z})|\vec{p} \rangle_{P_1^c}, \   t_E<0 \ . 
\end{align}
However, at $t_E=0$, the constructions above guarantee that for $|t|<|\vec{z}|$, one has $G_>^{(1)}(t,\vec{z})=G_<^{(1)}(t, \vec{z})$. This is due to the fact that in the one-particle completion, we have added all the required multi-particle cuts to guarantee the locality. As such, we can write
\begin{align}
f(t,\vec{z})=2Z\cos E_pt+\langle \vec{p}|\phi(t,\vec{z})\phi(0)|\vec{p}\rangle_{P_1^c}+\langle \vec{p}|\phi(t,\vec{z})\phi(0)|\vec{p}\rangle_{P_2''} \ , 
\end{align}
where each term is separately local. Moreover, the $P_2''$ term remains to be at least two particles in the sense of spectral condition, thus can be bounded at $\vec{p}=0$ by $e^{-\Delta m' z}$ using the previous method. Here 
\begin{align}
\Delta m'=\min \left({\Delta m, \sqrt{s_0}m},M+m\right)  , 
\end{align}
where the $M+m$ and $\sqrt{s_0}m$ are from the two and three particle cuts in the $I_t$. $\Delta m$ can be found at the beginning of this section. Thus, it is sufficient to estimate the 
\begin{align}
\langle &\vec{p}|\phi(t=0,\vec{z})\phi(0)|\vec{p}\rangle_{P_1^c} \nonumber \\ 
&=\int_{s_0}^{\infty} ds_1ds_2 \rho(s_1)\rho(s_2) \bigg(I_t(s_1,s_2,(0,\vec{z}),\vec{p})+I_t(s_1,s_2,(0,-\vec{z}),\vec{p}) \bigg) \ . 
\end{align}
However, since $y^2=\frac{M^2}{m^2} \ge 1$, one can see from the parametric representations of the one-loop integrals as will be shown later, that the slowest decay at large $\vec{p}$ can be bounded as ($C_i$ are bounded by polynomials)
\begin{align}\label{eq:oneparbound}
&\langle\vec{p}|\phi(0,\vec{z})\phi(0)|\vec{p}\rangle_{P_1^c}\rightarrow C_1e^{-M|\vec{z}|} \cos \vec{p}\cdot \vec{z}+C_2e^{-\sqrt{s_0}m |\vec{z}|}+C_3e^{-\sqrt{1+\frac{\vec{p}^2}{m^2}}m_c(s_0,m,M)} \ , \\
& m_c(s_0,m,M)=m^2\sqrt{y^2-\frac{(s_0-1-y^2)^2}{4}} , \  y^2-1<s_0<y^2+1  \ , \\
&m_c(s_0,m,M)= \sqrt{M^2-m^2}, \  0<s_0<y^2-1 \ . 
\end{align}
If $s_0>y^2+1$, the $m_c$ term would be absent. In no cases, there will be ``real singularities'' that could be contained in the original one-particle form factor Eq.~(\ref{eq:onepform}). As such, we have shown that locality also leads to the cancellation of the final-state form factor singularities.  Moreover, the locality completion of the one-particle contribution decays exponentially even at large $\vec{p}$, a pattern which we believe should hold for the $P_2''$ terms as well. 

Here we provide a detailed analysis of the one-loop integral $I_t(s_1,s_2,(0,\vec{z}),\vec{p})$ leading to the bounds in Eq.~(\ref{eq:oneparbound}). First, we can write
\begin{align}
&\frac{1}{(k-p)^2-s_1m^2+i0}\frac{1}{(k-p)^2-s_2m^2+i0} \nonumber \\ 
&=\frac{1}{(s_1-s_2)m^2} \bigg(\frac{1}{(k-p)^2-s_1m^2+i0}-\frac{1}{(k-p)^2-s_2m^2+i0}
\bigg) \ . 
\end{align}
In the presence of a discrete spectrum in the $\rho(s)$, cancellation of singularities between the two terms is required and could lead to polynomial enhancement, but will not change the exponential decay speed. So, to read the exponential speed, we can focus on one of them. Then, after introducing the Schwinger parameterization, one has
\begin{align}
&\int\frac{d^4k}{(2\pi)^4}\frac{i}{k^2-M^2}\frac{-e^{i\vec{k}\cdot \vec{z}-i\vec{p}\cdot \vec{z}}}{(k-p)^2-s_1m^2+i0}\nonumber \\ 
&=\frac{1}{(4\pi)^2} e^{-i\vec{p}\cdot \vec{z}}\int_0^{\infty}\frac{d\rho}{\rho} \int_0^1 dx e^{-\frac{z^2}{4\rho}+ix\vec{z}\cdot \vec{p}-\rho\left((1-x)M^2+xs_1m^2-x(1-x)m^2\right)} \ . 
\end{align}
The decay is controlled by the polynomial 
\begin{align}
{\cal P}(x,m^2,M^2,s_1)=(1-x)M^2+xs_1m^2-x(1-x)m^2 \ .
\end{align}
Denoting $y^2=\frac{M^2}{m^2}$, the global minimal is achieved at 
\begin{align}
&x_c=\frac{y^2-(s_1-1)}{2} \ , \\
&{\cal P}(x_c,m^2,M^2,s_1)=\frac{m^2}{4}\left(2y+y^2-(s_1-1)\right) \left(2y-y^2+(s_1-1)\right) \ . 
\end{align}
Here, we only consider the situation where $y\ge 1$. In this situation,  the condition $0\le x_c\le 1$ requires that
\begin{align}
&y^2-1\le s_1\le y^2+1 \ , 
\end{align}
Under this condition, the ${\cal P}(x_c,m^2,M^2,s_1)$ is always positive. More general, denote
\begin{align}
    m_c(s_1,m,M)=\sqrt{{\rm min}_{0<x<1} {\cal P}(x,m^2,M^2,s_1)} \ , 
\end{align}
one has
\begin{align}
m_c(s_1,m,M)=& \nonumber M \ , s_1\ge y^2+1  \\
=&m\sqrt{y^2-\frac{(s_1-1-y^2)^2}{4}} \ge \sqrt{M^2-m^2} \ ,  \  y^2-1<s_1\le y^2+1 \ , \\
=&\sqrt{s_1} m, \   s_1\le y^2-1 \ , 
\end{align}
which leads to the bounds in Eq.~(\ref{eq:oneparbound}).

\subsection{Generalization to $\vec{p}\ne 0$: all order perturbative proof through parametric representations}
Finally, we justify the existence of exponential decay for the non-zero external momentum $\vec{p}\ne 0$, considering all the possible intermediate states. The exponential decay arises from the spectral representation with a finite spectral gap $s_0>0$. The existence of $s_0>0$ as well as its relation to exponential decay are justified up to all-order perturbation theory in $\phi^3$ or $\phi^4$.

The generalization is based on the following spectral representation
\begin{align}
f(\vec{z},\vec{p})=2Z\cos \vec{z}\cdot \vec{p}+\int_{-1}^{1} dx \int_{s(x)}^{\infty} ds \rho(s,x) e^{ix \vec{z}\cdot \vec{p}}K_0(s|\vec{z}|) \ ,  \label{eq:spectralcoorgen}
\end{align}
where $\rho(s,x)$ is a two-dimensional tempered {\it spectral distribution}, $s(x)\ge 0$ is the spectral gap at $x$. The support in $x$ is between $(-1,1)$, which is a result of locality. Then, the crucial quantity controlling the exponential decay is 
\begin{align}
s_0={\rm min}_{x\in [-1,1]}s(x) \ . 
\end{align}
If $s_0>0$, then there will be exponential decay for any $\vec{p}$. On the other hand, if $s_0=0$, then at $\vec{p}=0$, there will be algebraic decay, unless the spectral function vanishes faster than any polynomial speed at the minimum of $s(x)$, which is very unlikely in QFT with power-law bounds. Therefore, the $s_0=0$ case is inconsistent with discussions in the previous two subsections and should be excluded. Moreover, as $\vec{p}$ increases, generically the oscillations in the phase will also increase the decay speed, unless the spectral distribution $\rho(s,x)$ is also highly oscillating and cancels exactly with the phase introduced by the momentum, which is also unlikely. 

We should also notice that, if $s_0$ is achieved at $-1<x_0<1$ within $(-1,1)$, with the following Taylor expansion $s(x)-s_0=a(x-x_0)^2+..$ nearby, then in the presence of the phase, the contribution near $x_0$ will be severely suppressed after the saddle point integrals. This explains the suppression of the ``non-analytic terms'' at large momentum, as they are exactly achieved away from the end-points.  On the other hand, the contributions near $s(1)$, $s(-1)$, $s(0)$ will not be canceled, since they are either due to the end-points, or due to possible singularities/discontinuities at $x=0$. As such,  they will dominate the large $\vec{z}$ expansion at large $\vec{p}$, explaining the three phases $e^{i\vec{p}\cdot \vec{z}}$, $e^{-i\vec{p}\cdot \vec{z}}$ and $1$ in the asymptotic formulas in all the known examples. 

\begin{figure}
    \centering
    \includegraphics[width=0.5\linewidth]{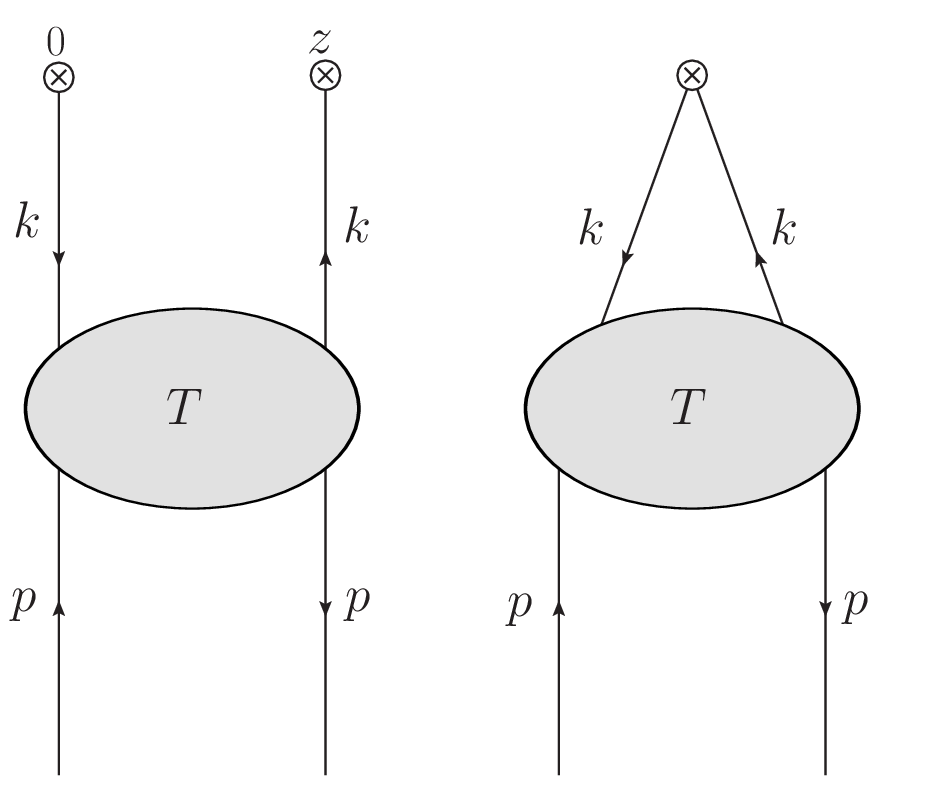}
    \caption{The four-point function graph $G$ (left) and the contracted graph $\tilde G$ (right). The coordinates space correlator is obtained by contracting two external legs with momentum $\pm k$ to the operators. Notice that the $G$ includes the two external legs with momentum $\pm k$ as well as the two-particle amputated kernel $T$.}
    \label{fig:fourpoints}
\end{figure}

The above can be proven to all orders in perturbation theory in $\phi^3$ or $\phi^4$, using parametric representations and properties of graph polynomials. In the rest of the subsection, we use the Euclidean signature for all the momenta, such as $p^2=p_t^2+\vec{p}^2$. To start, we first consider a momentum-space 4-point function connected graph $G$ with forward kinematics, and write the $\alpha$-parametric representation as (see~\cite{Krajewski:2008fa,Bogner:2010kv,Rivasseau:2014zpm} for reviews of parametric representations and graph polynomials)
\begin{align}
I_G(k,p)&=\frac{1}{(4\pi)^{2l}}\int_0^{\infty}\frac{d\alpha}{U_G^{2}}e^{-\frac{V_G}{U_G}-m^2\sum_{i=1}^n\alpha_i} \ \nonumber \\ 
&\equiv \frac{1}{(4\pi)^{2l}}\int_0^{\infty}\frac{d\alpha}{U_G^2}e^{-p^2P_1(\alpha)-k^2P_2(\alpha)-(p+k)^2P_3(\alpha)-(p-k)^2P_4(\alpha)-m^2\sum_{i=1}^n\alpha_i} \ .
\end{align}
In the graph $G$, two external legs with momentum $\pm p$ are amputated, while another two legs with momentum $\pm k$ are not amputated. See Fig.~\ref{fig:fourpoints} for a depiction. The number of propagators in $G$ is $n$, while the number of loops is $l$. The polynomial
\begin{align}\label{eq:U_G}
U_G=\sum_{{\cal T} \in T_1(G)}\prod_{i\notin {\cal T}}\alpha_i \ ,
\end{align}
is the standard first Zymanzik polynomial of the Feynman graph $G$ as a summation over the set of one-trees $T_1(G)$ of $G$, where the one-trees are connected subdiagrams of $G$ containing all its vertices without loops. On the other hand, 
\begin{align}\label{eq:V_G}
V_G&=\sum_{{\cal T}  \in T_2(G)} (\sum_{e\in {\cal T}_e} p_e)^2\prod_{i\notin {\cal T}}\alpha_i\nonumber \\ 
&\equiv \left(p^2P_1(\alpha)+k^2P_2(\alpha)+(p+k)^2P_3(\alpha)+(p-k)^2P_4(\alpha) \right)\times U_G \ , 
\end{align}
is the second Zymanzik polynomial as a summation over the two-trees $T_2(G)$. For any ${\cal T} \in T_2(G)$, one can write it as a union of two disjoint connected components ${\cal T}_1 \bigcup {\cal T}_2$ obtained by cutting a line in a one-tree, and the total incoming momentum-squared flowing into the cut is denoted as $(\sum_{e\in {\cal T}_e} p_e)^2$. For the four-point function graph $G$ in the forward limit, there are four kinds of cuts with four possible momentum squared $k^2$, $p^2$, $(k\pm p)^2$, giving rise to the four {\it positive} rational functions $P_1$ to $P_4$ after dividing the $U_G$. 

Then, given the four-point function in the momentum space, after integrating the $k$, one obtains the following parametric representation for the correlator in the coordinate space
\begin{align}\label{eq:fGI}
f_{G}(\vec{z},\vec{p})  = \int \frac{d^4 k}{(2\pi)^4} e^{i \vec{k}\cdot \vec{z}} I_{G}(k,p)
=\frac{1}{(4\pi)^{2l+2}}\int_{0}^{+\infty}\frac{d\alpha}{U^2_{\tilde  G}(\alpha)}e^{-\frac{z^2 U_G}{4U_{\tilde G}}+ix(\alpha)\vec{z}\cdot \vec{p}-p^2\frac{{\cal V}_{\tilde G}}{U_{\tilde G}}-m^2\sum_{i=1}^n\alpha_i} \ . 
\end{align}
The graph $\tilde G$ is the forward form-factor graph formed by contracting the two external legs of the four-point function graph $G$ connected to the operators, see Fig.~\ref{fig:fourpoints} for a depiction. Due to the forwardness, the only momentum squared in the second Zymanzik polynomial $V_{\tilde G}$ of  $\tilde G$ is $p^2$, and we have separated out the $p^2$ dependency explicitly
\begin{align}
\frac{V_{\tilde G}}{U_{\tilde G}} \equiv p^2\frac{{\cal V}_{\tilde G}}{U_{\tilde G}} \ . 
\end{align}
Finally, the momentum fraction ratio 
\begin{align}
x(\alpha)=\frac{P_4(\alpha)-P_3(\alpha)}{P_2(\alpha)+P_3(\alpha)+P_4(\alpha)} \ , 
\end{align}
is a homogeneous function in $\alpha$ of degree $0$ and satisfies the inequality $-1<x(\alpha)<1$.

By introducing the parameters
\begin{align}
\alpha_i=\rho x_i, \ \sum_{i=1}^nx_i=1 \ , 
\end{align}
One can write at $p^2=-m^2$
\begin{align}
f_G(\vec{z},\vec{p})=\int_0^{\infty} d\rho \rho^{N}\int_0^1 d x_i \delta(1-\sum_ix_i) \frac{e^{-\frac{z^2 U_G(x)}{4\rho U_{\tilde G}(x)}+i\vec{z}\cdot \vec{p}x(x_i)-m^2\rho(1-\frac{{\cal V}_{\tilde G}(x)}{U_{\tilde G}(x)})}}{(4\pi)^{2l+2}U^2_{\tilde G}(x)} \ .
\end{align}
Now, the polynomials $U(x)$ and $V(x)$ are obtained from the $\alpha$ dependent versions by removing the overall factors of $\rho$. The number $N$ depends on the theory. In $\phi^3$, one has combination $g^{2N+2}\rho^{N}$, where $N+1$ is the order of $g^2$, while in $\phi^4$, $N$ is always $-1$ due to the renormalizability.  In this form, we can integrate over $\rho$ and obtain the desired spectral representation
\begin{align}
&f_G(\vec{z},\vec{p})=\int_0^1 dx_i \frac{\delta(1-\sum_ix_i)}{(4\pi)^{2l+2}U^2_{\tilde G}(x)} e^{i\vec{z}\cdot \vec{p}x(x_i)} \left(\frac{U_G}{U_{\tilde G}\left(1-\frac{{\cal V}_{\tilde G}}{U_{\tilde G}}\right)}\right)^{\frac{N+1}{2}} \nonumber \\ 
&\times \frac{1}{2^N}\left(\frac{|\vec{z}|}{m}\right)^{N+1}K_{N+1}\bigg(m|\vec{z}|\sqrt{\frac{U_G}{U_{\tilde G}}}\sqrt{1-\frac{{\cal V}_{\tilde G}}{U_{\tilde G}}}\bigg) \ .
\end{align}
The integral is absolutely convergent, as far as the loop integrals inside the four-point function graph $G$ are convergent. Furthermore, being a tempered, Lorentz-invariant two-point correlator with a branch cut along the time-like axis and with a power-law bound, it can always be cast to Eq.~(\ref{eq:spectralcoorgen}). The corresponding spectral minimal for the graph $G$ is then
\begin{align}
s_0(G)=m\times {\rm min}_{x_i}\bigg(\sqrt{\frac{U_G}{U_{\tilde G}}}\sqrt{1-\frac{{\cal V}_{\tilde G}}{U_{\tilde G}}}\bigg) \ .
\end{align}
To provide a lower bound,  first notice the following inequality
\begin{align}
\frac{U_G(x)}{U_{\tilde G}(x)} \ge 1 . 
\end{align}
This is equivalent to 
\begin{align}
U_{\tilde G}(\alpha) \le \sum_{i=1}^n\alpha_i \times U_G(\alpha) \ .
\end{align}
Indeed, since $\tilde G$ is formed by contracting two external points in $G$, any one-tree in $T_1(\tilde G)$ must be formed by deleting one of the edges in another one-tree in $T_1(G)$, which establishes this inequality. Another set of more non-trivial inequalities are the so called {\it Nakanish inequalities}~\cite{Nakanishi1961,Jaffe1965}
\begin{align}
&4m^2\frac{{\cal V}_{\tilde G}}{U_{\tilde G}}-m^2 \le 0 \ ,  \phi^3 \ , \\
&9m^2\frac{{\cal V}_{\tilde G}}{U_{\tilde G}}-m^2 \le 0,  \ \phi^4 \ . 
\end{align}
The above inequality has a clear physical interpretation related to threshold production. When the external time-like momentum $-p^2$ is below the threshold for particle productions, the term $(-p^2\frac{{\cal V}_{\tilde G}}{U_{\tilde G}}-m^2\sum_{i=1}^n\alpha_i)<0$ on the exponential in Eq.~(\ref{eq:fGI}) to provide an analytic function after the integral. The momentum $-p^2$ needs to increase to $4m^2$ in $\phi^3$ and $9m^2$ in $\phi^4$, in order for the Feynman integrals to start hitting the threshold of two-particle or three-particle production, related to $(-p^2\frac{{\cal V}_{\tilde G}}{U_{\tilde G}}-m^2\sum_{i=1}^n\alpha_i) \leq 0$. As such, we have shown to all orders in perturbation theory that 
\begin{align}
&\sqrt{1-\frac{{\cal V}_{\tilde G}}{U_{\tilde G}}} \ge \sqrt{1-\frac{1}{4}}= \frac{\sqrt{3}}{2} \ ,  \  \phi^3 \ ,  \\
&\sqrt{1-\frac{{\cal V}_{\tilde G}}{U_{\tilde G}}} \ge \sqrt{1-\frac{1}{9}}= \frac{2\sqrt{2}}{3}  \ ,  \ \phi^4 \ . 
\end{align}
As such, we have obtained the following {\it stability bounds} for the spectral minimal
\begin{align}
&s_0(G) \ge \frac{\sqrt{3}m}{2},  \ \phi^3 \ , \\
&s_0(G) \ge \frac{2\sqrt{2}m}{3} \ , \ \phi^4 \ ,
\end{align}
that apply uniformly to {\it an arbitrary} connected four-point function Graph's contribution to the correlator. Notice that the lower bound $\frac{\sqrt{3}}{2}$ in $\phi^3$ is {\it optimal}, although suppressed in the large $\vec{p}$ limit. However, the stability bound in the $\phi^4$ is slightly lower than the optimal locality bound $m$, because the $\frac{U_G}{U_{\tilde G}} \ge 1$ and the stability inequality can not be saturated at the same point. 
\begin{figure}
    \centering
    \includegraphics[width=0.5\linewidth]{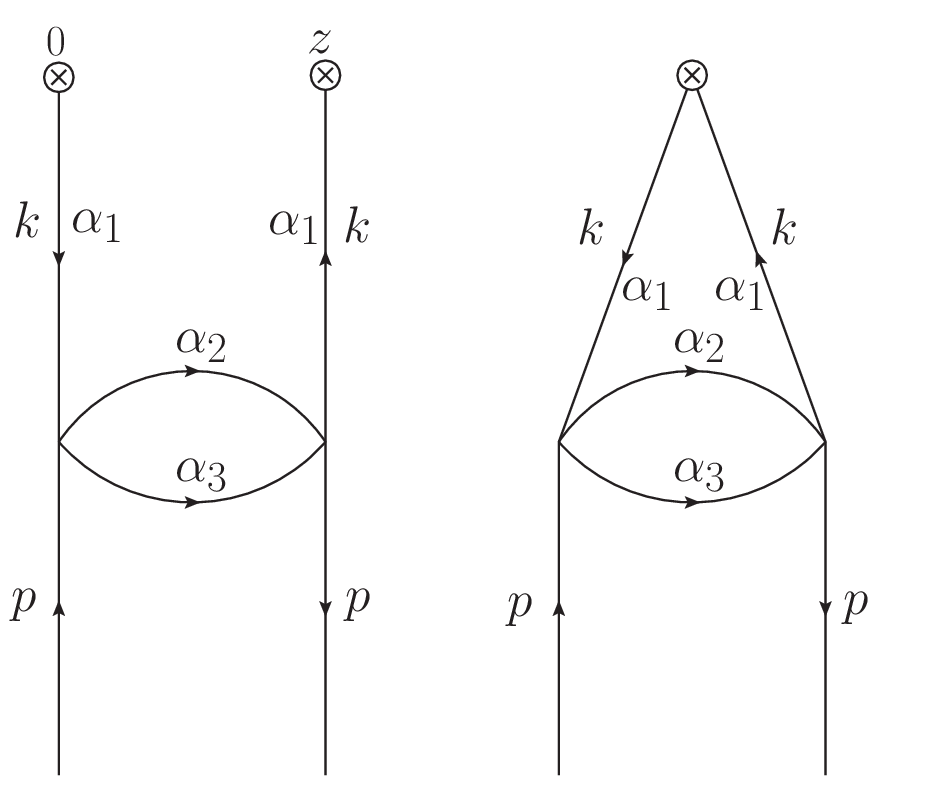}
    \caption{The two-loop $t$ channel four-point function graph $G$ (left) and the contracted graph $\tilde G$ (right) in the scalar $\phi^4$. The $\alpha$ parameters are labeled. }
    \label{fig:fourpointsphi4}
\end{figure}

To show this, it is instructive to consider the $t$-channel two-loop diagram in $\phi^4$ shown in Fig.~\ref{fig:fourpointsphi4}. The  crucial combination is 
\begin{align}
\sqrt{\frac{U_G}{U_{\tilde G}}}\sqrt{1-\frac{{\cal V}_{\tilde G}}{U_{\tilde G}}}=\frac{\sqrt{(x_1+x_2)(x_1+x_3)}(x_2+x_3)}{x_1 x_2+x_2 x_3+x_3 x_1}\ .
\end{align}
The minimal is achieved at $x_2=x_3=0$ and equals to $1$, consistent with the causality bound Eq.~(\ref{eq:Causalitybound}). Why the stability bound is not optimal? The reason is that, separately, one has
\begin{align}
\frac{ U_G}{U_{\tilde G}}=\frac{x_2+x_3}{x_1x_2+x_1x_3+x_2x_3} \ge 1 \ ,
\end{align}
and
\begin{align}
1-\frac{{\cal V}_{\tilde G}}{U_{\tilde G}}=\frac{(x_1+x_2)(x_2+x_3)(x_3+x_1)}{x_1x_2+x_2x_3+x_3x_1} \ge \frac{8}{9} \ . 
\end{align}
And each of them are optimal. However, the stability inequality is saturated at $x_1=x_2=x_3=\frac{1}{3}$. At this place, $\frac{U_G}{U_{\tilde G}}=2$ and makes the stability lower bound an under-estimation. On the other hand, the inequality $U_G\ge U_{\tilde G}$ is saturated at $x_2+x_3=0$, at this place $1-\frac{{\cal  V}_{\tilde G}}{U_{\tilde G}}$ also equals $1$. This configuration exactly corresponds to the ``forward singularity cut'', which also appears in Eq.~(\ref{eq:MYoff}). 

To see the structure of the forward singularity and to derive a closed formula, we can further separate the 4-point function graph $G$ into the two-to-two connected amputated kernel $T$ and two external legs with momentum $\pm k$, see Fig.~\ref{fig:fourpoints} for a depiction. We show the following: the forward singularity contribution can be reproduced using the parametric representations and is proportional to the two-particle amplitude $T$ evaluated at specific $k$. In $\phi^3$, we can further decompose the $T$ into one-particle irreducible parts in the $t$ and $u$ channels, and use the $e^{i(k^2+m^2+i0)\alpha}$ form for the one-particle propagators connecting different 1PI blocks that are locked to momentum $p\pm k$, to avoid above-threshold one-particle propagators. All the propagators within the 1PI blocks are with Euclidean parametric representations, as one needs a higher $4m^2$ threshold to make the 1PI parts on-shell. In this way, we can write 
\begin{align}
&(4\pi)^{2l+2}f_G(\vec{z},\vec{p})\nonumber \\
&=(4\pi)^2 \int_0^{+\infty} \frac{\beta d\beta d\alpha}{U_T^2(\alpha)} \int \frac{d^4 k}{(2\pi)^4} e^{i \vec{k}\cdot \vec{z} - k^2 \beta - p^2 P_1(\alpha) - k^2 P_2(\alpha) - (p+k)^2 P_3(\alpha) - (p-k)^2 P_4(\alpha) - m^2 (\sum_i \alpha_i + \beta) } \nonumber \\
&=\int_{0}^{+\infty}\frac{\beta d\beta d\alpha}{U_T^2(\alpha)(P_{234}+\beta)^2}e^{-\frac{z^2}{4(P_{234}+\beta)}+i\vec{z}\cdot\vec{p} \frac{P_4-P_3}{P_{234}+\beta}-m^2\left(\sum_i\alpha_i+\beta-P_{134}+\frac{(P_3-P_4)^2}{P_{234}+\beta}\right)} \ . 
\end{align}
In the above, the parameter for the external legs with momentum $\pm k$ is $\beta$. All the $\alpha$ parameters are for $T$. The polynomials $P_1$ to $P_4$ are only for the amputated kernel $T$ and is $\beta$ independent. Also, $P_{134}=P_1+P_3+P_4$ and $P_{234}=P_2+P_3+P_4$. We now consider the $\beta\gg \alpha_i$ region's contribution to the large $z$ asymptotics. This region exactly leads to the ``forward singularity''. For this purpose, we define 
\begin{align}
&\beta=\rho (1-\delta) \ , \\
&\alpha_i=\rho \delta y_i, \  \sum_iy_i=1 \ . 
\end{align}
In this manner, one has
\begin{align}
\frac{ \beta d\beta d\alpha} {U_T^2(\alpha)(P_{234}+\beta)^2} \rightarrow \frac{\rho^Nd\rho \delta^N(1-\delta)d\delta dy}{U_T^2(y)(1-\delta+\delta P_{234}^2)} \ . 
\end{align}
Then, one has
\begin{align}
&f_G\left(\vec{z},\vec{p}\right)=\int_0^1 \delta^N(1-\delta)d\delta\int_0^1\frac{dy \, \delta(\sum_i y_i-1)}{(4\pi)^{2l+2}U_T^2}\int_0^{+\infty}\frac{\rho^Nd\rho}{(1-\delta+\delta P_{234})^2} \nonumber \\ 
&\times \exp\bigg[ -\frac{z^2}{4\rho(1-\delta+\delta P_{234})}+\frac{i\vec{z}\cdot\vec{p}\delta(P_4-P_3)}{\delta P_{234}+1-\delta}-m^2\rho\left(1-\delta P_{134}+\frac{\delta^2(P_3-P_4)^2}{1-\delta+\delta P_{234}}\right)\bigg] \ . 
\end{align}
Notice that in the above, the overall $\rho$ and $\delta$ dependencies are factorized out from the rational functions $P_1(y)$ to $P_4(y)$, as well as in $U_T$.
Now, to perform the asymptotic expansion in the $\delta \rightarrow 0$ and $|\vec{z}| \rightarrow \infty$ limit, we integrate over $\rho$ and expand at large $z$ and small $\delta$, leading to
\begin{align}\label{eq:fGLz&Sd}
f_G\left(\vec{z},\vec{p}\right)  \rightarrow &\int_0^1 d\delta \, \delta^N \int_0^1\frac{dy \, \delta(\sum_i y_i-1)}{(4\pi)^{2l+2}U_T^2} \frac{\sqrt{\pi}e^{-m|\vec{z}|}}{2^N\sqrt{2m|\vec{z}|}}\left(\frac{|\vec{z}|}{m}\right)^{N+1} \nonumber\\
&\times \exp \bigg(-\frac{\delta m|\vec{z}|}{2}(1-P_{234}-P_{134})+izp^z \delta(P_4-P_3)\bigg) \ . 
\end{align} 
An important fact is that
\begin{align}
{\rm Re} \left(1-P_{134}-P_{234}\right)>0  \ , 
\end{align}
if the one-particle reducible ``bridges'' in $\phi^3$ are exponentiated using Minkowski parametric representations. The integral over $\delta$, in the large $|\vec{z}|$ limit, will then receive contribution from the region $\delta \sim (m|\vec{z}|)^{-1}$ and lead to polynomial corrections in the large $z$ expansion. 

To compute the leading polynomial dependency and the over-all coefficient,  we change the integral variable $\delta \rightarrow\frac{2m}{|z|}\rho$ in Eq.~(\ref{eq:fGLz&Sd}) and obtain the following  leading large $z$ asymptotics in the region $\beta \gg \alpha_i$,
\begin{align}
&f_G(\vec{z},\vec{p})\rightarrow\frac{\sqrt{\pi}e^{-m|\vec{z}|}}{8\pi^2\sqrt{2m|\vec{z}|}}\int_{0}^{+\infty} d\rho \, \rho^N \int_{0}^{1}\frac{dy \, \delta\left(\sum_i y_i-1\right)}{(4\pi)^{2l}U_T^2(y)} e^{-\rho m^2(1-P_{134}-P_{234})+i2mp^z\rho (P_4-P_3)} \ . 
\end{align} 
Here, there are two observations. First,  the integral (in the second line, we introduced back the $\alpha_i=\rho y_i$ for $T$)
\begin{align}
&I_T=\int\frac{d\rho dy \, \rho^N\delta\left(\sum_i y_i-1\right)}{(4\pi)^{2l}U_T^2(y)} e^{-\rho m^2(1-P_{134}-P_{234})+i2mp^z\rho (P_4-P_3)} \nonumber \\ 
&=\int \frac{d\alpha}{(4\pi)^{2l}U_T^2(\alpha)}e^{m^2P_1+m^2P_2-(-2m^2+2mip^z)P_3-(-2m^2-2mip^z)P_4-m^2\sum_i\alpha_i} \ , 
\end{align}
is nothing but the original amputated four-point function denoted as $T$ in Fig.~\ref{fig:fourpoints}, evaluated at the following invariants in the Euclidean signature
\begin{align}
k^2=p^2=-m^2, \  (p\pm k)^2=-2m^2 \pm 2imp^z \ .
\end{align}
This exactly corresponds to
\begin{align}
k=(0,0,0,im) \ , 
\end{align}
analytically continued to the imaginary momentum. Second, this configuration is still below the two-particle thresholds in the one-particle irreducible parts (loops) of $T$. As such, $T$ is well-defined without any ambiguity.  In fact, it is possible to show that, under this configuration, $T$ is purely real. In conclusion, we have derived from the parametric representation that the ``forward-singularity'' contribution takes the following general form 
\begin{align}
f_{\rm forward}(\vec{z},\vec{p})=\frac{1}{8\pi\sqrt{2\pi m|\vec{z}|}}e^{-m|\vec{z}|} \times T(2m^2-2imp^z, 2m^2+2imp^z) \ , 
\end{align}
where we have changed back to the Minkowskian signature in the amplitude. The crucial quantity $T(s,u)$ is the two-particle on-shell amplitude at $s=(p+k)^2$, $u=(p-k)^2$ and $t=0$. Our configuration lies exactly on the imaginary axis halfway to the two-particle thresholds starting at $s=4m^2, \ u=0$ or $u=4m^2, \ s=0$.

\section{Conclusion and discussion}\label{sec:conclusion}
In this paper, we use HQET reduction, dispersive analysis, and Lorentz symmetry to derive the large distance asymptotic forms of quasi-correlators, where many aspects are confirmed by two-loop calculations in $\phi^3$ theory as well as all-order justifications from the perspective of locality. We demonstrate the controlled precision of asymptotic analysis, which is a few percent with current lattice data and is expected to improve in the near future. Our results are valuable for precision calculations under LaMET. 

A natural question is whether the parameters appearing in the asymptotic forms can be calculated directly rather than determined by fitting. It is possible to calculate the mass gaps in Eq.~(\ref{eq:qPDFLzGen}) from lattice QCD. The physical masses in Eq.~(\ref{eq:qPDFLzGen}) can be directly extracted from two-point functions of heavy quark systems on the lattice. By subtracting off the heavy quark pole mass defined under the same renormalon regularization scheme as the linear divergence, one obtains the binding energies that can be directly used in the extrapolated forms. However, it is challenging to calculate the coefficients $a_{X,l}$, $a_{Y,l}$, and $a_{Z,l}$ in Eq.~(\ref{eq:hexp}) for a spacelike $x$. Although they have matrix element definitions that depend on the momenta $k_i$ of inserted states, the spacelike correlator requires the analytical continuation of $k_i$ to a spatial direction with imaginary numbers, which is difficult to perform with numerical values. Since the FT uncertainties in the moderate $y$ range mostly depend on the error of the magnitude at the truncation point $z_L$, fixing the binding energies but not the coefficients does not improve the precision there. 


\acknowledgments
We thank Yong Zhao, Rui Zhang, Chen Yang, Jinchen He, Fangcheng He, Jinghong Yang, Fei Yao, Jun Hua, Haoyang Bai, Muhua Zhang, and Jia-Lu Zhang for useful discussions and comments. We thank the ANL/BNL collaboration and the Lattice Parton Collaboration (LPC) for sharing lattice data with us. Y. S. is partially supported by the Quark-Gluon Tomography (QGT) collaboration, which is supported by the U.S. Department of Energy (DOE) topical collaboration program (DE-SC0023646). Y. L. is supported in part by a Priority Research Area DigiWorld grant under the Strategic Program Excellence Initiative at the Jagiellonian University (Kraków, Poland).

\appendix

\section{Matching between dynamical heavy quark and gauge link}\label{sec:MatQGL}
This section aims to justify the coordinate space HQET reduction in Eqs.~(\ref{eq:MatQGLorg}) and~(\ref{eq:MatQGL}), through the expansion by region analysis, one-loop perturbative calculations, and complex analysis. 

\subsection{Expansion by region analysis in the timelike separation}\label{sec:MatQGLexp}
This subsection justifies Eqs.~(\ref{eq:MatQGLorg}) and~(\ref{eq:MatQGL}) for the timelike case using the expansion-by-region analysis.

\begin{figure}
    \centering
    \includegraphics[width=0.618\linewidth]{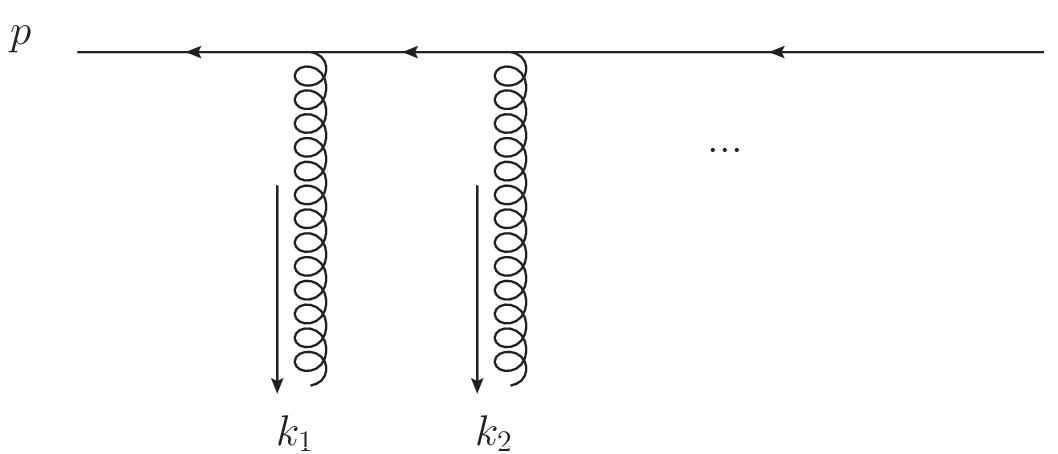}
    \caption{An on-shell external heavy quark inserted by multiple gluons.}
    \label{fig:HQET_on_shell}
\end{figure}
First, let's recall the HQET reduction for an on-shell external heavy quark $p=(m_{Q},0,0,0)$ and define the velocity vector $v \equiv (1,0,0,0)$. Consider multi-gluon insertions on the quark line with momenta $k_1,k_2,...$, as shown in Fig.~\ref{fig:HQET_on_shell},
\begin{align}
I = \bar{u}(p) \gamma^{\mu_1} t^{a_1} \frac{i(\slashed{p}-\slashed{k_1}+m_Q)}{(p+k_1)^2-m^2_Q} \gamma^{\mu_2} t^{a_2} \frac{i(\slashed{p}+\slashed{k_1}+\slashed{k_2}+m_Q)}{(p+k_1+k_2)^2-m^2_Q} ...
\end{align}
In the heavy quark limit $m_{Q} \rightarrow +\infty$, the gluon momenta are counted as $k_{i} \sim \lambda m_{Q}(1,1,1,1)$ where $\lambda \ll 1$. Therefore, the quark propagators can be reduced as,
\begin{align}
&I \approx \bar{u}(p) \gamma^{\mu_1} t^{a_1} \frac{i(\slashed{v}+1)}{2\left(p^{0}+k_1^{0}-\frac{(\vec{p}+\vec{k}_1)^2}{2 m_{Q}}-m_{Q}\right)} \gamma^{\mu_2} t^{a_2} \frac{i(\slashed{v}+1)}{2\left(p^{0}+k_1^{0}+k_2^{0}-\frac{(\vec{p}+\vec{k}_1+\vec{k}_2)^2}{2 m_{Q}}-m_{Q}\right)} ... \nonumber\\
&\approx \bar{u}(p) v^{\mu_1} t^{a_1} \frac{i(\slashed{v}+1)}{2 v \cdot k_1} v^{\mu_2} t^{a_2} \frac{i(\slashed{v}+1)}{2 v \cdot k_2} ... \nonumber\\
&= \bar{u}(p) \frac{i v^{\mu_1}  t^{a_1}}{v \cdot k_1}  \frac{i v^{\mu_2} t^{a_2}}{v \cdot k_2} ...
\end{align}
In the first line, the non-relativistic expansion is performed in the denominator. For a heavy quark at rest, the ``kinematic energy terms" $(\vec{p}+\vec{k}_1)^2/(2 m_{Q})$ can be ignored, which leads to the second line. The projection of Dirac gamma matrices has also been applied here $\gamma^{\mu_1} = \slashed{v} v^{\mu_1} + ...$, where ``..." does not contribute to the leading term. The final result is eikonalized as a gauge link. 

For a coordinate-space heavy quark propagator $Q(t)\bar{Q}(0)$ of interest in Eqs.~(\ref{eq:MatQGLorg}) and~(\ref{eq:MatQGL}), the HQET reduction becomes slightly different in the way that the heavy quark momentum $p$ has quantum fluctuations around the timelike direction and the ``kinematic energy terms" should be kept in the power counting. To understand this, we first consider the tree-level propagator under non-relativistic expansion,
\begin{align}
&I'_{\rm tree} = \int \frac{d^4 p}{(2\pi)^4} e^{-i p^0 t} \frac{i (\slashed{p}+m_{Q})}{p^2-m_{Q}^2} \nonumber\\
&\approx \int \frac{d^4 p}{(2\pi)^4} e^{-i p^0 t} \frac{1}{\left(p^0 - m_{Q} - \frac{\vec{p}^2}{2 m_{Q}} + i 0^{+}\right)} \frac{i(\slashed{v}+1)}{2}
\end{align}
In this case, the kinematic term $\vec{p}^2/(2 m_{Q})$ can not be ignored otherwise the integral becomes divergent. Pick up the residue pole of $p^0$ and integrate out $\vec{p}$,
\begin{align}\label{eq:I'tree}
&I'_{\rm tree} \approx \int \frac{d^3 \vec{p}}{(2\pi)^3} e^{-i m_{Q} t} e^{-i \frac{\vec{p}^2}{2 m_{Q}} t} \frac{\slashed{v}+1}{2}   
= \frac{m_Q^3 e^{-i m_Q t}}{2 \sqrt{2} \pi ^{3/2} \left(i m_Q t\right)^{3/2}} \frac{\slashed{v}+1}{2} \ .
\end{align}
In this calculation, the momentum mode $p \sim \left(m_{Q},\sqrt{2m_{Q}/t},\sqrt{2m_{Q}/t},\sqrt{2m_{Q}/t}\right)$ plays a crucial role regarding the width of the Gaussian factor $e^{-i \frac{\vec{p}^2}{2 m_{Q}} t}$. The overall polynomial dependence $m_{Q}^3/\left(m_Q t\right)^{3/2}$ is related to the size of quantum fluctuations along the spatial directions.

\begin{figure}
    \centering
    \includegraphics[width=0.618\linewidth]{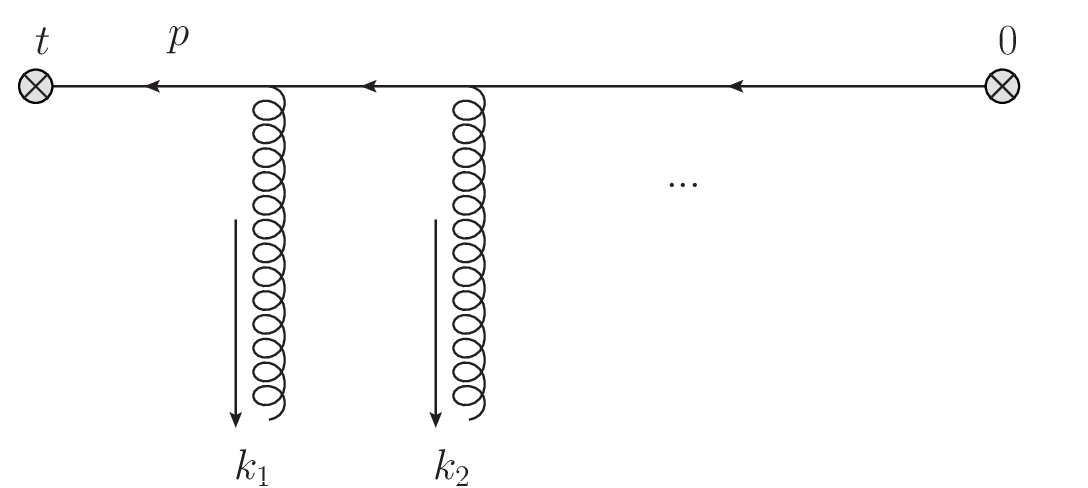}
    \caption{A coordinate-space heavy quark propagator $Q(t)\bar{Q}(0)$ inserted by multiple gluons.}
    \label{fig:HQET_prop}
\end{figure}
The next step is to check if HQET reduction holds for the momentum mode $p \sim \left(m_{Q},\sqrt{2m_{Q}/t},\sqrt{2m_{Q}/t},\sqrt{2m_{Q}/t}\right)$. Here $p$ is almost on-shell, and the off-shellness is treated as higher orders in non-relativistic expansion, $p^{0}-m_{Q}-\frac{\vec{p}^2}{2m_{Q}} \sim O\left(\vec{p}^4/m_{Q}^3\right)$. And the power counting for the gluon momenta will be $k_{i} \sim \lambda m_{Q}(1,1,1,1)$, where $\vec{p}^4/m_{Q}^4\ll \lambda \ll 1$. Start with a coordinate space heavy quark propagator with multi-gluon insertions in Fig.~\ref{fig:HQET_prop},
\begin{align}
I' =  \int \frac{d^4 p}{(2\pi)^4} e^{-i p^0 t} \frac{i (\slashed{p}+m_{Q})}{p^2-m_{Q}^2} \gamma^{\mu_1} t^{a_1} \frac{i(\slashed{p}-\slashed{k_1}+m_Q)}{(p+k_1)^2-m^2_Q} \gamma^{\mu_2} t^{a_2} \frac{i(\slashed{p}+\slashed{k_1}+\slashed{k_2}+m_Q)}{(p+k_1+k_2)^2-m^2_Q} ...
\end{align}
Perform the non-relativistic expansion up to the kinematic energy term, 
\begin{align}
I' \approx & \int \frac{d^4 p}{(2\pi)^4} e^{-i p^0 t} \frac{i(\slashed{v}+1)}{2 \left(p^0 - m_{Q} - \frac{\vec{p}^2}{2 m_{Q}} + i 0^{+}\right)} \nonumber\\
&\times v^{\mu_1} t^{a_1} \frac{i(\slashed{v}+1)}{2\left(p^{0}+k_1^{0}-\frac{(\vec{p}+\vec{k}_1)^2}{2 m_{Q}}-m_{Q}\right)} v^{\mu_2} t^{a_2} \frac{i(\slashed{v}+1)}{2\left(p^{0}+k_1^{0}+k_2^{0}-\frac{(\vec{p}+\vec{k}_1+\vec{k}_2)^2}{2 m_{Q}}-m_{Q}\right)} ...  
\end{align}
The key point here is the kinematic terms $(\vec{p}+\vec{k}_1)^2/(2 m_{Q}), (\vec{p}+\vec{k}_1+\vec{k}_2)^2/(2 m_{Q})$ cannot be ignored anymore. They can be reduced by power counting,
\begin{align}
k_1^0 - \frac{(\vec{p}+\vec{k}_1)^2}{2 m_{Q}} = k_1^0 -  \frac{\vec{p}^2}{2 m_{Q}} - \frac{\vec{k}_{1}^2}{2 m_{Q}} - \frac{\vec{p}\cdot \vec{k}_1}{m_{Q}} \approx  k_1^0 -  \frac{\vec{p}^2}{2 m_{Q}} + O\left(\lambda^2 m_{Q}\right) + O\left(\frac{\lambda m_{Q}}{\sqrt{t m_{Q}}}\right) \ , 
\end{align}
where the $\vec{k}_{1}^2/(2 m_{Q}) \sim \lambda^2 m_Q$ and $\vec{p}\cdot \vec{k}_1/m_{Q} \sim \lambda m_Q /\sqrt{t m_Q}$ are ignored since they are more suppressed than $k_1^0 \sim \lambda m_Q$. The kinematic term $\vec{p}^2/(2m_{Q})\sim 1/t$ is kept since it can be similarly large as $k_1^0 \sim \lambda m_Q$ as we do not require the hierarchy between $1/(m_Q t)$ and $\lambda$. Therefore, the integral becomes,
\begin{align}
I' \approx & \int \frac{d^4 p}{(2\pi)^4} e^{-i p^0 t} \frac{i(\slashed{v}+1)}{2 \left(p^0 - m_{Q} - \frac{\vec{p}^2}{2 m_{Q}} + i 0^{+}\right)} \nonumber\\
&\times v^{\mu_1} t^{a_1} \frac{i(\slashed{v}+1)}{2\left(p^{0}+k_1^{0}-\frac{\vec{p}^2}{2 m_{Q}}-m_{Q}\right)} v^{\mu_2} t^{a_2} \frac{i(\slashed{v}+1)}{2\left(p^{0}+k_1^{0}+k_2^{0}-\frac{\vec{p}^2}{2 m_{Q}}-m_{Q}\right)} ...    
\end{align}
By redefining $p'^0 = p^0-\frac{\vec{p}^2}{2 m_{Q}}-m_{Q} \sim O\left(\vec{p}^4/m_{Q}^3\right)$, one obtains,
\begin{align}
I' \approx & \int \frac{d^3 \vec{p}}{(2\pi)^3} \frac{d p'^0}{2\pi} e^{-i p'^0 t} e^{-i m_{Q} t} e^{- i \frac{\vec{p}^2}{2 m_{Q}} t} \frac{i(\slashed{v}+1)}{2 \left(p'^0 + i 0^{+}\right)}  v^{\mu_1} t^{a_1} \frac{i(\slashed{v}+1)}{2\left(p'^0+k_1^{0}\right)} v^{\mu_2} t^{a_2} \frac{i(\slashed{v}+1)}{2\left(p'^{0}+k_1^{0}+k_2^{0}\right)} ...  \nonumber\\
\approx & \int \frac{d^3 \vec{p}}{(2\pi)^3} \frac{d p'^0}{2\pi} e^{-i m_{Q} t} e^{- i \frac{\vec{p}^2}{2 m_{Q}} t} \frac{i(\slashed{v}+1)}{2 \left(p'^0 + i 0^{+}\right)}  v^{\mu_1} t^{a_1} \frac{i(\slashed{v}+1)}{2 k_1^{0}} v^{\mu_2} t^{a_2} \frac{i(\slashed{v}+1)}{2\left(k_1^{0}+k_2^{0}\right)} ... \nonumber\\
= & \frac{m_Q^3 e^{-i m_Q t}}{2 \sqrt{2} \pi ^{3/2} \left(i m_Q t\right)^{3/2}} \frac{\slashed{v}+1}{2} \frac{i v^{\mu_1} t^{a_1}}{v \cdot k_1} \frac{i v^{\mu_2} t^{a_2}}{v \cdot (k_1 +k_2)} ...
\end{align}
where the dynamical, Dirac and color structures are fully factorized. The color structure is the same as a gauge link along the timelike direction $v$. The above region analysis justifies Eqs.~(\ref{eq:MatQGLorg}) and~(\ref{eq:MatQGL}) for the timelike case in the main text.

\subsection{One-loop verification for the timelike case}\label{sec:MatQGL1loop}
This subsection targets at justifying Eq.~(\ref{eq:MatQGL}) in the timelike separation $x=(t,0,0,0)$ up to one-loop in a massless on-shell quark external state with $P=(P^z,0,0,P^z)$. In order to regulate the UV and IR divergences, we implement the dimensional regularization for the loop momentum of gluon, while the momentum of heavy quark propagator is fixed at 4D for simplicity. $m_Q$ is treated as a pole mass in our calculation and the other UV divergences are renormalized to $\overline{\rm MS}$ scheme.  

The quantities are expanded in terms of QCD coupling $\alpha(\mu)$,
\begin{align}
M_{\Gamma_1 \Gamma_2}(x,P,\mu) = M^{(0)}_{\Gamma_1 \Gamma_2}(x,P,\mu) + \frac{\alpha(\mu) C_F}{2\pi} M^{(1)}_{\Gamma_1 \Gamma_2}(x,P,\mu) + ...
\end{align}
\begin{align}
\tilde{h}_{\Gamma_1\Gamma_2}\left(x,P,\mu\right) = \tilde{h}^{(0)}_{\Gamma_1 \Gamma_2}(x,P,\mu) + \frac{\alpha(\mu) C_F}{2\pi} \tilde{h}^{(1)}_{\Gamma_1 \Gamma_2}(x,P,\mu) + ...
\end{align}
\begin{align}
H_{\Gamma_1\Gamma_2,\hat{x}}\left(\frac{m_{Q}^2} {\mu^2}\right) = H^{(0)}_{\Gamma_1 \Gamma_2,\hat{x}}\left(\frac{m_{Q}^2} {\mu^2}\right) + \frac{\alpha(\mu) C_F}{2\pi} H^{(1)}_{\Gamma_1 \Gamma_2,\hat{x}}\left(\frac{m_{Q}^2} {\mu^2}\right) + ...
\end{align}
The tree level results are,
\begin{align}
M^{(0)}_{\Gamma_1 \Gamma_2}(x,P,\mu) =  \frac{m_Q^3 e^{-i m_Q t}}{2 \sqrt{2} \pi ^{3/2} \left(i m_Q t\right)^{3/2}} e^{i \,t \, P^t} \bar{u}(P) \Gamma_1 \frac{\gamma^t + 1}{2} \Gamma_2 u(P) \left[ 1+O\left(\frac{1}{m_Q t}\right)\right]  \ ,
\end{align}
\begin{align}
\tilde{h}^{(0)}_{\Gamma_1 \Gamma_2}(x,P,\mu) = e^{i \,t \, P^t} \bar{u}(P) \Gamma_1 \frac{\gamma^t + 1}{2} \Gamma_2 u(P) \ .
\end{align}
Therefore, the hard kernel at tree-level is obtained by comparing the above results with Eq.~(\ref{eq:MatQGL}),
\begin{align}
H^{(0)}_{\Gamma_1\Gamma_2,\hat{x}}\left(\frac{m_{Q}^2} {\mu^2}\right) = 1 \ ,
\end{align}
which does not have $P$ or $x$ dependences as expected. 

The one-loop diagrams are calculated under Feynman gauge and decomposed based on diagram topology,
\begin{align}
&M^{(1)}_{\Gamma_1 \Gamma_2}(x,P,\mu) = \left[ M^{\rm vertex}_{\Gamma_1 \Gamma_2}(x,P,\mu) + M^{\rm sail}_{\Gamma_1 \Gamma_2}(x,P,\mu) \right.\nonumber\\
&\left.+ M^{\rm t.p.}_{\Gamma_1 \Gamma_2}(x,P,\mu)  + M^{\rm w.fn.}_{\Gamma_1 \Gamma_2}(x,P,\mu) \right] M^{(0)}_{\Gamma_1 \Gamma_2}(x,P,\mu)  \ ,
\end{align}
\begin{align}
\tilde{h}^{(1)}_{\Gamma_1 \Gamma_2}(x,P,\mu) = \left[ \tilde{h}^{\rm vertex}_{\Gamma_1 \Gamma_2}(x,P,\mu) + \tilde{h}^{\rm sail}_{\Gamma_1 \Gamma_2}(x,P,\mu) \right.\nonumber\\
\left.+ \tilde{h}^{\rm t.p.}_{\Gamma_1 \Gamma_2}(x,P,\mu)  + \tilde{h}^{\rm w.fn.}_{\Gamma_1 \Gamma_2}(x,P,\mu) \right] \tilde{h}^{(0)}_{\Gamma_1 \Gamma_2}(x,P,\mu) \ ,
\end{align}
where the names of Feynman diagrams are the same as Ref.~\cite{Izubuchi:2018srq}. The overall tree-level factors are factorized out. 

First, it is obvious that
\begin{align}
M^{\rm vertex}_{\Gamma_1 \Gamma_2}(x,P,\mu) =  \tilde{h}^{\rm vertex}_{\Gamma_1 \Gamma_2}(x,P,\mu) +O\left(\frac{1}{m_Q t}\right) \ ,
\end{align}
\begin{align}
M^{\rm w.fn.}_{\Gamma_1 \Gamma_2}(x,P,\mu) =  \tilde{h}^{\rm w.fn.}_{\Gamma_1 \Gamma_2}(x,P,\mu) +O\left(\frac{1}{m_Q t}\right)  \ ,
\end{align}
because the gluon and heavy quark momentum integrals can be studied independently in those diagrams. One can integrate out the heavy quark momentum and perform the large $m_Q$ expansion, the remaining gluon momentum integral is the same as the corresponding contribution from $\tilde{h}^{(1)}_{\Gamma_1 \Gamma_2}(x,P,\mu)$. 

The perturbative results from the heavy quark self energy are,
\begin{align}
M^{\rm t.p.}_{\Gamma_1 \Gamma_2}(x,P,\mu) = \frac{3}{2}\ln \frac{m_Q^2}{\mu^2} + \ln \frac{-t^2 \mu^2}{4 e^{-2\gamma_E}}  +O\left(\frac{1}{m_Q t}\right)  \ ,
\end{align}
\begin{align}
\tilde{h}^{\rm t.p.}_{\Gamma_1 \Gamma_2}(x,P,\mu) = \ln \frac{-t^2 \mu^2}{4 e^{-2\gamma_E}} + 2   \ , 
\end{align}
where $M^{\rm t.p.}_{\Gamma_1 \Gamma_2}(x,P,\mu)$ is calculated using the cut diagram techniques and non-relativistic expansion of heavy quark momentum similar as Eq.~(\ref{eq:It1calc}).

The perturbative results from the sail diagrams are,
\begin{align}
\tilde{h}^{\rm sail}_{\Gamma_1 \Gamma_2}(x,P,\mu) 
= &   2 \left(\frac{1}{\epsilon_{\rm IR}} - \ln\frac{4P_z^2}{\mu^2} + 2\ln(-i \lambda e^{\gamma_E}) + 1 \right)  \nonumber\\
& \times \left(\ln \left(-i \lambda  e^{\gamma_E }\right)+\frac{i}{\lambda}-\frac{i e^{i \lambda }}{\lambda }+\Gamma (0,-i \lambda )-1\right) \nonumber\\
& - \ln\frac{4P_z^2}{\mu^2} + 2\ln(-i \lambda  e^{\gamma_E }) + 4 i \lambda  \, _3F_3(1,1,1;2,2,2;i \lambda ) 
\end{align}
\begin{align}
M^{\rm sail}_{\Gamma_1 \Gamma_2}(x,P,\mu) =  \tilde{h}^{\rm sail}_{\Gamma_1 \Gamma_2}(x,P,\mu) + V_{\Gamma_1,\hat{x}}\left(\frac{m_Q^2}{\mu^2}\right) + V_{\Gamma_2,\hat{x}}\left(\frac{m_Q^2}{\mu^2}\right) +O\left(\frac{1}{m_Q t}\right) \ ,
\end{align}
where dimensionless distance $\lambda \equiv - P^t t$. The correction $V$ is,
\begin{align}
V_{\Gamma_1,\hat{x}}\left(\frac{m_Q^2}{\mu^2}\right)   = \begin{cases}
-\frac{3}{2} \ln\frac{m_Q^2}{\mu^2} + 2, &\Gamma_1 = I \\
\frac{i}{4\sqrt{-x^2}}{\rm Tr}[\slashed{x} \gamma^{\mu}] - 1, &\Gamma_1 = \gamma^{\mu} 
\end{cases} \ ,
\end{align}
which has been calculated for $\Gamma_1,\Gamma_2 \in \{I,\gamma^t,\gamma^x,\gamma^y,\gamma^z\}$ as examples. 

By comparing the above expressions, one validates the matching relation Eq.~(\ref{eq:MatQGL}) in the timelike separation up to one-loop. The hard kernel can be extracted,
\begin{align}
H^{(1)}_{\Gamma_1 \Gamma_2,\hat{x}}\left(\frac{m_{Q}^2} {\mu^2}\right) = \frac{3}{2}\ln \frac{m_Q^2}{\mu^2} - 2 + V_{\Gamma_1,\hat{x}}\left(\frac{m_Q^2}{\mu^2}\right) + V_{\Gamma_2,\hat{x}}\left(\frac{m_Q^2}{\mu^2}\right) \ ,
\end{align}
which only depends on the hard scale $m_Q$ and irrelevant to other physical scales. 

\subsection{Gauge link reduction at spacelike separations}\label{sec:reductionspace}
This subsection generalizes Eq.~(\ref{eq:MatQGL}) to the spacelike case using the methods of complex analysis. 

To start, we first introduce the hadronic matrix elements, in which the exponential factor of the pole mass and certain other factors are removed
\begin{align}\label{eq:hatM}
\hat M_{\Gamma_1\Gamma_2}(t,\vec{z})=\frac{2\sqrt{2}\pi^{\frac{3}{2}}}{m_Q^3}(m_Q\sqrt{-t^2+z^2})^{\frac{3}{2}}e^{m_Q\sqrt{-t^2+z^2}}M_{\Gamma_1\Gamma_2}(t,\vec{z}) \ . 
\end{align}
Here, $M_{\Gamma_1\Gamma_2}$ is defined in Eq.~(\ref{eq:MqQQq}) if one sets $x=\left(t,\vec{z}\right)$. The matrix element with gauge link is defined in the main text as Eq.~(\ref{eq:hqUq}),
\begin{align}
\tilde h_{\Gamma_1\Gamma_2}(t,\vec{z})\equiv   \langle H(P) | \bar{\psi}(x) \Gamma_1 U(x,0) \frac{i \slashed{x} + \sqrt{-x^2}}{2 \sqrt{-x^2}} \Gamma_2 \psi(0) | H(P) \rangle_c \ .  
\end{align}
The function variables $P$ and $\mu$ are omitted for simplicity. It is convenient to define the following difference function
\begin{align}
f(t+it_E,\vec{z})=\hat M_{\Gamma_1\Gamma_2}(t+it_E,\vec{z})- H_{\Gamma_1 \Gamma_2}(m_Q) \times  \tilde h_{\Gamma_1\Gamma_2}(t+it_E,\vec{z}) \ .
\end{align}
Our goal is to prove that $f(t+it_E,\vec{z})$ at $ t + i t_E = 0$ and non-zero $|\vec{z}|$ is at least $O\left(1/m_Q\right)$ suppressed in the large $m_Q$ expansion.

To show the main idea of the proof, we first neglect the subtleties due to the short distance region and consider a simplified scenario. We assume that the HQET reduction at the timelike separation can be bounded uniformly by 
\begin{align}
\bigg|f(t\pm i0,\vec{z})\bigg|_{|t|>|\vec{z}|}\le \frac{C}{m_Q }  \ , \label{eq:HEQEreduapp}
\end{align}
where the $|\vec{z}|$ and $P$ dependencies in $C=C(P,|\vec{z}|)$ are bound by power laws, and the matrix element in the lower and upper half-planes allows the following bounds
\begin{align}
\bigg|f(t+it_E,\vec{z})\bigg|\le C'e^{m_0|t_E|}\ , \label{eq:timecondition}
\end{align}
where $m_0$ is {\it $m_Q$ independent}, which can be either positive or negative. Then we can prove, that at the spacelike separation, the reduction must hold as well
\begin{align}
\bigg|f(0,\vec{z})\bigg|\le \frac{C''}{m_Q} e^{m_0|\vec{z}|}\ . \label{eq:redspace}
\end{align}
Next, we will show the detailed derivations for Eq.~(\ref{eq:redspace}). The crucial regularity property is, both of the matrix element $\tilde h_{\Gamma_1\Gamma_2}(t,\vec{z})$ with gauge link and the hadron matrix element $\hat M_{\Gamma_1\Gamma_2}(t,\vec{z})$, as functions of $t$, can be analytically continued to the upper and lower half complex planes of $t$, with a branch cut on the real-time axis along $(-\infty,-|\vec{z}|) \bigcup (|\vec{z}|,\infty)$. Again, locality is important in gluing the upper and lower half planes together, which is similar to the case in Sec.~\ref{sec:decayzerowoone}. The upper and lower boundary values along the branch cuts are exactly the time-like matrix elements, which have been shown to obey the HQET reduction Eq.~(\ref{eq:HEQEreduapp}). Now, the condition Eq.~(\ref{eq:timecondition}) implies that the function, 
\begin{align}\label{eq:gwz}
g(w,\vec{z})=\frac{f(w,\vec{z})}{(-w^2+|\vec{z}|^2)^{\beta}}e^{-(m_0+\epsilon)\sqrt{-w^2+z^2}}, w=t+it_E \ , 0<\beta<1 \ , \epsilon>0 \ ,
\end{align}
decays to zero sufficiently fast in the imaginary time direction and has the same singularity structure as $f$. As such, one can establish a dispersive representation for $g(0,\vec{z})$ as
\begin{align} \label{eq:siperg}
g(0,\vec{z})=\int_{|\vec{z}|}^{\infty}\frac{dt}{2\pi i t}\bigg(g(t+i0,\vec{z})-g(t-i0,\vec{z})\bigg)\nonumber \\ +\int^{-|\vec{z}|}_{-\infty}\frac{dt}{2\pi i t}\bigg(g(t+i0,\vec{z})-g(t-i0,\vec{z})\bigg) \ .
\end{align}
The values along the branch cuts are exactly bounded by the HQET reduction Eq.~(\ref{eq:HEQEreduapp}), leading to 
\begin{align}
\bigg|g(0,\vec{z})\bigg|\le \frac{2C}{\pi m_Q}\int_{|\vec{z}|}\frac{dt}{t(t^2-|\vec{z}|^2)^{\beta}}=\frac{C''}{ m_Q |\vec{z}|^{\beta}} \ , 
\end{align}
which, combined with Eq.~(\ref{eq:gwz}), gives the desired reduction Eq.~(\ref{eq:redspace}). 

We now come to the major subtle point in the simplified scenario: the convergence to the HQET correlator can not be uniform for all the separations. The point is that if $t^2-z^2\rightarrow0$, the matrix elements are dominated instead by the short distance limits,
\begin{align}
\hat M_{\Gamma_1\Gamma_2}(t,\vec{z})\rightarrow\frac{C}{m_Q^{\frac{3}{2}}\left(\sqrt{-t^2+\vec{z}^2}\right)^{\frac{3}{2}}} \ , \label{eq:hatMSD} \\
\tilde h_{\Gamma_1\Gamma_2}(t,\vec{z})\rightarrow C'\ln^{\gamma}(-t^2+\vec{z}^2) \ ,  
\end{align}
where $\gamma$ depends on the one-loop anomalous dimensions and the beta functions. Note that $M_{\Gamma_1 \Gamma_2}$ itself scales as $1/\left(\sqrt{-t^2+\vec{z}^2}\right)^3$ at short distance by power counting, which, combined with Eq.~(\ref{eq:hatM}), gives the scaling behavior of Eq.~(\ref{eq:hatMSD}). The separation line between the large distance and short distance limit is around
\begin{align}
m_Q\sqrt{t^2-|\vec{z}|^2} \sim 1 \ .
\end{align}
As such, we can now bound more realistically as 
\begin{align}
&\bigg|f(t\pm i0,\vec{z})\bigg|\le \frac{C_1}{m_Q^{\frac{3}{2}}(t^2-|\vec{z}|^2)^{\frac{3}{4}}} +C_1|\ln^{\gamma}(t^2-|\vec{z}|^2)| \ , \sqrt{t^2-|\vec{z}|^2} \le \frac{1}{m_Q} \ , \label{eq:neq1} \\
& \bigg|f(t\pm i0,\vec{z})\bigg| \le \tilde C_2 \ , \frac{1}{m_Q}\le \sqrt{t^2-|\vec{z}|^2}\le |\vec{z}|\left(\frac{1}{m_Q |\vec{z}|}\right)^{\frac{1}{2}-\eta} \ , \label{eq:neq2} \\
&\bigg|f(t \pm i0,\vec{z})\bigg|\le \frac{C_2}{m_Q\sqrt{t^2-|\vec{z}|^2}} , \  |\vec{z}|\left(\frac{1}{m_Q |\vec{z}|}\right)^{\frac{1}{2}-\eta}\le \sqrt{t^2-|\vec{z}|^2}\le |\vec{z}| \ , \label{eq:neq3} \\
&\bigg|f(t \pm i0,\vec{z})\bigg|\le \frac{C_3}{m_Q}, |\vec{z}|\le \sqrt{t^2-|\vec{z}|^2} \ . \label{eq:neq4}
\end{align}
In those inequalities, $|\vec{z}|$ is treated as a distance much larger than $1/m_Q$, which means HQET reduction works for $\sqrt{t^2-|\vec{z}|^2} \sim |\vec{z}|$ or larger. For $\sqrt{t^2-|\vec{z}|^2} \geq |\vec{z}|$, the power correction in Eq.~(\ref{eq:MatQGL}) can be estimated as $O(\Lambda_{\rm QCD}/m_Q)$, which provides Eq.~(\ref{eq:neq4}). For $|\vec{z}|\left(m_Q |\vec{z}|\right)^{\eta-\frac{1}{2}} \le \sqrt{t^2-|\vec{z}|^2}\le |\vec{z}|$, one needs to consider the distance dependence in the power correction, ending up with Eq.~(\ref{eq:neq3}). Note that $|\vec{z}|\left(m_Q |\vec{z}|\right)^{\eta-\frac{1}{2}}$ is a semi-hard scale appearing in perturbative calculations, which denotes the smallest distance for HQET reduction to hold. For the $\sqrt{t^2-|\vec{z}|^2} \le 1/m_Q$, the HQET reduction does not hold, and the difference function is bounded by the short-distance scaling behaviors as shown in Eq.~(\ref{eq:neq1}). Within the gap of the two scales $1/m_Q$ and $|\vec{z}|\left(m_Q |\vec{z}|\right)^{\eta-\frac{1}{2}}$, we choose the maximum value $\tilde{C_2}$ in this region, shown as the bound in Eq.~(\ref{eq:neq2}). In the above, $C_i$ are constants whose dependencies on $\vec{z}$ and $P$ are bounded by power laws, but depend on $m_Q$ only up to a finite number of logarithms. Moreover, to avoid collision of singularities, we can introduce 
\begin{align}
g(w,\vec{z})=\frac{f(w,\vec{z})}{\sqrt{-w^2+2|\vec{z}|^2}}e^{-(m_0+\epsilon)\sqrt{-w^2+|z|^2}} \ , 
\end{align}
without changing the dispersion relation. Then, one has
\begin{align}
&\bigg|g(0,\vec{z})\bigg| \le \frac{2C_1}{\pi}\int_0^{\frac{1}{m_Q}} \frac{sds}{(s^2+|\vec{z}|^2)\sqrt{|\vec{z}|^2-s^2}} \bigg(\frac{1}{m_Q^{\frac{3}{2}}s^{\frac{3}{2}}}+\ln^{\gamma}\frac{1}{s}\bigg) \nonumber \\ 
&+\frac{2\tilde C_2}{\pi}\int_{\frac{1}{m_Q}}^{ |\vec{z}|\left(\frac{1}{m_Q |\vec{z}|}\right)^{\frac{1}{2}-\eta}} \frac{sds} {(s^2+|\vec{z}|^2)\sqrt{|\vec{z}|^2-s^2}}+\frac{2C_2}{\pi}\int_{|\vec{z}|\left(\frac{1}{m_Q |\vec{z}|}\right)^{\frac{1}{2}-\eta}}^{|\vec{z}|} \frac{sds} {(s^2+|\vec{z}|^2)\sqrt{|\vec{z}|^2-s^2}} \frac{1}{m_Qs}\nonumber \\ 
&+\frac{2C_3 }{\pi m_Q}\int_{|\vec{z}|}^{\infty} \frac{sds} {(s^2+|\vec{z}|^2)\sqrt{-|\vec{z}|^2+s^2}}  \ ,
\end{align}
leading to
\begin{align}
|f(0,\vec{z})| \le \bigg(\frac{C_1'}{|\vec{z}|^2m_Q^2}+\frac{C_2'}{m_Q|\vec{z}|}+\frac{C_3'}{m_Q}\bigg)e^{m_0|\vec{z}|} + \hat C\left(\frac{1}{m_Q|\vec{z}|}\right)^{1-2\eta} e^{m_0|\vec{z}|} \ .  \label{eq:boundsimprov}
\end{align}
As far as $\eta<\frac{1}{2}$, the reduction still holds. In fact, the exponent $\frac{1}{2}$ corresponds to the scaling of the semi-hard scale, related to the fluctuations in the transverse directions. If the distance becomes much softer than the semi-hard scale, there should be no reason to expect violations to the reduction. As such, we expect that the $\eta$ can be made at least arbitrarily close to $\frac{1}{2}$. 
In fact, in one-loop perturbation theory, we found that $\eta$ can be chosen arbitrarily close to $-\frac{1}{2}$, suggesting negligible effects of the transition region.  

Finally, we should comment that possible subtleties in the large time region are easier to deal with: the condition in the $\sqrt{t^2-|\vec{z}|^2}>|\vec{z}|$ region can be further relaxed to include {\it power-law growths} 
\begin{align}
\bigg|f(t\pm i0,\vec{z})\bigg|\le \frac{C_3}{m_Q} (t^2-|\vec{z}|^2)^{\frac{n_0}{2}}, |\vec{z}|\le \sqrt{t^2-|\vec{z}|^2} \ , 
\end{align}
where $n_0\ge 0$ is a finite integer. To handle the possible large $t$ growth, we introduce the dressing function
\begin{align}
&D_{n_0}(w,\vec{z})=\prod_{i=1}^{n_0+1}\frac{1}{\sqrt{-w^2+(i+1)|\vec{z}|^2 }} \ , \\
&g_{n_0}(w,\vec{z})=D_{n_0}(w,\vec{z})f(w,\vec{z})e^{-(m_0+\epsilon)\sqrt{-w^2+|z|^2}} \ ,
\end{align}
and proceed with $g_{n_0}(w,|\vec{z}|)$. As far as the bounds in the small-distance region remain unchanged, the bound Eq.~(\ref{eq:boundsimprov}) remains essentially the same form (off-course, constants will be modified).

\section{Large distance behavior for correlator in $\phi^3$ theory}\label{sec:phi3}
In this section, we study the large-distance asymptotic behavior of the ``quasi-PDF matrix element" in $\phi^3$ theory, using perturbative calculations up to two-loop order. These results provide supporting evidence for multiple features of dispersive analysis in Secs.~\ref{sec:Disana} and~\ref{sec:Lorentz}, such as the physical origins of mass gaps, the impacts of form factor connectedness on the phases, power counting for polynomial dependences, cancellation and impact of forward singularity, and Lorentz symmetry to connect timelike and spacelike correlators. It also demonstrates the controlled precision of the large distance expansion truncated to a certain order, supporting the discussions in Sec.~\ref{sec:fitcriteria}.

\subsection{Definitions and notations}
The Lagrangian for $\phi^3$ theory,
\begin{align}
\mathcal{L} = \frac{1}{2} (\partial^{\mu} \phi)^2 - \frac{1}{2} m^2 \phi^2 - \frac{g}{3!} \phi^3 + \frac{1}{2} \delta_Z (\partial_\mu \phi)^2 - \frac{1}{2} \delta_m \phi^2 - \frac{\delta_g}{3!} \phi^3 \ ,
\end{align}
where $\delta_Z$, $\delta_m$ and $\delta_g$ are renormalization counterterms. $\delta_m$ is defined for $m$ to be a pole mass, and $\delta_Z, \delta_g$ are in $\overline{\rm MS}$ scheme. 

To mimic the quasi-PDF matrix element in QCD, the following correlator is constructed,
\begin{align}\label{eq:Mdefphi3}
M\left(\sqrt{-x^2}, \frac{i P\cdot x}{\sqrt{-x^2}}\right) \equiv Z^{-1}_{\phi} \langle P, {\rm out} |  \phi(x) \phi(0)  | P, {\rm in} \rangle_c \ ,
\end{align}
where the $| P \rangle$ is a large momentum single-particle state, with $P=(\sqrt{P_z^2+m^2},0,0,P^z)$ and $P^z \gg m$. They are defined as ``in" and ``out" states respectively so that the amplitude is a forward amplitude. $\phi(x) \phi(0)$ are scalar fields separated by $x$, which can be either timelike or spacelike. The disconnected diagrams are subtracted off as indicated by the lower script ``$c$" defined in Eq.~(\ref{eq:connected}), and the wavefunction diagram $Z_{\phi} = \langle P | \phi |\Omega \rangle \langle \Omega | \phi  | P \rangle$ is eliminated out. The vacuum bubbles are not considered in the calculations.

The large distance ($x^2 \rightarrow \infty$ with fixed $P\cdot x/\sqrt{-x^2}$) asymptotic forms can be obtained using form factor representation\footnote{For the timelike case, we choose $t>0$ to do the dispersive analysis.},
\begin{align}\label{eq:Mdisphi3}
M\left(\sqrt{-x^2}, \frac{i P\cdot x}{\sqrt{-x^2}}\right) = Z^{-1}_{\phi} \sum_{n} \int d \, \Gamma_n(k) \, e^{i (P - k) \cdot x} \langle P, {\rm in} | \phi | n(k), {\rm out} \rangle \langle n(k), {\rm out} | \phi | P, {\rm in} \rangle |_{c} \ ,
\end{align}
where the intermediate particles are chosen as ``out" states. The external state on the left is switched from ``out" state to ``in" state, which causes no difference for a stable particle but makes the discussions more convenient~\cite{Collins_2023}. $\Gamma_n(k)$ is the multi-particle phase space and $k$ is the total momentum as defined in or around Eq.~(\ref{eq:PhaseSpace}). The form factor representation can be classified based on the connectedness of form factors,
\begin{align}\label{eq:Mclaphi3}
&M\left(\sqrt{-x^2}, \frac{i P\cdot x}{\sqrt{-x^2}}\right) 
= 
{\color{red} M_{X}\left(\sqrt{-x^2}, \frac{i P\cdot x}{\sqrt{-x^2}}\right)} \nonumber\\
&+ {\color{blue} M_{Y}\left(\sqrt{-x^2}, \frac{i P\cdot x}{\sqrt{-x^2}}\right)} 
+ {\color{orange} M_{Z}\left(\sqrt{-x^2}, \frac{i P\cdot x}{\sqrt{-x^2}}\right)} \ ,
\end{align}
which follows the same convention as Eq.~(\ref{eq:Mcla}).

\subsection{Perturbative results up to two-loop order}\label{sec:phi3pert}
In this subsection, we present perturbative results on the large-distance asymptotic behavior up to two-loop order, which provide supporting evidence for Lorentz symmetry in connecting timelike and spacelike correlators, for the methods to determine the physical origins of mass gaps, for the relation between form factor connectedness and phases, and for power counting of polynomial dependences.

The perturbative results for a timelike separation $x=(t,0,0,0)$ in the $t\rightarrow +\infty$ limit with fixed large $P^t$ up to two-loop order,
\begin{align}\label{eq:MLtphi3}
M\left(i t, P^t\right)_{t\rightarrow+\infty} 
= & \left[ M^{(0)}\left(i t, P^t\right) 
+ \frac{g^2}{(2\pi)^2} M^{(1)}\left(i t, P^t\right) 
+ \frac{g^4}{(2\pi)^4}  M^{(2)}\left(i t, P^t\right) \right] \nonumber\\
&\times \left[1+O\left(\frac{1}{i m t}\right)\right] 
+ O\left(e^{-\frac{\sqrt{3}}{2} | P^z t |}\right) +  O\left(g^6\right) \ ,
\end{align}
where the tree-level result is,
\begin{align}
M^{(0)}\left(i t, P^t\right) = e^{i t P^{t}} + e^{-i t P^t} \ .
\end{align}
The one-loop result is (colors distinguish the contributions from different form factor representations: {\color{red} red for $M_{X}$}, {\color{blue} blue for $M_{Y}$} and {\color{orange} orange for $M_{Z}$}),
\begin{align}\label{eq:MLt(1)phi3}
&M^{(1)}\left(i t, P^t\right) = 
{\color{red} \frac{e^{i t P^{t}} e^{-i m t}}{(i m t)^{3/2}} \frac{\sqrt{\pi/2}}{(m-2 P^t)^2} } \nonumber\\
&+ {\color{blue} \frac{e^{-i m t}}{(i m t)^{1/2}} \frac{\sqrt{\pi/2}(4P_t^2+m^2)}{2m^2(4P_t^2-m^2)} } 
+ {\color{orange} \frac{e^{-i t P^{t}} e^{-i m t}}{(i m t)^{3/2}} \frac{\sqrt{\pi/2}}{(m+2 P^t)^2} } \ ,
\end{align}
which is evaluated in both covariant perturbation theory and form factor representation, and the results are consistent between the two methods. 
The two-loop result is calculated using the form factor representation under stationary phase analysis,
\begin{align}\label{eq:MLt(2)phi3}
&M^{(2)}\left(i t, P^t\right) \nonumber\\
&= {\color{red} \frac{e^{i t P^{t}} e^{-i m t}}{(i m t)^{3/2}} \frac{2 \sqrt{2} \pi ^{5/2}}{\left(m-2 P^t\right)^2} \left[ 2 I_{\rm tr}(2m(m-P^t)) - 2 I_{\rm se}(2m(m-P^t)) - I_{\rm se}(m^2) \right] } \nonumber\\
&{\color{red}+\frac{e^{i t P^{t}} e^{-i 2 m t}}{(i m t)^{3}} \frac{\left(7 m-2 P^t\right)^2 \pi }{576 m^2 \left(m-2P^t\right)^2 \left(m-P^t\right)^2}}  \nonumber\\
&{\color{blue}+\frac{e^{-i m t}}{(i m t)^{1/2}} \sqrt{2} \pi^{5/2} \, C(m P^t) +\frac{e^{-i 2 m t}}{(i m t)^{2}}\frac{\left(4 P_t^2+m^2\right)\pi }{36 m^4 \left(4P_t^2-m^2\right)} }  \nonumber\\
&{\color{orange}+\frac{e^{-i t P^{t}} e^{-i m t}}{(i m t)^{3/2}} \frac{2 \sqrt{2} \pi ^{5/2}}{\left(m+2 P^t\right)^2} \left[ 2 I_{\rm tr}(2m(m+P^t)) - 2 I_{\rm se}(2m(m+P^t)) - I_{\rm se}(m^2) \right] } \nonumber\\
&{\color{orange}+\frac{e^{-i t P^{t}} e^{-i 2 m t}}{(i m t)^{3}} \frac{\left(7 m+2 P^t\right)^2 \pi }{576 m^2 \left(m+2P^t\right)^2 \left(m+P^t\right)^2}} \ ,
\end{align}
where the coefficient $C(k\cdot P)$ is,
\begin{align}\label{eq:C}
C(k\cdot P) = &
- I_{\rm cbox}(2m^2+2 k\cdot P) - I_{\rm cbox}(2m^2-2 k\cdot P) \nonumber\\
&- I_{\rm box}(2m^2+2 k\cdot P) - I_{\rm box}(2m^2-2 k\cdot P) \nonumber\\
&-\frac{2 I_{\rm tr}(2m^2+2 k \cdot P)}{m^2+2 k \cdot P} -\frac{2 I_{\rm tr}(2m^2-2 k \cdot P)}{m^2-2 k \cdot P} - \frac{2I_{\rm tr0}(m^2)}{m^2} \nonumber\\
& +\frac{I_{\rm se}(2m^2+2 k\cdot P)}{m^2+2 k\cdot P} +\frac{I_{\rm se}(2m^2-2 k\cdot P)}{m^2-2 k\cdot P} +\frac{4 m^2 I_{\rm se}(m^2)}{m^4-4 (k\cdot P)^2} -\frac{I_{\rm se}(0)+2I_{\rm se}(m^2)}{m^2} \ ,
\end{align}
where the master integrals are,
\begin{align}\label{eq:Icbox}
&I_{\text{cbox}}(s=(P+k)^2) =  i  \int \frac{d^4 q}{(2\pi)^4} \frac{1}{q^2-m^2} \frac{1}{(q-P)^2-m^2} 
\nonumber\\
&\times\frac{1}{(q-k)^2-m^2} \frac{1}{(q-k-P)^2-m^2} \bigg|_{k^2=m^2,P^2=m^2} \ ,
\end{align}

\begin{align}\label{eq:Ibox}
I_{\rm box}(s=(P+k)^2) = i \int \frac{d^4 q}{(2\pi)^4}  \frac{1}{(P-q+k)^2-m^2} \left(\frac{1}{\left(P-q\right)^2-m^2}\right)^2 \frac{1}{q^2-m^2} \bigg|_{k^2=m^2,P^2=m^2} \ ,
\end{align}

\begin{align}\label{eq:Itr}
I_{\rm tr}(t=(P-k)^2)
= i \int \frac{d^4 q}{(2\pi)^4} \frac{1}{(q-k)^2-m^2}\frac{1}{q^2-m^2} \frac{1}{(q-P)^2-m^2} \bigg|_{k^2=m^2,P^2=m^2} \ ,
\end{align}

\begin{align}\label{eq:Itr0}
I_{\rm tr0}(k^2) = \int \frac{d^4 q}{(2\pi)^4} \frac{i}{(k-q)^2-m^2} \left(\frac{i}{q^2-m^2}\right)^2  \ ,
\end{align}
\begin{align}\label{eq:Isedef}
I_{\rm se}(k^2) = \frac{i}{k^2-m^2} \frac{1}{2} \left( \frac{\mu^2 e^{\gamma_E}}{4\pi} \right)^{\epsilon} \int \frac{d^d q}{(2\pi)^d} \frac{i}{(k-q)^2-m^2} \frac{i}{q^2-m^2} - \frac{k^2 \delta^{(1)}_Z-\delta^{(1)}_m}{k^2-m^2} \ ,
\end{align}
where the renormalization contourterms at one-loop is,
\begin{align}
\delta^{(1)}_Z = 0, \quad
\delta^{(1)}_{m} = \frac{1}{32\pi^2} \left(- \frac{1}{\epsilon_{\rm UV}} + \ln\frac{m^2}{\mu^2} - 2 + \frac{\pi}{\sqrt{3}}\right) \ .
\end{align}
The three-point and four-point scalar integrals $I_{\rm tr}$, $I_{\rm box}$, and $I_{\rm cbox}$ have been studied in Refs.~\cite{THOOFT1979365,DENNER1991637,Denner:1991kt,Denner:2010tr}. In Eq.~(\ref{eq:MLtphi3}), for each type of exponential behavior, we only keep the leading polynomial dependence and the omitted higher order terms are denoted as $O\left(1/(imt)\right)$. The exponential decaying terms regarding the single-particle on-shell production in the form factors are ignored because they are at $O\left(e^{-\frac{\sqrt{3}}{2} |P^z t|}\right)$ which are much more suppressed compared to other terms for a large momentum external state. All the propagators in the master integrals are Feynman propagators with $+ i \epsilon $ prescription. 

The perturbative calculations have also been performed for a spacelike separation $x=(0,0,0,z)$ in the $z\rightarrow +\infty$ limit with fixed large $P^z$,
\begin{align}\label{eq:MLzphi3}
M\left(z, -i P^z\right)_{z\rightarrow+\infty} 
= & \left[ M^{(0)}\left(z, -i P^z\right) 
+ \frac{g^2}{(2\pi)^2} M^{(1)}\left(z, -i P^z\right) 
+ \frac{g^4}{(2\pi)^4}  M^{(2)}\left(z, -i P^z\right) \right] \nonumber\\
&\times \left[1+O\left(\frac{1}{m z}\right)\right] 
+ O\left(e^{-\frac{\sqrt{3}}{2} \sqrt{P_z^2+m^2} |z| }\right) +  O\left(g^6\right) \ ,
\end{align}
where the one-loop result is,
\begin{align}\label{eq:MLz(1)phi3}
&M^{(1)}\left(z, -i P^z\right) =  {\color{red} \frac{e^{- i P^z z - m z}}{ (m z)^{3/2}} \frac{\sqrt{\pi/2}}{\left(m+2 i P^z\right)^2} } \nonumber\\  
&+ {\color{blue} \frac{e^{-m z}}{\sqrt{m z}} \frac{\sqrt{\pi/2} \left(-m^2+4 P_z^2\right)}{2 m^{2}\left(m^2+4 P_z^2\right)}} 
+  {\color{orange} \frac{e^{i P^z z} e^{-m z}}{(m z)^{3/2}} \frac{\sqrt{\pi/2}}{ \left(m-2 i P^z\right)^2} }  \ ,
\end{align}
and the two-loop result is,
\begin{align}\label{eq:MLz(2)phi3}
&M^{(2)}\left(z, -i P^z\right) \nonumber\\
&=  {\color{red} \frac{e^{-i z P^z} e^{-m z}}{(z m)^{3/2}} \frac{2 \sqrt{2} \pi ^{5/2}}{\left(m+2 i P^z\right)^2} \left( 2 I_{\rm tr}(2m(m+iP^z))-2I_{\rm se}(2m(m+iP^z)) - I_{\rm se}(m^2)  \right)  }
\nonumber\\
&{\color{red} + \frac{e^{-i z P^z} e^{-2 m z}}{(z m)^3} \frac{\pi  \left(7 m+2 i P^z\right)^2}{576 m^2 \left(m^2+3 i m P^z-2P_z^2\right){}^2} } \nonumber\\
&{\color{blue} + \frac{e^{-m z}}{\sqrt{z m}} \sqrt{2} \pi^{5/2} C(-i m P^z) 
 + \frac{e^{-2 m z}}{(z m)^2} \frac{\pi  \left(-m^2+4 P_z^2\right)}{36 m^4 \left(m^2+4 P_z^2\right)} } \nonumber\\
&{\color{orange}+\frac{e^{i z P^z} e^{-m z}}{(z m)^{3/2}} \frac{2 \sqrt{2} \pi ^{5/2}}{ \left(m-2i P^z\right)^2} \left( 2 I_{\rm tr}(2m(m-iP^z))-2I_{\rm se}(2m(m-iP^z)) - I_{\rm se}(m^2)  \right) }\nonumber\\
&{\color{orange} +\frac{e^{i z P^z} e^{-2 m z}}{(zm)^3} \frac{\pi  \left(7 m-2 i P^z\right)^2}{576 m^2 \left(m^2-3 i m P^z-2P_z^2\right){}^2}  } \ .
\end{align}
Please check the definitions for $I_{\rm tr}$ in Eq.~(\ref{eq:Itr}), $I_{\rm se}$ in Eq.~(\ref{eq:Isedef}), and $C$ in Eq.~(\ref{eq:C}).

The timelike and spacelike results in Eqs.~(\ref{eq:MLt(1)phi3}),~(\ref{eq:MLt(2)phi3}),~(\ref{eq:MLz(1)phi3}) and~(\ref{eq:MLz(2)phi3}) are consistent with the following Lorentz invariant forms,
\begin{align}\label{eq:MLx(1)phi3}
&M^{(1)}\left(\sqrt{-x^2}, \mathcal{P}_x \equiv \frac{i x \cdot P}{\sqrt{-x^2}} \right) = 
{\color{red} \frac{e^{i x \cdot P} e^{-m \sqrt{-x^2}}}{(m \sqrt{-x^2})^{3/2}} \frac{\sqrt{\pi/2}}{(m-2 \mathcal{P}_x)^2} } \nonumber\\
&+ {\color{blue} \frac{e^{-m \sqrt{-x^2}}}{(m \sqrt{-x^2})^{1/2}} \frac{\sqrt{\pi/2}(4 \mathcal{P}_x^2+m^2)}{2m^2(4\mathcal{P}_x^2-m^2)} } 
+ {\color{orange} \frac{e^{-i x \cdot P} e^{-m \sqrt{-x^2}}}{(m \sqrt{-x^2})^{3/2}} \frac{\sqrt{\pi/2}}{(m+2 \mathcal{P}_x)^2} } \ ,
\end{align}
\begin{align}\label{eq:MLx(2)phi3}
&M^{(2)}\left(\sqrt{-x^2}, \mathcal{P}_x \equiv \frac{i x \cdot P}{\sqrt{-x^2}} \right) \nonumber\\
&= {\color{red} \frac{e^{i x \cdot P} e^{-m \sqrt{-x^2}}}{(m \sqrt{-x^2})^{3/2}} \frac{2 \sqrt{2} \pi ^{5/2}}{\left(m-2 \mathcal{P}_x\right)^2} \left[ 2 I_{\rm tr}(2m(m-\mathcal{P}_x)) - 2 I_{\rm se}(2m(m-\mathcal{P}_x)) - I_{\rm se}(m^2) \right] } \nonumber\\
&{\color{red}+\frac{e^{i x \cdot P} e^{-2 m \sqrt{-x^2}}}{(m \sqrt{-x^2})^{3}} \frac{\left(7 m-2 \mathcal{P}_x\right)^2 \pi }{576 m^2 \left(m-2\mathcal{P}_x\right)^2 \left(m-\mathcal{P}_x\right)^2}}  \nonumber\\
&{\color{blue}+\frac{e^{-m \sqrt{-x^2}}}{(m \sqrt{-x^2})^{1/2}} \sqrt{2} \pi^{5/2} \, C(m \mathcal{P}_x) +\frac{e^{-2m \sqrt{-x^2}}}{(m \sqrt{-x^2})^{2}}\frac{\left(4 \mathcal{P}_x^2+m^2\right)\pi }{36 m^4 \left(4\mathcal{P}_x^2-m^2\right)} }  \nonumber\\
&{\color{orange}+\frac{e^{-i x \cdot P} e^{-m \sqrt{-x^2}}}{(m \sqrt{-x^2})^{3/2}} \frac{2 \sqrt{2} \pi ^{5/2}}{\left(m+2 \mathcal{P}_x\right)^2} \left[ 2 I_{\rm tr}(2m(m+\mathcal{P}_x)) - 2 I_{\rm se}(2m(m+\mathcal{P}_x)) - I_{\rm se}(m^2) \right] } \nonumber\\
&{\color{orange}+\frac{e^{-i x \cdot P} e^{-2 m \sqrt{-x^2}}}{(m \sqrt{-x^2})^{3}} \frac{\left(7 m+2 \mathcal{P}_x\right)^2 \pi }{576 m^2 \left(m+2\mathcal{P}_x\right)^2 \left(m+\mathcal{P}_x\right)^2}} \ ,
\end{align}
which justifies the usage of Lorentz symmetry in Eq.~(\ref{eq:qPDFLzGen}) in the main text. Also see the discussions below Eq.~(\ref{eq:DQ}) on the prescription of $\sqrt{-x^2}$. 

In the above results, the mass gaps in the exponential factors are $m$ and $2m$, which are the threshold masses of one-particle and two-particle intermediate states, respectively. They are covered by the spectra of the quantum numbers $\phi |P\rangle$ for {\color{red}$M_{X}$}, $\phi |\Omega\rangle$ for {\color{blue}$M_{Y}$}, and $\phi |\bar{P} \rangle$ for {\color{orange}$M_{Z}$}, that appear at one loop and two loop. Therefore, these results substantiate the methods to identify the physical origins of mass gaps in Eqs.~(\ref{eq:MXSP}),~(\ref{eq:MY}),~(\ref{eq:MZSP}), and~(\ref{eq:Mexp}). 

Consider the overall phase factors associated with the external momentum $P$. The factor {\color{red}$\sim e^{i x \cdot P}$} appears in the fully connected form factor representation {\color{red}$M_{X}$}. If one form factor is connected and the other one is not, as in {\color{blue}$M_{Y}$}, there is no overall phase factor involving the external momentum. The factor {\color{orange}$\sim e^{-i x \cdot P}$} corresponds to {\color{orange}$M_{Z}$}, where both form factors are disconnected. Consequently, these perturbative calculations provide supporting evidence for the relation between form factor connectedness ({\color{red}$M_{X}$},{\color{blue}$M_{Y}$},{\color{orange}$M_{Z}$}) and phases ({\color{red}$e^{i x \cdot P}$}, {\color{blue}$1$} or {\color{orange}$e^{-i x \cdot P}$}) in Eqs.~(\ref{eq:MXSP}),~(\ref{eq:MY}),~(\ref{eq:MZSP}), and~(\ref{eq:Mexp}). 

Regarding the polynomial dependences on $\sqrt{-x^2}$, the asymptotic behaviors of {\color{red}$M_{X}$} and {\color{orange}$M_{Z}$}, including one-particle and two-particle insertions, agree with the naive power counting $k_i\sim \left(m_i,\sqrt{m_i/t},\sqrt{m_i/t},\sqrt{m_i/t}\right)$ explained below Eq.~(\ref{eq:MXSP}). In the case of {\color{blue}$M_{Y}$} here, the polynomial dependence is linearly enhanced as a consequence of the forward singularity, while this enhancement does not appear in Eq.~(\ref{eq:MY}) in the main text because it is forbidden by the spin quantum numbers as explained around Eq.~(\ref{eq:MYsfLt}). Further discussions on the forward singularity will be provided in the next subsection.

\subsection{Cancellation and impact of forward singularity}\label{sec:phi3FWSG}
In this subsection, we explicitly show how the forward singularity is canceled in a representative one-loop diagram, thereby substantiating the discussion surrounding Eq.~(\ref{eq:MYsgcancel}). We further interpret the resulting enhanced polynomial dependence as the remnant of the forward singularity, providing additional support for Eq.~(\ref{eq:MYsgLint}).

\begin{figure}
    \centering
    \includegraphics[width=0.3\linewidth]{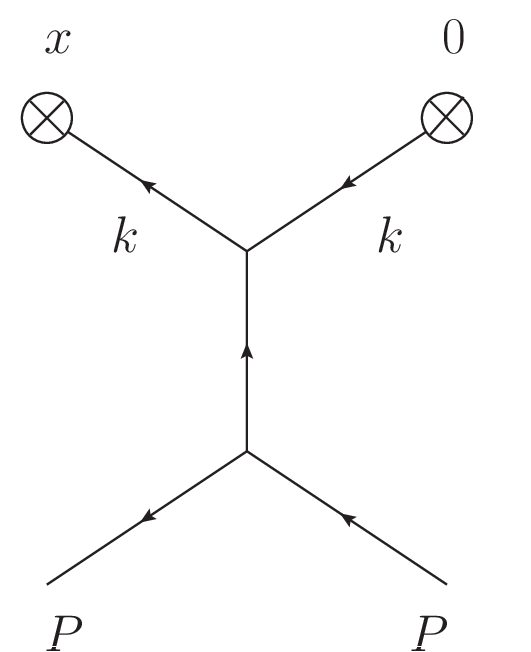}
    \caption{The one-loop example $I_{t}$ in covariant perturbation theory for ``quasi-PDF matrix element" in $\phi^3$ theory. }
    \label{fig:phi3_Is_co}
\end{figure}

\begin{figure}
    \centering
    \includegraphics[width=0.6\linewidth]{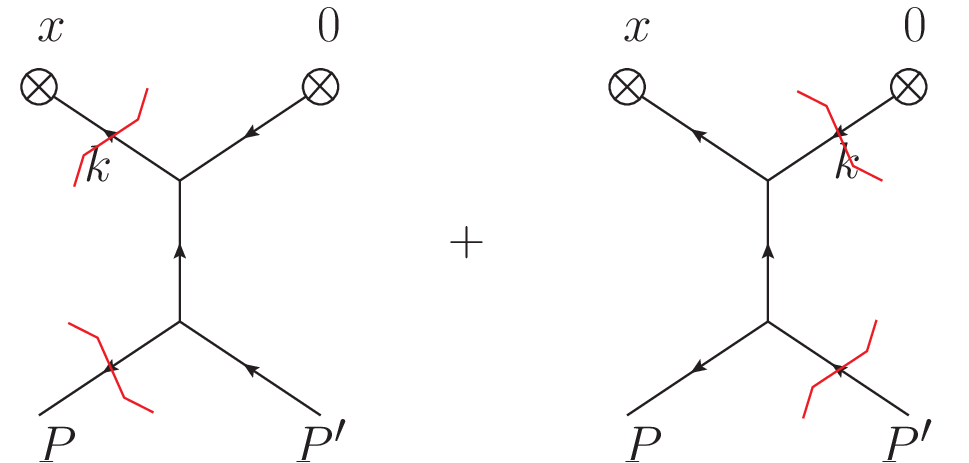}
    \caption{The form factor representation of Fig.~\ref{fig:phi3_Is_co}, which belongs to $M_{Y}$ in Eq.~(\ref{eq:Mclaphi3}), where one form factor is connected and the other one is disconnected. }
    \label{fig:phi3_Is_form}
\end{figure}
Consider the ``$t$-channel" diagram at one-loop displayed in Fig.~\ref{fig:phi3_Is_co}. In covariant perturbation theory, the corresponding integral is
\begin{align}\label{eq:IsCP}
&I_{t}\left(\sqrt{-x^2}\right) = - g^2 \int \frac{d^4 k}{(2\pi)^4} e^{-i k \cdot x} \left(\frac{i}{k^2-m^2+ i \epsilon} \right)^2 \frac{i}{- m^2+ i \epsilon} = \frac{g^2 K_0\left(m \sqrt{-x^2}\right)}{m^2 \left(8 \pi^2\right)} \ ,
\end{align}
which can be evaluated directly in a Lorentz-invariant manner using Schwinger parameters. The final result is finite and can be written in terms of the modified Bessel function of the second kind, $K_0$.

However, its form factor representation contains a potential singularity originating from the quadratic term $\left[i/(k^2-m^2+ i \epsilon) \right]^2$ due to the symmetric structure of the Feynman diagram. When one attempts to derive the form-factor representation by cutting one of the propagators using the Cutkosky cutting rule, the other one becomes singular. Such singularity has also been studied in Ref.~\cite{Collins_2023}. 

To regulate this singularity, one introduces off-forwardness, which takes the external momenta to be different,
\begin{align}
\langle P, {\rm out} | \phi(x) \phi(0) | P, {\rm in} \rangle_c \rightarrow \langle P, {\rm out} | \phi(x) \phi(0) | P', {\rm in} \rangle_c \ .
\end{align}
Then, the form factor representation, in Fig.~\ref{fig:phi3_Is_form}, can be obtained by directly applying Feynman rules of form factors, 
\begin{align}\label{eq:It}
&I_{t}\left(\sqrt{-x^2}\right) =  \frac{g^2}{(P'-P)^2-m^2} \int \frac{d^{3} \vec{k}}{(2\pi)^{3}}  \frac{e^{-i k \cdot x}}{2 E_{k}}  \frac{1}{(E_{P'}-E_{P}-E_{k})^2-E_{k+P-P'}^2+i\epsilon} \nonumber\\
& +  \frac{g^2 e^{i (P-P') \cdot x}}{(P'-P)^2-m^2} \int \frac{d^{3} \vec{k}}{(2\pi)^{3}}  \frac{e^{-i k \cdot  x}}{2 E_{k}}  \frac{1}{(E_{P}-E_{P'}-E_{k})^2-E_{k+P'-P}^2-i\epsilon}  \ ,
\end{align}
where $E_{k} \equiv \sqrt{\vec{k}^2+m^2}$. The above expression contains two terms, and each of them becomes singular in the forward limit $P' \rightarrow P$, which is referred to as forward singularity in our terminology. However, we will demonstrate the cancellation of this forward singularity between the two terms. Also note that the signs of the $i \epsilon$ prescription originate from the time-ordering regarding the ``in" and ``out" states in Eq.~(\ref{eq:Mdisphi3}), which are essential for verifying this cancellation.

We begin by examining the timelike case $x=(t,0,0,0)$. After changing integral variables and recombining terms, one obtains
\begin{align}
&I_{t}\left(i t\right) = \frac{g^2}{(P'-P)^2-m^2} \int \frac{d^{3} \vec{k}}{(2\pi)^{3}} \frac{e^{-i E_k t}}{2 E_{k}} 
\left[ \frac{e^{-i\left(E_{k+P'-P}-E_k\right)t}-e^{i\left(E_P-E_{P'}\right) t}}{2 E_{k+P'-P}\left(E_P-E_{P'}+E_{k+P'-P}-E_k+i \epsilon \right)}  \right.\nonumber\\
&\left.+\frac{1}{2 E_{P'-P-k}\left(E_{P'}-E_P-E_k-E_{P'-P-k}\right)} + \frac{e^{i\left(E_P-E_{P'}\right) t}}{2 E_{P-P'-k}\left(E_P-E_{P'}-E_k-E_{P-P'-k}\right)} \right] \ ,
\end{align}
where, in the first term in the square brackets, the forward singularity is canceled between the numerator and the denominator. The forward limit $P'\rightarrow P$ can then be smoothly taken,
\begin{align}\label{eq:Istl}
&I_{t}\left(i t\right)  = \frac{g^2}{-m^2} \int \frac{d^{3} \vec{k}}{(2\pi)^{3}} \frac{e^{-i E_k t}}{2 E_{k}} 
\left[ \frac{-i t}{2 E_{k}}  - \frac{1}{2 E^2_{k}} \right] \ ,
\end{align}
where the factor $-i t$ originates from the expansion of the phase factor $e^{-i\left(E_{k+P'-P}-E_k\right)t}-e^{i\left(E_P-E_{P'}\right) t}$, which cancels the singularity from the denominator and enhances the polynomial dependence compared to naive power counting.  

The cancellation of forward singularity can also be studied in a spacelike case $x=(0,0,0,z)$, where the expression can be simplified by redefining the integral variables,
\begin{align}
&I_{t}\left(z\right) =  \frac{g^2}{(P'-P)^2-m^2} \int \frac{d^{3} \vec{k}}{(2\pi)^{3}}  \frac{e^{-i P^z z} e^{i k^z z}}{2 E_{k-P} 2 E_{k-P'}} \nonumber\\
&\times  \left[ \frac{1}{E_{P'}-E_P-E_{k-P}-E_{k-P'}}-\frac{1}{E_{P'}-E_P-E_{k-P}+E_{k-P'}-i \epsilon } \right.\nonumber\\
&\left. \quad + \frac{1}{E_P-E_{P'}-E_{k-P'}-E_{k-P}}-\frac{1}{E_P-E_{P'}-E_{k-P'}+E_{k-P}+i \epsilon } \right] ,
\end{align}
in which the forward singularity is exactly canceled between the second and fourth terms in the square bracket. One can then take the forward limit $P'\rightarrow P$,
\begin{align}\label{eq:Issl}
&I_{t}\left(z\right)   =  \frac{g^2}{-m^2} \int \frac{d^{3} \vec{k}}{(2\pi)^{3}}  \frac{ e^{i k^z z}}{2 E_{k}}  \left[ \frac{-1}{2 E^2_{k}} \right] \ ,
\end{align}
where the energy denominator exhibits a more singular behavior at $|\vec{k}|= \pm i m$ compared to the $M_{X}$ and $M_{Z}$ cases, which causes the enhancement of the polynomial dependence, as we will show in detail as the last example in the next subsection.

\subsection{Calculation methods under form factor representation}\label{sec:CalMets}
In this subsection, we present our methods for calculating the large distance asymptotic forms in Eqs.~(\ref{eq:MLtphi3}) and~(\ref{eq:MLzphi3}), using selected one-loop diagrams as examples. This subsection is not just a collection of technical details but also a pedagogical journey deepening our understanding of the origins of asymptotic forms.

\begin{figure}
    \centering
    \includegraphics[width=0.3\linewidth]{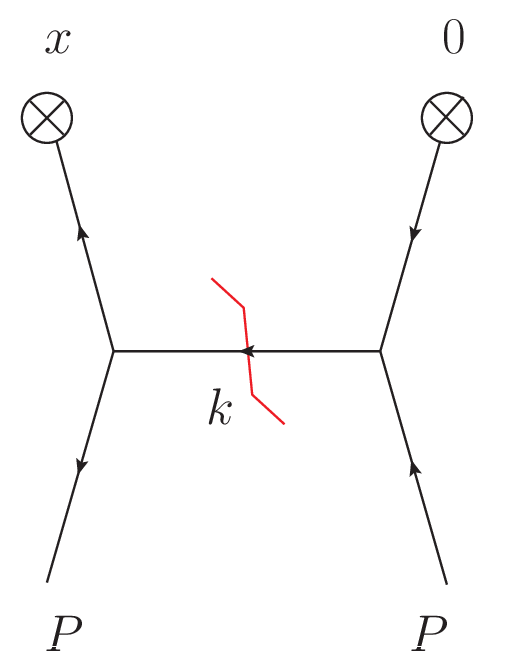}
    \caption{The one-loop example $I_{s,1}$ belonging to $M_{X}$ in Eq.~(\ref{eq:Mclaphi3}), where both form factors are connected.}
    \label{fig:phi3_It1_form}
\end{figure}
We first consider a one-loop example associated with $M_{X}$ in Eq.~(\ref{eq:Mclaphi3}),
\begin{align}\label{eq:Is1}
I_{s,1}\left(\sqrt{-x^2}, \frac{i P\cdot x}{\sqrt{-x^2}}\right)
= g^2 \int \frac{d^3 \vec{k}}{(2\pi)^3} \frac{e^{i (P-k)\cdot x}}{2 E_k} \left[ \frac{1}{(E_k-E_P)^2-E^2_{k-P}} \right]^2 \ ,
\end{align}
with the corresponding Feynman diagram displayed in Fig.~\ref{fig:phi3_It1_form}. The quantity $E_k$ is defined below Eq.~(\ref{eq:It}).

For a timelike separation $x=(t,0,0,0)$, the integral becomes,
\begin{align}
I_{s,1}\left(i t, P^t\right)
= \frac{g^2 e^{i P^t t}}{(2\pi)^2} \int \frac{d\Omega}{4\pi}\int_{0}^{+\infty} d k \, 2k^2 \frac{e^{-i E_k t}}{2 E_k} \left[ \frac{1}{(E_k-E_P)^2-E^2_{k-P}} \right]^2    \ .  
\end{align}
In the large time limit, the leading asymptotics can be obtained using the stationary phase analysis,
\begin{align}\label{eq:It1calc}
I_{s,1}\left(i t, P^t\right)_{t\rightarrow+\infty} = & \frac{g^2 e^{i P^t t}}{(2\pi)^2} \int \frac{d\Omega}{4\pi}\int_{0}^{+\infty} d k \, 2k^2 \frac{e^{-i m t - i \frac{k^2}{2m} t}}{2 m} \left[ \frac{1}{(m-P^t)^2-P_t^2} \right]^2 \nonumber\\
&\times \left[1+O\left(k P^z \cos(\theta)\right) + O\left(k^2\right) \right] \ ,
\end{align}
where, in the exponent, we have retained the non-relativistic expansion up to the kinetic energy term $k^2/(2m)$. Here $\theta$ denotes the angle between $\vec{k}$ and $\vec{P}$. Before doing this integral explicitly, we anticipate an overall exponential factor of the form $\sim e^{i P^t t} e^{-i m t}$, related to the energy $m$ at the stationary point. Spatial fluctuations around this stationary point scale as $\sim\left(\sqrt{m/t},\sqrt{m/t},\sqrt{m/t}\right)$, which determines the polynomial dependence on $t$; by simple power counting in $k$, the leading contribution scales as $1/t^{3/2}$.  The remaining $k$-dependence in the integrand is expanded order by order, which corresponds to various higher-order polynomial contributions in the large $t$ expansion. Integrating out $k$ and $\Omega$ gives,
\begin{align}\label{eq:Is1timelike}
I_{s,1}\left(i t, P^t\right)_{t\rightarrow+\infty} = \frac{g^2}{(2\pi)^2} \frac{e^{i P^t t}e^{-i m t}}{(i m t)^{3/2}} \frac{\sqrt{\pi/2}}{(m-2 P^t)^2} \left[1+O\left(\frac{1}{i m t}\right)\right] \ ,
\end{align}
where the leading exponential and polynomial dependences agree with our expectations. The subleading term is suppressed by $O(1/(imt))$ instead of $O(1/\sqrt{i m t})$ because the $O(k P^z \cos{\theta})$ term vanishes after we integrate out the solid angle, a consequence of rotational symmetry.

\begin{figure}
    \centering
    \includegraphics[width=0.475\linewidth]{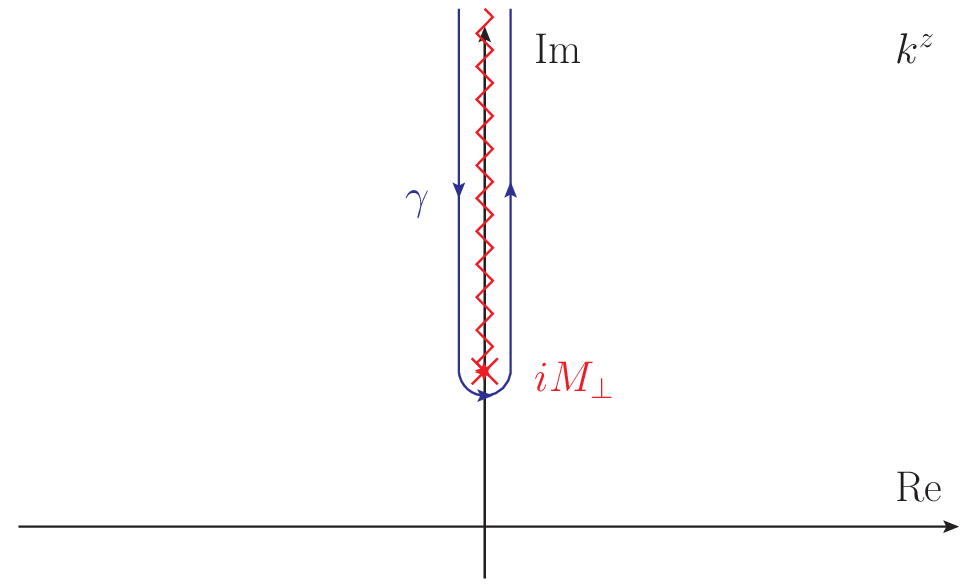}
    \quad
    \includegraphics[width=0.475\linewidth]{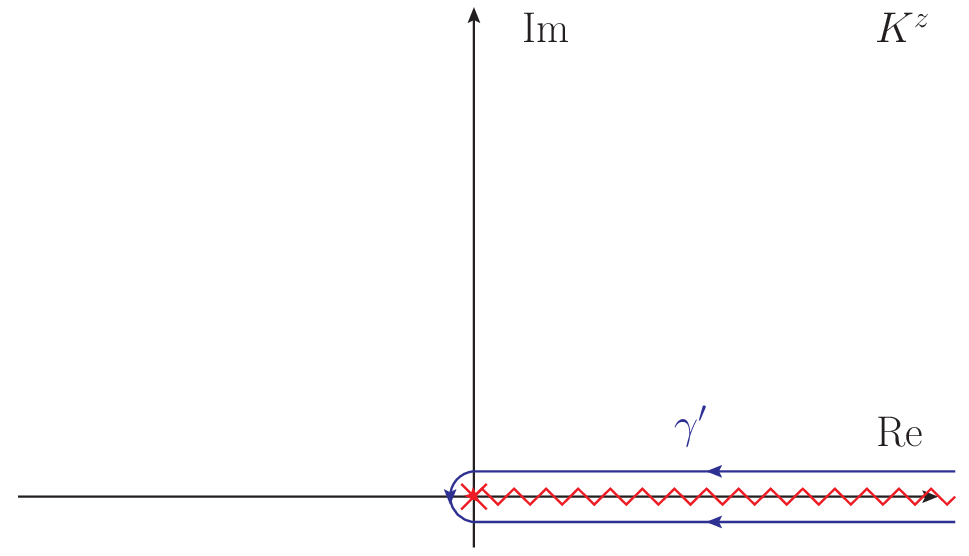}
    \caption{Integral contour in the complex plane of $k^z$ or $K^z$ for evaluating the large distance asymptotic form. }
    \label{fig:phi3_It1_sl}
\end{figure}
For a spacelike separation $x=(0,0,0,z)$, Eq.~(\ref{eq:Is1}) becomes,
\begin{align}\label{eq:It1sl}
I_{s,1}\left(z, -i P^z\right)
= \frac{g^2 e^{-i P^z z}}{(2\pi)^2} \int_{m}^{+\infty} d M_{\perp} \int_{-\infty}^{+\infty} d k^z \, M_{\perp}  e^{i k^z z} f_{s,1}(k^z,M_{\perp})  \ ,  
\end{align}
where
\begin{align}
f_{s,1}(k^z,M_{\perp}) \equiv \frac{1}{2 E_k} \left[ \frac{1}{(E_k-E_P)^2-E^2_{k-P}} \right]^2 \ ,
\end{align}
and $M_{\perp} \equiv \sqrt{m^2 + k_x^2 + k_y^2}$ represents the effective mass associated with the transverse momentum. To evaluate the results for $z \rightarrow +\infty$, we deform the $k^z$ integral contour upward in the complex plane around the non-holomorphic structures of the integrand. The energy denominator $1/E_k$ has a branch cut starting at $k^z=i M_{\perp}$ and extending upward along the imaginary axis, which leads to the exponential behavior $\sim e^{-i P^z z} e^{- m z}$. In addition, the factor $\left[ 1/(E_k-E_P)^2-E^2_{k-P} \right]^2$ contains a second order residue pole located at $k^z=\frac{P^z}{2} + i \sqrt{\frac{M^2_{\perp}}{m^2}-\frac{1}{4}} P^t$, yielding the exponential factor $\sim e^{-i \frac{P^z}{2} z-\frac{\sqrt{3}}{2}P^t z}$. For $P^z \gg m$, the latter one is highly suppressed compared to the first one. Therefore, we focus on the first one, 
\begin{align}
I_{s,1}\left(z, -i P^z\right)_{z\rightarrow +\infty}
= \frac{g^2 e^{-i P^z z}}{(2\pi)^2} \int_{m}^{+\infty} d M_{\perp} \int_{\gamma} d k^z \, M_{\perp}  e^{i k^z z} f_{s,1}(k^z,M_{\perp}) + O\left(e^{-\frac{\sqrt{3}}{2}P^t z}\right) \ ,
\end{align}
where $\gamma$ is a contour around the branch cut of the energy denominator $1/\sqrt{k_z^2+M^2_{\perp}}$ as depicted in the left panel of Fig.~\ref{fig:phi3_It1_sl}. We now perform the change of variables $k^z \rightarrow i \left( \frac{K^z}{z} + M_{\perp}\right)$ followed by $M_{\perp} \rightarrow \frac{M_{\perp}}{z} + m$,
\begin{align}
I_{s,1}\left(z, -i P^z\right)_{z\rightarrow +\infty}
= \frac{g^2 e^{-i P^z z} e^{-m z}}{(2\pi)^2} \int_{0}^{+\infty} d M_{\perp} \int_{\gamma'} d K^z \, \frac{i}{z^2} \left(\frac{M_{\perp}}{z} + m\right)  \nonumber\\
\times e^{-K^z-M_{\perp}} f_{s,1}\left(i \left( \frac{K^z}{z} + \frac{M_{\perp}}{z} + m\right),\frac{M_{\perp}}{z} + m\right) + O\left(e^{-\frac{\sqrt{3}}{2}P^t z}\right) \ ,
\end{align}
where $\gamma'$ now denotes a contour along the real axis shown in the right panel of Fig.~\ref{fig:phi3_It1_sl}. The exponential behavior $\sim e^{-i P^z z} e^{-m z}$ has already been factorized out and is consistent with our expectation from the start of the branch cut, $\vec{k}\sim (0,0,im)$. To calculate the leading polynomial $z$ dependence, we perform the large $z$ expansion for the integrand,
\begin{align}\label{eq:Is1spacelike}
&I_{s,1}\left(z, -i P^z\right)_{z\rightarrow +\infty}
= \frac{g^2 e^{-i P^z z} e^{-m z}}{(2\pi)^2} \int_{0}^{+\infty} d M_{\perp} \int_{\gamma'} d K^z \, e^{-K^z-M_{\perp}} \nonumber\\
&\times \frac{i m}{z^2} \frac{\sqrt{z}}{2 \sqrt{2} \sqrt{-K^z m}} \frac{1}{\left(m^2+2 i m P^z\right)^2} \left[1+O\left(\frac{1}{m z}\right)\right] + O\left(e^{-\frac{\sqrt{3}}{2}P^t z}\right) \nonumber\\
&= \frac{g^2 e^{-i P^z z} e^{-m z}}{(2\pi)^2} \int_{0}^{+\infty} d M_{\perp} \int_{0}^{+\infty} d K^z \, e^{-K^z-M_{\perp}} \nonumber\\
&\times \frac{i m}{z^2} \frac{\sqrt{z}}{\sqrt{2} i \sqrt{K^z m}} \frac{1}{\left(m^2+2 i m P^z\right)^2} \left[1+O\left(\frac{1}{m z}\right)\right] + O\left(e^{-\frac{\sqrt{3}}{2}P^t z}\right) \nonumber\\
&= \frac{g^2}{(2\pi)^2} \frac{e^{-i P^z z} e^{-m z}}{(m z)^{3/2}} \frac{\sqrt{\pi/2}}{\left(m+2 i P^z\right)^2} \left[1+O\left(\frac{1}{m z}\right)\right] + O\left(e^{-\frac{\sqrt{3}}{2}P^t z}\right) \ ,
\end{align}
where the 3D phase space provides $1/z^2$ from spatial fluctuations $\sim \left(\sqrt{m/z},\sqrt{m/z},1/z\right)$ and the energy denominator $1/E_k$ yields $\sqrt{z/m}$ due to its singular structure at $k^z \sim i M_{\perp}$ or $K^z \sim 0$. 

Although the mathematical origins of exponential and polynomial behaviors seem quite different between the timelike and spacelike cases, the final results for the leading asymptotics can be cast in a Lorentz invariant format,
\begin{align}
I_{s,1}\left(\sqrt{-x^2}, \mathcal{P}_x \equiv \frac{i x \cdot P}{\sqrt{-x^2}} \right)_{\sqrt{-x^2}\rightarrow\infty} = \frac{g^2}{(2\pi)^2} \frac{e^{i P \cdot x}e^{-m\sqrt{-x^2}}}{(m\sqrt{-x^2})^{3/2}} \frac{\sqrt{\pi/2}}{(m-2 \mathcal{P}_x)^2} \left[1+O\left(\frac{1}{m \sqrt{-x^2}}\right)\right] \ .
\end{align}
Moreover, even the power counting in the two cases can be related by a complex Lorentz transformation~\cite{Streater:1989vi}. In the timelike result in Eq.~(\ref{eq:Is1timelike}), the exponential and polynomial $t$ dependences can be reproduced by power counting $k\sim\left(m+C/it,\sqrt{m/it},\sqrt{m/it},\sqrt{m/it}\right)$ for the integrand and $d k\sim\left(C/it,\sqrt{m/it} ,\sqrt{m/it},\sqrt{m/it}\right)$ for the phase space $\int d^3 \vec{k}$ in Eq.~(\ref{eq:Is1}). We also impose $k^2 - m^2 \sim 1/t^2$, implying that the momentum is nearly on-shell. The $O(1/t)$ in the time component does not affect the leading asymptotics. On the other hand, the exponential and polynomial $z$ dependences in the spacelike result in Eq.~(\ref{eq:Is1spacelike}) follows from power counting $k\sim\left(i\sqrt{m/z},\sqrt{m/z},\sqrt{m/z},i m +iC/z\right)$ and $d k\sim\left(i\sqrt{m/z},\sqrt{m/z},\sqrt{m/z},iC/z\right)$ in Eq.~(\ref{eq:Is1}). In this case, the $O\left(1/z\right)$ term does contribute to the leading asymptotics. The momentum modes in the spacelike case can be obtained from the timelike case via the following complex Lorentz transformation\footnote{This is to relate the Lorentz vector structures of the momentum modes. The Lorentz scalars $i t$ and $z$ are linked through the invariant $\sqrt{-x^2}$.},
\begin{align}
    \Lambda(\zeta) = \left(
\begin{array}{cccc}
 \cosh (\zeta) & 0 & 0 & \sinh (\zeta ) \\
 0 & 1 & 0 & 0 \\
 0 & 0 & 1 & 0 \\
 \sinh (\zeta ) & 0 & 0 & \cosh (\zeta ) \\
\end{array}
\right) \ ,
\end{align}
with $\zeta= i \pi/2$, which implies $\cosh (\zeta)=0$ and $\sinh (\zeta )=i$.

To gain further insight into how the forward singularity influences the polynomial dependence, we present the calculations for $I_{t}$ defined in Eq.~(\ref{eq:IsCP}), associated with $M_{Y}$ in Eq.~(\ref{eq:Mclaphi3}). The corresponding form factor representation in the timelike region is given in Eq.~(\ref{eq:Istl}). Its large time asymptotic behavior can be calculated using stationary phase analysis,
\begin{align}
&I_{t}\left(i t\right)_{t\rightarrow +\infty}  = \frac{g^2}{-m^2 (2\pi)^2} \int \frac{d\Omega}{4\pi}\int_{0}^{+\infty} d k \, 2k^2 \frac{e^{-i m t - i \frac{k^2}{2m} t}}{2 m} \frac{(-i t)}{2 m} \left[1+O\left(\frac{1}{i t}\right)+O(k^2)\right] \nonumber\\
&= \frac{g^2}{(2\pi)^2}  \frac{e^{-i m t}}{\sqrt{i m t}} \frac{\sqrt{\pi}}{2 \sqrt{2} m^2} \left[1+O\left(\frac{1}{i m t}\right)\right] \ ,
\end{align}
where the naive power counting of phase space provides $1/t^{3/2}$, and the remnant of forward singularity gives $t$. As a result, the polynomial $t$ dependence becomes $1/t^{1/2}$. 

The expression for the spacelike case has been obtained in Eq.~(\ref{eq:Issl}), 
\begin{align}
&I_{t}\left(z\right)   =  \frac{g^2}{(2\pi)^2 m^2} \int_{m}^{+\infty} d M_{\perp} \int_{-\infty}^{+\infty} d k^z \, M_{\perp} e^{i k^z z}  \frac{1}{4 \left(\sqrt{k_z^2+M_{\perp}^2}\right)^3} \nonumber\\
&=\frac{g^2}{(2\pi)^2 m^2} \int_{m}^{+\infty} d M_{\perp} \int_{\gamma} d k^z \, M_{\perp} e^{i k^z z}  \frac{1}{4 \left(\sqrt{k_z^2+M_{\perp}^2}\right)^3} \ ,
\end{align}
where the contour $\gamma$ is depicted in the left panel of Fig.~\ref{fig:phi3_It1_sl}. Change the integral variable $k^z \rightarrow i \left( \frac{K^z}{z} + M_{\perp}\right)$ followed by $M_{\perp} \rightarrow \frac{M_{\perp}}{z} + m$, and perform the large $z$ expansion,
\begin{align}
I_{t}\left(z\right)_{z\rightarrow+\infty} =  \frac{g^2 e^{-m z}}{(2\pi)^2 m^2} \int_{0}^{+\infty} d M_{\perp} \int_{\gamma'} d K^z \, e^{-K^z-M_{\perp}} \frac{i m}{z^2} \frac{z^{3/2}}{\left(-K^z\right)^{3/2} 8 \sqrt{2}m^{3/2}} \left[1+O\left(\frac{1}{m z}\right)\right] \ ,
\end{align}
where $\gamma'$ denotes a contour illustrated in the right panel of Fig.~\ref{fig:phi3_It1_sl}. The phase space produces $1/z^2$, and the singularity from $1/E_k^3$ generates $z^{3/2}$, which leads to an overall scaling of $1/z^{1/2}$. 

To complete the calculation, we apply the derivative trick,
\begin{align}
\frac{\partial }{\partial K^z} \frac{2}{(-K^z)^{1/2}} = \frac{1}{(-K^z)^{3/2}} \ ,
\end{align}
together with integration by parts,
\begin{align}\label{eq:Itz}
&I_{t}\left(z\right)_{z\rightarrow+\infty} =  \frac{g^2 e^{-m z}}{(2\pi)^2 m^2} \int_{0}^{+\infty} d M_{\perp} \int_{\gamma'} d K^z \, e^{-K^z-M_{\perp}} \frac{i m}{z^2} \frac{2 z^{3/2}}{\left(-K^z\right)^{1/2} 8 \sqrt{2}m^{3/2}} \left[1+O\left(\frac{1}{m z}\right)\right] \nonumber\\
&= \frac{g^2 e^{-m z}}{(2\pi)^2 m^2} \int_{0}^{+\infty} d M_{\perp} \int_{0}^{+\infty} d K^z \, e^{-K^z-M_{\perp}} \frac{i m}{z^2} \frac{4 z^{3/2}}{i\left(K^z\right)^{1/2} 8 \sqrt{2}m^{3/2}} \left[1+O\left(\frac{1}{m z}\right)\right] \nonumber\\
&=\frac{g^2}{(2\pi)^2} \frac{e^{-m z}}{\sqrt{z m}} \frac{\sqrt{\pi }}{2 \sqrt{2} m^2} \left[1+O\left(\frac{1}{m z}\right)\right] \ .
\end{align}
Although the mechanism for the enhanced polynomial dependence originating from the forward singularity seems quite different between the timelike and spacelike cases, the final results, including the polynomial dependences, are connected by Lorentz symmetry. 

\subsection{Forward singularity under split mass regularization}\label{sec:phi3FWSGsplit}
This subsection studies the forward singularity under split mass regularization, which is a different method compared to the off-forward regulator adopted in Eq.~(\ref{eq:MYoff}), Appendices~\ref{sec:phi3FWSG} and~\ref{sec:CalMets}. It provides a complementary perspective on the forward singularity and further justifies its cancellation and impact. 
\begin{figure}
    \centering
    \includegraphics[width=0.7\linewidth]{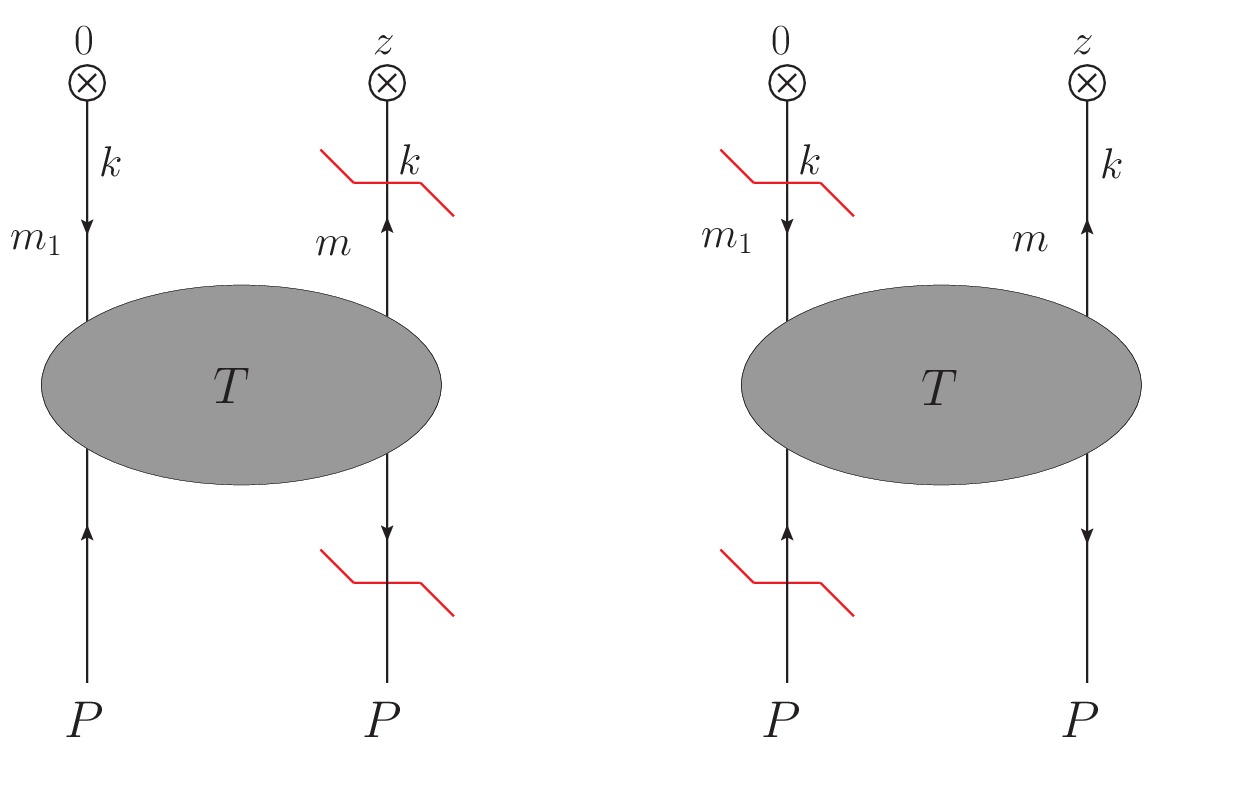}
    \caption{Forward singularity with split mass regularization. The cuts are shown in red.}
    \label{fig:forwardmass}
\end{figure}

The leading diagrammatic structure for the forward singularity is shown in the Fig.~\ref{fig:forwardmass}.\footnote{The general diagrammatic structure is shown in Fig.~\ref{fig:Corre_DC_C_C_DC_sg}, where the leading structure corresponds to $Y(k)=n(q)$, which means $Y(k)$ becomes a single particle state and there is no particle in $Y(k)/n(q)$.} The diagram consists of an amputated $2$ to $2$ connected kernel $T$ shown as the gray blobs, connected by the initial and final momentum, as well as another two legs connecting to the operators, which we call the operator legs.\footnote{The wavefunction diagrams of the operator legs are omitted in this subsection for simplicity.} It is the two operator legs, when participating in cuts as in the Fig.~\ref{fig:forwardmass}, that lead to the ``forward singularity''.  Clearly, since both of the operator legs can be on the cut, the forward singularity cut-diagrams always come in pairs, like in  Fig.~\ref{fig:forwardmass}.

To regulate this singularity, in each of such pairs, we change the mass for one of the operator legs to $m_1$, but all other mass parameters are unchanged. We chose $|m_1-m|<m$ to avoid additional singularities in the $s$ channel. Then, in one of the two cut diagrams, the $2$ to $2$ kernel is on the conjugate amplitude side and has a different $i0$ prescription. To combine the two diagrams and also to avoid conflict of the singularities when shifting the contour, we need to change the $-i0$ to $i0$ in the conjugating amplitudes, or change $i0$ to $-i0$ in the amplitudes. This will lead to additional discontinuities proportional to $T^{\dagger}-T$, generated when virtual propagators within $T$ become on-shell. By expressing $T^{\dagger}-T$ as a summation over cuts, one can see that such discontinuities cancel with discontinuities in the ``real-diagrams'', in which the cuts are within the $2 $ to $2$ kernel $T$, while the discontinuities are generated when the virtual operator legs become on-shell.\footnote{ The cancellation of those discontinuities has been checked at two-loop in $\phi^3$ theory. On the other hand, those discontinuities must cancel between different cuts due to locality as discussed in Sec.~\ref{sec:spacelike}.}

After switching to the same $i0$ prescriptions, we sum up the two diagrams in the pair.  This way, for the space-like correlator, one has 
\begin{align}
&M_{T}(z,P,m_1)=\int \frac{d^3\vec{k}}{(2\pi)^3}\bigg(\frac{e^{i\vec{k}\cdot \vec{z}} }{m^2-m_1^2}\frac{T(k,P) }{2\sqrt{m^2+\vec{k}{^2}}}+\frac{e^{i\vec{k}\cdot \vec{z}} }{m_1^2-m^2}\frac{T(k_1,P) }{2\sqrt{m_1^2+\vec{k}^2}} \bigg) \  , \label{eq:spaceinni} \\
&k=\left(\sqrt{m^2+\vec{k}^2},\vec{k}\right)=(E_k,\vec{k}) \ , \ k_1=\left(\sqrt{m_1^2+\vec{k}^2},\vec{k}\right) \ . 
\end{align}
Here, momentum of the operator legs is denoted as $k$, which depends on the $m_1$ through the $k^0$ component. The $2$ to $2$ kernel as a function of $k,P$ is denoted by $T(k,P)$. For example, for the one-loop cut-diagrams shown in the Fig.~\ref{fig:phi3_Is_form},
\begin{align}
T(k,P)\bigg|_{\rm Fig.~\ref{fig:phi3_Is_form}}=\frac{g^2}{-m^2} \ , 
\end{align}
which is a constant. 

We now take the $m_1\rightarrow m$ limit by Taylor expanding the $m_1$ at $m$. The only dependence on $m_1$ in the $T(k,P)$ is through the $k^0(m_1)$. Thus, after the expansion and using $m_1=m$, the singularity is canceled, and one obtains the regular limit,
\begin{align}
\lim_{m_1\rightarrow m} M_{T}(z,P,m_1)=\int\frac{d^3\vec{k}}{(2\pi)^3}\left(-\frac{T_0}{4E_k^3}+\frac{T_0'}{4E_k^2}\right)e^{i\vec{k}\cdot \vec{z}}\ , 
\end{align}
where one denotes 
\begin{align}
T_0\equiv T(k,P) \ , \ T_0'=\frac{d T(k,P)}{dk^0}\bigg|_{k=(E_k,\vec{k})} \ . 
\end{align}
In particular, the part with the slowest decay is given by
\begin{align}
M_{T;L}(z,P)=\int\frac{d^3\vec{k}}{(2\pi)^3}\left(-\frac{T_0}{4E_k^3}\right)e^{i\vec{k}\cdot \vec{z}} \ .
\end{align}
This way, after the asymptotic expansion, one has the general expression for the leading large $z$ decay
\begin{align}
M_{T;L}(z,P)=-T(k=i \, {\rm sign}(z) \, m \, n^z,P)\times\frac{e^{-m|z|}}{8\sqrt{2}\pi^{\frac{3}{2}} \sqrt{zm}} \ , 
\end{align}
for the space-like correlator. $n^z=(0,0,0,1)$ is a unit vector along $z$ direction. Notice that it is proportional to the on-shell two-particle amplitude at $t=0, u=2m^2\pm 2imP^z, s=2m^2\mp2imP^z$ and is purely real. Clearly, the picture is that the operator legs remain soft, while the large incoming momentum flows in the $s$ or $u$ channels to the final state, similar to the ``small-$x$'' kinematics. As such, there is a connection between the Regge-type limit and the large $z$ asymptotics of large-$P$ space-like correlator. Also note that the above result is consistent with Eq.~(\ref{eq:Itz}).

\subsection{Non-convergent expansion and controlled precision}
In the previous subsections, we have performed the large-$z$ expansion when evaluating the integrals. This naturally leads to the question: Does this large-$z$ expansion converge? The answer is No. The large $z$ asymptotic expansion is divergent because the point $z=\infty$ is non-analytic. Nonetheless, the expansion is Borel summable, which means it yields an unambiguous result if regularized in the Borel plane. Furthermore, truncating the series at a relatively low order already provides an accurate approximation of the original function at large distances; For instance, truncation at the minimum point leads to an exponentially small error. 

We demonstrate the above idea in the simplest case $I_{t}$, whose full result has been evaluated in Eq.~(\ref{eq:IsCP}),
\begin{align}
I_{t}(z) = \frac{g^2 K_0( m |z|)}{8 \pi ^2 m^2} \ ,
\end{align}
where $K_0$ is a modified second kind Bessel function. For convenience, we introduce the notation $Z \equiv m|z|$. The large $Z$ asymptotic series of $K_0(Z)$ has been worked out by Hankel~\cite{DLMF_Bessel_Asy,Watson_Bessel_1944},
\begin{align}\label{eq:K0Lz}
K_0(Z) = \sqrt{\frac{\pi}{2 Z}} e^{-Z} \left[ 1 + \sum_{l=0} (-1)^{l+1} \frac{((2l+1)!!)^2}{(l+1)! (8Z)^{l+1}} \right] \ .    
\end{align}
In the large $l$ limit, the coefficient behaves as
\begin{align}
(-1)^{l+1} \frac{((2l+1)!!)^2}{(l+1)! 8^{l+1}} 
\overset{l \rightarrow +\infty}{\rightarrow} 
(-1)^{l+1} \frac{l!}{2^{l+1} \pi} \ ,
\end{align}
where the factorial growth $l!$ indicates that the series is divergent. 

Since the series is alternating, it is Borel summable, which means there is no ambiguity if regularized in the Borel plane. Applying the Borel transform gives
\begin{align}
&K_0(Z) = \sqrt{\frac{\pi}{2 Z}} e^{-Z} \left[ 1 + \int_{C} dt \, e^{-t Z} \sum_{l=0} \frac{t^{l}}{l!} (-1)^{l+1} \frac{((2l+1)!!)^2}{(l+1)! 8^{l+1}} \right] \nonumber\\
&=\sqrt{\frac{\pi}{2 Z}} e^{-Z} \left[ 1 - \frac{1}{8} \int_{C} dt \, e^{-t Z}  \, _2F_1\left(\frac{3}{2},\frac{3}{2};2;-\frac{t}{2}\right) \right] \ ,
\end{align}
where $C$ is a contour in the complex $t$-plane from $0$ to $+\infty$. The function $_2F_1\left(\frac{3}{2},\frac{3}{2};2;-\frac{t}{2}\right)$ is a hypergeometric function whose only branch cut lies along the negative real axis for $t < -2$. Consequently, different choices of contours around the positive axis of $t$ do not introduce any ambiguity. 

One can also perform a truncation near the minimum point,
\begin{align}
\frac{\partial}{\partial l_{\rm min}}\frac{l_{\rm min}!}{2^{l_{\rm min}+1} \pi \, Z^{l_{\rm min}+1}}  = 0 
\quad \Rightarrow  \quad
l_{\rm min} \sim 2 Z \ ,
\end{align}
where Stirling's approximation, $n! \sim \sqrt{2\pi n} \left(n/e\right)^n$, has been employed. 
The truncation error is then of the order of the minimal term,
\begin{align}
\sim \frac{l_{\rm min}!}{2^{l_{\rm min}+1} \pi Z^{l_{\rm min}+1}} \sim \frac{e^{-2 Z}}{\sqrt{\pi } \sqrt{Z}} \ ,
\end{align}
which is exponentially suppressed in the large $Z$ limit. 

If the series is truncated at $l < l_{\rm min}$, the truncation error can be approximated by a higher-order polynomial dependence.

\begin{figure}
    \centering
    \includegraphics[width=0.49\linewidth]{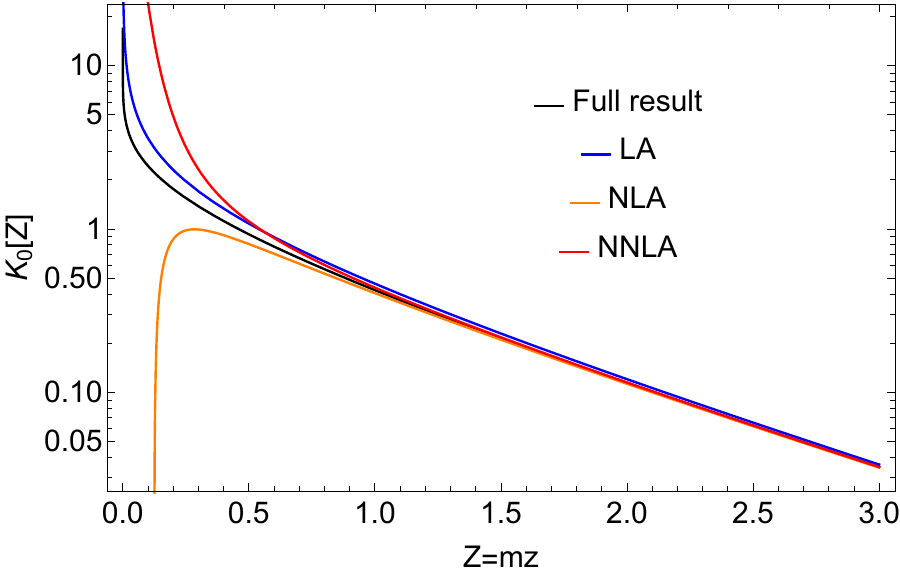}
    \includegraphics[width=0.49\linewidth]{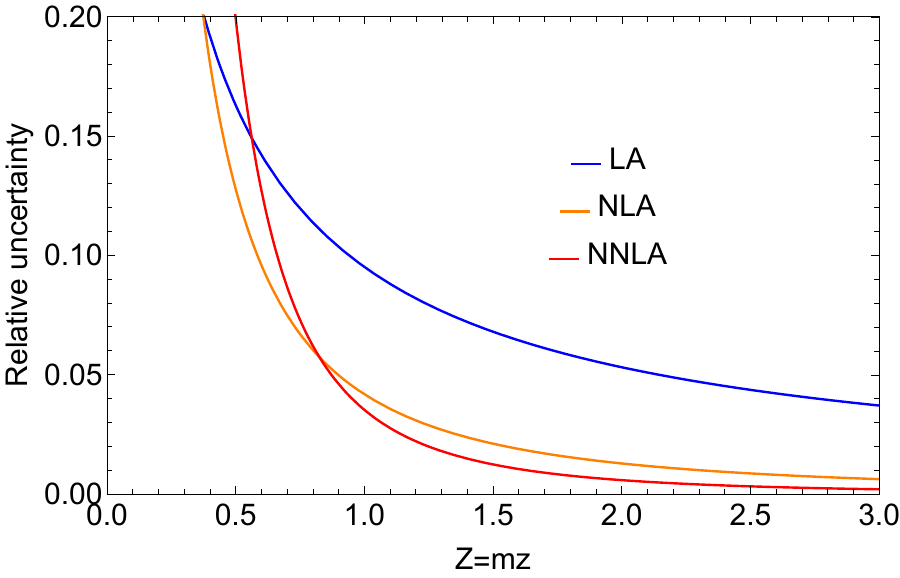}
    \caption{Various truncation orders of the large distance asymptotic forms for $K_0[Z]$, compared with the full result. The left panel displays LA (blue), NLA (orange), and NNLA (red), which correspond to truncations up to $l$ = $-1$, $0$, and $1$ in Eq.~(\ref{eq:K0Lz}), respectively. The black curve denotes the full result. The right panel displays the relative uncertainty of each order, $| {\rm trunc. \, res.} - {\rm full \, res.} | / {\rm full \, res.} $ }
    \label{fig:AEphi3}
\end{figure}
As shown in Fig.~\ref{fig:AEphi3}, the full result is well described by large distance asymptotic expansions for $z \gtrsim 1/m $. At $z=1/m$, LA only deviates from the full result by 9.5\%, NLA by 4.2\%, and NNLA by 3.5\%. At $z=2/m$, LA by 5.3\%, NLA by 1.3\%, and NNLA by 0.6\%.  

Notice that if $z$ becomes an imaginary time or real time with $1\pm i0$ prescriptions, then the threshold-type terms with phases $e^{\pm izP^z}$ become exponentially enhanced or suppressed, while the ``small-$x$'' type terms discussed here are exponentially suppressed and acquire phase ambiguities. These phase ambiguities cancel with the non Borel-summabilities of the leading threshold-type terms, due to poles and branch cuts along the integration paths, which are absent in case of space-like correlators.

\section{Nucleon quark transversity and helicity PDFs}\label{sec:PolPDF}
The nucleon quark transversity quasi-PDF matrix element is constructed as~\cite{Alexandrou:2021oih,LatticeParton:2022xsd},
\begin{align}
\langle N(P S_{\perp}) | \bar{\psi}(z) \gamma^t \gamma^{\perp} \gamma^5 U(z,0) \psi(0) | N(P S_{\perp}) \rangle_{c} \ ,
\end{align}
where $| N(P S_{\perp}) \rangle$ denotes a proton with momentum $P=\left(\sqrt{m_N^2+P_z^2},0,0,P^z\right)$ with $P^z \gg m_N$ and transverse polarization $S_{\perp}$ in $x$ or $y$ direction.  
Following the similar logic as Sec.~\ref{sec:theory} but replacing Eq.~(\ref{eq:Gammax}) into 
\begin{align}
\Gamma_x \equiv 
\begin{cases}
\gamma^z \gamma^{\perp} \gamma^5 \ , \quad x=(t,0,0,0) \\
\gamma^t \gamma^{\perp} \gamma^5 \ , \quad x=(0,0,0,z)
\end{cases} \ ,
\end{align}
we obtain this correlator's large distance expansion up to all orders in $\overline{\rm MS}$ or hybrid scheme,
\begin{align}\label{eq:qPDFPLzGen}
&\langle N(P S_{\perp}) | \bar{\psi}(z) \gamma^t \gamma^{\perp} \gamma^5 U(z,0) \psi(0) | N(P S_{\perp}) \rangle_{c}   \nonumber\\
&{\overset{z\rightarrow \infty}{=}}   
 \sum_{X,l=0} \frac{e^{-i P^z z} e^{-\Lambda_X |z| }}{|z|^{\frac{3}{2}(N_X-1)+l}} \, A_{X,l} \, e^{i \, {\rm sign}(z) \, \phi_{X,l}} \nonumber\\
&\quad + \sum_{Y,l=0} \frac{e^{-\Lambda_Y |z|}}{|z|^{\frac{3}{2}(N_Y-1)+l}} \, A_{Y,l} \, e^{i \, {\rm sign}(z) \, \phi_{Y,l}} \nonumber\\
&\quad + \sum_{Z,l=0} \frac{e^{i P^z z} e^{-\Lambda_Z |z|}}{|z|^{\frac{3}{2}(N_Z-1)+l}} \, A_{Z,l} \, e^{i \, {\rm sign}(z) \, \phi_{Z,l}} \ ,
\end{align}
where $\Lambda_X=m_X-m_Q$ is the binding energy of the state $X$ projected by the common quantum numbers of $\bar{Q} \psi | N(P S_{\perp}) \rangle$ and $\bar{Q} \gamma^z \gamma^{\perp} \gamma^5 \psi | N(P S_{\perp}) \rangle$, $\Lambda_Y=m_Y-m_Q$ is for the state $Y$ with quantum numbers of $\bar{Q}  \psi | \Omega \rangle$ or $\bar{Q} \gamma^z \gamma^{\perp} \gamma^5 \psi | \Omega \rangle$, and $\Lambda_Z=m_Z-m_Q$ is for the state $Z$ with common quantum numbers of $\bar{Q} \psi | \bar{N}(P S_{\perp}) \rangle$ and $\bar{Q} \gamma^z \gamma^{\perp} \gamma^5 \psi | \bar{N}(P S_{\perp}) \rangle$. $N_X$, $N_Y$, and $N_Z$ denote the numbers of single-particles in the multi-particle states $X$, $Y$ and $Z$, respectively. The $P$ and $S_{\perp}$ dependences in the coefficients and phases are omitted for simplicity. 

We repeat this logic for nucleon quark helicity quasi-PDF~\cite{Alexandrou:2021oih},
\begin{align}\label{eq:qPDFPH}
\langle N(P S_{L}) | \bar{\psi}(z) \gamma^z \gamma^5  U(z,0) \psi(0) | N(P S_{L}) \rangle_{c} \ ,
\end{align}
except for replacing Eq.~(\ref{eq:Gammax}) into 
\begin{align}
\Gamma_x \equiv 
\begin{cases}
\gamma^t \gamma^5 \ , \quad x=(t,0,0,0) \\
\gamma^z \gamma^5 \ , \quad x=(0,0,0,z)
\end{cases} \ .
\end{align}
The same mathematical structure as Eq.~(\ref{eq:qPDFPLzGen}) can be obtained, although the values of mass gaps, coefficients, and phases could differ. In this case, the state $X$ is projected by the common quantum numbers of $\bar{Q} \psi | N(P S_{L}) \rangle$ and $\bar{Q} \gamma^t \gamma^5 \psi | N(P S_{L}) \rangle$, the state $Y$ with quantum numbers of $\bar{Q}  \psi | \Omega \rangle$ or $\bar{Q} \gamma^t \gamma^5 \psi | \Omega \rangle$, and the state $Z$ with common quantum numbers of $\bar{Q} \psi | N(P S_{L}) \rangle$ and $\bar{Q} \gamma^t \gamma^5 \psi | N(P S_{L}) \rangle$.

Similar to the unpolarized case in Eq.~(\ref{eq:qPDFLzGen}), the polarized case in Eq.~(\ref{eq:qPDFPLzGen}) does not receive the linear $z$ enhancement from the forward singularity as well. This is because the forward singularity in the large distance limit is also forbidden by the quantum numbers of the operators for the polarized case. See the discussions between and around Eqs.~(\ref{eq:MYorg}) and~(\ref{eq:replace}) on the origin and impact of forward singularity.

Following the accuracy-counting principles in Sec.~\ref{sec:AccuCount}, we truncate the large distance expansion in Eq.~(\ref{eq:qPDFPLzGen}) up to LA and NLA,
\begin{align}
\tilde{h}^{\rm LA}\left(z,P^z\right) = 
A e^{i \phi \, {\rm sign}(z)} e^{-\Lambda |z|}  \ ,
\end{align}
\begin{align}
\tilde{h}^{\rm NLA}\left(z,P^z\right) = 
\left[A e^{i \phi \, {\rm sign}(z)} +  \frac{A' e^{i \phi' \, {\rm sign}(z)}}{|z|} \right] e^{-\Lambda |z|}  \ ,
\end{align}
which apply to both transversity and helicity. $\Lambda$ is the binding energy of a vector (for transversity) or pseudo-scalar (for helicity) heavy-light meson if $\psi$ is a light quark, or a heavy-heavy meson if $\psi$ is a heavy quark.

\section{Meson DA}\label{sec:DA}
The quasi-DA matrix element for a pseudo scalar meson~\cite{LatticeParton:2022zqc,Zhang:2017bzy},
\begin{align}
\langle M(P) | \bar{\psi}_1(z) U(z,0) \gamma^z \gamma^5 \psi_{2}(0) | \Omega \rangle_c \ ,
\end{align}
where $\psi_1$ and $\psi_2$ are quark fields under different flavors. Following the methods in Sec.~\ref{sec:theory}, we obtain the all-order large distance asymptotic form in $\overline{\rm MS}$ or hybrid scheme,
\begin{align}\label{eq:qDALzGen}
&\langle M(P) | \bar{\psi}_1(z) U(z,0) \gamma^z \gamma^5 \psi_{2}(0) | \Omega \rangle_c  \nonumber\\
&{\overset{z\rightarrow \infty}{=}}   
 \sum_{X,l=0} \frac{e^{-i P^z z} e^{-\Lambda_X |z| }}{|z|^{\frac{3}{2}(N_X-1)+l}} \, A_{X,l} \, e^{i \, {\rm sign}(z) \, \phi_{X,l}} \nonumber\\
&\quad + \sum_{Y,l=0} \frac{e^{-\Lambda_Y |z|}}{|z|^{\frac{3}{2}(N_Y-1)+l}} \, A_{Y,l} \, e^{i \, {\rm sign}(z) \, \phi_{Y,l}}  \ ,
\end{align}
where $\Lambda_X=m_X-m_Q$ is the binding energy of the state $X$ with the quantum numbers of $\bar{Q} \psi_2 |\Omega\rangle$ or $\bar{Q} \gamma^t \gamma^5 \psi_2 |\Omega\rangle$, and $\Lambda_{Y}=m_Y-m_Q$ is for the state $Y$ with the quantum numbers of $\bar{Q} \psi_1 |\Omega\rangle$ or $\bar{Q} \gamma^t \gamma^5 \psi_1 |\Omega\rangle$. See the definitions of $N_X$ and $N_Y$ below Eq.~(\ref{eq:qPDFPLzGen}). Compared with the quasi-PDF in Eq.~(\ref{eq:qPDFLzGen}), the DA lacks the $Z$ term in which both form factors are disconnected. 

Based on the accuracy counting principles in Sec.~\ref{sec:AccuCount}, we truncate the large distance expansion in Eq.~(\ref{eq:qDALzGen}) up to LA and NLA,
\begin{align}
\tilde{h}^{\rm LA}\left(z,P^z\right) = 
\left[A_1 e^{i \phi_1 \, {\rm sign}(z)} e^{-i z P^z} + A_2 e^{i \phi_2 \, {\rm sign}(z)} \right]e^{-\Lambda |z|}  \ ,
\end{align}
\begin{align}
&\tilde{h}^{\rm NLA}\left(z,P^z\right) = 
\left[ \left(A_1 e^{i \phi_1 \, {\rm sign}(z)} e^{-i z P^z} + A_2 e^{i \phi_2 \, {\rm sign}(z)} \right) \right. \nonumber\\
&\left.+ \left(A'_1 e^{i \phi'_1 \, {\rm sign}(z)} e^{-i z P^z} + A'_2 e^{i \phi'_2 \, {\rm sign}(z)} \right) \frac{1}{|z|} \right] e^{-\Lambda |z|}  \ .
\end{align}
The above asymptotic forms apply to light meson DAs. If both $\psi_1$ and $\psi_2$ are light quarks with similar masses, all terms should be kept. $\Lambda$ corresponds to the binding energy of a pseudoscalar heavy-light meson. In particular, $A_1=A_2$, $\phi_1=-\phi_2$, $A'_1=A'_2$ and $\phi'_1=-\phi'_2$ for pion quasi-DA. They can be generalized to heavy-light meson DAs~\cite{Xu:2022guw,Hu:2023bba,LatticeParton:2024zko,Han:2024cht,Deng:2024dkd,Han:2024fkr,Guo:2025obm}, which have been hot topics recently. If $\psi_1$ is a light quark and $\psi_2$ is a heavy quark, the state $X$ is heavier than $Y$ and the leading exponential decay arises from the $Y$ term, which means $A_1 = A'_1 = 0$. If $\psi_1$ is a heavy quark and $\psi_2$ is a light quark, $A_2 = A'_2 = 0$.

\section{Quark GPD}\label{sec:GPD}
The quark unpolarized quasi-GPD~\cite{Ji:2015qla,Bhattacharya:2023jsc},
\begin{align}
\langle H(P) | \bar{\psi}(z) U(z,0) \gamma^t  \psi(0) | H(P') \rangle_c \ ,
\end{align}
whose all-order large distance asymptotic expansion in $\overline{\rm MS}$ or hybrid scheme can be obtained following a similar derivation process as Sec.~\ref{sec:theory},
\begin{align}\label{eq:qGPDLzGen}
&\langle H(P) | \bar{\psi}(z) U(z,0) \gamma^t  \psi(0) | H(P') \rangle_c   \nonumber\\
&{\overset{z\rightarrow \infty}{=}}   
 \sum_{X,l=0} \frac{e^{-i P^z z} e^{-\Lambda_X |z| }}{|z|^{\frac{3}{2}(N_X-1)+l}} \, A_{X,l} \, e^{i \, {\rm sign}(z) \, \phi_{X,l}} \nonumber\\
&\quad + \sum_{Y,l=0} \frac{e^{-\Lambda_Y |z|}}{|z|^{\frac{3}{2}(N_Y-1)+l}} \, \left[ A_{Y,l} \, e^{i \, {\rm sign}(z) \, \phi_{Y,l}} +  \tilde{A}_{Y,l} \, e^{i \, {\rm sign}(z) \, \tilde{\phi}_{Y,l}} e^{-i (P^z-P'^z) z}\right]\nonumber\\
&\quad + \sum_{Z,l=0} \frac{e^{i P'^z z} e^{-\Lambda_Z |z|}}{|z|^{\frac{3}{2}(N_Z-1)+l}} \, A_{Z,l} \, e^{i \, {\rm sign}(z) \, \phi_{Z,l}} \ ,
\end{align}
where the binding energies have the same physical origins as Eq.~(\ref{eq:qPDFLzGen}). See the definitions of $N_X$, $N_Y$ and $N_Z$ below Eq.~(\ref{eq:qPDFPLzGen}). The forward singularity discussed between and around Eqs.~(\ref{eq:MYorg}) and~(\ref{eq:replace}) does not exist in the off-forward GPD, which means there is no enhanced polynomial $z$-dependence in the $Y$ term. 

For pion external state, we truncate the large distance expansion in Eq.~(\ref{eq:qGPDLzGen}) up to LA and NLA following Sec.~\ref{sec:AccuCount},
\begin{align}
\tilde{h}^{\rm LA}\left(z,P^z,P'^z\right) =
&\left[A_1 e^{i \phi_1 \, {\rm sign}(z)} e^{-i z P^z} + A_3 e^{i \phi_3 \, {\rm sign}(z)} e^{i z P'^z}  \right. \nonumber\\
&\left.+A_2 e^{i \phi_2 \, {\rm sign}(z)} + \tilde{A}_2 e^{i \tilde{\phi}_2 \, {\rm sign}(z)} e^{-i(P^z-P'^z)z}  \right] e^{-\Lambda |z|} \ ,
\end{align}
\begin{align}
\tilde{h}^{\rm NLA}\left(z,P^z,P'^z\right) =
&\left[A_1 e^{i \phi_1 \, {\rm sign}(z)} e^{-i z P^z}  + A_3 e^{i \phi_3 \, {\rm sign}(z)} e^{i z P'^z} \right. \nonumber\\
&\left.+A_2 e^{i \phi_2 \, {\rm sign}(z)} + \tilde{A}_2 e^{i \tilde{\phi}_2 \, {\rm sign}(z)} e^{-i(P^z-P'^z)z}\right]e^{-\Lambda |z|} \nonumber\\ 
&+\left[A'_1 e^{i \phi'_1 \, {\rm sign}(z)} e^{-i z P^z} + A'_3 e^{i \phi'_3 \, {\rm sign}(z)} e^{i z P'^z} \right. \nonumber\\
&\left.+A'_2 e^{i \phi'_2 \, {\rm sign}(z)} + \tilde{A}'_2 e^{i \tilde{\phi}'_2 \, {\rm sign}(z)} e^{-i(P^z-P'^z)z} \right] \frac{e^{-\Lambda |z|}}{|z|} \ .
\end{align}
For valence combination, all terms are preserved for the same reason as the pion valence quasi-PDF discussed in Sec.~\ref{sec:AccuCount}. For sea quarks (including light and heavy quarks), $A_1 = A_3 = A'_1 = A'_3 = 0$ since their mass gaps are associated with four-quark states that have the combined quantum numbers of a light meson and a heavy meson and are therefore larger than those of a single heavy meson.

For proton external state, the LA and NLA of Eq.~(\ref{eq:qGPDLzGen}) can be obtained following Sec.~\ref{sec:AccuCount},
\begin{align}\label{eq:protonqGPDLA}
\tilde{h}^{\rm LA}\left(z,P^z,P'^z\right) = 
\left[ A_2 e^{i \phi_2 \, {\rm sign}(z)} + \tilde{A}_2 e^{i \tilde{\phi}_2 \, {\rm sign}(z)} e^{-i(P^z-P'^z)z} \right] e^{-\Lambda |z|}  \ ,
\end{align}
\begin{align}\label{eq:protonqGPDNLA}
&\tilde{h}^{\rm NLA}\left(z,P^z,P'^z\right) = 
\left[ A_2 e^{i \phi_2 \, {\rm sign}(z)} + \tilde{A}_2 e^{i \tilde{\phi}_2 \, {\rm sign}(z)} e^{-i(P^z-P'^z)z} \right.\nonumber\\ 
&\left.+ \left( A'_2 e^{i \phi'_2 \, {\rm sign}(z)} + \tilde{A}'_2 e^{i \tilde{\phi}'_2 \, {\rm sign}(z)} e^{-i(P^z-P'^z)z} \right) \frac{1}{|z|}  \right] e^{-\Lambda |z|}  \ ,
\end{align}
where the leading exponential decay arises from the $Y$ term for the same reason as the proton quark unpolarized quasi-PDF discussed in Sec.~\ref{sec:AccuCount}.

As for the proton polarized quasi-GPD~\cite{Ji:2015qla,Bhattacharya:2023jsc},
\begin{align}
\langle N(P S)| \bar{\psi}(z) U(z,0) \gamma^z \gamma^5 \psi(0) | N(P' S')\rangle_c \ ,    
\end{align}
we follow the methods in Sec.~\ref{sec:theory} to derive its large distance expansion, which has the same mathematical structure as Eq.~(\ref{eq:qGPDLzGen}) though the values of mass gaps, coefficients and phases can be different. The physical origins of binding energies are the same as the proton helicity quasi-PDF in Eq.~(\ref{eq:qPDFPH}). Following Sec.~\ref{sec:AccuCount}, the LA and NLA have the same mathematical formats as Eqs.~(\ref{eq:protonqGPDLA}) and~(\ref{eq:protonqGPDNLA}), respectively.

\section{Gluon PDF}\label{sec:Gluon}

In this section, we derive the large distance expansion up to all orders for the gluon quasi-PDF matrix element using the adjoint heavy quark. However, its truncations up to LA and NLA are presented as conjectures because the spectra of the exotic hadrons have not been fully understood\footnote{at least by the authors}.

Consider the gluon quasi-PDF matrix element in the adjoint representation~\cite{Zhang:2018diq},
\begin{align}
\langle H(P) | F^{z\mu}_{a}(z) L_{ab}(z,0) F^{z}_{b,\mu}(0) | H(P) \rangle_c \ .
\end{align}
For the convenience of dispersive analysis, we introduce the auxiliary heavy quark in the adjoint representation $\mathcal{Q}$ with trivial spin degrees of freedom, following a similar method as Ref.~\cite{Zhang:2018diq},
\begin{align}
\mathcal{L} = \mathcal{L}_{\rm QCD} + \bar{\mathcal{Q}}(x) \left(i n \cdot D - m_{\mathcal{Q}} - \frac{D_{\perp}^2}{2m_{\mathcal{Q}}}\right) \mathcal{Q}(x) \ ,
\end{align}
where the kinematic energy term $D_{\perp}^2/(2m_{\mathcal{Q}})$ is retained to ensure the correct polynomial dependence. Then, the adjoint gauge link can be represented by adjoint heavy quarks, 
\begin{align}
\langle H(P) | F^{z\mu}_{a}(x) \mathcal{Q}_{a}(x) \bar{\mathcal{Q}}_{b}(0) F^{z}_{b,\mu}(0) | H(P) \rangle_c \ .
\end{align}
We follow a similar logic as Sec.~\ref{sec:theory} to derive the all-order large distance expansion for the gluon quasi-PDF matrix element in $\overline{\rm MS}$ or hybrid scheme,
\begin{align}\label{eq:GluonqPDFLzGen}
&\langle H(P) | F^{z\mu}_{a}(z) L_{ab}(z,0) F^{z}_{b,\mu}(0) | H(P) \rangle_c   \nonumber\\
&{\overset{z\rightarrow \infty}{=}}   
 \sum_{X,l=0} \frac{e^{-i P^z z} e^{-\Lambda_X |z| }}{|z|^{\frac{3}{2}(N_X-1)+l}} \, A_{X,l} \, e^{i \, {\rm sign}(z) \, \phi_{X,l}} \nonumber\\
&\quad + \sum_{Y,l=0} \frac{e^{-\Lambda_Y |z|}}{|z|^{\frac{3}{2}(N_Y-1)+l-1}} \, A_{Y,l} \, e^{i \, {\rm sign}(z) \, \phi_{Y,l}} \nonumber\\
&\quad + \sum_{Z,l=0} \frac{e^{i P^z z} e^{-\Lambda_Z |z|}}{|z|^{\frac{3}{2}(N_Z-1)+l}} \, A_{Z,l} \, e^{i \, {\rm sign}(z) \, \phi_{Z,l}} \ ,
\end{align}
where $\Lambda_X=m_X-m_{\mathcal{Q}}$ is the binding energy of the state $X$ with the quantum numbers of $\bar{\mathcal{Q}}_{b} F^{t}_{b,\mu} | H(P) \rangle$, $\Lambda_Y=m_Y-m_{\mathcal{Q}}$ is for the state $Y$ of $\bar{\mathcal{Q}}_{b} F^{t}_{b,\mu}  | \Omega \rangle$, and $\Lambda_Z=m_Z-m_{\mathcal{Q}}$ for $\bar{\mathcal{Q}}_{b} F^{t}_{b,\mu} | \bar{H}(P) \rangle$. According to Ref.~\cite{Bruschini2025JLabSeminar}, an adjoint heavy quark has the same color quantum number as a pair of fundamental heavy quark and anti-quark under repulsive color-Coulomb potential, which means $\bar{\mathcal{Q}}_{b} F^{t}_{b,\mu}  | \Omega \rangle$ can produce the quarkonium-hybrid ($\bar{Q}Q g$). Also see the definitions of $N_X$, $N_Y$ and $N_Z$ below Eq.~(\ref{eq:qPDFPLzGen}). Since the gluon quasi-PDF matrix element is purely real, the above expression can be further simplified,
\begin{align}\label{eq:GluonqPDFLzSim}
&\langle H(P) | F^{z\mu}_{a}(z) L_{ab}(z,0) F^{z}_{b,\mu}(0) | H(P) \rangle_c   \nonumber\\
&{\overset{z\rightarrow \infty}{=}}   
 \sum_{X,l=0} \frac{e^{-\Lambda_X |z| }}{|z|^{\frac{3}{2}(N_X-1)+l}} \, A_{X,l} \, \cos\left[-P^z z + {\rm sign}(z) \phi_{X,l}\right]  \nonumber\\
&\quad + \sum_{Y,l=0} \frac{e^{-\Lambda_Y |z|}}{|z|^{\frac{3}{2}(N_Y-1)+l-1}} \, A_{Y,l}  \ ,
\end{align}
where the $X$ and $Z$ terms are combined into a single term, and the phase of the $Y$ term does not exist. Compared to the unpolarized quark quasi PDF case in Eq.~(\ref{eq:qPDFLzGen}), the $Y$ terms in the gluon PDF manifest the linear $z$ enhancement as the remnant of the forward singularity. This is because the symmetric structure of the gluon quasi-correlator has the desired quantum numbers to preserve the forward singularity diagram in the large time limit, according to the discussions around Eqs.~(\ref{eq:GOdis}) and~(\ref{eq:replace}).

The next thing is to truncate the large distance expansion in Eq.~(\ref{eq:GluonqPDFLzSim}) up to LA and NLA following the accuracy-counting principles in Sec.~\ref{sec:AccuCount}, which requires ranking the spectrum of $\{\bar{\mathcal{Q}}_{b} F^{t}_{b,\mu} | H(P) \rangle$, $\bar{\mathcal{Q}}_{b} F^{t}_{b,\mu} | \Omega \rangle \}$ and finding the lowest mass gap. However, it is challenging to rank this spectrum because it involves hybrids ($\bar{Q}Q g$), tetraquarks ($\bar{Q}Q \bar{q}q$), and pentaquarks ($\bar{Q}Q qqq$), which are exotic hadrons under investigation from both experimental and theoretical perspectives~\cite{Guo:2017jvc,Ali:2017jda,Olsen:2017bmm,Brambilla:2019esw,Liu:2019zoy,Chen:2022asf,Berwein:2024ztx}. Moreover, the introduced auxiliary heavy-quark field could alter the original QCD spectrum, further complicating this problem.

Here, we propose the following conjecture regarding the spectrum of $\{\bar{\mathcal{Q}}_{b} F^{t}_{b,\mu} | H(P) \rangle$, $\bar{\mathcal{Q}}_{b} F^{t}_{b,\mu} | \Omega \rangle \}$ that the lightest hybrid ($\bar{Q}Q g$) and tetraquark ($\bar{Q}Q \bar{q}q$) share similar masses, which are smaller than the lightest pentaquark ($\bar{Q}Q qqq$). Based on this conjecture, we derive the LA and NLA. 

For light baryon external states, $\bar{\mathcal{Q}}_{b} F^{t}_{b,\mu} | H(P) \rangle$ corresponds to a heavy pentaquark ($\bar{Q}Q qqq$), which has higher energy than a tetraquark ($\bar{Q}Q \bar{q}q$) or hybrid ($\bar{Q}Q g$) generated by $\bar{\mathcal{Q}}_{b} F^{t}_{b,\mu} | \Omega \rangle$. Therefore, the leading exponential decay comes from the $Y$ term, 
\begin{align}\label{eq:gluonLA}
\tilde{h}^{\rm LA}\left(z,P^z\right) = 
A  |z| e^{-\Lambda |z|}  \ ,
\end{align}
\begin{align}\label{eq:gluonNLA}
\tilde{h}^{\rm NLA}\left(z,P^z\right) = 
\left[ A  |z| +  A'  \right] e^{-\Lambda |z|}  \ ,
\end{align}
where $A$, $A'$, and $\Lambda$ are real parameters to be fitted with lattice data. 

For light meson external states, $\bar{\mathcal{Q}}_{b} F^{t}_{b,\mu} | H(P) \rangle$ corresponds to a tetraquark ($\bar{Q}Q \bar{q}q$), whose mass is similar to a hybrid ($\bar{Q}Q g$) produced by $\bar{\mathcal{Q}}_{b} F^{t}_{b,\mu} | \Omega \rangle$. Therefore, all three types of terms could potentially contribute to the leading exponential decay. At LA, one only considers the $Y$ term due to the enhancement of polynomial-$z$ dependence, while at NLA, one should consider all three types, 
\begin{align}
\tilde{h}^{\rm LA}\left(z,P^z\right) = 
A_2 |z| e^{-\Lambda |z|}  \ ,
\end{align}
\begin{align}
\tilde{h}^{\rm NLA}\left(z,P^z\right) 
= \left[ A_{2} |z| + A'_2 + 2 A_1 \cos\left(\phi-z P^z\right) \right] e^{-\Lambda |z|} \ ,
\end{align} 
where the real parameters $A_2$, $\Lambda$, $A'_2$, $A_1$, and $\phi$ can be fitted from lattice data. 

For heavy baryon or meson external states, the LA and NLA forms should share the same mathematical structures as Eqs.~(\ref{eq:gluonLA}) and~(\ref{eq:gluonNLA}), respectively. 

There are two ways to examine the conjecture presented above. The first is to perform a detailed analysis of the spectra of hybrids, tetraquarks, and pentaquarks in lattice QCD, considering the auxiliary field. The second is to validate the fit formulas directly using gluon quasi-PDF data.

\bibliographystyle{apsrev4-1}
\bibliography{bibliography}

@article{Kravchuk:2021kwe,
    author = "Kravchuk, Petr and Qiao, Jiaxin and Rychkov, Slava",
    title = "{Distributions in CFT. Part II. Minkowski space}",
    eprint = "2104.02090",
    archivePrefix = "arXiv",
    primaryClass = "hep-th",
    doi = "10.1007/JHEP08(2021)094",
    journal = "JHEP",
    volume = "08",
    pages = "094",
    year = "2021"
}

@article{Ji:2022ezo,
    author = "Ji, Xiangdong",
    title = "{Large-Momentum Effective Theory vs. Short-Distance Operator Expansion: Contrast and Complementarity}",
    eprint = "2209.09332",
    archivePrefix = "arXiv",
    primaryClass = "hep-lat",
    doi = "10.34133/research.0695",
    journal = "Research",
    volume = "8",
    pages = "0695",
    year = "2025"
}

@article{Bjorken:1969ja,
    author = "Bjorken, J. D. and Paschos, Emmanuel A.",
    title = "{Inelastic Electron Proton and gamma Proton Scattering, and the Structure of the Nucleon}",
    reportNumber = "SLAC-PUB-0572",
    doi = "10.1103/PhysRev.185.1975",
    journal = "Phys. Rev.",
    volume = "185",
    pages = "1975--1982",
    year = "1969"
}

@book{Jost:1965yxu,
    author = "Jost, Res",
    title = "{The general theory of quantized fields}",
    publisher = "American Mathematical Society",
    address = "Providence, RI",
    volume = "4",
    year = "1965"
}

@article{Bajnok:2006ze,
    author = "Bajnok, Z. and Palla, L. and Takacs, G.",
    title = "{On the boundary form-factor program}",
    eprint = "hep-th/0603171",
    archivePrefix = "arXiv",
    doi = "10.1016/j.nuclphysb.2006.05.019",
    journal = "Nucl. Phys. B",
    volume = "750",
    pages = "179--212",
    year = "2006"
}

@article{Berg:1978sw,
    author = "Berg, B. and Karowski, M. and Weisz, P.",
    title = "{Construction of Green Functions from an Exact S Matrix}",
    reportNumber = "FUB-HEP-78-16",
    doi = "10.1103/PhysRevD.19.2477",
    journal = "Phys. Rev. D",
    volume = "19",
    pages = "2477",
    year = "1979"
}

@article{Karowski:1978vz,
    author = "Karowski, M. and Weisz, P.",
    title = "{Exact Form-Factors in (1+1)-Dimensional Field Theoretic Models with Soliton Behavior}",
    reportNumber = "FUB-HEP 78/2",
    doi = "10.1016/0550-3213(78)90362-0",
    journal = "Nucl. Phys. B",
    volume = "139",
    pages = "455--476",
    year = "1978"
}

@article{Kirillov:1988zk,
    author = "Kirillov, A. N. and Smirnov, F. A.",
    title = "{Form-factors in O(3) Nonlinear $\sigma$ Model}",
    doi = "10.1142/S0217751X88000321",
    journal = "Int. J. Mod. Phys. A",
    volume = "3",
    pages = "731--741",
    year = "1988"
}

@article{Babujian:2013roa,
    author = "Babujian, Hrachya M. and Foerster, Angela and Karowski, Michael",
    title = "{Exact form factors of the O(N) $\sigma$-model}",
    eprint = "1308.1459",
    archivePrefix = "arXiv",
    primaryClass = "hep-th",
    doi = "10.1007/JHEP11(2013)089",
    journal = "JHEP",
    volume = "11",
    pages = "089",
    year = "2013"
}

@article{Balog:1994np,
    author = "Balog, J. and Hauer, T.",
    title = "{Polynomial form-factors in the O(3) nonlinear sigma model}",
    eprint = "hep-th/9406155",
    archivePrefix = "arXiv",
    reportNumber = "KFKI-1994-10-A",
    doi = "10.1016/0370-2693(94)91453-2",
    journal = "Phys. Lett. B",
    volume = "337",
    pages = "115--121",
    year = "1994"
}

@book{Smirnov:1992vz,
    author = "Smirnov, F. A.",
    title = "{Form-factors in completely integrable models of quantum field theory}",
    volume = "14",
    year = "1992"
}

@article{Babujian:1998uw,
    author = "Babujian, Hratchya M. and Fring, A. and Karowski, M. and Zapletal, A.",
    title = "{Exact form-factors in integrable quantum field theories: The Sine-Gordon model}",
    eprint = "hep-th/9805185",
    archivePrefix = "arXiv",
    doi = "10.1016/S0550-3213(98)00737-8",
    journal = "Nucl. Phys. B",
    volume = "538",
    pages = "535--586",
    year = "1999"
}

@article{Jaffe1965,
author = {Arthur Jaffe},
title = {{Divergence of perturbation theory for bosons}},
volume = {1},
journal = {Communications in Mathematical Physics},
number = {2},
publisher = {Springer},
pages = {127 -- 149},
year = {1965},
}

@article{Nakanishi1961,
    author = {Nakanishi, Noboru},
    title = {Parametric Integral Formulas and Analytic Properties in Perturbation Theory},
    journal = {Progress of Theoretical Physics Supplement},
    volume = {18},
    pages = {1-81},
    year = {1961},
    month = {02},
    issn = {0375-9687},
    doi = {10.1143/PTPS.18.1},
    url = {https://doi.org/10.1143/PTPS.18.1},
    eprint = {https://academic.oup.com/ptps/article-pdf/doi/10.1143/PTPS.18.1/5384081/18-1.pdf},
}

@article{Krajewski:2008fa,
    author = "Krajewski, T. and Rivasseau, V. and Tanasa, A. and Wang, Zhituo",
    title = "{Topological Graph Polynomials and Quantum Field Theory, Part I: Heat Kernel Theories}",
    eprint = "0811.0186",
    archivePrefix = "arXiv",
    primaryClass = "math-ph",
    reportNumber = "LPT-ORSAY-08-87",
    journal = "J. Noncommut. Geom.",
    volume = "4",
    pages = "29--82",
    year = "2010"
}

@book{Rivasseau:2014zpm,
    author = "Rivasseau, V.",
    title = "{From Perturbative to Constructive Renormalization}",
    isbn = "978-0-691-60835-8",
    publisher = "Princeton University Press",
    month = "7",
    year = "2014"
}

@article{Fredenhagen:1984bz,
    author = "Fredenhagen, K.",
    title = "{A Remark on the Cluster Theorem}",
    reportNumber = "CPT-84/P-1622",
    doi = "10.1007/BF01213409",
    journal = "Commun. Math. Phys.",
    volume = "97",
    pages = "461",
    year = "1985"
}

@article{delRio:2023pse,
    author = "del R\'\i{}o, \'Oscar and Vladimirov, Alexey",
    title = "{Quasi Transverse Momentum Dependent Distributions at Next-to-Next-to-Leading order}",
    eprint = "2304.14440",
    archivePrefix = "arXiv",
    primaryClass = "hep-ph",
    month = "4",
    year = "2023"
}

@article{Ji:2021znw,
    author = "Ji, Xiangdong and Liu, Yizhuang",
    title = "{Computing light-front wave functions without light-front quantization: A large-momentum effective theory approach}",
    eprint = "2106.05310",
    archivePrefix = "arXiv",
    primaryClass = "hep-ph",
    doi = "10.1103/PhysRevD.105.076014",
    journal = "Phys. Rev. D",
    volume = "105",
    number = "7",
    pages = "076014",
    year = "2022"
}

@article{Ji:2020ect,
    author = "Ji, Xiangdong and Liu, Yu-Sheng and Liu, Yizhuang and Zhang, Jian-Hui and Zhao, Yong",
    title = "{Large-momentum effective theory}",
    eprint = "2004.03543",
    archivePrefix = "arXiv",
    primaryClass = "hep-ph",
    doi = "10.1103/RevModPhys.93.035005",
    journal = "Rev. Mod. Phys.",
    volume = "93",
    number = "3",
    pages = "035005",
    year = "2021"
}

@article{Chen:2020ody,
    author = "Chen, Long-Bin and Wang, Wei and Zhu, Ruilin",
    title = "{Next-to-Next-to-Leading Order Calculation of Quasiparton Distribution Functions}",
    eprint = "2006.14825",
    archivePrefix = "arXiv",
    primaryClass = "hep-ph",
    doi = "10.1103/PhysRevLett.126.072002",
    journal = "Phys. Rev. Lett.",
    volume = "126",
    number = "7",
    pages = "072002",
    year = "2021"
}

@article{Ji:2013dva,
    author = "Ji, Xiangdong",
    title = "{Parton Physics on a Euclidean Lattice}",
    eprint = "1305.1539",
    archivePrefix = "arXiv",
    primaryClass = "hep-ph",
    doi = "10.1103/PhysRevLett.110.262002",
    journal = "Phys. Rev. Lett.",
    volume = "110",
    pages = "262002",
    year = "2013"
}

@article{Ji:2014gla,
    author = "Ji, Xiangdong",
    title = "{Parton Physics from Large-Momentum Effective Field Theory}",
    eprint = "1404.6680",
    archivePrefix = "arXiv",
    primaryClass = "hep-ph",
    doi = "10.1007/s11433-014-5492-3",
    journal = "Sci. China Phys. Mech. Astron.",
    volume = "57",
    pages = "1407--1412",
    year = "2014"
}

@article{Gao:2021hxl,
    author = "Gao, Xiang and Lee, Kyle and Mukherjee, Swagato and Shugert, Charles and Zhao, Yong",
    title = "{Origin and resummation of threshold logarithms in the lattice QCD calculations of PDFs}",
    eprint = "2102.01101",
    archivePrefix = "arXiv",
    primaryClass = "hep-ph",
    doi = "10.1103/PhysRevD.103.094504",
    journal = "Phys. Rev. D",
    volume = "103",
    number = "9",
    pages = "094504",
    year = "2021"
}

@article{Xiong:2013bka,
    author = "Xiong, Xiaonu and Ji, Xiangdong and Zhang, Jian-Hui and Zhao, Yong",
    title = "{One-loop matching for parton distributions: Nonsinglet case}",
    eprint = "1310.7471",
    archivePrefix = "arXiv",
    primaryClass = "hep-ph",
    doi = "10.1103/PhysRevD.90.014051",
    journal = "Phys. Rev. D",
    volume = "90",
    number = "1",
    pages = "014051",
    year = "2014"
}

@book{Streater:1989vi,
    author = "Streater, R. F. and Wightman, A. S.",
    title = "{PCT, spin and statistics, and all that}",
    isbn = "978-0-691-07062-9",
    year = "1989"
}

@article{Ji:2014hxa,
    author = "Ji, Xiangdong and Sun, Peng and Xiong, Xiaonu and Yuan, Feng",
    title = "{Soft factor subtraction and transverse momentum dependent parton distributions on the lattice}",
    eprint = "1405.7640",
    archivePrefix = "arXiv",
    primaryClass = "hep-ph",
    doi = "10.1103/PhysRevD.91.074009",
    journal = "Phys. Rev. D",
    volume = "91",
    pages = "074009",
    year = "2015"
}

@article{Cichy:2018mum,
    author = "Cichy, Krzysztof and Constantinou, Martha",
    title = "{A guide to light-cone PDFs from Lattice QCD: an overview of approaches, techniques and results}",
    eprint = "1811.07248",
    archivePrefix = "arXiv",
    primaryClass = "hep-lat",
    doi = "10.1155/2019/3036904",
    journal = "Adv. High Energy Phys.",
    volume = "2019",
    pages = "3036904",
    year = "2019"
}

@article{Izubuchi:2018srq,
    author = "Izubuchi, Taku and Ji, Xiangdong and Jin, Luchang and Stewart, Iain W. and Zhao, Yong",
    title = "{Factorization Theorem Relating Euclidean and Light-Cone Parton Distributions}",
    eprint = "1801.03917",
    archivePrefix = "arXiv",
    primaryClass = "hep-ph",
    reportNumber = "MIT-CTP-4960, MIT-CTP 4960",
    doi = "10.1103/PhysRevD.98.056004",
    journal = "Phys. Rev. D",
    volume = "98",
    number = "5",
    pages = "056004",
    year = "2018"
}

@article{Alekhin:2017kpj,
    author = {Alekhin, S. and Bl\"umlein, J. and Moch, S. and Placakyte, R.},
    title = "{Parton distribution functions, $\alpha_s$, and heavy-quark masses for LHC Run II}",
    eprint = "1701.05838",
    archivePrefix = "arXiv",
    primaryClass = "hep-ph",
    reportNumber = "DESY-16-179, DO-TH-16-13",
    doi = "10.1103/PhysRevD.96.014011",
    journal = "Phys. Rev. D",
    volume = "96",
    number = "1",
    pages = "014011",
    year = "2017"
}

@article{Hou:2019efy,
    author = "Hou, Tie-Jiun and others",
    title = "{New CTEQ global analysis of quantum chromodynamics with high-precision data from the LHC}",
    eprint = "1912.10053",
    archivePrefix = "arXiv",
    primaryClass = "hep-ph",
    reportNumber = "MSUHEP-19-025, PITT-PACC-1911, SMU-HEP-19-03",
    doi = "10.1103/PhysRevD.103.014013",
    journal = "Phys. Rev. D",
    volume = "103",
    number = "1",
    pages = "014013",
    year = "2021"
}

@article{H1:2015ubc,
    author = "Abramowicz, H. and others",
    collaboration = "H1, ZEUS",
    title = "{Combination of measurements of inclusive deep inelastic ${e^{\pm }p}$ scattering cross sections and QCD analysis of HERA data}",
    eprint = "1506.06042",
    archivePrefix = "arXiv",
    primaryClass = "hep-ex",
    reportNumber = "DESY-15-039",
    doi = "10.1140/epjc/s10052-015-3710-4",
    journal = "Eur. Phys. J. C",
    volume = "75",
    number = "12",
    pages = "580",
    year = "2015"
}

@article{Jimenez-Delgado:2014twa,
    author = "Jimenez-Delgado, Pedro and Reya, Ewald",
    title = "{Delineating parton distributions and the strong coupling}",
    eprint = "1403.1852",
    archivePrefix = "arXiv",
    primaryClass = "hep-ph",
    reportNumber = "DO-TH-14-03, JLAB-THY-14-1853",
    doi = "10.1103/PhysRevD.89.074049",
    journal = "Phys. Rev. D",
    volume = "89",
    number = "7",
    pages = "074049",
    year = "2014"
}

@article{Lin:2014zya,
    author = "Lin, Huey-Wen and Chen, Jiunn-Wei and Cohen, Saul D. and Ji, Xiangdong",
    title = "{Flavor Structure of the Nucleon Sea from Lattice QCD}",
    eprint = "1402.1462",
    archivePrefix = "arXiv",
    primaryClass = "hep-ph",
    reportNumber = "NT@UW-14-03, INT-PUB-14-002",
    doi = "10.1103/PhysRevD.91.054510",
    journal = "Phys. Rev. D",
    volume = "91",
    pages = "054510",
    year = "2015"
}

@article{Alexandrou:2015rja,
    author = "Alexandrou, Constantia and Cichy, Krzysztof and Drach, Vincent and Garcia-Ramos, Elena and Hadjiyiannakou, Kyriakos and Jansen, Karl and Steffens, Fernanda and Wiese, Christian",
    title = "{Lattice calculation of parton distributions}",
    eprint = "1504.07455",
    archivePrefix = "arXiv",
    primaryClass = "hep-lat",
    reportNumber = "SFB-CPP-14-124, DESY-15-059, CP3-ORIGINS-2015-013, DIAS-2015-13",
    doi = "10.1103/PhysRevD.92.014502",
    journal = "Phys. Rev. D",
    volume = "92",
    pages = "014502",
    year = "2015"
}

@article{Chen:2016utp,
    author = "Chen, Jiunn-Wei and Cohen, Saul D. and Ji, Xiangdong and Lin, Huey-Wen and Zhang, Jian-Hui",
    title = "{Nucleon Helicity and Transversity Parton Distributions from Lattice QCD}",
    eprint = "1603.06664",
    archivePrefix = "arXiv",
    primaryClass = "hep-ph",
    reportNumber = "MIT-CTP-4776, INT-PUB-16-009",
    doi = "10.1016/j.nuclphysb.2016.07.033",
    journal = "Nucl. Phys. B",
    volume = "911",
    pages = "246--273",
    year = "2016"
}

@article{Alexandrou:2016jqi,
    author = "Alexandrou, Constantia and Cichy, Krzysztof and Constantinou, Martha and Hadjiyiannakou, Kyriakos and Jansen, Karl and Steffens, Fernanda and Wiese, Christian",
    title = "{Updated Lattice Results for Parton Distributions}",
    eprint = "1610.03689",
    archivePrefix = "arXiv",
    primaryClass = "hep-lat",
    reportNumber = "DESY-16-192, DESY-16---192",
    doi = "10.1103/PhysRevD.96.014513",
    journal = "Phys. Rev. D",
    volume = "96",
    number = "1",
    pages = "014513",
    year = "2017"
}

@article{Alexandrou:2018pbm,
    author = "Alexandrou, Constantia and Cichy, Krzysztof and Constantinou, Martha and Jansen, Karl and Scapellato, Aurora and Steffens, Fernanda",
    title = "{Light-Cone Parton Distribution Functions from Lattice QCD}",
    eprint = "1803.02685",
    archivePrefix = "arXiv",
    primaryClass = "hep-lat",
    doi = "10.1103/PhysRevLett.121.112001",
    journal = "Phys. Rev. Lett.",
    volume = "121",
    number = "11",
    pages = "112001",
    year = "2018"
}

@article{Chen:2018xof,
    author = "Chen, Jiunn-Wei and Jin, Luchang and Lin, Huey-Wen and Liu, Yu-Sheng and Yang, Yi-Bo and Zhang, Jian-Hui and Zhao, Yong",
    title = "{Lattice Calculation of Parton Distribution Function from LaMET at Physical Pion Mass with Large Nucleon Momentum}",
    eprint = "1803.04393",
    archivePrefix = "arXiv",
    primaryClass = "hep-lat",
    reportNumber = "MSUHEP-18-003, MIT-CTP/4991, MIT-CTP-4991",
    month = "3",
    year = "2018"
}

@article{Lin:2018pvv,
    author = "Lin, Huey-Wen and Chen, Jiunn-Wei and Ji, Xiangdong and Jin, Luchang and Li, Ruizi and Liu, Yu-Sheng and Yang, Yi-Bo and Zhang, Jian-Hui and Zhao, Yong",
    title = "{Proton Isovector Helicity Distribution on the Lattice at Physical Pion Mass}",
    eprint = "1807.07431",
    archivePrefix = "arXiv",
    primaryClass = "hep-lat",
    reportNumber = "MSUHEP-18-013, MIT-CTP/5032",
    doi = "10.1103/PhysRevLett.121.242003",
    journal = "Phys. Rev. Lett.",
    volume = "121",
    number = "24",
    pages = "242003",
    year = "2018"
}

@article{LatticeParton:2018gjr,
    author = "Liu, Yu-Sheng and others",
    collaboration = "Lattice Parton",
    title = "{Unpolarized isovector quark distribution function from lattice QCD: A systematic analysis of renormalization and matching}",
    eprint = "1807.06566",
    archivePrefix = "arXiv",
    primaryClass = "hep-lat",
    doi = "10.1103/PhysRevD.101.034020",
    journal = "Phys. Rev. D",
    volume = "101",
    number = "3",
    pages = "034020",
    year = "2020"
}

@article{Alexandrou:2018eet,
    author = "Alexandrou, Constantia and Cichy, Krzysztof and Constantinou, Martha and Jansen, Karl and Scapellato, Aurora and Steffens, Fernanda",
    title = "{Transversity parton distribution functions from lattice QCD}",
    eprint = "1807.00232",
    archivePrefix = "arXiv",
    primaryClass = "hep-lat",
    doi = "10.1103/PhysRevD.98.091503",
    journal = "Phys. Rev. D",
    volume = "98",
    number = "9",
    pages = "091503",
    year = "2018"
}

@article{Liu:2018hxv,
    author = "Liu, Yu-Sheng and Chen, Jiunn-Wei and Jin, Luchang and Li, Ruizi and Lin, Huey-Wen and Yang, Yi-Bo and Zhang, Jian-Hui and Zhao, Yong",
    title = "{Nucleon Transversity Distribution at the Physical Pion Mass from Lattice QCD}",
    eprint = "1810.05043",
    archivePrefix = "arXiv",
    primaryClass = "hep-lat",
    reportNumber = "MSUHEP-18-019, MIT-CTP/5033",
    month = "10",
    year = "2018"
}

@article{Chen:2018fwa,
    author = {Zhang, Jian-Hui and Chen, Jiunn-Wei and Jin, Luchang and Lin, Huey-Wen and Sch\"afer, Andreas and Zhao, Yong},
    title = "{First direct lattice-QCD calculation of the $x$-dependence of the pion parton distribution function}",
    eprint = "1804.01483",
    archivePrefix = "arXiv",
    primaryClass = "hep-lat",
    doi = "10.1103/PhysRevD.100.034505",
    journal = "Phys. Rev. D",
    volume = "100",
    number = "3",
    pages = "034505",
    year = "2019"
}

@article{Izubuchi:2019lyk,
    author = "Izubuchi, Taku and Jin, Luchang and Kallidonis, Christos and Karthik, Nikhil and Mukherjee, Swagato and Petreczky, Peter and Shugert, Charles and Syritsyn, Sergey",
    title = "{Valence parton distribution function of pion from fine lattice}",
    eprint = "1905.06349",
    archivePrefix = "arXiv",
    primaryClass = "hep-lat",
    doi = "10.1103/PhysRevD.100.034516",
    journal = "Phys. Rev. D",
    volume = "100",
    number = "3",
    pages = "034516",
    year = "2019"
}

@inproceedings{Shugert:2020tgq,
    author = "Shugert, Charles and Gao, Xiang and Izubichi, Taku and Jin, Luchang and Kallidonis, Christos and Karthik, Nikhil and Mukherjee, Swagato and Petreczky, Peter and Syritsyn, Sergey and Zhao, Yong",
    title = "{Pion valence quark PDF from lattice QCD}",
    booktitle = "{37th International Symposium on Lattice Field Theory}",
    eprint = "2001.11650",
    archivePrefix = "arXiv",
    primaryClass = "hep-lat",
    month = "1",
    year = "2020"
}

@article{Chai:2020nxw,
    author = "Chai, Yahui and others",
    title = "{Parton distribution functions of $\Delta^+$ on the lattice}",
    eprint = "2002.12044",
    archivePrefix = "arXiv",
    primaryClass = "hep-lat",
    doi = "10.1103/PhysRevD.102.014508",
    journal = "Phys. Rev. D",
    volume = "102",
    number = "1",
    pages = "014508",
    year = "2020"
}

@article{Lin:2020ssv,
    author = "Lin, Huey-Wen and Chen, Jiunn-Wei and Fan, Zhouyou and Zhang, Jian-Hui and Zhang, Rui",
    title = "{Valence-Quark Distribution of the Kaon and Pion from Lattice QCD}",
    eprint = "2003.14128",
    archivePrefix = "arXiv",
    primaryClass = "hep-lat",
    reportNumber = "MSUHEP-20-006",
    doi = "10.1103/PhysRevD.103.014516",
    journal = "Phys. Rev. D",
    volume = "103",
    number = "1",
    pages = "014516",
    year = "2021"
}

@article{Fan:2020nzz,
    author = "Fan, Zhouyou and Gao, Xiang and Li, Ruizi and Lin, Huey-Wen and Karthik, Nikhil and Mukherjee, Swagato and Petreczky, Peter and Syritsyn, Sergey and Yang, Yi-Bo and Zhang, Rui",
    title = "{Isovector parton distribution functions of the proton on a superfine lattice}",
    eprint = "2005.12015",
    archivePrefix = "arXiv",
    primaryClass = "hep-lat",
    doi = "10.1103/PhysRevD.102.074504",
    journal = "Phys. Rev. D",
    volume = "102",
    number = "7",
    pages = "074504",
    year = "2020"
}

@article{Gao:2021dbh,
    author = "Gao, Xiang and Hanlon, Andrew D. and Mukherjee, Swagato and Petreczky, Peter and Scior, Philipp and Syritsyn, Sergey and Zhao, Yong",
    title = "{Lattice QCD Determination of the Bjorken-x Dependence of Parton Distribution Functions at Next-to-Next-to-Leading Order}",
    eprint = "2112.02208",
    archivePrefix = "arXiv",
    primaryClass = "hep-lat",
    doi = "10.1103/PhysRevLett.128.142003",
    journal = "Phys. Rev. Lett.",
    volume = "128",
    number = "14",
    pages = "142003",
    year = "2022"
}

@article{Gao:2022iex,
    author = "Gao, Xiang and Hanlon, Andrew D. and Karthik, Nikhil and Mukherjee, Swagato and Petreczky, Peter and Scior, Philipp and Shi, Shuzhe and Syritsyn, Sergey and Zhao, Yong and Zhou, Kai",
    title = "{Continuum-extrapolated NNLO valence PDF of the pion at the physical point}",
    eprint = "2208.02297",
    archivePrefix = "arXiv",
    primaryClass = "hep-lat",
    doi = "10.1103/PhysRevD.106.114510",
    journal = "Phys. Rev. D",
    volume = "106",
    number = "11",
    pages = "114510",
    year = "2022"
}

@article{Gao:2022uhg,
    author = "Gao, Xiang and Hanlon, Andrew D. and Holligan, Jack and Karthik, Nikhil and Mukherjee, Swagato and Petreczky, Peter and Syritsyn, Sergey and Zhao, Yong",
    title = "{Unpolarized proton PDF at NNLO from lattice QCD with physical quark masses}",
    eprint = "2212.12569",
    archivePrefix = "arXiv",
    primaryClass = "hep-lat",
    doi = "10.1103/PhysRevD.107.074509",
    journal = "Phys. Rev. D",
    volume = "107",
    number = "7",
    pages = "074509",
    year = "2023"
}

@article{Su:2022fiu,
    author = "Su, Yushan and Holligan, Jack and Ji, Xiangdong and Yao, Fei and Zhang, Jian-Hui and Zhang, Rui",
    title = "{Resumming quark's longitudinal momentum logarithms in LaMET expansion of lattice PDFs}",
    eprint = "2209.01236",
    archivePrefix = "arXiv",
    primaryClass = "hep-ph",
    doi = "10.1016/j.nuclphysb.2023.116201",
    journal = "Nucl. Phys. B",
    volume = "991",
    pages = "116201",
    year = "2023"
}

@article{LatticeParton:2022xsd,
    author = "Yao, Fei and others",
    collaboration = "Lattice Parton",
    title = "{Nucleon Transversity Distribution in the Continuum and Physical Mass Limit from Lattice QCD}",
    eprint = "2208.08008",
    archivePrefix = "arXiv",
    primaryClass = "hep-lat",
    month = "8",
    year = "2022"
}

@book{stein2003complex,
  title     = {Complex Analysis},
  author    = {Stein, Elias M. and Shakarchi, Rami},
  series    = {Princeton Lectures in Analysis, vol. II},
  year      = {2003},
  publisher = {Princeton University Press},
  address   = {Princeton, NJ},
  isbn      = {978-0-691-11385-2},
}

@article{Lin:2021umz,
    author = "Lin, Yong-Hui and Hammer, Hans-Werner and Mei{\ss}ner, Ulf-G.",
    title = "{Dispersion-theoretical analysis of the electromagnetic form factors of the nucleon: Past, present and future}",
    eprint = "2106.06357",
    archivePrefix = "arXiv",
    primaryClass = "hep-ph",
    doi = "10.1140/epja/s10050-021-00562-0",
    journal = "Eur. Phys. J. A",
    volume = "57",
    number = "8",
    pages = "255",
    year = "2021"
}

@online{DLMF_Bessel_Asy,
  author = {{NIST Digital Library of Mathematical Functions}},
  title  = {Bessel Functions — Asymptotic Expansions},
  url    = {https://dlmf.nist.gov/10.40},
  note   = {Accessed: 2025-08-21}
}

@book{Watson_Bessel_1944,
  author    = {Watson, G. N.},
  title     = {A Treatise on the Theory of Bessel Functions},
  edition   = {2},
  publisher = {Cambridge University Press},
  address   = {Cambridge},
  year      = {1944}
}

@article{Denner:2010tr,
    author = "Denner, A. and Dittmaier, S.",
    title = "{Scalar one-loop 4-point integrals}",
    eprint = "1005.2076",
    archivePrefix = "arXiv",
    primaryClass = "hep-ph",
    reportNumber = "FR-PHENO-2010-020, PSI-PR-10-10",
    doi = "10.1016/j.nuclphysb.2010.11.002",
    journal = "Nucl. Phys. B",
    volume = "844",
    pages = "199--242",
    year = "2011"
}

@article{THOOFT1979365,
title = {Scalar one-loop integrals},
journal = {Nuclear Physics B},
volume = {153},
pages = {365-401},
year = {1979},
issn = {0550-3213},
doi = {https://doi.org/10.1016/0550-3213(79)90605-9},
url = {https://www.sciencedirect.com/science/article/pii/0550321379906059},
author = {G. {'t Hooft} and M. Veltman}
}

@article{Denner:1991kt,
    author = "Denner, Ansgar",
    title = "{Techniques for calculation of electroweak radiative corrections at the one loop level and results for W physics at LEP-200}",
    eprint = "0709.1075",
    archivePrefix = "arXiv",
    primaryClass = "hep-ph",
    reportNumber = "PRINT-91-0349 (WURZBURG)",
    doi = "10.1002/prop.2190410402",
    journal = "Fortsch. Phys.",
    volume = "41",
    pages = "307--420",
    year = "1993"
}

@article{DENNER1991637,
title = {A compact expression for the scalar one-loop four-point function},
journal = {Nuclear Physics B},
volume = {367},
number = {3},
pages = {637-656},
year = {1991},
issn = {0550-3213},
doi = {https://doi.org/10.1016/0550-3213(91)90011-L},
url = {https://www.sciencedirect.com/science/article/pii/055032139190011L},
author = {A. Denner and U. Nierste and R. Scharf}
}

@article{JAFFE1983205,
title = {Parton distribution functions for twist 4},
journal = {Nuclear Physics B},
volume = {229},
number = {1},
pages = {205-230},
year = {1983},
issn = {0550-3213},
doi = {https://doi.org/10.1016/0550-3213(83)90361-9},
url = {https://www.sciencedirect.com/science/article/pii/0550321383903619},
author = {R.L. Jaffe}
}

@article{Neubert:1996wg,
    author = "Neubert, Matthias",
    editor = "Zichichi, A.",
    title = "{Heavy quark effective theory}",
    eprint = "hep-ph/9610266",
    archivePrefix = "arXiv",
    reportNumber = "CERN-TH-96-281",
    journal = "Subnucl. Ser.",
    volume = "34",
    pages = "98--165",
    year = "1997"
}

@article{Mannel:1991mc,
    author = "Mannel, Thomas and Roberts, Winston and Ryzak, Zbigniew",
    title = "{A Derivation of the heavy quark effective Lagrangian from QCD}",
    reportNumber = "HUTP-91-A017",
    doi = "10.1016/0550-3213(92)90204-O",
    journal = "Nucl. Phys. B",
    volume = "368",
    pages = "204--217",
    year = "1992"
}

@book{Collins_2023, place={Cambridge}, series={Cambridge Monographs on Particle Physics, Nuclear Physics and Cosmology}, title={Foundations of Perturbative QCD}, publisher={Cambridge University Press}, author={Collins, John}, year={2023}, collection={Cambridge Monographs on Particle Physics, Nuclear Physics and Cosmology}}

@article{ParticleDataGroup:2024cfk,
    author = "Navas, S. and others",
    collaboration = "Particle Data Group",
    title = "{Review of particle physics}",
    doi = "10.1103/PhysRevD.110.030001",
    journal = "Phys. Rev. D",
    volume = "110",
    number = "3",
    pages = "030001",
    year = "2024"
}

@article{Pineda:2001zq,
    author = "Pineda, Antonio",
    title = "{Determination of the bottom quark mass from the Upsilon(1S) system}",
    eprint = "hep-ph/0105008",
    archivePrefix = "arXiv",
    reportNumber = "TTP-01-12",
    doi = "10.1088/1126-6708/2001/06/022",
    journal = "JHEP",
    volume = "06",
    pages = "022",
    year = "2001"
}

@article{Ayala:2019hkn,
    author = "Ayala, Cesar and Lobregat, Xabier and Pineda, Antonio",
    title = "{Hyperasymptotic approximation to the top, bottom and charm pole mass}",
    eprint = "1909.01370",
    archivePrefix = "arXiv",
    primaryClass = "hep-ph",
    doi = "10.1103/PhysRevD.101.034002",
    journal = "Phys. Rev. D",
    volume = "101",
    number = "3",
    pages = "034002",
    year = "2020"
}

@article{Zhang:2023bxs,
    author = "Zhang, Rui and Holligan, Jack and Ji, Xiangdong and Su, Yushan",
    title = "{Leading power accuracy in lattice calculations of parton distributions}",
    eprint = "2305.05212",
    archivePrefix = "arXiv",
    primaryClass = "hep-lat",
    doi = "10.1016/j.physletb.2023.138081",
    journal = "Phys. Lett. B",
    volume = "844",
    pages = "138081",
    year = "2023"
}

@article{Gao:2020ito,
    author = "Gao, Xiang and Jin, Luchang and Kallidonis, Christos and Karthik, Nikhil and Mukherjee, Swagato and Petreczky, Peter and Shugert, Charles and Syritsyn, Sergey and Zhao, Yong",
    title = "{Valence parton distribution of the pion from lattice QCD: Approaching the continuum limit}",
    eprint = "2007.06590",
    archivePrefix = "arXiv",
    primaryClass = "hep-lat",
    doi = "10.1103/PhysRevD.102.094513",
    journal = "Phys. Rev. D",
    volume = "102",
    number = "9",
    pages = "094513",
    year = "2020"
}

@article{Gao:2022ytj,
    author = "Gao, Xiang and Hanlon, Andrew D. and Mukherjee, Swagato and Petreczky, Peter and Scior, Philipp and Syritsyn, Sergey and Zhao, Yong",
    title = "{Lattice QCD Determination of the Bjorken-{\ensuremath{\boldsymbol{\mathit{x}}}} Dependence of PDFs at NNLO}",
    doi = "10.22323/1.430.0104",
    journal = "PoS",
    volume = "LATTICE2022",
    pages = "104",
    year = "2023"
}

@article{Ji:2024hit,
    author = "Ji, Xiangdong and Liu, Yizhuang and Su, Yushan and Zhang, Rui",
    title = "{Effects of threshold resummation for large-x PDF in large momentum effective theory}",
    eprint = "2410.12910",
    archivePrefix = "arXiv",
    primaryClass = "hep-ph",
    doi = "10.1007/JHEP03(2025)045",
    journal = "JHEP",
    volume = "03",
    pages = "045",
    year = "2025"
}

@article{ATLAS:2021vod,
    author = "Aad, Georges and others",
    collaboration = "ATLAS",
    title = "{Determination of the parton distribution functions of the proton using diverse ATLAS data from $pp$ collisions at $\sqrt{s} = 7$, 8 and 13~TeV}",
    eprint = "2112.11266",
    archivePrefix = "arXiv",
    primaryClass = "hep-ex",
    reportNumber = "CERN-EP-2021-239",
    doi = "10.1140/epjc/s10052-022-10217-z",
    journal = "Eur. Phys. J. C",
    volume = "82",
    number = "5",
    pages = "438",
    year = "2022"
}

@article{NNPDF:2021njg,
    author = "Ball, Richard D. and others",
    collaboration = "NNPDF",
    title = "{The path to proton structure at 1{\%} accuracy}",
    eprint = "2109.02653",
    archivePrefix = "arXiv",
    primaryClass = "hep-ph",
    reportNumber = "Edinburgh 2021/12, Nikhef-2021-013, TIF-UNIMI-2021-11",
    doi = "10.1140/epjc/s10052-022-10328-7",
    journal = "Eur. Phys. J. C",
    volume = "82",
    number = "5",
    pages = "428",
    year = "2022"
}

@article{Bailey:2020ooq,
    author = "Bailey, S. and Cridge, T. and Harland-Lang, L. A. and Martin, A. D. and Thorne, R. S.",
    title = "{Parton distributions from LHC, HERA, Tevatron and fixed target data: MSHT20 PDFs}",
    eprint = "2012.04684",
    archivePrefix = "arXiv",
    primaryClass = "hep-ph",
    reportNumber = "IPPP/20/58",
    doi = "10.1140/epjc/s10052-021-09057-0",
    journal = "Eur. Phys. J. C",
    volume = "81",
    number = "4",
    pages = "341",
    year = "2021"
}

@article{Ji:2024oka,
    author = "Ji, Xiangdong",
    title = "{Euclidean effective theory for partons in the spirit of Steven Weinberg}",
    eprint = "2408.03378",
    archivePrefix = "arXiv",
    primaryClass = "hep-ph",
    doi = "10.1016/j.nuclphysb.2024.116670",
    journal = "Nucl. Phys. B",
    volume = "1007",
    pages = "116670",
    year = "2024"
}

@article{Li:2020xml,
    author = "Li, Zheng-Yang and Ma, Yan-Qing and Qiu, Jian-Wei",
    title = "{Extraction of Next-to-Next-to-Leading-Order Parton Distribution Functions from Lattice QCD Calculations}",
    eprint = "2006.12370",
    archivePrefix = "arXiv",
    primaryClass = "hep-ph",
    reportNumber = "JLAB-THY-20-3214",
    doi = "10.1103/PhysRevLett.126.072001",
    journal = "Phys. Rev. Lett.",
    volume = "126",
    number = "7",
    pages = "072001",
    year = "2021"
}

@article{Bogner:2010kv,
    author = "Bogner, Christian and Weinzierl, Stefan",
    title = "{Feynman graph polynomials}",
    eprint = "1002.3458",
    archivePrefix = "arXiv",
    primaryClass = "hep-ph",
    doi = "10.1142/S0217751X10049438",
    journal = "Int. J. Mod. Phys. A",
    volume = "25",
    pages = "2585--2618",
    year = "2010"
}

@article{Ji:2020brr,
    author = {Ji, Xiangdong and Liu, Yizhuang and Sch\"afer, Andreas and Wang, Wei and Yang, Yi-Bo and Zhang, Jian-Hui and Zhao, Yong},
    title = "{A Hybrid Renormalization Scheme for Quasi Light-Front Correlations in Large-Momentum Effective Theory}",
    eprint = "2008.03886",
    archivePrefix = "arXiv",
    primaryClass = "hep-ph",
    doi = "10.1016/j.nuclphysb.2021.115311",
    journal = "Nucl. Phys. B",
    volume = "964",
    pages = "115311",
    year = "2021"
}

@article{Braun:2018brg,
    author = "Braun, Vladimir M. and Vladimirov, Alexey and Zhang, Jian-Hui",
    title = "{Power corrections and renormalons in parton quasidistributions}",
    eprint = "1810.00048",
    archivePrefix = "arXiv",
    primaryClass = "hep-ph",
    doi = "10.1103/PhysRevD.99.014013",
    journal = "Phys. Rev. D",
    volume = "99",
    number = "1",
    pages = "014013",
    year = "2019"
}

@article{Ji:2023pba,
    author = "Ji, Xiangdong and Liu, Yizhuang and Su, Yushan",
    title = "{Threshold resummation for computing large-x parton distribution through large-momentum effective theory}",
    eprint = "2305.04416",
    archivePrefix = "arXiv",
    primaryClass = "hep-ph",
    doi = "10.1007/JHEP08(2023)037",
    journal = "JHEP",
    volume = "08",
    pages = "037",
    year = "2023"
}

@article{Chen:2019lcm,
    author = "Chen, Jiunn-Wei and Lin, Huey-Wen and Zhang, Jian-Hui",
    title = "{Pion generalized parton distribution from lattice QCD}",
    eprint = "1904.12376",
    archivePrefix = "arXiv",
    primaryClass = "hep-lat",
    doi = "10.1016/j.nuclphysb.2020.114940",
    journal = "Nucl. Phys. B",
    volume = "952",
    pages = "114940",
    year = "2020"
}

@article{Alexandrou:2019dax,
    author = "Alexandrou, Constantia and Cichy, Krzysztof and Constantinou, Martha and Hadjiyiannakou, Kyriakos and Jansen, Karl and Scapellato, Aurora and Steffens, Fernanda",
    title = "{Quasi-PDFs with Twisted Mass Fermions}",
    eprint = "1910.13229",
    archivePrefix = "arXiv",
    primaryClass = "hep-lat",
    doi = "10.22323/1.363.0036",
    journal = "PoS",
    volume = "LATTICE2019",
    pages = "036",
    year = "2019"
}

@article{Lin:2020rxa,
    author = "Lin, Huey-Wen",
    title = "{Nucleon Tomography and Generalized Parton Distribution at Physical Pion Mass from Lattice QCD}",
    eprint = "2008.12474",
    archivePrefix = "arXiv",
    primaryClass = "hep-ph",
    reportNumber = "MSUHEP-20-014, MSUHEP-20-014",
    doi = "10.1103/PhysRevLett.127.182001",
    journal = "Phys. Rev. Lett.",
    volume = "127",
    number = "18",
    pages = "182001",
    year = "2021"
}

@article{Alexandrou:2020zbe,
    author = "Alexandrou, Constantia and Cichy, Krzysztof and Constantinou, Martha and Hadjiyiannakou, Kyriakos and Jansen, Karl and Scapellato, Aurora and Steffens, Fernanda",
    title = "{Unpolarized and helicity generalized parton distributions of the proton within lattice QCD}",
    eprint = "2008.10573",
    archivePrefix = "arXiv",
    primaryClass = "hep-lat",
    reportNumber = "DESY-20-150",
    doi = "10.1103/PhysRevLett.125.262001",
    journal = "Phys. Rev. Lett.",
    volume = "125",
    number = "26",
    pages = "262001",
    year = "2020"
}

@article{Lin:2021brq,
    author = "Lin, Huey-Wen",
    title = "{Nucleon helicity generalized parton distribution at physical pion mass from lattice QCD}",
    eprint = "2112.07519",
    archivePrefix = "arXiv",
    primaryClass = "hep-lat",
    reportNumber = "MSUHEP-21-024",
    doi = "10.1016/j.physletb.2021.136821",
    journal = "Phys. Lett. B",
    volume = "824",
    pages = "136821",
    year = "2022"
}

@article{Scapellato:2022mai,
    author = "Scapellato, Aurora and Alexandrou, Constantia and Cichy, Krzysztof and Constantinou, Martha and Hadjiyiannakou, Kyriakos and Jansen, Karl and Steffens, Fernanda",
    title = "{Proton generalized parton distributions from lattice QCD}",
    eprint = "2201.06519",
    archivePrefix = "arXiv",
    primaryClass = "hep-lat",
    doi = "10.31349/SuplRevMexFis.3.0308104",
    journal = "Rev. Mex. Fis. Suppl.",
    volume = "3",
    number = "3",
    pages = "0308104",
    year = "2022"
}

@article{Zhang:2017bzy,
    author = "Zhang, Jian-Hui and Chen, Jiunn-Wei and Ji, Xiangdong and Jin, Luchang and Lin, Huey-Wen",
    title = "{Pion Distribution Amplitude from Lattice QCD}",
    eprint = "1702.00008",
    archivePrefix = "arXiv",
    primaryClass = "hep-lat",
    doi = "10.1103/PhysRevD.95.094514",
    journal = "Phys. Rev. D",
    volume = "95",
    number = "9",
    pages = "094514",
    year = "2017"
}

@article{Chen:2017gck,
    author = {Zhang, Jian-Hui and Jin, Luchang and Lin, Huey-Wen and Sch\"afer, Andreas and Sun, Peng and Yang, Yi-Bo and Zhang, Rui and Zhao, Yong and Chen, Jiunn-Wei},
    collaboration = "LP3",
    title = "{Kaon Distribution Amplitude from Lattice QCD and the Flavor SU(3) Symmetry}",
    eprint = "1712.10025",
    archivePrefix = "arXiv",
    primaryClass = "hep-ph",
    reportNumber = "MSUHEP-17-023",
    doi = "10.1016/j.nuclphysb.2018.12.020",
    journal = "Nucl. Phys. B",
    volume = "939",
    pages = "429--446",
    year = "2019"
}

@article{Zhang:2020gaj,
    author = "Zhang, Rui and Honkala, Carson and Lin, Huey-Wen and Chen, Jiunn-Wei",
    title = "{Pion and kaon distribution amplitudes in the continuum limit}",
    eprint = "2005.13955",
    archivePrefix = "arXiv",
    primaryClass = "hep-lat",
    reportNumber = "MSUHEP-20-010",
    doi = "10.1103/PhysRevD.102.094519",
    journal = "Phys. Rev. D",
    volume = "102",
    number = "9",
    pages = "094519",
    year = "2020"
}

@article{Hua:2020gnw,
    author = "Hua, Jun and Chu, Min-Huan and Sun, Peng and Wang, Wei and Xu, Ji and Yang, Yi-Bo and Zhang, Jian-Hui and Zhang, Qi-An",
    collaboration = "Lattice Parton",
    title = "{Distribution Amplitudes of K* and \ensuremath{\phi} at the Physical Pion Mass from Lattice QCD}",
    eprint = "2011.09788",
    archivePrefix = "arXiv",
    primaryClass = "hep-lat",
    doi = "10.1103/PhysRevLett.127.062002",
    journal = "Phys. Rev. Lett.",
    volume = "127",
    number = "6",
    pages = "062002",
    year = "2021"
}

@article{Shanahan:2019zcq,
    author = "Shanahan, Phiala and Wagman, Michael L. and Zhao, Yong",
    title = "{Nonperturbative renormalization of staple-shaped Wilson line operators in lattice QCD}",
    eprint = "1911.00800",
    archivePrefix = "arXiv",
    primaryClass = "hep-lat",
    reportNumber = "MIT/CTP-5153",
    doi = "10.1103/PhysRevD.101.074505",
    journal = "Phys. Rev. D",
    volume = "101",
    number = "7",
    pages = "074505",
    year = "2020"
}

@article{Shanahan:2020zxr,
    author = "Shanahan, Phiala and Wagman, Michael and Zhao, Yong",
    title = "{Collins-Soper kernel for TMD evolution from lattice QCD}",
    eprint = "2003.06063",
    archivePrefix = "arXiv",
    primaryClass = "hep-lat",
    reportNumber = "FERMILAB-PUB-20-102-T, MIT/CTP-5184",
    doi = "10.1103/PhysRevD.102.014511",
    journal = "Phys. Rev. D",
    volume = "102",
    number = "1",
    pages = "014511",
    year = "2020"
}

@article{Zhang:2020dbb,
    author = "Zhang, Qi-An and others",
    collaboration = "Lattice Parton",
    title = "{Lattice-QCD Calculations of TMD Soft Function Through Large-Momentum Effective Theory}",
    eprint = "2005.14572",
    archivePrefix = "arXiv",
    primaryClass = "hep-lat",
    doi = "10.22323/1.396.0477",
    journal = "Phys. Rev. Lett.",
    volume = "125",
    number = "19",
    pages = "192001",
    year = "2020"
}

@article{Chou:2022drv,
    author = "Chou, Chien-Yu and Chen, Jiunn-Wei",
    title = "{One-loop hybrid renormalization matching kernels for quasiparton distributions}",
    eprint = "2204.08343",
    archivePrefix = "arXiv",
    primaryClass = "hep-lat",
    doi = "10.1103/PhysRevD.106.014507",
    journal = "Phys. Rev. D",
    volume = "106",
    number = "1",
    pages = "014507",
    year = "2022"
}

@article{Gao:2022vyh,
    author = "Gao, Xiang and Hanlon, Andrew D. and Karthik, Nikhil and Mukherjee, Swagato and Petreczky, Peter and Scior, Philipp and Syritsyn, Sergey and Zhao, Yong",
    title = "{Pion distribution amplitude at the physical point using the leading-twist expansion of the quasi-distribution-amplitude matrix element}",
    eprint = "2206.04084",
    archivePrefix = "arXiv",
    primaryClass = "hep-lat",
    reportNumber = "JLAB-THY-22-3626",
    doi = "10.1103/PhysRevD.106.074505",
    journal = "Phys. Rev. D",
    volume = "106",
    number = "7",
    pages = "074505",
    year = "2022"
}

@article{Bhattacharya:2022aob,
    author = "Bhattacharya, Shohini and Cichy, Krzysztof and Constantinou, Martha and Dodson, Jack and Gao, Xiang and Metz, Andreas and Mukherjee, Swagato and Scapellato, Aurora and Steffens, Fernanda and Zhao, Yong",
    title = "{Generalized parton distributions from lattice QCD with asymmetric momentum transfer: Unpolarized quarks}",
    eprint = "2209.05373",
    archivePrefix = "arXiv",
    primaryClass = "hep-lat",
    doi = "10.1103/PhysRevD.106.114512",
    journal = "Phys. Rev. D",
    volume = "106",
    number = "11",
    pages = "114512",
    year = "2022"
}

@article{Xu:2022guw,
    author = "Xu, Ji and Zhang, Xi-Ruo",
    title = "{Matching the B-meson quasidistribution amplitude in the RI/MOM scheme}",
    eprint = "2209.10719",
    archivePrefix = "arXiv",
    primaryClass = "hep-ph",
    doi = "10.1103/PhysRevD.106.114019",
    journal = "Phys. Rev. D",
    volume = "106",
    number = "11",
    pages = "114019",
    year = "2022"
}

@article{Deng:2023csv,
    author = "Deng, Zhi-Fu and Han, Chao and Wang, Wei and Zeng, Jun and Zhang, Jia-Lu",
    title = "{Light-cone distribution amplitudes of a light baryon in large-momentum effective theory}",
    eprint = "2304.09004",
    archivePrefix = "arXiv",
    primaryClass = "hep-ph",
    doi = "10.1007/JHEP07(2023)191",
    journal = "JHEP",
    volume = "07",
    pages = "191",
    year = "2023"
}

@article{Han:2023xbl,
    author = "Han, Chao and Su, Yushan and Wang, Wei and Zhang, Jia-Lu",
    title = "{Hybrid renormalization for quasi distribution amplitudes of a light baryon}",
    eprint = "2308.16793",
    archivePrefix = "arXiv",
    primaryClass = "hep-ph",
    doi = "10.1007/JHEP12(2023)044",
    journal = "JHEP",
    volume = "12",
    pages = "044",
    year = "2023"
}

@article{Han:2023hgy,
    author = "Han, Chao and Zhang, Jialu",
    title = "{Light baryon spatial correlators at short distances}",
    eprint = "2311.02669",
    archivePrefix = "arXiv",
    primaryClass = "hep-ph",
    month = "11",
    year = "2023"
}

@article{Zhang:2022xuw,
    author = "Zhang, Kuan and Ji, Xiangdong and Yang, Yi-Bo and Yao, Fei and Zhang, Jian-Hui",
    collaboration = "[Lattice Parton Collaboration (LPC)]",
    title = "{Renormalization of Transverse-Momentum-Dependent Parton Distribution on the Lattice}",
    eprint = "2205.13402",
    archivePrefix = "arXiv",
    primaryClass = "hep-lat",
    doi = "10.1103/PhysRevLett.129.082002",
    journal = "Phys. Rev. Lett.",
    volume = "129",
    number = "8",
    pages = "082002",
    year = "2022"
}

@article{Deng:2022gzi,
    author = "Deng, Zhi-Fu and Wang, Wei and Zeng, Jun",
    title = "{Transverse-momentum-dependent wave functions and soft functions at one-loop in large momentum effective theory}",
    eprint = "2207.07280",
    archivePrefix = "arXiv",
    primaryClass = "hep-th",
    doi = "10.1007/JHEP09(2022)046",
    journal = "JHEP",
    volume = "09",
    pages = "046",
    year = "2022"
}

@article{Rodini:2022wic,
    author = "Rodini, Simone and Vladimirov, Alexey",
    title = "{Factorization for quasi-TMD distributions of sub-leading power}",
    eprint = "2211.04494",
    archivePrefix = "arXiv",
    primaryClass = "hep-ph",
    doi = "10.1007/JHEP09(2023)117",
    journal = "JHEP",
    volume = "09",
    pages = "117",
    year = "2023"
}

@article{Shu:2023cot,
    author = {Shu, Hai-Tao and Schlemmer, Maximilian and Sizmann, Tobias and Vladimirov, Alexey and Walter, Lisa and Engelhardt, Michael and Sch\"afer, Andreas and Yang, Yi-Bo},
    title = "{Universality of the Collins-Soper kernel in lattice calculations}",
    eprint = "2302.06502",
    archivePrefix = "arXiv",
    primaryClass = "hep-lat",
    doi = "10.1103/PhysRevD.108.074519",
    journal = "Phys. Rev. D",
    volume = "108",
    number = "7",
    pages = "074519",
    year = "2023"
}

@article{LatticePartonLPC:2023pdv,
    author = "Chu, Min-Huan and others",
    collaboration = "Lattice Parton (LPC)",
    title = "{Lattice calculation of the intrinsic soft function and the Collins-Soper kernel}",
    eprint = "2306.06488",
    archivePrefix = "arXiv",
    primaryClass = "hep-lat",
    doi = "10.1007/JHEP08(2023)172",
    journal = "JHEP",
    volume = "08",
    pages = "172",
    year = "2023"
}

@article{Avkhadiev:2023poz,
    author = "Avkhadiev, Artur and Shanahan, Phiala and Wagman, Michael and Zhao, Yong",
    title = "{Collins-Soper kernel from lattice QCD at the physical pion mass}",
    eprint = "2307.12359",
    archivePrefix = "arXiv",
    primaryClass = "hep-lat",
    reportNumber = "MIT-CTP/5587, FERMILAB-PUB-23-375-T",
    month = "7",
    year = "2023"
}

@article{Zhao:2023ptv,
    author = "Zhao, Yong",
    title = "{Transverse Momentum Distributions from Lattice QCD without Wilson Lines}",
    eprint = "2311.01391",
    archivePrefix = "arXiv",
    primaryClass = "hep-ph",
    doi = "10.1103/PhysRevLett.133.241904",
    journal = "Phys. Rev. Lett.",
    volume = "133",
    number = "24",
    pages = "241904",
    year = "2024"
}

@article{Gao:2023lny,
    author = "Gao, Xiang and Liu, Wei-Yang and Zhao, Yong",
    title = "{Parton Distributions from Boosted Fields in the Coulomb Gauge}",
    eprint = "2306.14960",
    archivePrefix = "arXiv",
    primaryClass = "hep-ph",
    month = "6",
    year = "2023"
}

@article{Gao:2023ktu,
    author = "Gao, Xiang and Hanlon, Andrew D. and Mukherjee, Swagato and Petreczky, Peter and Shi, Qi and Syritsyn, Sergey and Zhao, Yong",
    title = "{Transversity PDFs of the proton from lattice QCD with physical quark masses}",
    eprint = "2310.19047",
    archivePrefix = "arXiv",
    primaryClass = "hep-lat",
    month = "10",
    year = "2023"
}

@article{Baker:2024zcd,
    author = {Baker, Ethan and Bollweg, Dennis and Boyle, Peter and Clo\"et, Ian and Gao, Xiang and Mukherjee, Swagato and Petreczky, Peter and Zhang, Rui and Zhao, Yong},
    title = "{Lattice QCD calculation of the pion distribution amplitude with domain wall fermions at physical pion mass}",
    eprint = "2405.20120",
    archivePrefix = "arXiv",
    primaryClass = "hep-lat",
    doi = "10.1007/JHEP07(2024)211",
    journal = "JHEP",
    volume = "07",
    pages = "211",
    year = "2024"
}

@article{Cloet:2024vbv,
    author = "Cloet, Ian and Gao, Xiang and Mukherjee, Swagato and Syritsyn, Sergey and Karthik, Nikhil and Petreczky, Peter and Zhang, Rui and Zhao, Yong",
    title = "{Lattice QCD Calculation of $x$-dependent Meson Distribution Amplitudes at Physical Pion Mass with Threshold Logarithm Resummation}",
    eprint = "2407.00206",
    archivePrefix = "arXiv",
    primaryClass = "hep-lat",
    month = "6",
    year = "2024"
}

@article{Han:2024ucv,
    author = "Han, Chao and Wang, Wei and Zeng, Jun and Zhang, Jia-Lu",
    title = "{Lightcone and quasi distribution amplitudes for light octet and decuplet baryons}",
    eprint = "2404.04855",
    archivePrefix = "arXiv",
    primaryClass = "hep-ph",
    doi = "10.1007/JHEP07(2024)019",
    journal = "JHEP",
    volume = "07",
    pages = "019",
    year = "2024"
}

@article{Han:2024cht,
    author = "Han, Chao and Wang, Wei and Zhang, Jia-Lu and Zhang, Jian-Hui",
    title = "{Power corrections to quasi-distribution amplitudes of a heavy meson}",
    eprint = "2408.13486",
    archivePrefix = "arXiv",
    primaryClass = "hep-ph",
    month = "8",
    year = "2024"
}

@article{Deng:2024dkd,
    author = "Deng, Zhi-Fu and Wang, Wei and Wei, Yan-Bing and Zeng, Jun",
    title = "{Probing heavy meson lightcone distribution amplitudes with heavy quark spin symmetry}",
    eprint = "2409.00632",
    archivePrefix = "arXiv",
    primaryClass = "hep-ph",
    month = "9",
    year = "2024"
}

@article{Chen:2024rgi,
    author = "Chen, Chen and Liu, Liuming and Sun, Peng and Yang, Yi-Bo and Geng, Yiqi and Yao, Fei and Zhang, Jian-Hui and Zhang, Kuan",
    title = "{Parton Distribution Function of a Deuteron-like Dibaryon System from Lattice QCD}",
    eprint = "2408.12819",
    archivePrefix = "arXiv",
    primaryClass = "hep-lat",
    month = "8",
    year = "2024"
}

@article{Bollweg:2024zet,
    author = "Bollweg, Dennis and Gao, Xiang and Mukherjee, Swagato and Zhao, Yong",
    title = "{Nonperturbative Collins-Soper kernel from chiral quarks with physical masses}",
    eprint = "2403.00664",
    archivePrefix = "arXiv",
    primaryClass = "hep-lat",
    doi = "10.1016/j.physletb.2024.138617",
    journal = "Phys. Lett. B",
    volume = "852",
    pages = "138617",
    year = "2024"
}

@article{Avkhadiev:2024mgd,
    author = "Avkhadiev, Artur and Shanahan, Phiala E. and Wagman, Michael L. and Zhao, Yong",
    title = "{Determination of the Collins-Soper Kernel from Lattice QCD}",
    eprint = "2402.06725",
    archivePrefix = "arXiv",
    primaryClass = "hep-lat",
    reportNumber = "FERMILAB-PUB-24-0037-T, MIT-CTP/5677",
    doi = "10.1103/PhysRevLett.132.231901",
    journal = "Phys. Rev. Lett.",
    volume = "132",
    number = "23",
    pages = "231901",
    year = "2024"
}

@article{Bhattacharya:2023jsc,
    author = "Bhattacharya, Shohini and others",
    title = "{Generalized parton distributions from lattice QCD with asymmetric momentum transfer: Axial-vector case}",
    eprint = "2310.13114",
    archivePrefix = "arXiv",
    primaryClass = "hep-lat",
    doi = "10.1103/PhysRevD.109.034508",
    journal = "Phys. Rev. D",
    volume = "109",
    number = "3",
    pages = "034508",
    year = "2024"
}

@article{Fan:2018dxu,
    author = "Fan, Zhou-You and Yang, Yi-Bo and Anthony, Adam and Lin, Huey-Wen and Liu, Keh-Fei",
    title = "{Gluon Quasi-Parton-Distribution Functions from Lattice QCD}",
    eprint = "1808.02077",
    archivePrefix = "arXiv",
    primaryClass = "hep-lat",
    doi = "10.1103/PhysRevLett.121.242001",
    journal = "Phys. Rev. Lett.",
    volume = "121",
    number = "24",
    pages = "242001",
    year = "2018"
}

@article{Good:2024iur,
    author = "Good, William and Hasan, Kinza and Lin, Huey-Wen",
    title = "{Toward the First Gluon Parton Distribution from the LaMET}",
    eprint = "2409.02750",
    archivePrefix = "arXiv",
    primaryClass = "hep-lat",
    month = "9",
    year = "2024"
}

@article{Holligan:2024wpv,
    author = "Holligan, Jack and Lin, Huey-Wen",
    title = "{Nucleon helicity parton distribution function in the continuum limit with self-renormalization}",
    eprint = "2405.18238",
    archivePrefix = "arXiv",
    primaryClass = "hep-lat",
    reportNumber = "MSUHEP-24-003",
    doi = "10.1016/j.physletb.2024.138731",
    journal = "Phys. Lett. B",
    volume = "854",
    pages = "138731",
    year = "2024"
}

@article{Holligan:2024umc,
    author = "Holligan, Jack and Lin, Huey-Wen",
    title = "{Pion valence quark distribution at physical pion mass of N $_{f}$ = 2 + 1 + 1 lattice QCD}",
    eprint = "2404.14525",
    archivePrefix = "arXiv",
    primaryClass = "hep-lat",
    reportNumber = "MSUHEP-23-032",
    doi = "10.1088/1361-6471/ad3162",
    journal = "J. Phys. G",
    volume = "51",
    number = "6",
    pages = "065101",
    year = "2024"
}

@article{Holligan:2023jqh,
    author = "Holligan, Jack and Lin, Huey-Wen",
    title = "{Systematic improvement of x-dependent unpolarized nucleon generalized parton distributions in lattice-QCD calculation}",
    eprint = "2312.10829",
    archivePrefix = "arXiv",
    primaryClass = "hep-lat",
    reportNumber = "MSUHEP-23-033",
    doi = "10.1103/PhysRevD.110.034503",
    journal = "Phys. Rev. D",
    volume = "110",
    number = "3",
    pages = "034503",
    year = "2024"
}

@article{Lin:2023gxz,
    author = "Lin, Huey-Wen",
    title = "{Pion valence-quark generalized parton distribution at physical pion mass}",
    eprint = "2310.10579",
    archivePrefix = "arXiv",
    primaryClass = "hep-lat",
    reportNumber = "MSUHEP-23-011",
    doi = "10.1016/j.physletb.2023.138181",
    journal = "Phys. Lett. B",
    volume = "846",
    pages = "138181",
    year = "2023"
}

@article{Holligan:2023rex,
    author = "Holligan, Jack and Ji, Xiangdong and Lin, Huey-Wen and Su, Yushan and Zhang, Rui",
    title = "{Precision control in lattice calculation of x-dependent pion distribution amplitude}",
    eprint = "2301.10372",
    archivePrefix = "arXiv",
    primaryClass = "hep-lat",
    doi = "10.1016/j.nuclphysb.2023.116282",
    journal = "Nucl. Phys. B",
    volume = "993",
    pages = "116282",
    year = "2023"
}

@article{Spanoudes:2024kpb,
    author = "Spanoudes, Gregoris and Constantinou, Martha and Panagopoulos, Haralambos",
    title = "{Renormalization of asymmetric staple-shaped Wilson-line operators in lattice and continuum perturbation theory}",
    eprint = "2401.01182",
    archivePrefix = "arXiv",
    primaryClass = "hep-lat",
    doi = "10.1103/PhysRevD.109.114501",
    journal = "Phys. Rev. D",
    volume = "109",
    number = "11",
    pages = "114501",
    year = "2024"
}

@article{Bhattacharya:2023nmv,
    author = "Bhattacharya, Shohini and Cichy, Krzysztof and Constantinou, Martha and Dodson, Jack and Metz, Andreas and Scapellato, Aurora and Steffens, Fernanda",
    title = "{Chiral-even axial twist-3 GPDs of the proton from lattice QCD}",
    eprint = "2306.05533",
    archivePrefix = "arXiv",
    primaryClass = "hep-lat",
    doi = "10.1103/PhysRevD.108.054501",
    journal = "Phys. Rev. D",
    volume = "108",
    number = "5",
    pages = "054501",
    year = "2023"
}

@article{Alexandrou:2023ucc,
    author = "Alexandrou, Constantia and others",
    title = "{Nonperturbative renormalization of asymmetric staple-shaped operators in twisted mass lattice QCD}",
    eprint = "2305.11824",
    archivePrefix = "arXiv",
    primaryClass = "hep-lat",
    doi = "10.1103/PhysRevD.108.114503",
    journal = "Phys. Rev. D",
    volume = "108",
    number = "11",
    pages = "114503",
    year = "2023"
}

@article{LatticeParton:2023xdl,
    author = "Chu, Min-Huan and others",
    collaboration = "Lattice Parton",
    title = "{Transverse-momentum-dependent wave functions of the pion from lattice QCD}",
    eprint = "2302.09961",
    archivePrefix = "arXiv",
    primaryClass = "hep-lat",
    doi = "10.1103/PhysRevD.109.L091503",
    journal = "Phys. Rev. D",
    volume = "109",
    number = "9",
    pages = "L091503",
    year = "2024"
}

@article{LatticePartonCollaborationLPC:2022myp,
    author = {He, Jin-Chen and Chu, Min-Huan and Hua, Jun and Ji, Xiangdong and Sch\"afer, Andreas and Su, Yushan and Wang, Wei and Yang, Yi-Bo and Zhang, Jian-Hui and Zhang, Qi-An},
    collaboration = "Lattice Parton Collaboration (LPC)",
    title = "{Unpolarized transverse momentum dependent parton distributions of the nucleon from lattice QCD}",
    eprint = "2211.02340",
    archivePrefix = "arXiv",
    primaryClass = "hep-lat",
    doi = "10.1103/PhysRevD.109.114513",
    journal = "Phys. Rev. D",
    volume = "109",
    number = "11",
    pages = "114513",
    year = "2024"
}

@article{LatticePartonLPC:2022eev,
    author = "Chu, Min-Huan and others",
    collaboration = "Lattice Parton (LPC)",
    title = "{Nonperturbative determination of the Collins-Soper kernel from quasitransverse-momentum-dependent wave functions}",
    eprint = "2204.00200",
    archivePrefix = "arXiv",
    primaryClass = "hep-lat",
    doi = "10.1103/PhysRevD.106.034509",
    journal = "Phys. Rev. D",
    volume = "106",
    number = "3",
    pages = "034509",
    year = "2022"
}

@article{LatticeParton:2022zqc,
    author = "Hua, Jun and others",
    collaboration = "Lattice Parton",
    title = "{Pion and Kaon Distribution Amplitudes from Lattice QCD}",
    eprint = "2201.09173",
    archivePrefix = "arXiv",
    primaryClass = "hep-lat",
    doi = "10.1103/PhysRevLett.129.132001",
    journal = "Phys. Rev. Lett.",
    volume = "129",
    number = "13",
    pages = "132001",
    year = "2022"
}

@article{Zhu:2022bja,
    author = "Zhu, Ruilin and Ji, Yao and Zhang, Jian-Hui and Zhao, Shuai",
    title = "{Gluon transverse-momentum-dependent distributions from large-momentum effective theory}",
    eprint = "2209.05443",
    archivePrefix = "arXiv",
    primaryClass = "hep-ph",
    reportNumber = "TUM-HEP-1417/22, JLAB-THY-22-3720, JLAB-THY-22-3720",
    doi = "10.1007/JHEP02(2023)114",
    journal = "JHEP",
    volume = "02",
    pages = "114",
    year = "2023"
}

@article{Liu:2022nnk,
    author = "Liu, Y.",
    title = "{Lecture Notes on Transverse-momentum-dependent Parton Distribution Function and Soft Functions in the Large-momentum Effective Theory}",
    doi = "10.5506/APhysPolB.53.4-A2",
    journal = "Acta Phys. Polon. B",
    volume = "53",
    number = "4",
    pages = "4-A2",
    year = "2022"
}

@article{Zhang:2023wea,
    author = "Zhang, Jian-Hui",
    title = "{Double Parton Distributions from Euclidean Lattice}",
    eprint = "2304.12481",
    archivePrefix = "arXiv",
    primaryClass = "hep-ph",
    month = "4",
    year = "2023"
}

@article{Jaarsma:2023woo,
    author = "Jaarsma, Max and Rahn, Rudi and Waalewijn, Wouter J.",
    title = "{Towards double parton distributions from first principles using Large Momentum Effective Theory}",
    eprint = "2305.09716",
    archivePrefix = "arXiv",
    primaryClass = "hep-ph",
    doi = "10.1007/JHEP12(2023)014",
    journal = "JHEP",
    volume = "12",
    pages = "014",
    year = "2023"
}

@article{Ayala:2019uaw,
    author = "Ayala, Cesar and Lobregat, Xabier and Pineda, Antonio",
    title = "{Superasymptotic and hyperasymptotic approximation to the operator product expansion}",
    eprint = "1902.07736",
    archivePrefix = "arXiv",
    primaryClass = "hep-th",
    doi = "10.1103/PhysRevD.99.074019",
    journal = "Phys. Rev. D",
    volume = "99",
    number = "7",
    pages = "074019",
    year = "2019"
}

@article{Liu:2023onm,
    author = "Liu, Yizhuang and Su, Yushan",
    title = "{Renormalon cancellation and linear power correction to threshold-like asymptotics of space-like parton correlators}",
    eprint = "2311.06907",
    archivePrefix = "arXiv",
    primaryClass = "hep-ph",
    doi = "10.1007/JHEP02(2024)204",
    journal = "JHEP",
    volume = "2024",
    pages = "204",
    year = "2024"
}

@article{Hoang:2008yj,
    author = "Hoang, Andre H. and Jain, Ambar and Scimemi, Ignazio and Stewart, Iain W.",
    title = "{Infrared Renormalization Group Flow for Heavy Quark Masses}",
    eprint = "0803.4214",
    archivePrefix = "arXiv",
    primaryClass = "hep-ph",
    reportNumber = "MIT-CTP-3940, MPP-2008-24",
    doi = "10.1103/PhysRevLett.101.151602",
    journal = "Phys. Rev. Lett.",
    volume = "101",
    pages = "151602",
    year = "2008"
}

@article{Hoang:2009yr,
    author = "Hoang, Andre H. and Jain, Ambar and Scimemi, Ignazio and Stewart, Iain W.",
    title = "{R-evolution: Improving perturbative QCD}",
    eprint = "0908.3189",
    archivePrefix = "arXiv",
    primaryClass = "hep-ph",
    reportNumber = "MIT-CTP-4062, MPP-2009-156",
    doi = "10.1103/PhysRevD.82.011501",
    journal = "Phys. Rev. D",
    volume = "82",
    pages = "011501",
    year = "2010"
}

@article{Zhao:2025oto,
    author = "Zhao, Yong",
    title = "{Improving the Precision of First-Principles Calculation of Parton Physics from Lattice QCD}",
    eprint = "2509.00247",
    archivePrefix = "arXiv",
    primaryClass = "hep-lat",
    month = "8",
    year = "2025"
}

@article{Bali:2016lva,
    author = {Bali, Gunnar S. and Lang, Bernhard and Musch, Bernhard U. and Sch{\"a}fer, Andreas},
    title = "{Novel quark smearing for hadrons with high momenta in lattice QCD}",
    eprint = "1602.05525",
    archivePrefix = "arXiv",
    primaryClass = "hep-lat",
    doi = "10.1103/PhysRevD.93.094515",
    journal = "Phys. Rev. D",
    volume = "93",
    number = "9",
    pages = "094515",
    year = "2016"
}

@article{Zhang:2025hyo,
    author = "Zhang, Rui and Grebe, Anthony V. and Hackett, Daniel C. and Wagman, Michael L. and Zhao, Yong",
    title = "{Kinematically enhanced interpolating operators for boosted hadrons}",
    eprint = "2501.00729",
    archivePrefix = "arXiv",
    primaryClass = "hep-lat",
    reportNumber = "FERMILAB-PUB-24-0968-T",
    doi = "10.1103/6dh4-6k4t",
    journal = "Phys. Rev. D",
    volume = "112",
    number = "5",
    pages = "L051502",
    year = "2025"
}

@article{Ji:2017oey,
    author = "Ji, Xiangdong and Zhang, Jian-Hui and Zhao, Yong",
    title = "{Renormalization in Large Momentum Effective Theory of Parton Physics}",
    eprint = "1706.08962",
    archivePrefix = "arXiv",
    primaryClass = "hep-ph",
    doi = "10.1103/PhysRevLett.120.112001",
    journal = "Phys. Rev. Lett.",
    volume = "120",
    number = "11",
    pages = "112001",
    year = "2018"
}

@article{Ishikawa:2017faj,
    author = "Ishikawa, Tomomi and Ma, Yan-Qing and Qiu, Jian-Wei and Yoshida, Shinsuke",
    title = "{Renormalizability of quasiparton distribution functions}",
    eprint = "1707.03107",
    archivePrefix = "arXiv",
    primaryClass = "hep-ph",
    doi = "10.1103/PhysRevD.96.094019",
    journal = "Phys. Rev. D",
    volume = "96",
    number = "9",
    pages = "094019",
    year = "2017"
}

@article{Green:2017xeu,
    author = "Green, Jeremy and Jansen, Karl and Steffens, Fernanda",
    title = "{Nonperturbative Renormalization of Nonlocal Quark Bilinears for Parton Quasidistribution Functions on the Lattice Using an Auxiliary Field}",
    eprint = "1707.07152",
    archivePrefix = "arXiv",
    primaryClass = "hep-lat",
    reportNumber = "DESY-17-109, DESY 17-109",
    doi = "10.1103/PhysRevLett.121.022004",
    journal = "Phys. Rev. Lett.",
    volume = "121",
    number = "2",
    pages = "022004",
    year = "2018"
}

@article{Constantinou:2017sej,
    author = "Constantinou, Martha and Panagopoulos, Haralambos",
    title = "{Perturbative renormalization of quasi-parton distribution functions}",
    eprint = "1705.11193",
    archivePrefix = "arXiv",
    primaryClass = "hep-lat",
    doi = "10.1103/PhysRevD.96.054506",
    journal = "Phys. Rev. D",
    volume = "96",
    number = "5",
    pages = "054506",
    year = "2017"
}

@article{Stewart:2017tvs,
    author = "Stewart, Iain W. and Zhao, Yong",
    title = "{Matching the quasiparton distribution in a momentum subtraction scheme}",
    eprint = "1709.04933",
    archivePrefix = "arXiv",
    primaryClass = "hep-ph",
    reportNumber = "MIT-CTP-4873, MIT--CTP-4873",
    doi = "10.1103/PhysRevD.97.054512",
    journal = "Phys. Rev. D",
    volume = "97",
    number = "5",
    pages = "054512",
    year = "2018"
}

@article{Alexandrou:2017huk,
    author = "Alexandrou, Constantia and Cichy, Krzysztof and Constantinou, Martha and Hadjiyiannakou, Kyriakos and Jansen, Karl and Panagopoulos, Haralambos and Steffens, Fernanda",
    title = "{A complete non-perturbative renormalization prescription for quasi-PDFs}",
    eprint = "1706.00265",
    archivePrefix = "arXiv",
    primaryClass = "hep-lat",
    reportNumber = "DESY-17-092",
    doi = "10.1016/j.nuclphysb.2017.08.012",
    journal = "Nucl. Phys. B",
    volume = "923",
    pages = "394--415",
    year = "2017"
}

@article{Chen:2017mzz,
    author = "Chen, Jiunn-Wei and Ishikawa, Tomomi and Jin, Luchang and Lin, Huey-Wen and Yang, Yi-Bo and Zhang, Jian-Hui and Zhao, Yong",
    title = "{Parton distribution function with nonperturbative renormalization from lattice QCD}",
    eprint = "1706.01295",
    archivePrefix = "arXiv",
    primaryClass = "hep-lat",
    reportNumber = "MSUHEP-17-007, MIT-CTP-4912",
    doi = "10.1103/PhysRevD.97.014505",
    journal = "Phys. Rev. D",
    volume = "97",
    number = "1",
    pages = "014505",
    year = "2018"
}

@article{LatticePartonLPC:2021gpi,
    author = "Huo, Yi-Kai and others",
    collaboration = "Lattice Parton (LPC)",
    title = "{Self-renormalization of quasi-light-front correlators on the lattice}",
    eprint = "2103.02965",
    archivePrefix = "arXiv",
    primaryClass = "hep-lat",
    doi = "10.1016/j.nuclphysb.2021.115443",
    journal = "Nucl. Phys. B",
    volume = "969",
    pages = "115443",
    year = "2021"
}

@article{Ji:2025mvk,
    author = "Ji, Yao and Yao, Fei and Zhang, Jian-Hui",
    title = "{Extracting Meson Distribution Amplitudes from Nonlocal Euclidean Correlations at Next-to-Next-to-Leading Order}",
    eprint = "2504.09367",
    archivePrefix = "arXiv",
    primaryClass = "hep-ph",
    month = "4",
    year = "2025"
}

@article{Jaffe:1983hp,
    author = "Jaffe, R. L.",
    title = "{Parton Distribution Functions for Twist Four}",
    reportNumber = "MIT-CTP-1085",
    doi = "10.1016/0550-3213(83)90361-9",
    journal = "Nucl. Phys. B",
    volume = "229",
    pages = "205--230",
    year = "1983"
}

@article{Alexandrou:2021oih,
    author = "Alexandrou, Constantia and Constantinou, Martha and Hadjiyiannakou, Kyriakos and Jansen, Karl and Manigrasso, Floriano",
    title = "{Flavor decomposition of the nucleon unpolarized, helicity, and transversity parton distribution functions from lattice QCD simulations}",
    eprint = "2106.16065",
    archivePrefix = "arXiv",
    primaryClass = "hep-lat",
    doi = "10.1103/PhysRevD.104.054503",
    journal = "Phys. Rev. D",
    volume = "104",
    number = "5",
    pages = "054503",
    year = "2021"
}

@article{Ji:2015qla,
    author = {Ji, Xiangdong and Sch{\"a}fer, Andreas and Xiong, Xiaonu and Zhang, Jian-Hui},
    title = "{One-Loop Matching for Generalized Parton Distributions}",
    eprint = "1506.00248",
    archivePrefix = "arXiv",
    primaryClass = "hep-ph",
    doi = "10.1103/PhysRevD.92.014039",
    journal = "Phys. Rev. D",
    volume = "92",
    pages = "014039",
    year = "2015"
}

@article{Zhang:2018diq,
    author = {Zhang, Jian-Hui and Ji, Xiangdong and Sch{\"a}fer, Andreas and Wang, Wei and Zhao, Shuai},
    title = "{Accessing Gluon Parton Distributions in Large Momentum Effective Theory}",
    eprint = "1808.10824",
    archivePrefix = "arXiv",
    primaryClass = "hep-ph",
    doi = "10.1103/PhysRevLett.122.142001",
    journal = "Phys. Rev. Lett.",
    volume = "122",
    number = "14",
    pages = "142001",
    year = "2019"
}

@article{CLQCD:2024yyn,
    author = "Du, Hai-Yang and others",
    collaboration = "CLQCD",
    title = "{Charmed meson masses and decay constants in the continuum limit from the tadpole improved clover ensembles}",
    eprint = "2408.03548",
    archivePrefix = "arXiv",
    primaryClass = "hep-lat",
    doi = "10.1103/PhysRevD.111.054504",
    journal = "Phys. Rev. D",
    volume = "111",
    number = "5",
    pages = "054504",
    year = "2025"
}

@article{CLQCD:2023sdb,
    author = "Hu, Zhi-Cheng and others",
    collaboration = "CLQCD",
    title = "{Quark masses and low-energy constants in the continuum from the tadpole-improved clover ensembles}",
    eprint = "2310.00814",
    archivePrefix = "arXiv",
    primaryClass = "hep-lat",
    doi = "10.1103/PhysRevD.109.054507",
    journal = "Phys. Rev. D",
    volume = "109",
    number = "5",
    pages = "054507",
    year = "2024"
}

@article{Bruno:2014jqa,
    author = "Bruno, Mattia and others",
    title = "{Simulation of QCD with N$_{f} =$ 2 $+$ 1 flavors of non-perturbatively improved Wilson fermions}",
    eprint = "1411.3982",
    archivePrefix = "arXiv",
    primaryClass = "hep-lat",
    reportNumber = "DESY-14-216, FTUAM-14-48, HIM-2014-01, HU-EP-14-51, MITP-14-091, SFB-CPP-14-89, IFT-UAM-CSIC-14-117",
    doi = "10.1007/JHEP02(2015)043",
    journal = "JHEP",
    volume = "02",
    pages = "043",
    year = "2015"
}

@article{Follana:2006rc,
    author = "Follana, E. and Mason, Q. and Davies, C. and Hornbostel, K. and Lepage, G. P. and Shigemitsu, J. and Trottier, H. and Wong, K.",
    collaboration = "HPQCD, UKQCD",
    title = "{Highly improved staggered quarks on the lattice, with applications to charm physics}",
    eprint = "hep-lat/0610092",
    archivePrefix = "arXiv",
    doi = "10.1103/PhysRevD.75.054502",
    journal = "Phys. Rev. D",
    volume = "75",
    pages = "054502",
    year = "2007"
}

@article{MILC:2012znn,
    author = "Bazavov, A. and others",
    collaboration = "MILC",
    title = "{Lattice QCD Ensembles with Four Flavors of Highly Improved Staggered Quarks}",
    eprint = "1212.4768",
    archivePrefix = "arXiv",
    primaryClass = "hep-lat",
    reportNumber = "FERMILAB-PUB-12-796-T",
    doi = "10.1103/PhysRevD.87.054505",
    journal = "Phys. Rev. D",
    volume = "87",
    number = "5",
    pages = "054505",
    year = "2013"
}

@misc{Bruschini2025JLabSeminar,
  author       = {Roberto Bruschini},
  title        = {Explaining Exotic Heavy Hadrons from QCD},
  howpublished = {Seminar slides, Jefferson Lab},
  year         = {2025},
  month        = feb,
  day          = {17},
  url          = {https://www.jlab.org/sites/default/files/theory/files/Seminar-Bruschini.pdf},
  note         = {Accessed: 2025-11-06}
}

@article{Berwein:2024ztx,
    author = "Berwein, Matthias and Brambilla, Nora and Mohapatra, Abhishek and Vairo, Antonio",
    title = "{Hybrids, tetraquarks, pentaquarks, doubly heavy baryons, and quarkonia in Born-Oppenheimer effective theory}",
    eprint = "2408.04719",
    archivePrefix = "arXiv",
    primaryClass = "hep-ph",
    reportNumber = "TUM-EFT 185/23",
    doi = "10.1103/PhysRevD.110.094040",
    journal = "Phys. Rev. D",
    volume = "110",
    number = "9",
    pages = "094040",
    year = "2024"
}

@article{IsgurWise1989,
  author       = {Isgur, Nathan and Wise, Mark B.},
  title        = {Weak Decays of Heavy Mesons in the Static Quark Approximation},
  journal      = {Phys. Lett. B},
  volume       = {232},
  pages        = {113--117},
  year         = {1989},
  doi          = {10.1016/0370-2693(89)90566-2}
}

@article{IsgurWise1990,
  author       = {Isgur, Nathan and Wise, Mark B.},
  title        = {Weak transition form-factors between heavy mesons},
  journal      = {Phys. Lett. B},
  volume       = {237},
  pages        = {527--530},
  year         = {1990},
  doi          = {10.1016/0370-2693(90)91219-2}
}

@article{EichtenHill1990,
  author       = {Eichten, E. and Hill, B. R.},
  title        = {An Effective Field Theory for the Motion of Heavy Quarks},
  journal      = {Phys. Lett. B},
  volume       = {234},
  pages        = {511--516},
  year         = {1990},
  doi          = {10.1016/0370-2693(90)92049-O}
}

@article{Georgi1990,
  author       = {Georgi, Howard},
  title        = {An Effective Field Theory for Heavy Quarks at Low Energies},
  journal      = {Phys. Lett. B},
  volume       = {240},
  pages        = {447--450},
  year         = {1990},
  doi          = {10.1016/0370-2693(90)91128-X}
}

@article{Grinstein1990,
  author       = {Grinstein, Benjamin},
  title        = {The Static Quark Effective Theory},
  journal      = {Nucl. Phys. B},
  volume       = {339},
  pages        = {253--268},
  year         = {1990},
  doi          = {10.1016/0550-3213(90)90349-I}
}

@article{Neubert1994,
  author       = {Neubert, Matthias},
  title        = {Heavy Quark Symmetry},
  journal      = {Phys. Rept.},
  volume       = {245},
  pages        = {259--396},
  year         = {1994},
  doi          = {10.1016/0370-1573(94)90091-4},
  eprint       = {hep-ph/9306320},
  archivePrefix= {arXiv},
  primaryClass = {hep-ph}
}

@book{ManoharWise2000,
  author       = {Manohar, Aneesh V. and Wise, Mark B.},
  title        = {Heavy Quark Physics},
  publisher    = {Cambridge University Press},
  year         = {2000},
  isbn         = {9780521642407}
}

@article{FalkLuke1992,
  author       = {Falk, Andrew F. and Luke, Michael E.},
  title        = {Strong decays of excited heavy mesons in chiral perturbation theory},
  journal      = {Phys. Lett. B},
  volume       = {292},
  pages        = {119--127},
  year         = {1992},
  doi          = {10.1016/0370-2693(92)90618-E}
}

@article{Bigi1993,
  author       = {Bigi, I. I. and Uraltsev, N. G. and Vainshtein, A. I.},
  title        = {Nonleptonic decays of beauty hadrons: from phenomenology to theory},
  journal      = {Phys. Rev. Lett.},
  volume       = {71},
  pages        = {496--499},
  year         = {1993},
  doi          = {10.1103/PhysRevLett.71.496}
}

@article{Orginos:2017kos,
    author = "Orginos, Kostas and Radyushkin, Anatoly and Karpie, Joseph and Zafeiropoulos, Savvas",
    title = "{Lattice QCD exploration of parton pseudo-distribution functions}",
    eprint = "1706.05373",
    archivePrefix = "arXiv",
    primaryClass = "hep-ph",
    reportNumber = "JLAB-THY-17-2494",
    doi = "10.1103/PhysRevD.96.094503",
    journal = "Phys. Rev. D",
    volume = "96",
    number = "9",
    pages = "094503",
    year = "2017"
}

@article{Dirac:1949cp,
    author = "Dirac, Paul A. M.",
    title = "{Forms of Relativistic Dynamics}",
    doi = "10.1103/RevModPhys.21.392",
    journal = "Rev. Mod. Phys.",
    volume = "21",
    pages = "392--399",
    year = "1949"
}

@article{Brodsky:1997de,
    author = "Brodsky, Stanley J. and Pauli, Hans-Christian and Pinsky, Stephen S.",
    title = "{Quantum chromodynamics and other field theories on the light cone}",
    eprint = "hep-ph/9705477",
    archivePrefix = "arXiv",
    reportNumber = "SLAC-PUB-7484, MPIH-V1-1997",
    doi = "10.1016/S0370-1573(97)00089-6",
    journal = "Phys. Rept.",
    volume = "301",
    pages = "299--486",
    year = "1998"
}

@article{Wilson:1974sk,
    author = "Wilson, Kenneth G.",
    editor = "Taylor, J. C.",
    title = "{Confinement of Quarks}",
    reportNumber = "CLNS-262",
    doi = "10.1103/PhysRevD.10.2445",
    journal = "Phys. Rev. D",
    volume = "10",
    pages = "2445--2459",
    year = "1974"
}

@article{Zhang:2024omt,
    author = "Zhang, Kuan and Huo, Yi-Kai and Ji, Xiangdong and Schaefer, Andreas and Shi, Chun-Jiang and Sun, Peng and Wang, Wei and Yang, Yi-Bo and Zhang, Jian-Hui",
    collaboration = "Lattice Parton",
    title = "{Impact of gauge fixing precision on the continuum limit of nonlocal quark-bilinear lattice operators}",
    eprint = "2405.14097",
    archivePrefix = "arXiv",
    primaryClass = "hep-lat",
    doi = "10.1103/PhysRevD.110.074505",
    journal = "Phys. Rev. D",
    volume = "110",
    number = "7",
    pages = "074505",
    year = "2024"
}

@article{Good:2025daz,
    author = "Good, William and Yao, Fei and Lin, Huey-Wen",
    title = "{First nucleon gluon PDF from large momentum effective theory}",
    eprint = "2505.13321",
    archivePrefix = "arXiv",
    primaryClass = "hep-lat",
    doi = "10.1016/j.physletb.2025.140067",
    journal = "Phys. Lett. B",
    volume = "872",
    pages = "140067",
    year = "2026"
}

@article{NieMiera:2025mwj,
    author = "NieMiera, Alex and Good, William and Lin, Huey-Wen and Yao, Fei",
    title = "{First Self-Renormalized Gluon PDF of Nucleon from Large-Momentum Effective Theory in the Continuum Limit}",
    eprint = "2510.17758",
    archivePrefix = "arXiv",
    primaryClass = "hep-lat",
    reportNumber = "MSUHEP-25-024",
    month = "10",
    year = "2025"
}

@article{Chen:2025xww,
    author = "Chen, Chen and Dong, Hongxin and Liu, Liuming and Sun, Peng and Xiong, Xiaonu and Yang, Yi-Bo and Yao, Fei and Zhang, Jian-Hui and Zeng, Chunhua and Zhong, Shiyi",
    title = "{Unpolarized gluon PDF of the nucleon from lattice QCD in the continuum limit}",
    eprint = "2510.26425",
    archivePrefix = "arXiv",
    primaryClass = "hep-lat",
    month = "10",
    year = "2025"
}

@article{NieMiera:2025vcx,
    author = "NieMiera, Alex and Good, William and Lin, Huey-Wen and Yao, Fei",
    title = "{Systematic Study of the Self-Renormalized Nucleon Gluon PDF in Large-Momentum Effective Theory}",
    eprint = "2511.14708",
    archivePrefix = "arXiv",
    primaryClass = "hep-lat",
    reportNumber = "MSUHEP-25-025",
    month = "11",
    year = "2025"
}

@article{Chu:2025jsi,
    author = "Chu, Min-Huan and Cichy, Krzysztof and Constantinou, Martha and Sznajder, Pawe{\l} and Wagner, Jakub",
    title = "{A Unified Neural-Network Framework for Nucleon Imaging from Numerical Simulations of QCD}",
    eprint = "2509.15931",
    archivePrefix = "arXiv",
    primaryClass = "hep-lat",
    month = "9",
    year = "2025"
}

@article{Zhang:2025mer,
    author = "Zhang, Jia-Lu",
    title = "{Renormalon effects on quasi-PDFs in the gradient flow formalism}",
    eprint = "2507.18233",
    archivePrefix = "arXiv",
    primaryClass = "hep-ph",
    doi = "10.1103/h5xf-pmww",
    journal = "Phys. Rev. D",
    volume = "112",
    number = "9",
    pages = "094504",
    year = "2025"
}

@article{Holligan:2025baj,
    author = "Holligan, Jack and Lin, Huey-Wen and Zhang, Rui and Zhao, Yong",
    title = "{Resummation for lattice QCD calculation of generalized parton distributions at nonzero skewness}",
    eprint = "2501.19225",
    archivePrefix = "arXiv",
    primaryClass = "hep-ph",
    reportNumber = "MSUHEP-24-023",
    doi = "10.1007/JHEP07(2025)241",
    journal = "JHEP",
    volume = "207",
    pages = "241",
    year = "2025"
}

@article{LatticeParton:2024vck,
    author = "Chu, Min-Huan and others",
    collaboration = "Lattice Parton",
    title = "{Light cone distribution amplitude for the {\ensuremath{\Lambda}} baryon from lattice QCD}",
    eprint = "2411.12554",
    archivePrefix = "arXiv",
    primaryClass = "hep-lat",
    doi = "10.1103/PhysRevD.111.034510",
    journal = "Phys. Rev. D",
    volume = "111",
    number = "3",
    pages = "034510",
    year = "2025"
}

@article{Xiong:2025obq,
    author = "Xiong, Ao-Sheng and Hua, Jun and Ling, Yu-Fei and Wei, Ting and Yu, Fu-Sheng and Zhang, Qi-An and Zheng, Yong",
    title = "{Ill-posedness in limited discrete Fourier inversion and regularization for quasi distributions in LaMET}",
    eprint = "2506.16689",
    archivePrefix = "arXiv",
    primaryClass = "hep-lat",
    doi = "10.1140/epjc/s10052-025-15130-9",
    journal = "Eur. Phys. J. C",
    volume = "85",
    number = "12",
    pages = "1409",
    year = "2025"
}

@article{Bollweg:2025ecn,
    author = "Bollweg, Dennis and Gao, Xiang and Mukherjee, Swagato and Zhao, Yong",
    title = "{Lattice QCD Benchmark of Proton Helicity and Flavor-Dependent Unpolarized Transverse Momentum-Dependent Parton Distribution Functions at Physical Quark Masses}",
    eprint = "2505.18430",
    archivePrefix = "arXiv",
    primaryClass = "hep-lat",
    doi = "10.1103/tb2w-3tks",
    journal = "Phys. Rev. Lett.",
    volume = "135",
    number = "20",
    pages = "201901",
    year = "2025"
}

@article{LatticeParton:2024zko,
    author = "Han, Xue-Ying and others",
    collaboration = "Lattice Parton",
    title = "{Calculation of heavy meson light-cone distribution amplitudes from lattice QCD}",
    eprint = "2410.18654",
    archivePrefix = "arXiv",
    primaryClass = "hep-lat",
    doi = "10.1103/PhysRevD.111.034503",
    journal = "Phys. Rev. D",
    volume = "111",
    number = "3",
    pages = "034503",
    year = "2025"
}

@article{Chowdhury:2024ymm,
    author = "Chowdhury, Talal Ahmed and Izubuchi, Taku and Kamruzzaman, Methun and Karthik, Nikhil and Khan, Tanjib and Liu, Tianbo and Paul, Arpon and Schoenleber, Jakob and Sufian, Raza Sabbir",
    title = "{Polarized and unpolarized gluon PDFs: Generative machine learning applications for lattice QCD matrix elements at short distance and large momentum}",
    eprint = "2409.17234",
    archivePrefix = "arXiv",
    primaryClass = "hep-lat",
    doi = "10.1103/PhysRevD.111.074509",
    journal = "Phys. Rev. D",
    volume = "111",
    number = "7",
    pages = "074509",
    year = "2025"
}

@article{Ling:2025olz,
    author = "Ling, Yu-Fei and Chu, Min-Huan and Liang, Jian and Hua, Jun and Xiong, Ao-Sheng and Zhang, Qi-An",
    title = "{Approaches to the Inverse Fourier Transformation with Limited and Discrete Data}",
    eprint = "2511.03593",
    archivePrefix = "arXiv",
    primaryClass = "hep-lat",
    month = "11",
    year = "2025"
}

@article{LPC:2025jvd,
    author = "Bai, Haoyang and others",
    collaboration = "LPC",
    title = "{Hybrid renormalization for distribution amplitude of a light baryon in large momentum effective theory}",
    eprint = "2508.08971",
    archivePrefix = "arXiv",
    primaryClass = "hep-lat",
    month = "8",
    year = "2025"
}

@article{Zhang:2024wyq,
    author = "Zhang, Jia-Lu",
    title = "{Conformal mapping in matching quark correlation functions to parton distribution functions}",
    eprint = "2411.16382",
    archivePrefix = "arXiv",
    primaryClass = "hep-ph",
    doi = "10.1103/PhysRevD.111.074016",
    journal = "Phys. Rev. D",
    volume = "111",
    number = "7",
    pages = "074016",
    year = "2025"
}

@article{Ding:2024saz,
    author = "Ding, Heng-Tong and Gao, Xiang and Mukherjee, Swagato and Petreczky, Peter and Shi, Qi and Syritsyn, Sergey and Zhao, Yong",
    title = "{Three-dimensional imaging of pion using lattice QCD: generalized parton distributions}",
    eprint = "2407.03516",
    archivePrefix = "arXiv",
    primaryClass = "hep-lat",
    doi = "10.1007/JHEP02(2025)056",
    journal = "JHEP",
    volume = "02",
    pages = "056",
    year = "2025"
}

@article{Guo:2025obm,
    author = "Guo, Tu and Han, Chao and Wang, Wei and Zhang, Jia-Lu",
    title = "{Power corrections in the determination of heavy meson light-cone distribution amplitudes: A renormalon-based estimation}",
    eprint = "2505.08611",
    archivePrefix = "arXiv",
    primaryClass = "hep-ph",
    doi = "10.1103/c97x-z7j9",
    journal = "Phys. Rev. D",
    volume = "112",
    number = "1",
    pages = "016013",
    year = "2025"
}

@article{Han:2024fkr,
    author = {Han, Xue-Ying and Hua, Jun and Ji, Xiangdong and L{\"u}, Cai-Dian and Wang, Wei and Xu, Ji and Zhang, Qi-An and Zhao, Shuai},
    title = "{Realistic method to access heavy meson light-cone distribution amplitudes from first-principle}",
    eprint = "2403.17492",
    archivePrefix = "arXiv",
    primaryClass = "hep-ph",
    doi = "10.1103/2t8s-w8t6",
    journal = "Phys. Rev. D",
    volume = "111",
    number = "11",
    pages = "L111503",
    year = "2025"
}

@article{Hu:2023bba,
    author = "Hu, Shu-Man and Wang, Wei and Xu, Ji and Zhao, Shuai",
    title = "{Accessing the subleading-twist B-meson light-cone distribution amplitude with large-momentum effective theory}",
    eprint = "2308.13977",
    archivePrefix = "arXiv",
    primaryClass = "hep-ph",
    doi = "10.1103/PhysRevD.109.034001",
    journal = "Phys. Rev. D",
    volume = "109",
    number = "3",
    pages = "034001",
    year = "2024"
}

@article{Zhang:2025npd,
    author = "Zhang, Jia-lu and Zhang, Mu-Hua",
    title = "{Hybrid Renormalization with Gradient Flow for Baryon Quasi-Distribution Amplitudes}",
    eprint = "2510.14325",
    archivePrefix = "arXiv",
    primaryClass = "hep-ph",
    month = "10",
    year = "2025"
}

@article{Tan:2025ofx,
    author = "Tan, Jin-Xin and others",
    title = "{Lattice QCD Determination of the Collins-Soper Kernel in the Continuum and Physical Mass Limits}",
    eprint = "2511.22547",
    archivePrefix = "arXiv",
    primaryClass = "hep-lat",
    month = "11",
    year = "2025"
}

@article{Alexandrou:2025xci,
    author = "Alexandrou, Constantia and Bacchio, Simone and Cichy, Krzysztof and Constantinou, Martha and Sen, Aniket and Spanoudes, Gregoris and Steffens, Fernanda and Tarello, Jacopo",
    title = "{Collins-Soper Kernel and Reduced Soft Function in Lattice QCD}",
    eprint = "2509.26316",
    archivePrefix = "arXiv",
    primaryClass = "hep-lat",
    month = "9",
    year = "2025"
}

@article{Bollweg:2025iol,
    author = "Bollweg, Dennis and Gao, Xiang and He, Jinchen and Mukherjee, Swagato and Zhao, Yong",
    title = "{Transverse-momentum-dependent pion structures from lattice QCD: Collins-Soper kernel, soft factor, TMDWF, and TMDPDF}",
    eprint = "2504.04625",
    archivePrefix = "arXiv",
    primaryClass = "hep-lat",
    doi = "10.1103/j3n6-8kxy",
    journal = "Phys. Rev. D",
    volume = "112",
    number = "3",
    pages = "034501",
    year = "2025"
}

@article{LatticeParton:2025eui,
    author = "Ma, Lingquan and others",
    collaboration = "Lattice Parton",
    title = "{Quark transverse spin-momentum correlation of the nucleon from lattice QCD: the Boer-Mulders function}",
    eprint = "2502.11807",
    archivePrefix = "arXiv",
    primaryClass = "hep-lat",
    doi = "10.1007/JHEP08(2025)086",
    journal = "JHEP",
    volume = "08",
    pages = "086",
    year = "2025"
}

@article{LatticeParton:2024mxp,
    author = "Walter, Lisa and others",
    collaboration = "Lattice Parton, LPC",
    title = "{Quark transverse spin-momentum correlation of the pion from lattice QCD: The Boer-Mulders function}",
    eprint = "2412.19988",
    archivePrefix = "arXiv",
    primaryClass = "hep-lat",
    doi = "10.1103/PhysRevD.111.094507",
    journal = "Phys. Rev. D",
    volume = "111",
    number = "9",
    pages = "094507",
    year = "2025"
}

@article{Gao:2024fbh,
    author = "Gao, Xiang and He, Jinchen and Zhang, Rui and Zhao, Yong",
    title = "{Systematic Uncertainties from Gribov Copies in Lattice Calculation of Parton Distributions in the Coulomb Gauge}",
    eprint = "2408.05910",
    archivePrefix = "arXiv",
    primaryClass = "hep-lat",
    doi = "10.1088/0256-307X/41/12/121201",
    journal = "Chin. Phys. Lett.",
    volume = "41",
    number = "12",
    pages = "121201",
    year = "2024"
}

@article{Xie:2025rrw,
    author = "Xie, Xiupeng and Lu, Zhun",
    title = "{Leading-twist gluon transverse momentum dependent distributions from large-momentum effective theory}",
    eprint = "2512.08292",
    archivePrefix = "arXiv",
    primaryClass = "hep-ph",
    month = "12",
    year = "2025"
}

@article{Chu:2025kew,
    author = "Chu, Min-Huan and Cola{\c{c}}o, Manuel and Bhattacharya, Shohini and Cichy, Krzysztof and Constantinou, Martha and Metz, Andreas and Steffens, Fernanda",
    title = "{Generalized parton distributions from lattice QCD with asymmetric momentum transfer: Unpolarized quarks at nonzero skewness}",
    eprint = "2508.17998",
    archivePrefix = "arXiv",
    primaryClass = "hep-lat",
    doi = "10.1103/ts5s-hvb1",
    journal = "Phys. Rev. D",
    volume = "112",
    number = "9",
    pages = "094510",
    year = "2025"
}

@article{Reitinger:2024ulw,
    author = {Reitinger, Daniel and Zimmermann, Christian and Diehl, Markus and Sch{\"a}fer, Andreas},
    title = "{Double parton distributions with flavor interference from lattice QCD}",
    eprint = "2401.14855",
    archivePrefix = "arXiv",
    primaryClass = "hep-lat",
    reportNumber = "DESY-24-014",
    doi = "10.1007/JHEP04(2024)087",
    journal = "JHEP",
    volume = "04",
    pages = "087",
    year = "2024"
}

@article{Chen:2025cxr,
    author = "Chen, Jiunn-Wei and others",
    title = "{LaMET's Asymptotic Extrapolation vs. Inverse Problem}",
    eprint = "2505.14619",
    archivePrefix = "arXiv",
    primaryClass = "hep-lat",
    month = "5",
    year = "2025"
}

@article{Bhattacharya:2025yba,
    author = "Bhattacharya, Shohini and Cichy, Krzysztof and Constantinou, Martha and Metz, Andreas and Miller, Joshua and Petreczky, Peter and Steffens, Fernanda",
    title = "{Generalized parton distributions from lattice QCD with asymmetric momentum transfer: Tensor case}",
    eprint = "2505.11288",
    archivePrefix = "arXiv",
    primaryClass = "hep-lat",
    doi = "10.1103/tlkb-ykgp",
    journal = "Phys. Rev. D",
    volume = "112",
    number = "11",
    pages = "114504",
    year = "2025"
}

@article{Miller:2025wgr,
    author = "Miller, Joshua and Torsiello, Joseph and Anderson, Isaac and Cichy, Krzysztof and Constantinou, Martha and Delmar, Joseph and Lampreich, Sarah",
    title = "{Pion and Kaon PDFs from Lattice QCD with Complementary Approaches}",
    eprint = "2512.06121",
    archivePrefix = "arXiv",
    primaryClass = "hep-lat",
    month = "12",
    year = "2025"
}

@article{Dutrieux:2025jed,
    author = "Dutrieux, Herv{\'e} and Karpie, Joe and Monahan, Christopher J. and Orginos, Kostas and Radyushkin, Anatoly and Richards, David and Zafeiropoulos, Savvas",
    title = "{Inverse problem in the LaMET framework}",
    eprint = "2504.17706",
    archivePrefix = "arXiv",
    primaryClass = "hep-lat",
    reportNumber = "JLAB-THY-25-4295",
    month = "4",
    year = "2025"
}

@article{Dutrieux:2025axb,
    author = "Dutrieux, Herv{\'e} and Karpie, Joe and Monahan, Christopher J. and Orginos, Kostas and Radyushkin, Anatoly and Richards, David and Zafeiropoulos, Savvas",
    title = "{Comment on ''LaMET's Asymptotic Extrapolation vs. Inverse Problem''}",
    eprint = "2506.24037",
    archivePrefix = "arXiv",
    primaryClass = "hep-lat",
    reportNumber = "JLAB-THY-25-4392",
    month = "6",
    year = "2025"
}

@article{Blossier:2009kd,
    author = "Blossier, Benoit and Della Morte, Michele and von Hippel, Georg and Mendes, Tereza and Sommer, Rainer",
    title = "{On the generalized eigenvalue method for energies and matrix elements in lattice field theory}",
    eprint = "0902.1265",
    archivePrefix = "arXiv",
    primaryClass = "hep-lat",
    reportNumber = "DESY-09-014, SFB-CPP-09-10, MKPH-T-09-01, LPT-ORSAY-09-05",
    doi = "10.1088/1126-6708/2009/04/094",
    journal = "JHEP",
    volume = "04",
    pages = "094",
    year = "2009"
}

@article{Barca:2025det,
    author = "Barca, Lorenzo",
    title = "{Current-enhanced excited states in lattice QCD three-point functions}",
    eprint = "2508.09006",
    archivePrefix = "arXiv",
    primaryClass = "hep-lat",
    reportNumber = "DESY-25-115",
    doi = "10.1103/69yc-d74z",
    journal = "Phys. Rev. D",
    volume = "112",
    number = "9",
    pages = "L091503",
    year = "2025"
}

@article{Wang:2025nsd,
    author = "Wang, Ji-Hao and Hu, Zhi-Cheng and Ji, Xiangdong and Jiang, Xiangyu and Su, Yushan and Sun, Peng and Yang, Yi-Bo",
    title = "{Precision determination of nucleon iso-vector scalar and tensor charges at the physical point}",
    eprint = "2511.02326",
    archivePrefix = "arXiv",
    primaryClass = "hep-lat",
    month = "11",
    year = "2025"
}

@article{Abbott:2025yhm,
    author = "Abbott, Ryan and Hackett, Daniel C. and Fleming, George T. and Pefkou, Dimitra A. and Wagman, Michael L.",
    title = "{Filtered Rayleigh-Ritz is all you need}",
    eprint = "2503.17357",
    archivePrefix = "arXiv",
    primaryClass = "hep-lat",
    reportNumber = "FERMILAB-PUB-25-0131-T, MIT-CTP/5849",
    month = "3",
    year = "2025"
}

@article{Guo:2017jvc,
    author = "Guo, Feng-Kun and Hanhart, Christoph and Mei{\ss}ner, Ulf-G. and Wang, Qian and Zhao, Qiang and Zou, Bing-Song",
    title = "{Hadronic molecules}",
    eprint = "1705.00141",
    archivePrefix = "arXiv",
    primaryClass = "hep-ph",
    doi = "10.1103/RevModPhys.90.015004",
    journal = "Rev. Mod. Phys.",
    volume = "90",
    number = "1",
    pages = "015004",
    year = "2018",
    note = "[Erratum: Rev.Mod.Phys. 94, 029901 (2022)]"
}

@article{Ali:2017jda,
    author = {Ali, Ahmed and Lange, Jens S{\"o}ren and Stone, Sheldon},
    title = "{Exotics: Heavy Pentaquarks and Tetraquarks}",
    eprint = "1706.00610",
    archivePrefix = "arXiv",
    primaryClass = "hep-ph",
    reportNumber = "DESY-17-071",
    doi = "10.1016/j.ppnp.2017.08.003",
    journal = "Prog. Part. Nucl. Phys.",
    volume = "97",
    pages = "123--198",
    year = "2017"
}

@article{Olsen:2017bmm,
    author = "Olsen, Stephen Lars and Skwarnicki, Tomasz and Zieminska, Daria",
    title = "{Nonstandard heavy mesons and baryons: Experimental evidence}",
    eprint = "1708.04012",
    archivePrefix = "arXiv",
    primaryClass = "hep-ph",
    doi = "10.1103/RevModPhys.90.015003",
    journal = "Rev. Mod. Phys.",
    volume = "90",
    number = "1",
    pages = "015003",
    year = "2018"
}

@article{Brambilla:2019esw,
    author = "Brambilla, Nora and Eidelman, Simon and Hanhart, Christoph and Nefediev, Alexey and Shen, Cheng-Ping and Thomas, Christopher E. and Vairo, Antonio and Yuan, Chang-Zheng",
    title = "{The $XYZ$ states: experimental and theoretical status and perspectives}",
    eprint = "1907.07583",
    archivePrefix = "arXiv",
    primaryClass = "hep-ex",
    reportNumber = "TUM-EFT 125/19",
    doi = "10.1016/j.physrep.2020.05.001",
    journal = "Phys. Rept.",
    volume = "873",
    pages = "1--154",
    year = "2020"
}

@article{Liu:2019zoy,
    author = "Liu, Yan-Rui and Chen, Hua-Xing and Chen, Wei and Liu, Xiang and Zhu, Shi-Lin",
    title = "{Pentaquark and Tetraquark states}",
    eprint = "1903.11976",
    archivePrefix = "arXiv",
    primaryClass = "hep-ph",
    doi = "10.1016/j.ppnp.2019.04.003",
    journal = "Prog. Part. Nucl. Phys.",
    volume = "107",
    pages = "237--320",
    year = "2019"
}

@article{Chen:2022asf,
    author = "Chen, Hua-Xing and Chen, Wei and Liu, Xiang and Liu, Yan-Rui and Zhu, Shi-Lin",
    title = "{An updated review of the new hadron states}",
    eprint = "2204.02649",
    archivePrefix = "arXiv",
    primaryClass = "hep-ph",
    doi = "10.1088/1361-6633/aca3b6",
    journal = "Rept. Prog. Phys.",
    volume = "86",
    number = "2",
    pages = "026201",
    year = "2023"
}

@article{Cheng:2017oqh,
    author = "Cheng, Hai-Yang and Yu, Fu-Sheng",
    title = "{Masses of Scalar and Axial-Vector B Mesons Revisited}",
    eprint = "1704.01208",
    archivePrefix = "arXiv",
    primaryClass = "hep-ph",
    doi = "10.1140/epjc/s10052-017-5252-4",
    journal = "Eur. Phys. J. C",
    volume = "77",
    number = "10",
    pages = "668",
    year = "2017"
}

@book{Gattringer:2010zz,
    author = "Gattringer, Christof and Lang, Christian B.",
    title = "{Quantum chromodynamics on the lattice}",
    doi = "10.1007/978-3-642-01850-3",
    isbn = "978-3-642-01849-7, 978-3-642-01850-3",
    publisher = "Springer",
    address = "Berlin",
    volume = "788",
    year = "2010"
}

@article{Lyu:2025lnd,
    author = "Lyu, Yan and Aoki, Sinya and Doi, Takumi and Hatsuda, Tetsuo and Murakami, Kotaro and Sugiura, Takuya",
    title = "{Decoding Two-Particle States in QCD with Spatial Wavefunctions}",
    eprint = "2507.09930",
    archivePrefix = "arXiv",
    primaryClass = "hep-lat",
    reportNumber = "RIKEN-iTHEMS-Report-25, FQSP-25-2, YITP-25-109",
    month = "7",
    year = "2025"
}

@book{Peskin:1995ev,
    author = "Peskin, Michael E. and Schroeder, Daniel V.",
    title = "{An Introduction to quantum field theory}",
    doi = "10.1201/9780429503559",
    isbn = "978-0-201-50397-5, 978-0-429-50355-9, 978-0-429-49417-8",
    publisher = "Addison-Wesley",
    address = "Reading, USA",
    year = "1995"
}

\end{document}